\documentclass[twocolumn]{aastex631}
\usepackage{multirow}
\usepackage[normalem]{ulem}
\useunder{\uline}{\ul}{}
\usepackage{savesym}
\savesymbol{tablenum}
\usepackage{siunitx}
\usepackage{cleveref}
\usepackage{natbib}
\usepackage{xcolor}
\usepackage[T1]{fontenc}

\usepackage{booktabs}
\restoresymbol{SIX}{tablenum}

\shorttitle{X-ray intraday variability of HBLs}
\shortauthors{Devanand et al.}

\begin{document}

\title{X-ray Intraday Variability of HBL Blazars with \emph{XMM-Newton}}

\author{P.\ U.\ Devanand}
\affiliation{Aryabhatta
Research Institute of Observational Sciences (ARIES), Manora Peak, Nainital 263001, India}

\author{Alok C.\ Gupta}
\affiliation{Aryabhatta
Research Institute of Observational Sciences (ARIES), Manora Peak, Nainital 263001, India}

\author{V.\ Jithesh}
\affiliation{Department of Physics and Electronics, CHRIST (Deemed to be University), Hosur Main Road, Bengaluru - 560029, India} 
\affiliation{Aryabhatta
Research Institute of Observational Sciences (ARIES), Manora Peak, Nainital 263001, India}

\author{Paul J.\ Wiita}
\affiliation{Department of Physics, The College of New Jersey, 2000 Pennington Rd., Ewing, NJ 08628, USA}

\begin{abstract}
\noindent
 We present an extensive study on  the X-ray intraday variability  of  ten TeV-emitting high synchrotron peaked blazars (HBLs): 1ES 0229+200, 1ES 0414+009, PKS 0548-322, 1ES 1101-232, 1H 1219+301, H 1426+428, Mrk 501, 1ES 1959+650, PKS 2005-489, and 1ES 2344+514 made with twenty-five \emph{XMM-Newton} pointed observations  during its operational period. Intraday variability has been estimated in three energy bands: soft (0.3--2 keV), hard (2--10 keV) and total (0.3--10 keV). Although seven out of these ten TeV HBLs exhibited some  intraday variability at three-sigma levels no major variations exceeding six percent were detected. We explored the spectral properties of the sample by extracting the hardness ratio from  the soft and hard  bands; no significant variations in the hardness ratio were observed in any source.  We performed power spectral density analyses on  the variable light-curves  by fitting power-laws, yielding slopes lying in the range from 1.11 to 2.93 for different HBLs. We briefly discuss possible  emission mechanisms and carry out  rough estimates for magnetic fields,  electron Lorentz factors and emission region sizes for seven of these  HBLs.
\end{abstract}


\keywords{general - HBL blazars: Individual (1ES 0229+200, 1ES 0414+009, PKS 0548-322, 1ES 1101-232, 1H 1219+301, H 1426+428, Mrk 501, 1ES 1959+650, PKS 2005-489, 1ES 2344+514)}

\section{Introduction} \label{sec:intro}
\noindent
Active galactic nuclei (AGN) are the  central regions of certain galaxies, which emit enormous amounts of energy  across the entire electromagnetic (EM) spectrum that   usually outshine all the stars in their host galaxies.
AGNs are universally believed to be powered by accreting supermassive black holes  (SMBH; $10^6-10^{10} \hspace{0.15cm}\textup{M}_\odot$), lying at their centers \citep[][]{Ree84}. About 10--15\% of AGNs clearly include well collimated jets of relativistic particles and these jetted AGNs emit significant radio emission \citep{2017A&ARv..25....2P}. These relativistic jets are particularly luminous at radio and $\gamma$-ray frequencies.  Types of jetted-AGN are distinguished by the angle of the jet with respect to the viewer's line of sight, with blazars being those in which one of their relativistic jets is aligned at a small angle \citep[$\lesssim $ 15\textdegree \hspace{0.1cm}to the observer;][]{Urr95}. Blazars have historically been classified into two categories: BL Lacertae (BL Lac) objects, which have nearly featureless spectra or very weak emission lines  \citep[equivalent width $\le$ 5{\AA};][]{Sto91,Mar96}, and flat spectrum radio quasars (FSRQs), which have broad and strong emissions lines in  their composite optical/UV spectra \citep{Bla78,Ghi97}. Some weaker AGN, particularly Narrow Line Seyfert 1 galaxies, also have relativistic jets pointing close to our line of sight \citep{2020Univ....6..136F}. Blazars exhibit flux and spectral variability across  all accessible EM bands. The observed emission, which is  predominantly non-thermal, is dominated by Doppler boosted radiation from relativistic jets. Other outstanding characteristics of blazars  are their core-dominated radio structures and strong polarization in radio and optical bands. \\
\\
 The multi-wavelength spectral energy distribution (SED) of blazars  is characterized by a double-humped structure  \citep[e.g.][]{Fos08}.
Synchrotron emission from relativistic electrons gyrating around  the magnetic field in relativistic jets  produces the low-energy hump which peaks somewhere between IR and X-ray band.  A recent examination of a large sample of jetted AGN \citep{2021MNRAS.505.4726K} has shown that the claimed anti-correlation between synchrotron peak frequency and peak luminosity, called the blazar sequence, is not significant. It appears to be preferable to classify these AGN by whether their jets are associated with efficient accretion (strong, `quasar-mode', or Type II jets) or inefficient accretion \citep[weak, Type I jets;][]{2021MNRAS.505.4726K}.\\ 
\\
The physical mechanisms responsible for  the high-energy hump, which peaks in GeV-TeV energies, can be broadly separated into leptonic and hadronic models.
 In leptonic models the high energy hump is due to inverse-Compton (IC) scattering of either  low energy synchrotron photons  \citep[Synchrotron Self-Compton (SSC); e.g.][]{Blo96} or low energy external photons  \citep[External Compton, EC; e.g.][]{Bla95} by the same electrons in relativistic jets which are responsible for  the synchrotron emission. In hadronic models,  the high energy hump is due to synchrotron emission from relativistic protons and/or proton-photon cascade processes. On the basis of  their peak synchrotron frequency ($\nu_s$), BL Lacs were classified as low energy peaked BL Lacs (LBLs),  intermediate energy peaked BL Lacs  (IBLs), and high energy peaked BL Lacs (HBLs). This classification was later modified to include FSRQs by \citet{Abd10},  giving us low synchrotron peaked blazars (LSPs; $\nu_s < 10^{14}$Hz), intermediate synchrotron peaked  blazars (ISPs;  $10^{14}{\rm Hz} < \nu_s < 10^{15}$Hz) and high synchrotron peaked blazars (HSPs; $\nu_s > 10^{15}$Hz).\\
\\
Blazars exhibit  flux variation across entire EM spectrum  in diverse timescales   down to timescales of hours or even a few minutes. Flux variations lasting from a few minutes to less than a day are termed as intraday variability  \citep[IDV; e.g.][]{Wag95} or intra-night variability \citep[e.g.][]{Goy09} or micro-variability \citep[e.g.][]{Mil89}. Variations in flux on timescales from  days to  a few weeks to even a few months are termed as short-term variability (STV), while flux variations  seen over greater timespans are termed as long-term variability \citep[LTV; e.g.][]{Gup04...422..505G}. Over the last decade, we have  studied the IDV of blazars in X-ray bands by utilizing timing data from various X-ray telescopes: \emph{Chandra} \citep[][]{Agg18}, \emph{NuSTAR} \citep[][]{Pan17,Pan18},\emph{ Suzaku} \citep[][]{Zha19,Zha21}, and \emph{XMM-Newton} \citep[][]{Gau10,Kal15,Bha16,Gup16,Kir20,Dhi21}. \\ 
\\
BL Lac objects are expected to be best candidates for TeV emission among blazars. This is based on the assumption that there is  much less TeV absorbing material in  the vicinity of their emission regions as their optical spectra contain  essentially no emission lines \citep[e.g.][]{Der94}. The terminology ``high-frequency-peaked BL Lac objects (HBLs)'' was introduced by \citet{Pad95} to describe BL Lac objects in which the lower energy hump peaks in the X-ray range.
 There were only six TeV HBLs (Markarian 421, H 1426+428, Markarian 501, 1ES 1959+650, PKS  2155-304 and 1ES 2344+514) known  until 2005.  The {\it Fermi} satellite and several  ground-based very-high energy (VHE) $\gamma$-ray facilities such as  the High Energy Spectroscopic System ({\it HESS}),  the Major Atmospheric Gamma-ray Imaging Cherenkov telescopes ({\it MAGIC}),  and the Very Energetic Radiation Imaging Telescope Array System ({\it VERITAS}), discovered  a substantial  number of HBLs in the last 17 years and revolutionized $\gamma$-ray blazar astronomy.  At the time of writing this paper, the total number of blazars in the TeV source catalogue\footnote{\url{http://tevcat.uchicago.edu/}} (TeVcat)   is  comprised of 55 HBLs, 10 IBLs, 2 LBLs, 9 FSRQs,  and 4 blazars and 2 BL Lacs with unclear classifications. \\ 
\\
X-ray IDV for blazars is an intrinsic phenomenon and may be related to some activity  in the innermost region near the central SMBH. IDV timescale studies can help in constraining the size of  the emitting region and in estimating a crude mass of the central SMBH. Our main motivation for this work is to understand X-ray variability properties on IDV-timescales of HBLs that show emission at highest energies. \\
\\
The paper is arranged as follows: Section \ref{sec2} shortly describes the \emph{XMM-Newton} satellite instrumentation,  the archival data of HBLs we selected and  our reduction methodology.  Various analysis techniques we used to study flux and spectral variations are discussed in section \ref{sec3}. Results and associated discussion are presented in section \ref{sec4} and \ref{sec5}, respectively. We report our conclusions in section \ref{sec6}.

\renewcommand\theHtable{\thetable}
\begin{deluxetable*}{cccccCccCCC}
\tablenum{1}
\tablecaption{Observation log for {\it XMM-Newton} data}
\tablewidth{0pt}
\tablehead
{\colhead{Source} &\colhead{$\alpha_{2000.0}$} &\colhead{$\delta_{2000.0}$}&\colhead{z} &\colhead{Obs Date} &  \colhead{Obs ID} & \colhead{Window} &
 \colhead{Pile-up} &   \colhead{Exposure}  &  \colhead{Good Exp }& \colhead{Bin Size}   \\
\colhead{} & \colhead{hh mm ss.ss} & \colhead{dd mm ss.ss} &\colhead{} &\colhead{} &\colhead{}&\colhead{Mode}&\colhead{}&\colhead{Time (ks)}&\colhead{Time (ks)}&\colhead{(s)}
}
\decimalcolnumbers
\startdata
1ES\hspace{0.1cm}0229+200 &02 32 48.61&+20 17 17.49&0.1400& 2009 Aug 21 &  0604210201 & Full & No & 23.61& 17.30& 100 \\
 &&&&2009 Aug 23&  0604210301&  Full & No & 27.71&21.20&100 \\
1ES\hspace{0.1cm}0414+009 & 04 16 52.49& +01 05 23.89 &0.2870 &2002 Aug 26 & 0094383101 & Small & No & 10.96 &10.40& 100 \\
&&& & 2003 Sep 01 &0161160101 & Small & No & 79.36&71.60 & 300\\
PKS\hspace{0.1cm}0548-322&05 50 40.57&-32 16 16.49 &0.0690& 2002 Oct 19 &  0142270101 & Full & Yes & 94.50 &80.00 & 400\\
&&&&2004 Oct 19&0205920501 & Timing & No & 40.92 &24.90 &100\\
1ES\hspace{0.1cm}1101-232&11 03 37.62& -23 29 31.20 &0.1860& 2009 Aug 23 &  0205920601 & Timing & No & 18.50 &17.30 & 100 \\
1H\hspace{0.1cm}1219+301&12 21 21.94&+30 10 37.16&0.1836 & 2001 Jun 11 &  0111840101 & Small & No & 29.94 &28.40& 200\\
H\hspace{0.1cm}1426+428 & 14 28 32.61&+42 40 21.05&0.1293&2001 Jun 16 &   0111850201  & Small & No& 68.57& 52.80&300	\\
  &&&&2004 Aug 04  & 	0165770101  & 	Small  & 	No  & 67.87  & 56.10&	300\\
&&&& 2004 Aug 06 &  0165770201 & Small & No &68.92&60.30& 300	\\
&&&  & 2005 Jan 24  & 	0212090201  &  	Small  & 	No  & 	30.41 &28.80 & 	200 \\	
&&&   & 2005 Jul 19  & 	0310190101  & 	Small  & 	Yes  & 	47.03 &33.40 & 	200 \\	
&&&   & 2005 Jun 25  & 	0310190201  & 	Small  & 	No  & 	49.50 &40.20& 	200	 \\
&&&   & 2005 Aug 04  & 	0310190501  & 	Small  & 	Yes  & 	47.54 & 35.20&	200	 \\
Mrk\hspace{0.1cm}501&16 53 52.22&+39 45 36.61&0.0330& 2010 Sep 08	& 0652570101 &	Small &	No &	44.91 &39.80&	200 \\
&&&& 2010 Sep 10	& 0652570201 &	Small &	No &	44.92 &44.80&	200\\
&&& & 2011 Feb 11 & 	0652570301 &	Small &	No &	40.91 &	40.80&200\\	
&&&& 2011 Feb 15	& 0652570401 &	Small &	No &	40.72 &	40.20&200	\\
1ES 1959+650&19 59 59.85& +65 08 54.65& 0.0470& 2019 Jul 05 &  0850980101 & Small & Yes & 44.00 &38.00& 200 \\
&&&  &2020 Jul 16 &  0870210101 & Small & Yes & 33.10 &31.40& 200 \\
PKS\hspace{0.1cm}2005-489 &20 09 25.34&-48 49 53.72& 0.0707 &	2004 Oct 04 & 	0205920401 &  Timing & 	No & 12.92 &11.70&	100	\\
&&& & 2005 Sep 26 &  0304080301 &  Timing & No & 27.92 & 14.60&100 \\
&&& & 2005 Sep 28	& 0304080401 &  Timing & No & 27.92&24.10 &200 \\
1ES\hspace{0.1cm}2344+514&23 47 04.84 &+51 42 17.88&0.0440&2020 Jul 22&0870400101&Small&No&28.90&26.80&200\\
\enddata
\tablecomments{Right Ascension ($\alpha_{2000.0}$), Declination ($\delta_{2000.0}$) and red-shift ($z$) are taken from the Simbad astronomical database ({\url{http://simbad.u-strasbg.fr/simbad/}}).}
\label{tab1}
\end{deluxetable*}

\section{INSTRUMENTATION,  DATA SELECTION, AND DATA REDUCTION\label{sec2}}
\subsection{Instrumentation}
\noindent
 The X-ray Multi-Mirror Newton (\emph{XMM-Newton}) mission is  a space observatory launched by  the European Space Agency on 1999 December 10 and  was, for the first time, capable of performing simultaneous imaging of sources in X-ray and optical (visible $\&$ UV)  bands \citep[][]{Mas01}. 
It  was placed in a 48-hour elliptical orbit at 40$^{\circ}$ inclination to the equator and carries both three Wolter type-1 X-ray telescopes and one UV/optical telescope. The three science instruments,  the European Photon Imaging  Camera (EPIC),  the Reflection Grating Spectrometer (RGS) and the Optical Monitor (OM)  enable  \emph{XMM-Newton} to do imaging  and spectrophotometry in X-ray and optical bands.  EPIC consists of 3 CCD cameras of two types: Metal oxide semiconductor (EPIC-MOS) and EPIC PN. These EPIC cameras  can perform extremely sensitive  imaging observations over a wide field-of-view of   30$^{'}$ (but only the inner 12$^{'}$ are efficiently corrected for vignetting)  over an energy range of  0.2--12 keV  with moderate angular resolution ($\SI{6}{\arcsecond}$ Full Width Half Maximum; ~$\SI{14}{\arcsecond} $ (MOS) and $\SI{15}{\arcsecond} $ (PN) Half Energy Width; see XMM Users Handbook 2021\footnote{\url{https://xmm-tools.cosmos.esa.int/external/xmm_user_support/documentation/uhb/}}, XMM-Newton Calibration Technical Note\footnote{\url{https://xmmweb.esac.esa.int/docs/documents/CAL-TN-0018.pdf}}). In our study we have considered only EPIC-PN data as it is more sensitive and less effected by photon pile-up effects \citep{Tur01}.
\subsection{Data Selection Criteria\label{dsc}}
\noindent
Among the 55 HBLs in TeVcat, 19  have been  observed by \emph{XMM-Newton}  since its launch until March 2022.  Of these, we excluded four of the sources previously studied by members of the group (e.g Mrk 421; \citet{Pri22}, PKS 2155-304; \citet{Bha14,Bha16}, H 2356-309; \citet{Kir20} and PG 1533+113; \citet{Dhi21}).
We also did not analyze the shorter observations with exposure time less than 10 ks and those  longer observations effected heavily by background flaring such that  the good exposure time is below 10 ks. Sources 1ES 0033+595, 1ES 0347-121, 1ES 0647+250, Mrk 180 and TXS 1515-273 were excluded due to this reason. In addition, some observations were ignored, where the  source fell between CCDs or where source was not detected at all. \\ 
\\
Applying our selection criteria, we were left with 10 HBLs (1ES 0299+200, 1ES 0414+009, PKS 0548-322, 1ES 1101-232, 1H 1219+301, H 1426+428, Mrk 501, 1ES 1959+650, PKS 2005-489,  and 1ES 2344+514) that involved  25 pointed observations of \emph{XMM-Newton}. The X-ray data of these HBLs taken by  EPIC was 
downloaded from the {\it XMM-Newton} public archive\footnote{\url{https://nxsa.esac.esa.int/nxsa-web/}}.  The observation log is provided in  \autoref{tab1}, which contains the name of  each source, its position and 
red-shift, date of observation, observation ID, window mode of observation,  whether there was pile-up, exposure time, good exposure time, and  the size of the temporal bin we used to produce X-ray light curves (LCs).

\subsection{Data Reduction}
\noindent
 EPIC-PN  takes extremely sensitive images of X-ray sources in  the energy range 0.15--15 keV \citep{Tur01}. However, soft proton flaring  often dominates over 10 keV so we restrict  our analysis to the 0.3--10 keV energy range. Observation Data Files (ODF) were reprocessed using the standard procedure of the {\it XMM-Newton} Science Analysis System (SAS) version 19.1.0 with the help of updated Calibration Current Files. SAS task \emph{epproc} was used to produce calibrated and concatenated EPIC-PN event  lists from uncalibrated event lists. In order to create clean event lists, we search for soft proton flares by examining LCs in  the 10--12 keV range. We then use SAS task \emph{tabtigen} to generate a Good Time Interval (GTI) file, which  contains information about  proton flare free timings.  Next we employ SAS task \emph{evselect}, which  uses GTI and uncalibrated event files as input  to produce cleaned event files.   Finally, SAS task \emph{epatplot} is used to detect pile-up, if and it is eliminated by removing a small region from the center of the source and thus provides an annular region from which we extract the source events. \\
 \\
 Out of  these 25 observations, three were in full window imaging mode, five were in timing mode and  the rest were in small window imaging mode. In imaging mode, source events are extracted using a circular aperture having $\SI{40}{\arcsecond}$ radius and background events are extracted using the same circular aperture but  placed far away from the source.
  One observation of PKS 0548-322 (0142270101)  and two observations each of H 1426+428 (0310190101, 0310190501) and 1ES 1959+650 (0850980101, 0870210101) were affected by pile-up.
  In those  cases we have discarded the central portion around source and selected  an annular region between $\SI{4}{\arcsecond}$ and $\SI{40}{\arcsecond}$ for extraction. In timing mode, source events are extracted from a $\SI{82}{\arcsecond}$ (RAW X=27--47) wide box  along RAWX centred on  the source's  vertical strip. Background events are extracted similarly from a box of $\SI{41}{\arcsecond}$  (RAWX=3--13)  width from a source free region.

\vspace*{0.1in}
\noindent
We obtained  LCs that are background and vignetting corrected using SAS task \emph{epiclccorr}.
High background flaring if detected at beginning or/and at the end of a LC were completely removed. However, when short periods of high background flaring  were detected  in middle of a LC, we removed  these points if they are  fewer than five  in number and used Lagrange's interpolation method  to replace the removed ones, as in \citet{Gon12}. We defined the resultant continuous time interval obtained by following the above method as good exposure time. LCs with good exposure times above 10, 25, 50 and 75 ks were binned  into segments of 100, 200, 300 and 400 s, respectively. Mean counts of the light  curves obtained in this fashion in the soft (0.3--2.0 keV), hard (2.0--10.0 keV) and total (0.3--10.0 keV) energy bands are given in  \autoref{tabhr}.

\section{ANALYSIS TECHNIQUEs\label{sec3}}
\noindent
In this section, we discuss  the  techniques that we have used  to analyze  our data.  The  results we  obtained by these techniques are discussed in section \ref{sec4}.

\subsection{Excess Variance and Fractional Variance}\label{fvar}
\noindent
We calculated excess variance $\sigma^2_{XS}$  and fractional variance $F_{var}$ 
\citep[e.g][]{Ede02}, which  are the   parameters commonly used to quantify the strength of X-ray variability. Due to measurement errors in the observation, finite  uncertainties 
will be present in the LC. These uncertainties will cause an additional variance on individual flux measurements. Excess variance,  a measure of the intrinsic variance of a source, removes this additional variance arising from measurement errors. Suppose  a LC consists of $N$ data points $x_j$ corresponding to times $t_j$ having associated  measurement errors $\sigma_{err,j}$, then the excess variance is given by following expression
\begin{equation}
  \sigma ^2_{XS}=S^2- \bar{\sigma }^2_{err}.
\end{equation}
\vspace{-0.02in}
Here $S^2$ is the sample variance of the LC, given by
\begin{equation}
    S^2=\frac{1}{N-1}{\sum_{j=1}^{N} (x_j-\bar{x})^2},
\end{equation}
 and $ \bar{\sigma}^2_{err} $ is the mean square error of uncertainties, given by

\begin{equation}
\bar{\sigma }^2_{err}=\frac{\sum_{j=1}^{N} \sigma ^2_{err,j}}{N}.
\end{equation}

\noindent
 The fractional rms variability amplitude, $F_{var}$ is the square root of normalized excess variance $\sigma ^2_{NXS}= {\sigma^2_{XS}}$/$ {\bar{x}^2}$ \citep[e.g.][]{Vau03} and is given by,
\begin{equation}
 F_{var}=\sqrt{\frac{S^2-\bar{\sigma} ^2_{err}}{\bar{x}^2}}.
\end{equation}

\noindent
The uncertainty associated with $F_{var}$  \citep[e.g.][]{Vau03} is given by,

\begin{equation}
   (F_{var})_{err}=\sqrt{\Big(\sqrt{\frac{1}{2N}}\frac{\bar{\sigma} ^2_{err}}{\bar{x}^2 F_{var}}\Big)^2+\Big(\sqrt{\frac{\bar{\sigma} ^2_{err}}{N}}\frac{1}{\bar{x}}}\Big)^2 .
\end{equation}

\noindent
We consider a LC to be variable when $F_{var}>3 \times (F_{var})_{err}$, following \citet{Dhi21}.

\begin{figure}[ht!]
   \begin{center}
        \vspace*{-0.7in}
        \epsscale{1.2}
        \hspace{-0.2in}\plotone{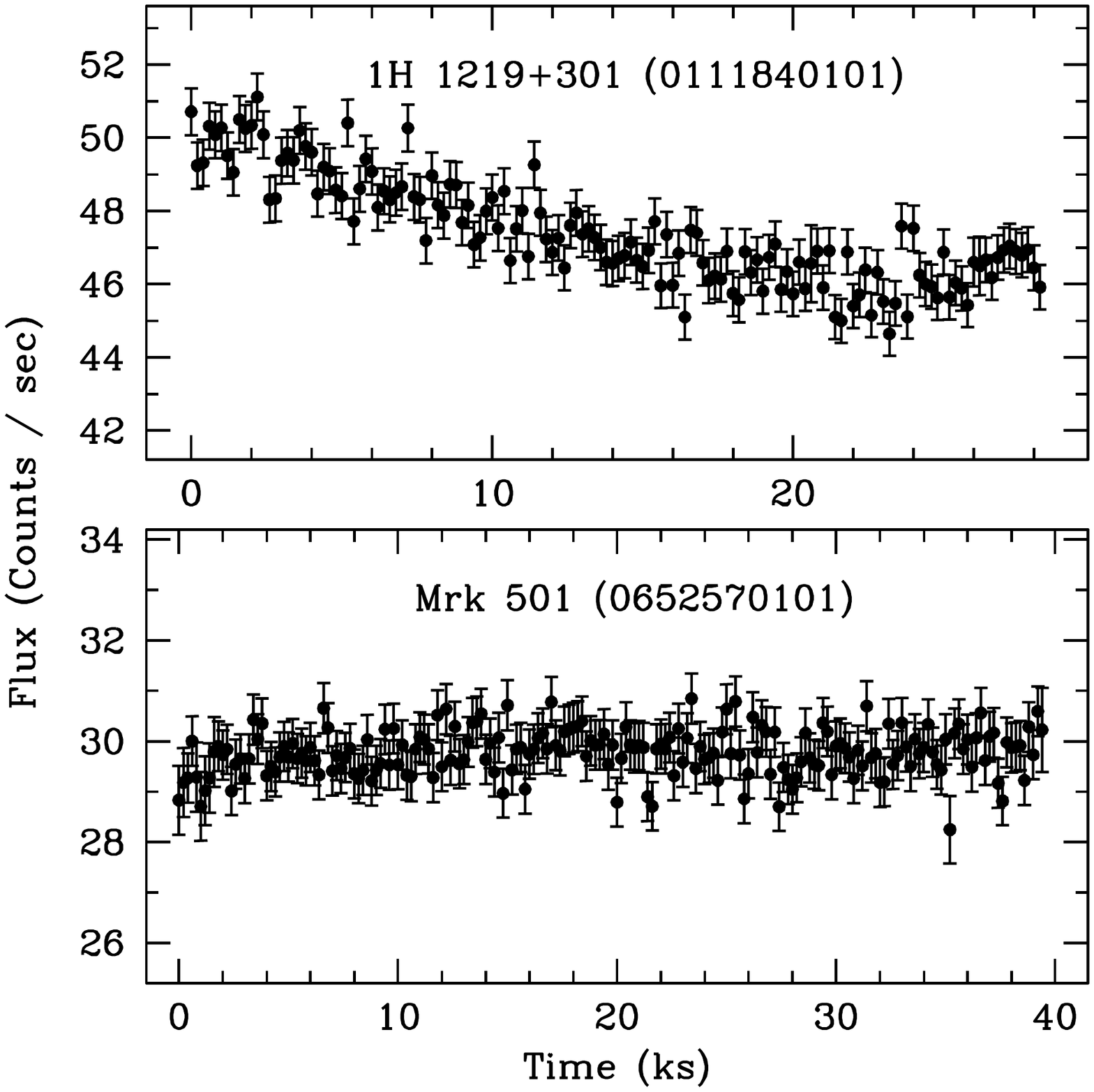}
    \end{center}
\vspace*{-0.9in}
    \caption{{\it XMM-Newton} LCs for 1H 1219+301 (Obs ID: 0111840101) and Mrk 501 (Obs ID:  0652570101) in the total energy range (0.3--10 keV). \label{TLC}}
\end{figure}

\begin{figure}[ht!]
   \begin{center}
        \vspace*{-0.7in}
        \epsscale{1.2}
        \hspace{-0.2in}\plotone{{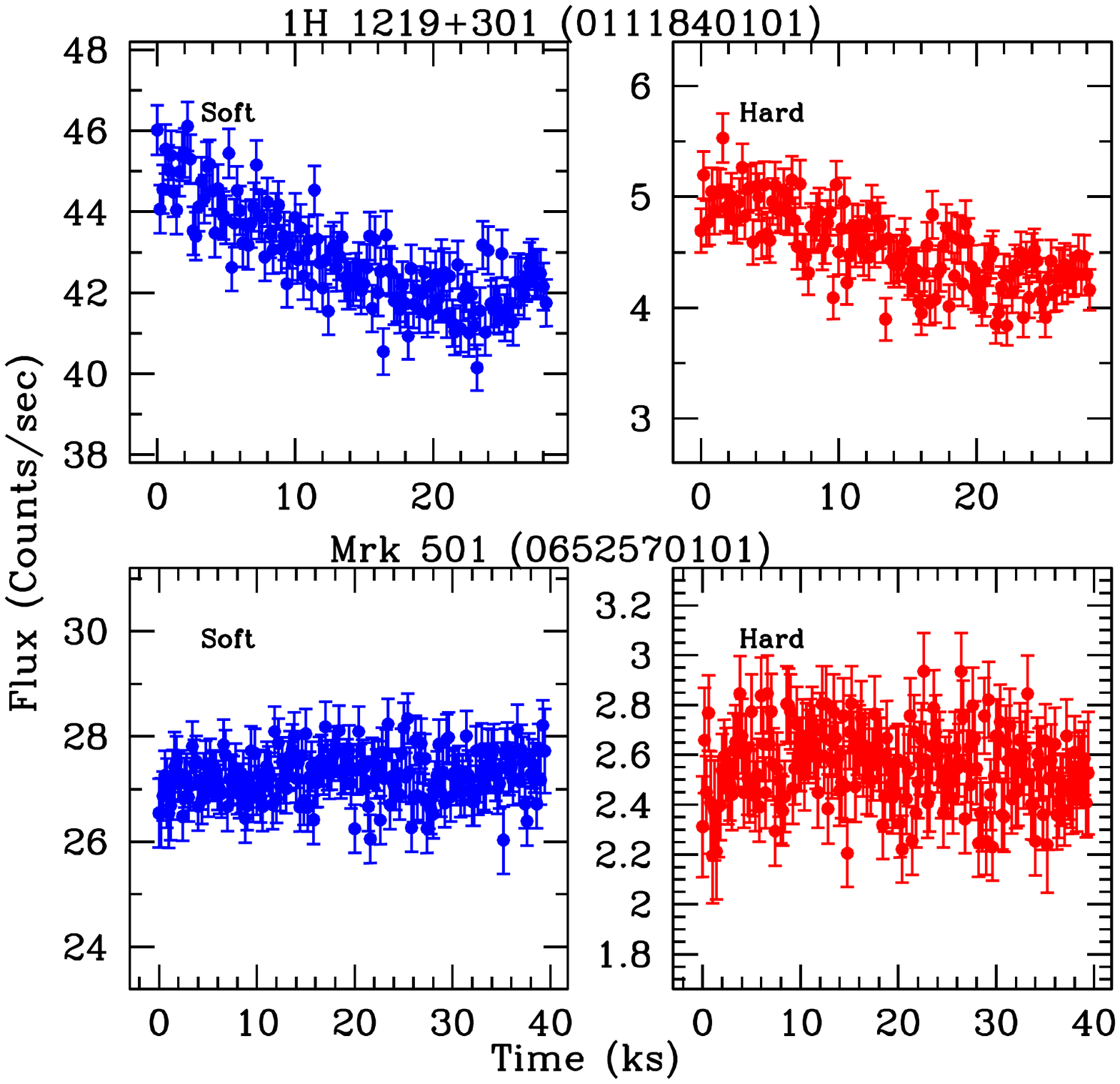}}
    \end{center}
\vspace*{-0.9in}
    \caption{ Soft (0.3--2 keV; denoted by blue filled circles) and hard (2--10 keV; denoted by red filled circles) LCs for 1H 1219+301 and Mrk 501. The observations used here are same as that of \autoref{TLC}. \label{SOFT-HARD-LC}}
\end{figure}

\renewcommand\theHtable{\thetable}
\begin{deluxetable*}{cccccCCCC}
\tablenum{2}
\tablecaption{Flux and hardness ratio of HBLs}
\tablewidth{0pt}
\tablehead
{\colhead{Source} &\colhead{Obs ID} & \multicolumn3C{ \mathrm{$ \mu $(counts/s)} } &
 \colhead{Mean HR} &   \colhead{$n$} &   \colhead{$\chi^2$} &   \colhead{$\chi^2_{0.99,n}$}\\
\colhead{} & \colhead{} & \colhead{Soft} &\colhead{Hard} & \colhead{Total} &
\colhead{} & \colhead{} & \colhead{} & \colhead{} \\
\nocolhead{}& \nocolhead{}& \colhead{0.3--2 keV}& \colhead{2--10 keV}& \colhead{0.3--10 keV}& \nocolhead{}& \colhead{} & \colhead{}
& \colhead{}
}
\decimalcolnumbers
\startdata
1ES\hspace{0.1cm}0229$+$200 &  0604210201 & 03.67 $\pm$ 0.22&01.22 $\pm$ 0.13&04.88 $\pm$ 0.26&$-$0.50 $\pm$ 0.07 &173 &71.72 & 219.20\\
  &  0604210301&03.67 $\pm$ 0.22&01.23 $\pm$ 0.13&04.88 $\pm$ 0.26&$-$0.50 $\pm$ 0.07 & 212& 100.57 &262.80\\
1ES\hspace{0.1cm}0414$+$009 & 0094383101 & 10.70 $\pm$ 0.42&01.45 $\pm$ 0.16&12.13 $\pm$ 0.45&$-$0.76 $\pm$ 0.06 & 104 & 18.50 & 140.50\\
 &0161160101 & 05.42 $\pm$ 0.17&00.54 $\pm$ 0.06&05.95 $\pm$ 0.19&$-$0.82 $\pm$ 0.05 & 238& 33.58& 291.70\\
PKS\hspace{0.1cm}0548$-$322 &0142270101&14.27 $\pm$ 0.27&03.03 $\pm$ 0.13&17.20 $\pm$ 0.30&$-$0.65$\pm$0.03&200&50.18&259.50\\
&0205920501&23.11 $\pm$ 0.52&04.79 $\pm$ 0.24&27.86 $\pm$ 0.57&$-$0.66$\pm$0.03&238&60.73&291.70\\
1ES\hspace{0.1cm}1101$-$232&   0205920601 & 33.09 $\pm$ 0.63&06.06 $\pm$ 0.27&39.09 $\pm$ 0.68& $-$0.69 $\pm$ 0.03 & 172 & 47.30 &218.10\\
1H\hspace{0.1cm}1219$+$301 & 0111840101  & 42.90 $\pm$ 0.59&04.54 $\pm$ 0.20&47.44 $\pm$ 0.63&$-$0.81$\pm$0.02 & 141 & 32.04 & 183.00\\
H\hspace{0.1cm}1426$+$428  &   0111850201 &15.34 $\pm$ 0.29&03.79 $\pm$ 0.14 &19.09 $\pm$ 0.32&$-$0.60 $\pm$ 0.02 & 175 & 59.08 & 221.40 \\
  & 0165770101&  19.67 $\pm$ 0.33&03.19 $\pm$ 0.13&22.82 $\pm$ 0.35&$-$0.72 $\pm$ 0.02&186&41.30&233.80\\
 &  0165770201 &19.55 $\pm$ 0.32& 03.20 $\pm$ 0.13&22.71 $\pm$ 0.35 &$-$0.72 $\pm$ 0.02 & 200 & 37.99 & 249.40 \\
&0212090201 & 24.57 $\pm$ 0.45& 03.92$ \pm$ 0.18&28.44 $\pm$ 0.48&$-$0.73 $\pm$ 0.03&143&25.31&185.30\\	
  & 0310190101  &35.29 $\pm$ 0.53 &06.91 $\pm$ 0.24& 42.12 $\pm$ 0.59&$-$0.67 $\pm$ 0.02&167&44.71&212.40\\	
 & 	0310190201&28.04 $\pm$ 0.48&  04.71 $\pm$ 0.20&32.69 $\pm$ 0.51&$-$0.71 $\pm$ 0.02&200&35.69&249.40\\
& 0310190501 &  28.13 $\pm$ 0.48&04.02 $\pm$ 0.18&32.11 $\pm$ 0.51& $-$0.75 $\pm$ 0.03 &175&32.81&221.40\\
Mrk\hspace{0.1cm}501& 0652570101  &27.25 $\pm$ 0.48&02.55 $\pm$ 0.15&29.77 $\pm$ 0.50&$-$0.83 $\pm$ 0.03&197&22.79&246.10\\
& 0652570201  &	27.61 $\pm$ 0.47 &02.47 $\pm$ 0.14&30.06 $\pm$ 0.49&$-$0.84 $\pm$ 0.03& 222 & 26.05 & 273.90\\
& 	0652570301 &27.02 $\pm$ 0.46&04.53 $\pm$ 0.19&31.50 $\pm$ 0.50 & $-$0.71 $\pm$ 0.03 & 202 & 39.04& 251.70 \\	
& 0652570401 &36.57 $\pm$ 0.54&06.10 $\pm$ 0.23&42.60 $\pm$ 0.59&$-$0.71 $\pm$ 0.02 &199 &34.50 &248.3\\
1ES 1959$+$650&0850980101 &172.14 $\pm$ 1.62&37.00 $\pm$ 0.76&207.98 $\pm$ 1.78 & $-$0.65 $\pm$ 0.01 & 189 & 67.15 & 237.10\\
&  0870210101 &123.55 $\pm$ 1.31&32.60 $\pm$ 0.68&155.17 $\pm$ 1.46 & $-$0.58 $\pm$ 0.01 & 156 & 51.85 & 200.00\\
PKS\hspace{0.1cm}2005$-$489 &	0205920401 &04.08 $\pm$ 0.23&00.21 $\pm$ 0.08&04.29 $\pm$ 0.25&$-$0.90 $\pm$ 0.10&116&14.12&154.3\\
&  0304080301 &23.80 $\pm$ 0.53& 03.05 $\pm$ 0.20&26.82 $\pm$ 0.56&$-$0.77 $\pm$ 0.03&144&18.82&186.40 \\
& 0304080401 &23.03 $\pm$ 0.52&02.87 $\pm $ 0.19&25.86 $\pm$ 0.55&-0.78 $\pm $ 0.04 &246 &41.80 & 299.40 \\
1ES\hspace{0.1cm}2344$+$514&0870400101&06.15 $\pm$ 0.22&01.53 $\pm$ 0.11&07.66 $\pm$ 0.25&$-$0.60 $\pm$0.05 & 134 & 66.37 &173.90\\
\enddata
\tablecomments{$\mathrm{ \mu } $ is mean count rate. HR is hardness ratio. n is the number of degrees of freedom. $\chi^2_{99,n}$ is is the $\chi^2$ at 99 per cent confidence level for n degrees of freedom}
\label{tabhr}
\end{deluxetable*}

\subsection{Variability timescale\label{vart}}
\noindent
We determine  the flux variability timescale  following the method described in \citet{Bha18}, where it is given by \citep{Bur74}
\begin{equation}
\tau_{var}=\Bigg|{\frac{\Delta t}{\Delta \ln F}}\Bigg|
\end{equation}
Here $\Delta t $ is time interval between measured flux values $F_1$ and $F_2$, with $F_1 >F_2$ since $\Delta \ln F = \ln F_1 - \ln F_2$. As described in \citet{Hag08}, we calculated all possible pairs of timescale $\tau_{ij}$ which satisfy the condition $\big| F_i - F_j \big| > \Delta F_i + \Delta F_j$   , where $\Delta F $ is the error associated with flux measurement. The shortest variability timescale is minimum of all such pairs
$\tau$ = min($\tau_{ij}$), where $i=1,2,...N-1, j=i,i+1,....N$, and $N$ is  the number of flux values. The uncertainty in $\tau_{var}$ is given by    \citep{Bha18}
\begin{equation}
\Delta\tau_{var} \simeq \sqrt{\frac{F_1^2\Delta F_2^2+F_2^2\Delta F_1^2}{F_1^2F_2^2(\ln[F_1/F_2])^4}}\Delta t .
\end{equation}

\subsection{Discrete Correlation Function\label{dcf_an}}
\noindent
 We use a discrete correlation function (DCF) analysis introduced by \citet{Ede88} and later modified by \citet{Huf92}  to search for the cross-correlation and possible  time lags  between LCs  in soft (0.3--2 keV) and hard (2--10 keV) energy bands. First we calculate  the set of unbinned $UDCF_{ij}$ discrete correlations between soft and hard energy bands using
 \begin{equation}
UDCF_{ij}=\frac{(a_i-\bar{a}) (b_j-\bar{b})}{\sqrt{\sigma_a^2\sigma_b^2}} .
\end{equation}
Here $a_i$ and $b_j$ are soft and hard data points, $\bar{a}$, $ \bar{b} $, $\sigma_a$ and $\sigma_b$ are means and  standard deviations of  the soft and hard data sets respectively. 
There excist a pairwise lag  $\Delta t_{ij}=t_j-t_i$ corresponding to each of these $UDCF_{ij}$  values.
 After binning the correlation function,  we calculate the DCF  for a time lag $\tau$ defined by $\tau-\frac{\Delta\tau}{2}\leq\Delta t_{ij}<\tau$+$\frac{\Delta\tau}{2}$, by averaging the $UDCF_{ij}$  values as,

\begin{longrotatetable}
\begin{deluxetable*}{cCCCcCCcCCccccc}
\tablenum{3}
\tablecaption{X-ray variability parameters for HBLs }
\tablewidth{600pt}
\tabletypesize{\scriptsize}
\tablehead{
\colhead{Source} & \colhead{Obs ID} & \multicolumn9C{F$_{var}$} & \multicolumn3C{$|\tau|$ (ks)} & \colhead{$|\tau|_{corr}$ (ks)}\\
\nocolhead{} & \nocolhead{} & \colhead{Soft} & \colhead{Sig} & \colhead{Var} &   \colhead{Hard} &  \colhead{Sig} & \colhead{Var} &   \colhead{Total} & \colhead{Sig} & \colhead{Var} & \colhead{Soft} & \colhead{Hard} & \colhead{Total} &\colhead{Total} \\
\nocolhead{} & \nocolhead{} & \colhead{0.3--2 keV} & \nocolhead{} &  \nocolhead{} & \colhead{2--10 keV} & \nocolhead{} &  \nocolhead{} &  \colhead{0.3--10 keV} & \nocolhead{} & \nocolhead{} &\colhead{0.3--2 keV} & \colhead{2--10 keV}&  \colhead{0.3--10 keV} &\colhead{0.3--10 keV}
}
\decimalcolnumbers
\startdata
1ES\hspace{0.1cm}0229$+$200 & 0604210201 &$-$&$-$&$-$&$-$&$-$&$-$ &$-$&$-$&$-$&$-$ &$-$&$-$&$-$\\
 &0604210301 &1.08\pm1.76&0.61&NV&$-$&$-$&$-$&$-$&$-$&$-$&$-$ &$-$&$-$&$-$\\
1ES\hspace{0.1cm}0414$+$009&0094383101&$-$&$-$&$-$&3.68 \pm 2.52&1.46&NV&0.64 \pm 1.51&0.42&NV&$-$ &$-$&$-$&$-$\\
 &0161160101&0.93 \pm 0.55&1.69&NV&2.98 \pm 2.19&1.36&NV&0.88 \pm 0.58&1.52&NV&$-$ &$-$&$-$&$-$\\
 PKS\hspace{0.1cm}0548$-$322&0142270101&0.64 \pm 0.31 & 2.06 & NV & 0.90 \pm 1.02 & 0.88& NV & 0.79 \pm 0.22 & 3.59 & V &$-$ & $-$& 04.75 $\pm$ 1.36 & 04.43 $\pm$ 1.27 \\
 &0205920501&0.76 $\pm$ 0.34 &2.24&NV& $-$&$-$&$-$& 0.40 $\pm$ 0.50&0.80&NV&$-$&$-$&$-$&$-$\\
1ES\hspace{0.1cm}1101$-$232	& 0205920601 & 1.30 \pm 0.21 &6.19& V &1.67 \pm 0.73 &2.29 &NV &1.05 \pm 0.20 &5.25&V&1.11 $\pm$ 0.35&$-$&1.10 $\pm$0.30 & 0.93 $\pm$ 0.26\\
1H \hspace{0.1cm}1219$+$301 & 0111840101 &2.53 \pm 0.13 &19.46&V & 6.09 \pm 0.42 &14.50 & V& 2.81 \pm 0.12 & 23.42 & V & 2.91 $ \pm $ 0.83 & 0.90 $ \pm $ 0.25 & 3.65 $ \pm $ 1.21 & 3.08 $ \pm $ 1.02\\
H\hspace{0.1cm}1426$+$428 & 0111850201 & 1.87 \pm 0.17 &11.00& V & 2.89 \pm 0.39&7.41& V &2.05 \pm 0.15 &13.66 & V &4.04 $ \pm $ 1.46 & 2.09 $ \pm $ 0.80 & 4.01 $ \pm $ 1.29 &  3.55 $ \pm $  1.14\\
 & 0165770101 & 0.91 \pm 0.20 &4.55& V&  1.68 \pm 0.61 &2.75&NV & 0.85 \pm 0.18 & 4.72 &	V  & 3.76 $ \pm $ 1.11 & $ -$ & 4.75 $ \pm $ 1.65 & 4.21 $ \pm $  1.46 \\
 & 0165770201 &  0.32 \pm 0.45 &0.26 &NV  &1.59 \pm 0.61 &2.61 &NV & 0.69 \pm 0.20&3.45&V & $ -$  & $ -$ & 4.84 $ \pm $ 1.72 & 4.29 $ \pm $  1.52 \\
 & 0212090201 &0.80\pm0.28 &2.85&NV& 0.76 \pm 1.67&0.43& NV &0.89 \pm 0.23&3.87&	V & $ -$ & $ -$ & 4.08 $ \pm $ 1.98 & 3.61 $ \pm $  1.75\\
 & 0310190101 &0.82 \pm 0.19 &4.32& V &2.38 \pm 0.38 &6.26 &V &1.10 \pm 0.14 &7.86 &V &3.63 $ \pm $ 1.42 & 1.62 $ \pm $ 0.63 & 3.34 $ \pm $ 1.10 & 2.96 $ \pm $  0.97\\
 & 0310190201 &0.94 \pm 0.19&4.95& V & $-$ & $-$ &$-$ &0.98 \pm 0.17&5.76 &V & 2.37 $ \pm $ 0.67 & $ -$ & 2.71 $ \pm $ 0.83 & 2.40 $ \pm $ 0.73 \\
 & 0310190501 &0.41 \pm 0.39&1.05& NV&2.07 \pm 0.63 &3.29&V&0.59 \pm 0.26 & 2.27&NV & $ -$ & 1.15 $ \pm $ 0.43 & $ -$& $ -$  \\
Mrk\hspace{0.1cm}501 & 0652570101 & $-$&$-$&$-$ &1.33 \pm 1.34&0.99&NV & $-$ & $-$ & $-$&$-$ &$-$&$-$&$-$\\
 &  0652570201 &0.22 \pm 0.63&0.34 & NV &2.34 \pm 0.76&3.08& V &0.20\pm0.63 &0.31 &NV& $ -$ & 0.86 $ \pm $ 0.29 & $ -$& $ -$ \\
 & 0652570301 &1.65 \pm 0.15&11.00&V &1.71 \pm 0.61&2.80&NV&1.70 \pm 0.13&13.08& V & 2.97 $ \pm $ 1.05 & $ -$ &3.38 $ \pm $ 1.28 & 3.27 $ \pm $  1.24 \\ 
 & 0652570401 & 0.79 \pm 0.17&4.65 &V &$-$&$-$ & $-$ &0.82 \pm 0.15&5.47&	V & 3.21 $ \pm $ 1.08 & $ -$ & 3.69 $ \pm $ 1.33 & 3.57 $ \pm $ 1.29 \\
1ES\hspace{0.1cm}1959$+$650 & 0850980101 &1.30 \pm 0.08&16.25&V&2.54\pm0.17&14.94&V&1.5\pm0.07&21.42&V & 4.99 $ \pm $ 1.68 & 2.73 $ \pm $ 1.10 & 5.60 $ \pm $ 1.92 & 5.35 $ \pm $ 1.83 \\
 & 0870210101 &1.72 \pm 0.09&19.11&V&2.03 \pm 0.21&9.66&V &1.8 \pm 0.08&22.50&V & 4.47 $ \pm $ 1.47 & 2.10 $ \pm $ 0.64 & 4.51 $ \pm $ 1.33 & 4.31  $ \pm $  1.27 \\
PKS\hspace{0.1cm}2005$-$489 & 0205920401&$-$&$-$&$-$	&$-$&$-$&$-$& $-$ & $-$ &$-$&$-$ &$-$&$-$&$-$\\
 & 0304080301 &0.28 \pm 1.03 &0.27&NV &$-$&$-$ & $-$ & 0.62 \pm 0.45 &1.37&	NV&$-$ &$-$&$-$&$-$ \\
 & 0304080401 & $-$&$-$&$-$ &1.79 \pm 1.24 &1.44 &NV & $-$ & $-$	& $-$ &$-$ &$-$&$-$&$-$ \\
1ES\hspace{0.1cm}2344$+$514& 0870400101&3.45 $\pm$ 0.39&08.85&V&4.27 $\pm$ 1.00&4.27&V&2.58 $\pm$ 0.38&6.79&V &1.45$ \pm $0.55&  0.74 $\pm $ 0.27 & 1.54 $\pm $0.56& 1.48 $\pm$ 0.54\\
\enddata
\tablecomments{
 Sig denotes significance and is calculated by $Sig=F_{var}/err(F_{var})$. 
\newline Var represents the variability and can take values V for variable LC and NV for non-variable LC.
\newline $|\tau|$ and $|\tau|_{corr}$ are the variability timescale and red shift corrected variability timescale, respectively.
\newline Dash represents fractional variance could not be calculated as the sample variance was less than mean square error.
}
\label{tab3}
\end{deluxetable*}
\end{longrotatetable}

\begin{equation}
    DCF(\tau)=\frac{\sum{UDCF_{ij}}}{M}
\end{equation}
where M is  the number of $UDCF_{ij}$  pairs over which DCF is averaged.
 The standard error $\sigma_{DCF}(\tau)$ associated with each bin is \citep{Ede88} 
\begin{equation}
\sigma_{DCF}(\tau)=\frac{\sqrt{\sum[UDCF_{ij}-DCF(\tau)]^2}}{M-1} .
\end{equation}

\begin{figure}[ht!]
   \begin{center}
        \vspace*{-0.7in}
        \epsscale{1.2}
        \hspace{-0.2in}\plotone{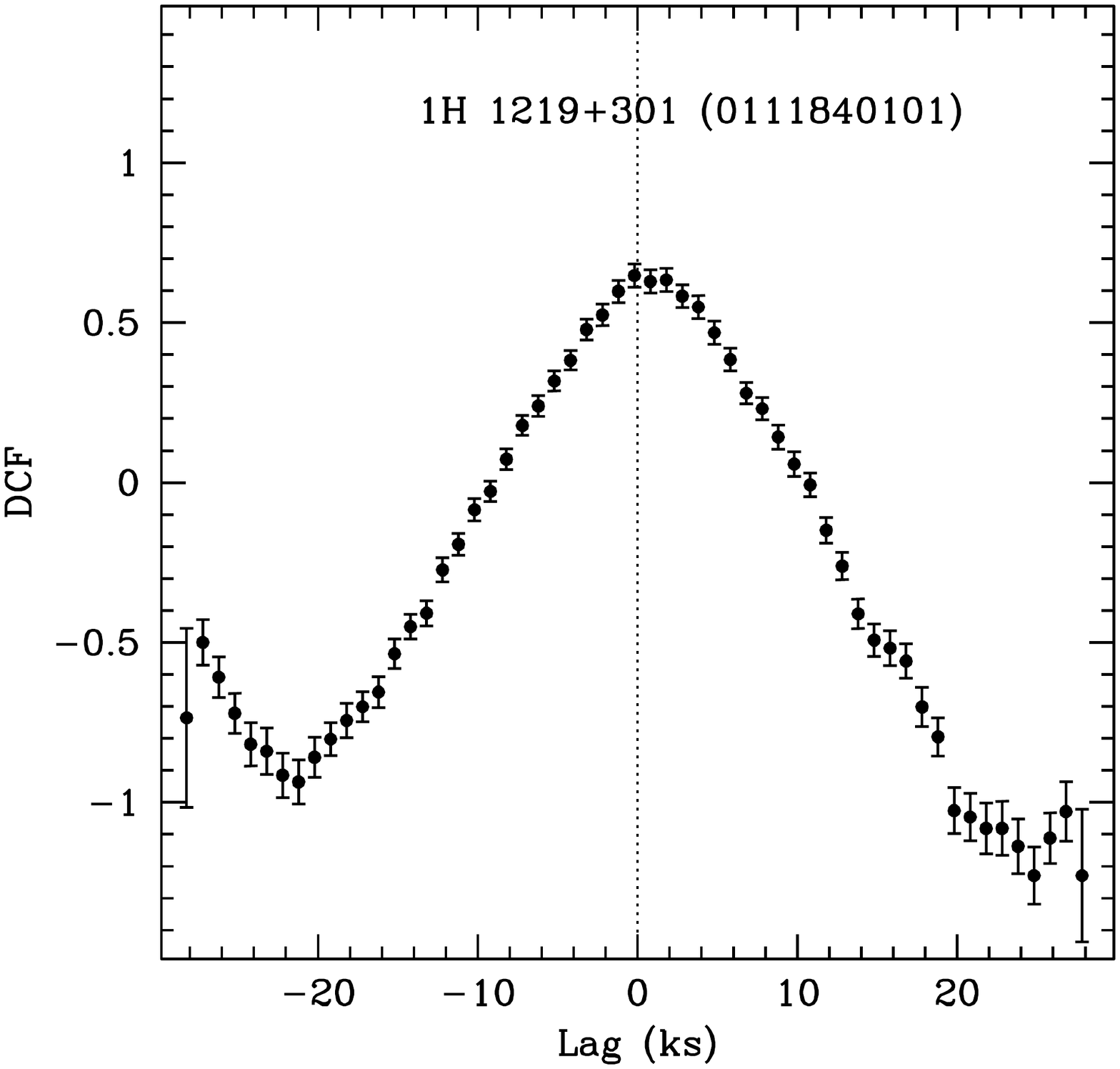}
     \end{center}
\vspace*{-0.9in}
    \caption{The DCF plot of 1H 1219+301 with observation ID 0111840101.\label{DCF}}
\end{figure}

\noindent
A DCF peak value $> 0$ implies soft and hard data sets are correlated  at that lag, while a DCF peak $ < 0$ implies they are anti-correlated. We can use the auto-correlation function (ACF, where $a = b$) to perform a crude search for periods in these astronomical time series. If an  observational time series contains periodic signals the ACF distribution would also show an oscillation at that period.

\begin{figure}[ht!]
    \begin{center}
        \vspace*{-0.7in}
        \epsscale{1.2}
        \hspace{-0.2in}\plotone{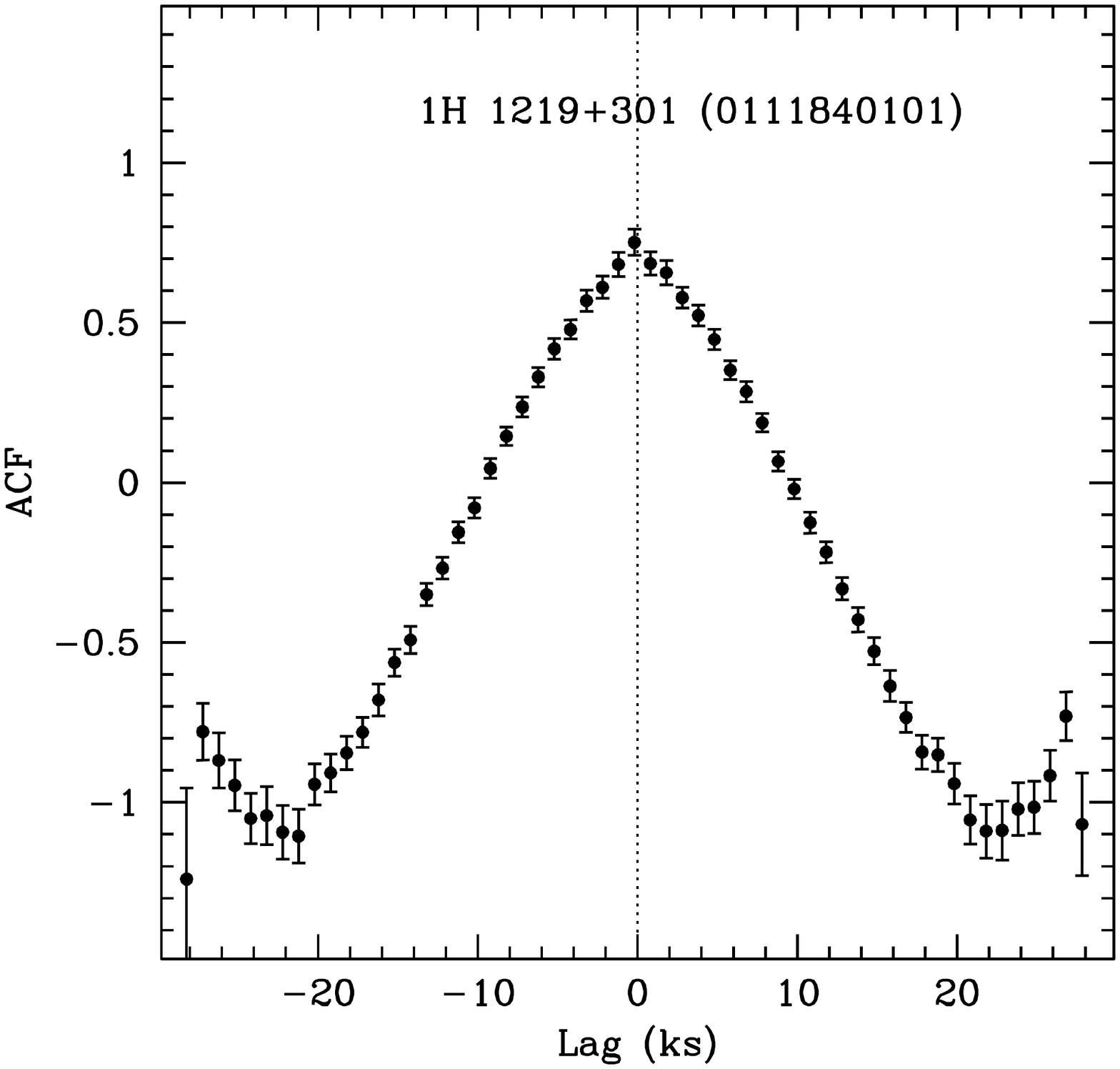}
     \end{center}
\vspace*{-0.9in}
        \caption{The ACF plot of 1H 1219+301 with observation ID 0111840101.\label{ACF}}
\end{figure}

\subsection{Hardness Ratio\label{hr_analysis}}
\noindent
 We can  investigate the X-ray spectral behavior in the coarse fashion prescribed by the relatively low counts in each temporal bin by computing the hardness ratio (HR) in ten of these blazars. Looking at any changes in HR for a given source lets us search for spectral variability; it is simply given by

\begin{equation}
    HR=\frac {H-S}{H+S}.
\end{equation}
\begin{figure}[ht!]
    \begin{center}
    \epsscale{1.2}
    \vspace*{-0.65in}
    \hspace{-0.2in}\plotone{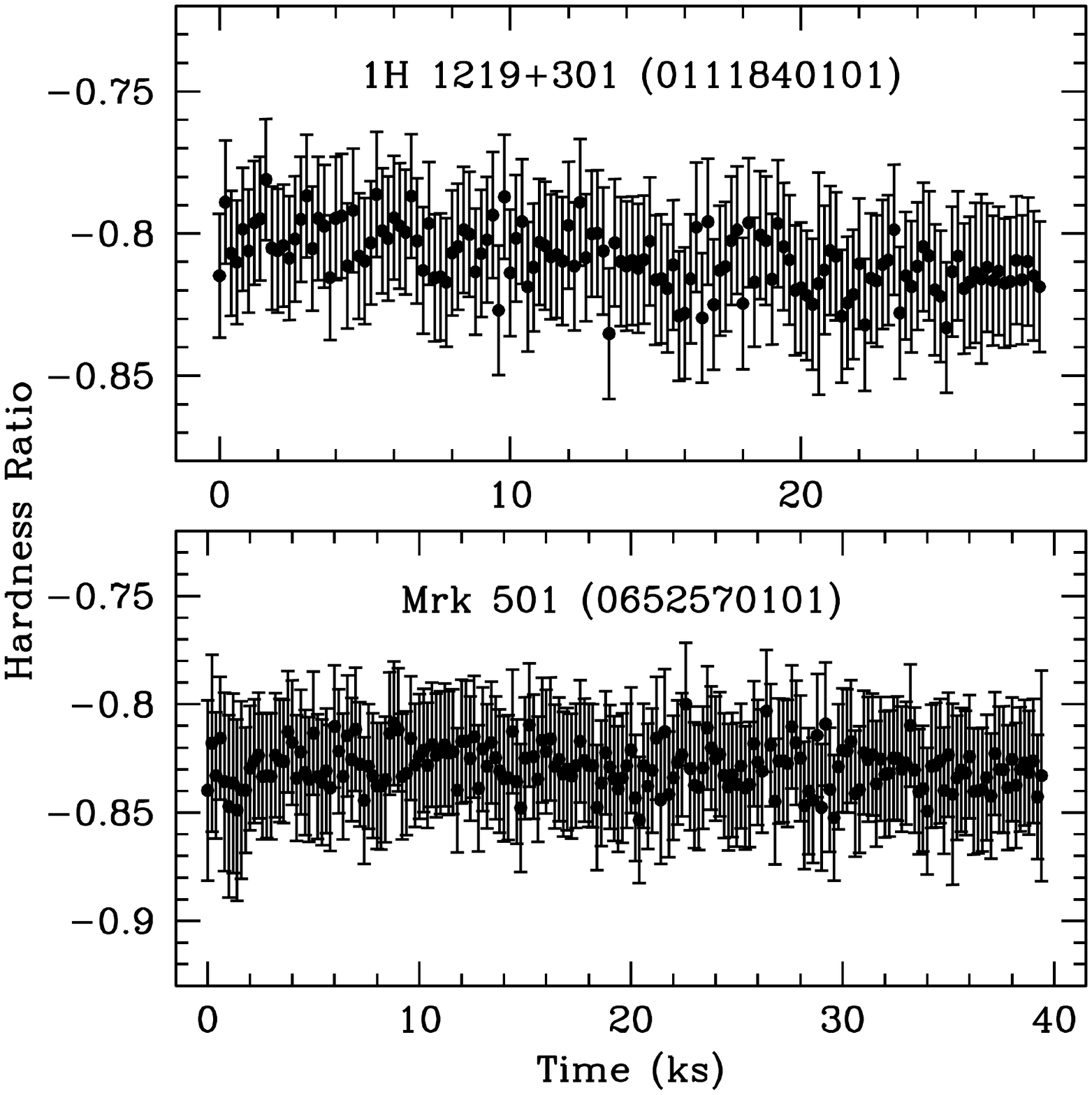}
    \end{center}
      \vspace*{-1.0in}
    \caption{ HR plots for 1ES 0414+009 and 1ES 1959+650 observed with {\it XMM-Newton}. The observations used here are same as that of \autoref{TLC}\label{HR}.}
\end{figure}
\noindent Here $H$ and $S$ denote the hard (2--10 keV) and soft (0.3--2 keV) net count rates. 
 The error associated with a HR measurement, $\sigma_{HR}$, is
\begin{equation}
\sigma_{HR}=\frac{2\sqrt{S^2\sigma_H^2+H^2\sigma_S^2}}{(H+S)^2} ,
\end{equation}
where $\sigma_H$ and $\sigma_S$ are the uncertainty in hard and soft band, respectively. We perform a standard $\chi^2$ test to investigate possible temporal  variations in $HR$:

\begin{equation}
  \chi^2=\sum_{j=1}^n\frac{x_j-\bar{x}}{\sigma^2_j}, 
\end{equation}
\noindent where $x_j$ and $\sigma_j$ denote the  HR value for the j$^{th}$ data point and its associated error, while $\bar{x}$ is mean value of all HR values.

 \subsection{Power Spectral Density\label{psd_an}}
 \noindent
 The power spectral density (PSD)  provides the amount of  variability power as a function of temporal frequency and is an useful tool to  search for presence of possible periodicities and quasi-periodic oscillations (QPOs) in a LC.  For AGNs, the PSD usually shows red noise behavior at lower frequency that changes to white noise behavior at higher frequencies where measurement errors prevail.  The standard procedure for evaluating a PSD is by calculating the periodogram \citep[][]{Vau03}. While computing periodogram function using scipy module, we normalize it in units of (rms/mean)$^2$.  We employ Bayesian statistics and maximum likelihood estimation in fitting the periodogram, as discussed in \citet{Vau10}.
 The best-fitting parameter $\theta$ for a particular parametric model $P(\nu,\theta)$
is assessed by maximizing the likelihood function, which is equivalent to
minimizing the following fit statistic,
 \begin{equation}
    S=2\sum_{j=1}^{N/2}\frac{I_j}{P_j} + \ln P_j .
\end{equation}
 Here $S$ is twice the minus log-likelihood,  $P_j$ and $ I_j$ are observed model spectral density and periodogram at Fourier frequency $f_j$, respectively. A significant QPO may be claimed if a peak rises at least 3 $\sigma$ (i.e 99.73 $\%$) above the red noise fit of the PSD. We adopt a simple power law plus a constant form to fit the PSDs of these X-ray LCs \citep[][]{Gon12,Moh15},  
 \begin{eqnarray}
    P(f)=Nf^{-\alpha}+C
\end{eqnarray}
The model uses three parameters: $N$, normalization constant for the power law fitting; $\alpha$, the spectral index for power law fitting; $C$, an additive constant  to take care of the Poisson noise.

\begin{figure}[ht!]
    \epsscale{1.2}
    \plotone{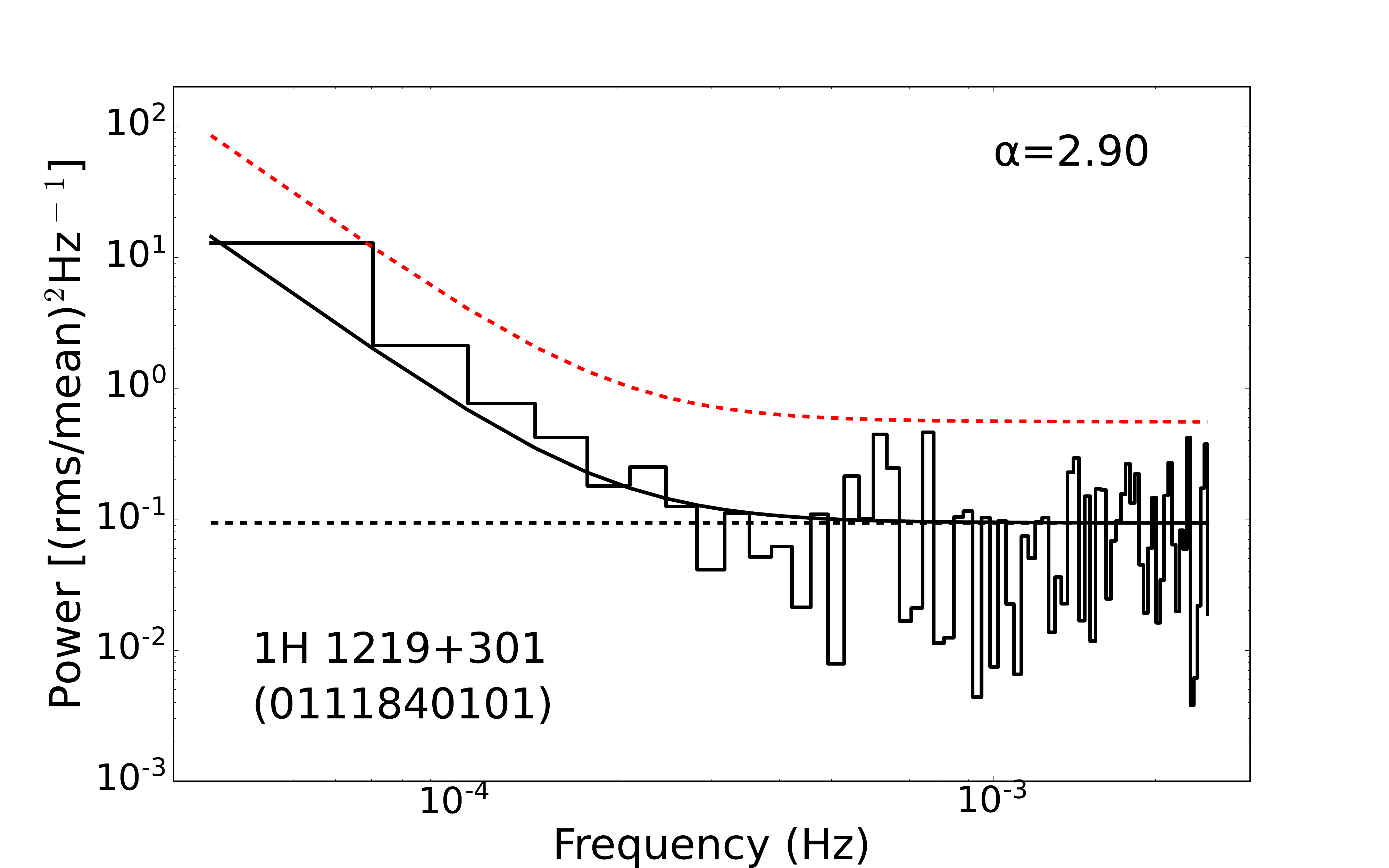}
        \plotone{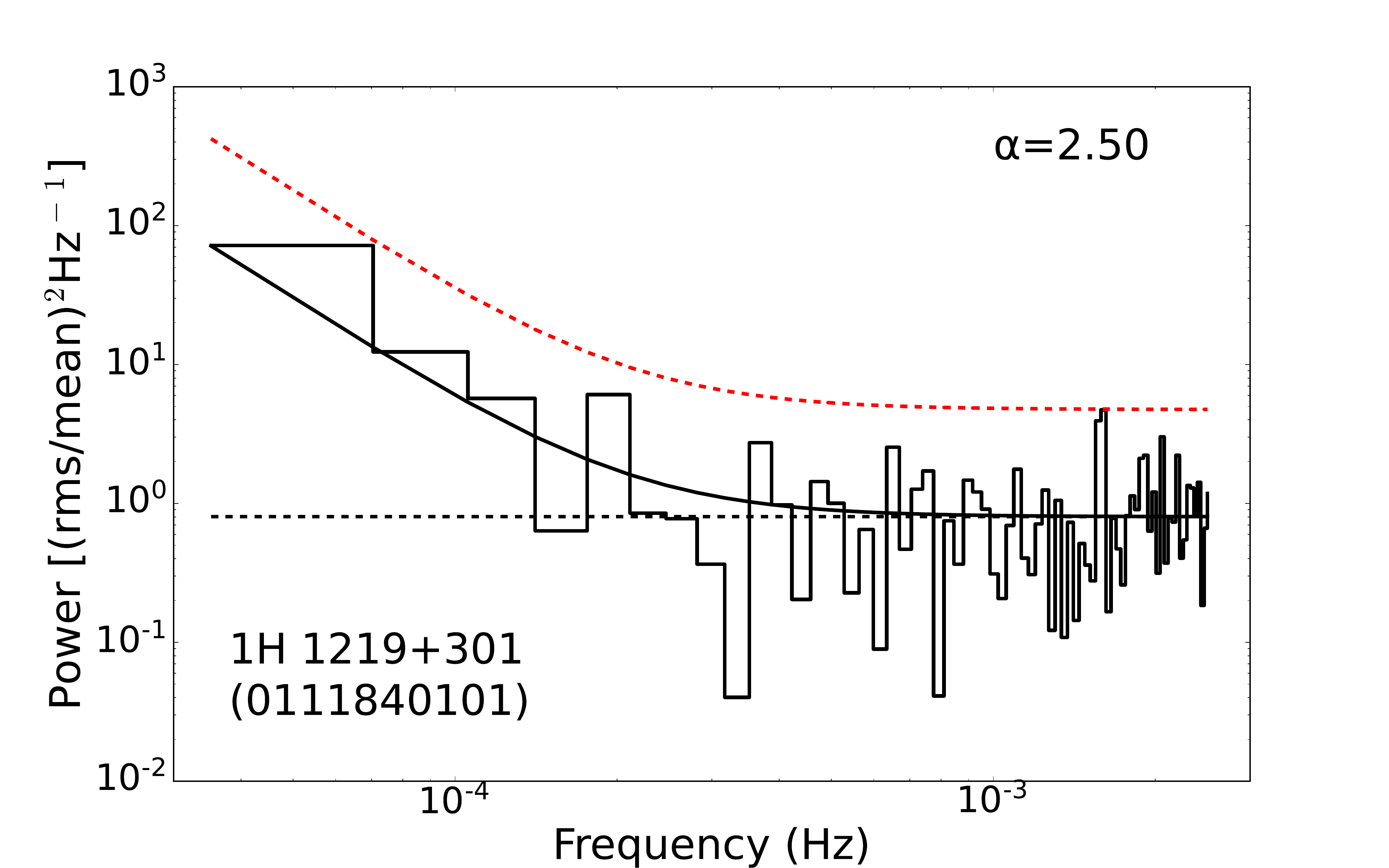}
            \plotone{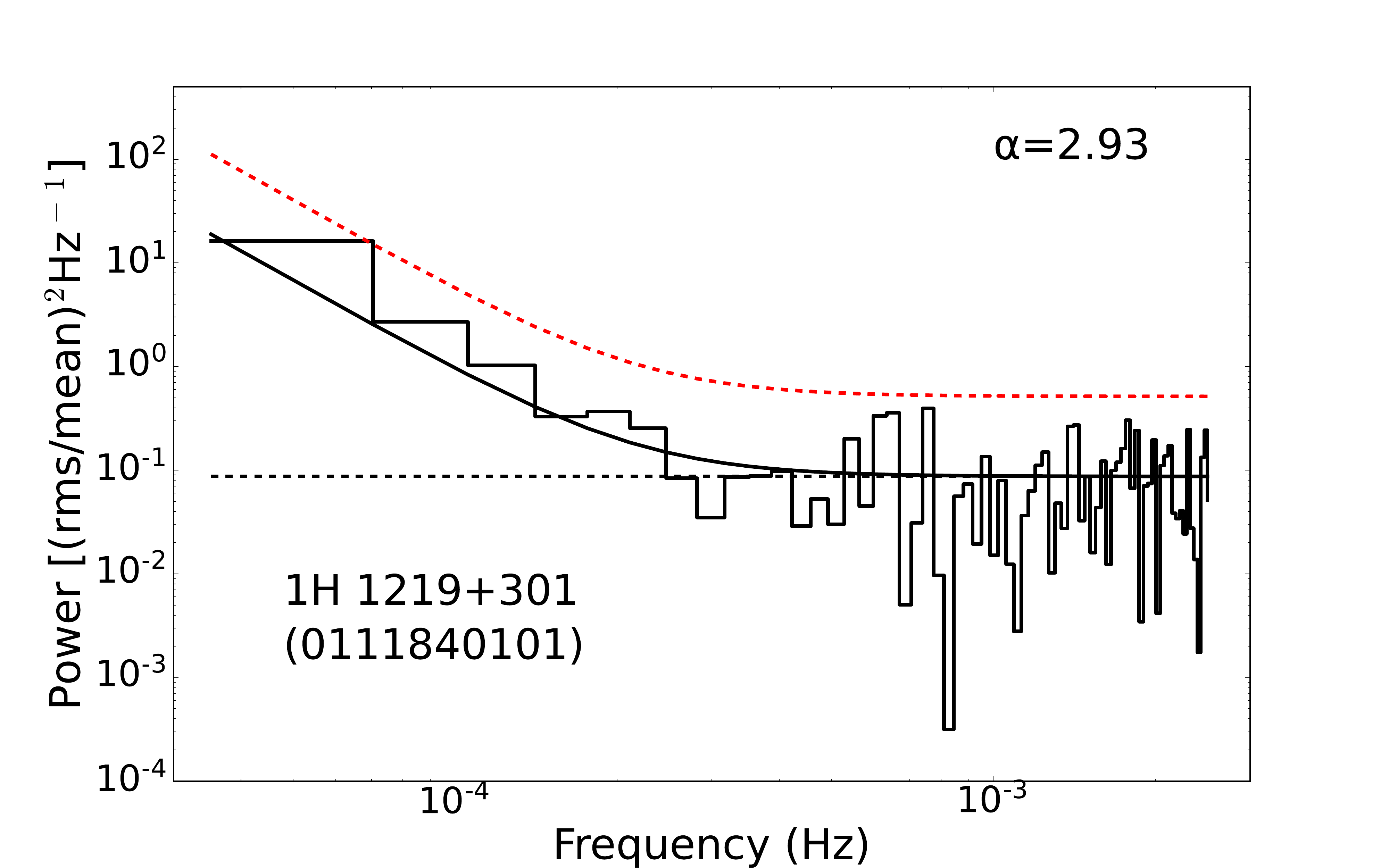}
    \caption{PSD plot for {\it XMM-Newton} observation of 1H 1219+301 with observation ID 0111840101 in the soft (top), hard (middle) and total (bottom) X-ray band. The obtained spectral index are also shown.\label{allPSD}}
\end{figure}

\begin{deluxetable*}{cCcccccc}
\tablenum{4}
\tablecaption{PSD parameters from the power law fit in soft, hard and total bands\label{tab4}}
\tablewidth{0pt}
\tablehead{
\colhead{Source} & \colhead{Obs ID} & \multicolumn2c{Soft (0.3--2 keV)} &\multicolumn2c{Hard (2--10 keV)} & \multicolumn2c{Total (0.3--10 keV)} \\
\nocolhead{} & \nocolhead{} & \colhead{$log_{10}N$} & \colhead{$\alpha$} &
\colhead{$log_{10}N$} & \colhead{$\alpha$} & \colhead{$log_{10}N$} & \colhead{$\alpha$} 
}
\decimalcolnumbers
\startdata
1ES\hspace{0.1cm}0229$+$200& 0604210201&	NV &NV &NV &NV &NV &NV\\
& 0604210301&	NV &NV &NV &NV &NV &NV\\
1ES\hspace{0.1cm}0414 $+$ 009& 0094383101 &	NV &NV &NV &NV &NV &NV\\
&0161160101 &  NV &NV &NV &NV&NV&NV\\
PKS\hspace{0.1cm}0548-322&0142270101&NV &NV &NV &NV&$-$&$-$\\
&0205920501&NV &NV &NV &NV&NV&NV\\
1ES\hspace{0.1cm}1101$-$232	& 0205920601 & 	$-$ &	$-$ &NV &NV &	$-$ &	$-$\\
1H\hspace{0.1cm}1219$+$301 & 0111840101 &$-$11.77 $\pm$0.64&2.90 $\pm$0.14& $-$9.26$\pm$ 0.45 & 2.5$\pm$ 0.10 &$-$11.79 $\pm$0.75&2.93 $\pm$0.17\\
H\hspace{0.1cm}1426$+$428 & 0111850201&$-$11.74 $\pm$0.96&2.72 $\pm$0.2&$-$9.79 $\pm$0.83&2.41 $\pm$0.18&$-$11.72 $\pm$1.18&2.75 $\pm$0.25\\
 & 0165770101 &	$-$ &	$-$ &NV &NV &	$-$& 	$-$\\
 & 0165770201 &	NV & NV &NV &NV &$-$10.14 $\pm$1.06&2.22 $\pm$0.22\\
 & 0212090201 &	NV &NV&NV&NV &$-$4.84 $\pm$0.54&1.11 $\pm$0.13\\
 & 0310190101 &	$-$6.98 $\pm$0.69 &	1.59 $\pm$0.16	&$-$ & $-$	&$-$6.63 $\pm$0.80 &1.52 $\pm$0.18\\
 & 0310190201 & $-$6.60 $\pm$0.62 & 1.49 $\pm$0.14 & NV&NV &$-$8.82 $\pm$0.7 & 2.01 $\pm$0.15\\
 & 0310190501 &	NV &NV &$-$6.27 $\pm$2.91 &1.51 $\pm$0.65&NV& NV\\
Mrk\hspace{0.1cm} 501& 0652570101 &NV &NV &NV &NV &NV& NV\\
 & 0652570201	&NV &NV &	$-$ &	$-$ &NV&NV\\
 & 0652570301	&$-$5.62 $\pm$0.27&1.36 $\pm$0.06&NV&NV &$-$6.97 $\pm$0.45&1.69 $\pm$0.1\\
 & 0652570401 &$-$12.17 $\pm$3.01 &2.67 $\pm$0.65& NV &NV & $-$12.38 $\pm$2.23 &	2.73 $\pm$0.48\\
1ES\hspace{0.1cm}1959$+$650 & 	0850980101 &$-$8.16 $\pm$0.51&1.89 $\pm$0.11&$-$7.04 $\pm$ 0.76&1.77 $\pm$ 0.17 &$-$7.96 $\pm$0.55&1.88 $\pm$0.12\\
 &0870210101&$-$11.02 $\pm$0.45&2.64 $\pm$0.1&$-$10.29 $\pm$1.19&2.54 $\pm$0.26&$-$11.33 $\pm$0.95&2.73 $\pm$0.21\\
PKS\hspace{0.1cm} 2005$-$489	& 0205920401&	NV &NV &NV &NV &NV& NV\\
	& 0304080301&	NV &NV &NV &NV &NV& NV\\
	& 0304080401 &NV &NV &	$-$ &	$-$ &NV&NV\\
1ES\hspace{0.1cm}2344$+$514&0870400101&$-$&$-$&$-$5.66 $\pm$ 1.47&1.59 $\pm$0.34&$-$&$-$\\
\enddata
\tablecomments{N and $\alpha$ denote normalisation and spectral index, respectively.\newline
NV denotes observation is non-variable.\newline
$-$ indicates variations were too small to compute a power spectral density.
}
\end{deluxetable*}

\section{RESULTS\label{sec4}}
\noindent
We now discuss the results obtained on applying  the analysis techniques discussed in section  \ref{sec3} to the \emph{XMM-Newton} data given in Table \ref{tab1}.

 \subsection{Intraday X-ray flux variability}
 \noindent
 We produced X-ray LCs for all 10 HBLs spanning over 25 observations observed by EPIC-PN instrument of \emph{XMM-Newton} in the  soft, hard, and total energy bands.
 Examples of a variable (top panel, 1H 1219+301) and a  non-variable (bottom panel, Mrk 501) total LC are shown in \autoref{TLC}. Corresponding LCs in the soft (left panel) and hard (right panel) energy bands are plotted in \autoref{SOFT-HARD-LC}. Similar LCs for all other observations IDs are given in  \autoref{A1} and \autoref{A2}. \\
 \\
 On visual inspection of  these plots, one observation of Mrk 501  and one of 1H 1219+301 appeared to exhibit variability on IDV timescales in the total energy band, while 1ES 1959+650 and H 1426+428 showed variability in multiple observations. To investigate the variability of all  of these blazars in IDV time scales  and to  quantify their variability amplitudes, we  used the excess variance method discussed in \autoref{fvar} and the obtained results are reported in \autoref{tab3}. Following \citet{Dhi21}, we consider a LC to be variable only when  its sample variance is greater than the mean square error and $F_{\rm var}>3 \times (F_{\rm var})_{\rm err}$. We calculated $F_{var}$ and associated error $(F_{\rm var})_{\rm err}$ in all three energy bands for 25 observations. \\ 
 \\
  Three sources (1ES 0229+200,1ES 0414+009 and PKS 2005-489) did not show variability in any energy bands in any of the observations. All observations of 1H 1219+301, 1ES 2344+514 and 1ES 1959+650 exhibited variability in all energy bands. Among the seven observations of H 1426+428, the source exhibited variability in all energy bands only in two observations (0111840101 and 0310190101), while in the rest of the observations the source is variable in one or two bands (see \autoref{tabhr}). PKS 0548-322 did not display any variability in one observation (0205920501), while it showed variability in the total band during the other observation (0142270101). Mrk 501 did not  show any variability in observation 0652570101, while the remaining three observations displayed variability in one or more bands. The single observation of 1ES 1101-232 revealed variability in both soft and total bands. Moreover, one observation each of  Mrk 501 (Obs ID 0652570201) and H 1426+428 (Obs ID  0310190501) showed variability in the hard band alone. However, in these cases the  ratio of $F_{var}$ to $(F_{\rm var})_{\rm err}$ barely exceeds 3. In summary, among the 25 observations considered in this work, 15 of them showed intraday variability in at least one band. These $F_{var}$ results, along with the corresponding variability time scales in the soft, hard and total energy  bands  are reported in \autoref{tabhr}. For further analysis, we consider  only these 15 observations  unless otherwise specified. \\
 \\
 We also performed auto-correlation function analysis to  see if there was any hint of periodicities  in the 15  variable LCs.
  An example of an ACF plot is shown in \autoref{ACF}. Similar ACF plots for other variable LCs are given in \autoref{A4}. Most of the ACF plots are noisy and those which are not noisy do not show any peak other than at zero lag,  so we see no evidence for any periods.\\
 \\
\subsection{Intraday cross-correlated variability}
\noindent
  We have followed discrete correlation function (DCF) analysis technique discussed in \autoref{dcf_an} to determine cross correlations and time lags between hard and soft X-ray bands. The DCF plot for 1H 1219+301 is shown as an example in \autoref{DCF}. Similar DCF plots for 15 variable LCs are given in \autoref{A3}.
  We note that the single observation of H 1219+301 and both observations of 1ES 1959+650 showed a significant DCF values (> 0.5) at non-zero  time lags between the hard and soft bands. In these three cases we fit the DCF with a Gaussian function of form:
  \begin{equation}
      DCF(\tau)= a\hspace{0.2cm}exp\frac{-(\tau-m)^2}{2\sigma^2},
  \end{equation}
  where $a$ is the DCF peak value, $m$ is  the time lag at which DCF peaks and $\sigma$ is  the width of  the Gaussian function.
   Through these fits we found positive time lags of 0.70 ks and 0.57 ks, respectively for 1H 1219+301 (0111840101) and 1ES 1959+650 (0870210101), which indicate soft energy emission precedes hard energy emission in these  cases, and a negative time lag of  $-1.23$ ks for  1ES 1959+650 (0850980101), which indicates hard energy emission precedes soft emission  for this observation.
   These results possibly indicate that the soft and hard X-ray emission emerge from somewhat different populations of leptons.
  For rest of the observations, the DCF plots are either noisy or show no visible lag because of the low count rate in the hard band. Interestingly, the  the single DCF plot of 1ES 2344+514 seems to  an anti-correlation, but the DCF value is not high enough  to allow for any claim that this is significant.

\subsection{Intraday spectral variability}
\noindent
We investigated X-ray spectral variations on IDV timescales  through measurements of the hardness ratio (HR).  The mean HR for each LC, numbers of degree of freedom, $\chi^2$ value, and $\chi^2$ at 99 per cent significance level are given in \autoref{tabhr}.
An example of HR plot for the variable (top panel) and non variable (bottom panel) LCs are given in \autoref{HR}. HR plots for the rest of the observations can be found in \autoref{A5}. We included all 25 observations in this analysis. On  visual inspection of  the HR plots, we did not find any spectral variations. To investigate the spectral variability quantitatively, we performed $\chi^2 $ tests as discussed in \autoref{hr_analysis}. If $\chi^2 > \chi^2_{99,n}$  (the 99 per cent confidence level for the number of degrees of freedom, $n$), the source is considered to have spectral variations. We did not find significant variations in the HR for any of the observations according to the $\chi^2$ test. These results are  not surprising in that we never observed large variability amplitude in the fluxes themselves. \\

\subsection{Intraday power spectral density analysis}
\noindent
 We performed PSD analyses on all 15 variable LCs to  characterize the type of noise present in the variations and to search for any QPOs present during those spans. Studies done on a large number of X-ray light-curves has confirmed the fact that PSDs are red noise dominated,  following a power-law P($\nu$) $\approx$ $\nu^{-\alpha}$ where $\nu$ is the temporal frequency and $\alpha$ is the spectral index, until they flatten into white noise.   \citet{Gon12} reported that X-ray variable LCs of  a large sample of AGN have $\alpha \approx 2$. Following \citet{Vau10}  and \citet{Pav22}, we fit power-law model to the variable X-ray LCs of the 15 observations as discussed in \autoref{psd_an}. The slope $\alpha $ and logarithm of normalization constant $\log_{10} N $  are tabulated in \autoref{tab4}. 
PSD plots for soft (0.3--2 keV), hard (2--10 keV) total (0.3--10 keV) variable LCs of  1H 1219+301 with  observation ID 0111840101 are given in \autoref{allPSD}.
Similar PSD plots for variable LCs in different energy bands are given in \autoref{A6}, \autoref{A7},and \autoref{A8}.

\section{Discussion\label{sec5}}
\noindent
The analysis of flux variations on diverse  timescales across  all EM bands will aid us in understanding emission mechanisms in blazars and other AGNs. Study of rapid flux variations in blazars can be used as a tool to estimate key features of emitting regions in jets such as their sizes, locations, and sometimes structures \citep[e.g.][]{Cip03}.
Two fundamental classes of  models can explain intrinsic AGN emission and flux variability: (a) relativistic-jet-based models \citep[e.g.][]{Mar85,Gop92,Mar14,Cal15};
(b) accretion-disk-based models \citep[e.g][]{Mang93,Cha93}. In  the case of blazars, in particular BL Lac objects, the relativistic jet emission dominates and  any contribution from  the accretion disk can be noticed only when  the BL Lac is observed in  a low flux state. IDV and STV seen in radio-quiet AGNs and blazars, particularly FSRQs  in low flux states can be explained by accretion-disk-based models.  For them brightness fluctuations on these timescales can be due to hot spots on or above the disk or arise from larger scale disk related instabilities  that might be caused by  a tilted disk or a dynamo \citep[e.g.][]{Cha93,Mang93,Hen12,Sad16}. \\
\\
In this sample, 1H 1219+301 has  the minimum weighted variability timescale $\tau_{var,m} = 3.65\pm1.21$ ks for the observation ID 0111840101. By utilizing the simplest causality argument, the  $\tau_{var,m}$  can be used  to estimate an upper limit for  the size of emitting region, $R$,  as
\begin{equation}
    R \le \frac{\delta}{1+z}{c\tau_{var,m}}.
\end{equation}
Here $\delta$ is the Doppler  factor; unfortunately, for 1H 1219+301 this value, estimated using leptonic models in different EM bands and in different flux states, covers a wide range between 20 and 80 \citep[][]{Sat08,Rug10,Cer15,Sin19,Sah20}.  We now assume  that the varying X-ray emission originates from a region in the relativistic jet. Taking $\tau_{var,m} =3.65$ ks, along with the complete range of Doppler factors ($\delta=20-80$), and $z=0.1836$ (see \autoref{tab1}) and making use of equation (17), we find that the upper limit  to the size of the emission region lies in the range  $(1.9-7.4) \times 10^{15}$ cm. \\
\\
 We can derive some other important parameters for these HBLs,  by continuing to consider the  scenario in which the emission arises from relativistic jets.
As already discussed, hard X-ray emission from HBLs is understood to be generated by synchrotron emission from relativistic electrons in the jet \citep[e.g][]{Pan18}.
A diffusive shock acceleration mechanism \citep[e.g.][]{Bla87} is  very likely to be responsible for electron acceleration within jets and in that case the acceleration timescale for electron of energy $E=\gamma m_ec^2$ in the observer's frame is given in \citet{Zha02} as 
 \begin{equation}
     t_{acc}(\gamma) \simeq 3.79 \times 10^{-7}\frac{1+z}{\delta }\frac{\xi\gamma}{B}\hspace{0.2cm} {\rm s}.
 \end{equation}
 Here  $\delta$ is the Doppler factor, $B$ is the magnetic field in Gauss,  $\gamma$ is the electron Lorentz factor, and $\xi$ {is the electron acceleration parameter which comes} from the relation between mean free path and electron Larmor radius: $\lambda(\gamma) = \gamma m_e c^2 \xi/(eB)$ \citep[][]{Kus10}. 
 The synchrotron cooling timescale for an individual electron in observer's frame is given in \citet{Rub85} as,
 \begin{equation}
     t_{cool}(\gamma) \simeq 7.74 \times 10^8\frac{1+z}{\delta}\frac{1}{ B^2 \gamma}\hspace{0.2cm} {\rm s}.
 \end{equation}
For a given $B$ and $\gamma$,  the characteristic X-ray frequency at which synchrotron emission  peaks, the critical synchrotron emission frequency, is \citep[][]{Pal15,Dhi21}

\begin{deluxetable*}{cCCCCCCC}
\tablenum{5}
\tablecaption{Model parameters for HBL blazars\label{tab5}}
\tablewidth{0pt}
\tablehead{
\colhead{Source$^{[1]}$} & \colhead{$\tau_{var,m}^{[2]}$} &\colhead{z$^{[3]}$} &\colhead{$\delta^{[4]}$} & \colhead{R$^{[5]}$}  &\colhead{B$^{[6]}$}&\colhead{$\gamma^{[7]}$}&\colhead{E$_{T,max}^{[8]}$}\\
\nocolhead{}&\colhead{(ks)}&\nocolhead{}& \nocolhead{}&\colhead{$10^{15}$cm} &\colhead{$\nu_{18}^{-1/3} $G} &\colhead{$10^5 \nu_{18}^{2/3}$}&\colhead{$\nu_{18}^{2/3}TeV$}
}
\decimalcolnumbers
\startdata
PKS 0548-322&4.75&0.0690&10-20$^{[a]}$&$<$1.3-2.7&$>$0.18-0.23&$<$2.7-3.3&$\simeq$1.6-2.6\\
1ES\hspace{0.1cm}1101$-$232&1.23&0.1860&10-60$^{[b]}$&$<$0.3-1.9&$>$0.32-0.59&$<$1.2-2.2&$\simeq$0.9-3.1\\
1H\hspace{0.1cm}1219$+$301&3.65&0.1836&20-80$^{[c]}$&$<$1.9-7.4&$>$0.14-0.22&$<$1.6-2.5&$\simeq$2.2-5.5\\
 H\hspace{0.1cm}1426$+$428&2.71&0.1293&11-27.3$^{[d]}$&$<$0.8-2.0&$>$0.24-0.33&$<$2.0-2.7&$\simeq$1.4-2.5\\
Mrk\hspace{0.1cm}501&3.38&0.0330&8.3-50$^{[e]}$&$<$0.8-4.9&$>$0.17-0.30&$<$1.7-3.1&$\simeq$1.3-4.2\\
 1ES\hspace{0.1cm}1959$+$650&4.51&0.0470&15-60$^{[f]}$&$<$1.9-7.8&$>$0.13-0.21&$<$1.8-2.8&$\simeq$2.1-5.3\\
 1ES\hspace{0.1cm}2344$+$514&1.51&0.0440&8.4-23$^{[g]}$&$<$0.4-1.0&$>$0.40-0.50&$<$1.6-2.4&$\simeq$1.0-1.8\\
\enddata
\tablecomments{[1] Name of the source; [2] Minimum variability timescale; [3] Red-shift; [4] Range of Doppler factor values; [5] Characteristic size of emitting region;  [6] Magnetic field [7] Electron Lorentz factor [8] Maximum electron energy in Thomson region.
[a] \citet{Sat08,Rug10}
[b] \citet{Zhe13,Abd10}
[c] \citet{Sat08,Sah20}
[d] \citet{Wol08,Pin08}
[e] \citet{Kin02,ALb07,Pan17}
[f] \citet{Pat18,Magic2020}
[g] \citet{Albe07,Rug11}
}
\end{deluxetable*}
\vspace{-0.5in}
 \begin{equation}
     \nu \simeq 4.2 \times 10^6 \frac{\delta}{1+z}B\gamma^2  \hspace{0.1cm}{\rm Hz} \simeq 10^{18} \nu_{18}\hspace{0.2cm}{\rm Hz},
 \end{equation}
 where $0.73 \le \nu_{18} \le 2.42$ for X-rays in {\emph{XMM-Newton's}} total energy range of 0.3 -- 10 keV.  Imposing the condition that the synchrotron cooling timescale 
 of electrons radiating in the \emph{XMM-Newton} range has to be  no greater than the observed minimum variability timescale,
 we have following inequality \citep[][]{Pal15}
 \begin{equation}
     t_{cool}(\gamma) \le \tau_{var,m}.
 \end{equation}
 We combine Eqns.\ (19) and (20) to come up with an expression for $t_{cool}(\gamma)$ without  an explicit dependence on $\gamma$
 and substitute it in above inequality Eqn\ (21) along with $\tau_{var,m}=3.65$ ks, we arrive at a  constraint on the magnetic field for 1H 1219+301 as,
 \begin{equation}
     B \ge 0.61\hspace{0.1cm} \delta^{-1/3} \nu_{18}^{-1/3}\hspace{0.2cm} {\rm G}.
 \end{equation}
 Using the complete range of Doppler factors (ie $\delta \sim 20-80$) we find  this lower limit for $B$ lies in the rather narrow range of ($0.14-0.22) \hspace{0.1cm} \nu_{18}^{-1/3}$ G. Previous  estimations of the magnetic field for 1H 1219+301 at different flux state and different epochs vary between $0.01-0.22 $ G \citep[e.g.][]{Sin19,Sah20}. As $\nu_{18}$ can have any value between 0.73 and 2.73, our magnetic field estimate is consistent with previous  values. Using Eqns.\ (20) and (22), we  now can set a  constraint on  the electron Lorentz factor for 1H 1219+301 as well,
 \begin{equation}
     \gamma \le 6.8 \times 10^5 \delta^{-1/3} \nu_{18}^{2/3}.
 \end{equation}
 Again taking the complete  range of Doppler factor, we find  upper limits to  $\gamma$   in the range $1.6 \times 10^5 \nu_{18}^{2/3} - 2.5 \times 10^5 \nu_{18}^{2/3}$. This is consistent with  a previous estimate of $\gamma \sim 5 \times 10^5$ \citep[][]{Rug10}, if we take into account uncertainty associated with $\tau_{var,m}$ as well as the  allowed range for  $\nu_{18}$. \\
 \\
 The maximum energy of $\gamma$-ray photons generated by  relativistic electrons through Compton scattering in Thomson regime  can be estimated by following expression \citep[e.g][]{Pan18}
 \begin{equation}
     E_{T,max} \simeq \frac{\delta}{1+z}\gamma_{max}m_e c^2 .
 \end{equation}
 Using Eqn.\ (23), $z=0.1836$, and  the same range of $\delta$ values, we see that  $2.2\hspace{0.1cm}  \nu_{18}^{2/3}\hspace{0.1cm} {\rm TeV} \le  E_{T,max} \le 5.5 \hspace{0.1cm} \nu_{18}^{2/3}\hspace{0.1cm} {\rm TeV} $. \\
 
 Similarly, we estimate all these parameters for the remaining 6  variable HBLs and report them in \autoref{tab5}.

\section{Conclusions\label{sec6}}

\noindent
We studied 25 LCs of ten TeV HBLs which  were observed by \emph{XMM-Newton} during its complete operational period. We searched for IDV and variability timescales, lags between soft and hard energy bands, spectral variations through  analysis of hardness ratio changes, and  also performed PSD analyses to characterize the type of noise present and  to search for the presence of possible QPOs. \\
\\
We summarize our conclusions as follows:
\begin{enumerate}
    \item The fractional variability amplitude analysis clearly shows 7 of 10 HBLs  exhibit IDV for at least one observation in total energy band (0.3$-$10 keV). 1ES 1101$-$232, 1H 1219$+$301 and 1ES 2344$+$514 showed IDV in their single observation.  1ES 1959$+$650 displayed IDV in both of its observations, while PKS 0548$-$322 displayed IDV only in one of two observations. Two of the four observations of Mrk 501 and six of the seven observations of H 1426$+$428 exhibited IDV. However, there  were no major  variations; the highest variability amplitude was just below 6 per cent.
    \item In general,  F$_{var}$ is lower in the soft band ($0.3-2$ keV) than in the hard band (2$-$10 keV) with  the exception of one observation, where  they are comparable (i.e., in the case of H 1426$+$428 and Obs ID 0212090201). We estimated variability  timescales  and determined a minimum variability timescale ($\tau_{var,m}$) for each of the seven HBLs.  Then, we used $\tau_{var,m}$ to estimate various parameters such as  the size of the emission region, the magnetic field in that region and the electron Lorentz factor for the ultra-relativistic electrons emitting X-rays in each of these HBLs.
    \item Most of the ACF plots were noisy and we did not find any  indication of a variability timescale from this approach.
    \item DCF plots for most of these TeV HBLs are almost flat, which indicates that there  could be no correlation between soft and hard energy bands. This  however, is more likely due to very low fluxes  in the hard band, which makes it difficult to detect correlations. However, for both observations of 1ES 1959+650 and the single observation of 1H 1219+301, DCF peaks at non-zero lags indicate the possibility that  much of the soft X-ray might originate from synchrotron  emission, while  the hard X-rays may be dominated by a SSC origin.
    \item We performed hardness ratio analysis to  attempt to study X-ray spectral variations of these 10 blazars, but found no significant variations.
    \item PSD  analyses in the soft, hard, and total  X-ray energy bands were performed for the 15 variable LCs. We found that PSDs are dominated by red noise and no evidence for a possible  QPOs was found in any of the PSD plots.  PSD  was fitted using power low model at lower frequencies and their slopes  range from $1.11$ to $2.93$ .

\end{enumerate}

\section*{Acknowledgements}
\noindent
We thank the anonymous referee for the constructive comments and suggestions that improved this manuscript.
This research is based on observations obtained with {\it XMM–Newton}, an ESA science
 mission with instruments and contributions directly funded by ESA member states and NASA.\\
\\
\software{ HEAsoft\citep{Hea14}, SAS \citep{Gab04}}

\bibliographystyle{aasjournal}
\bibliography{references}

\appendix
 \section{appendix section}
\restartappendixnumbering
\setcounter{figure}{0}
\begin{figure*}
\centering
  \vspace*{-1in}
\includegraphics[scale=0.4]{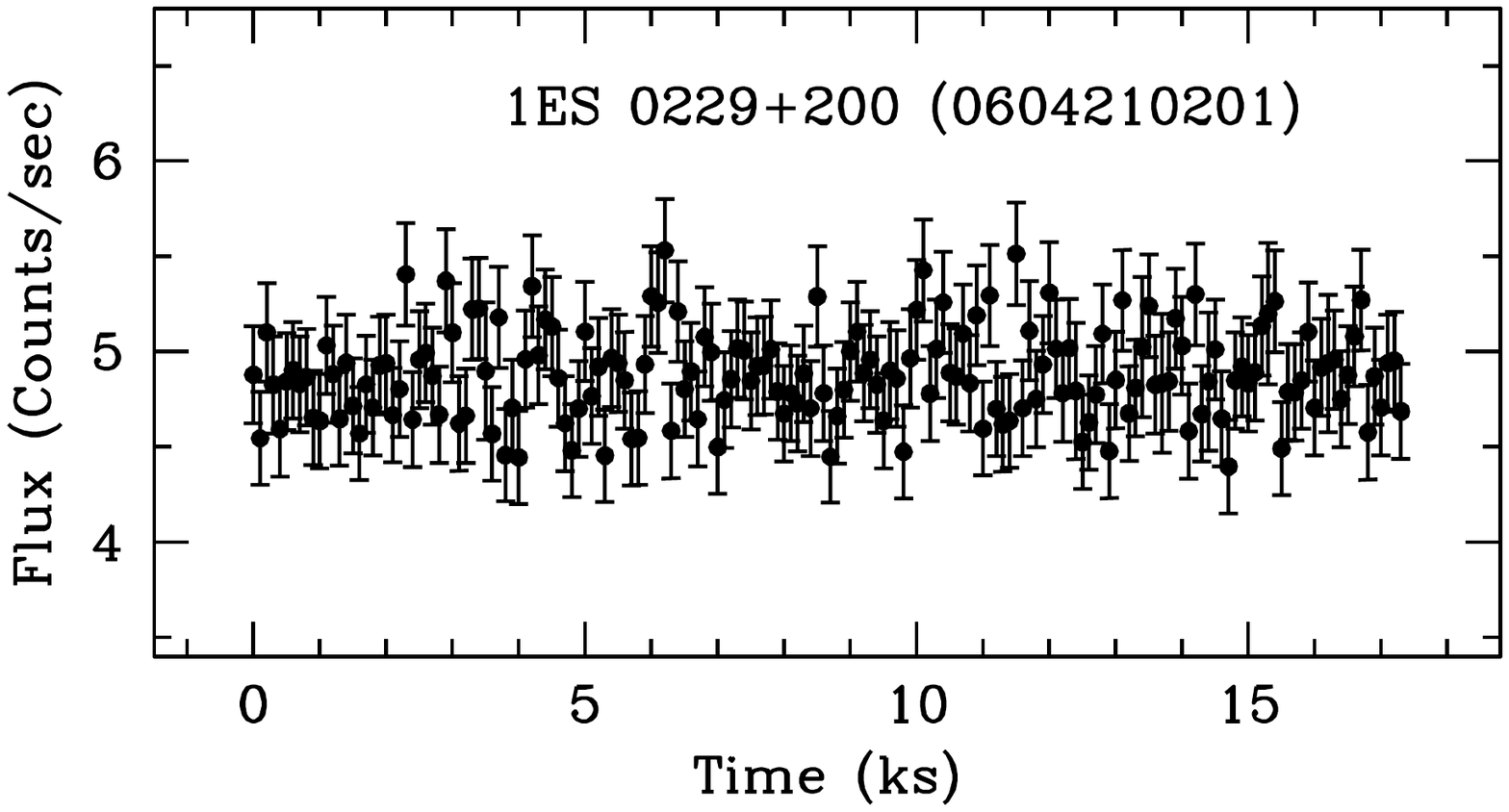}
\includegraphics[scale=0.4]{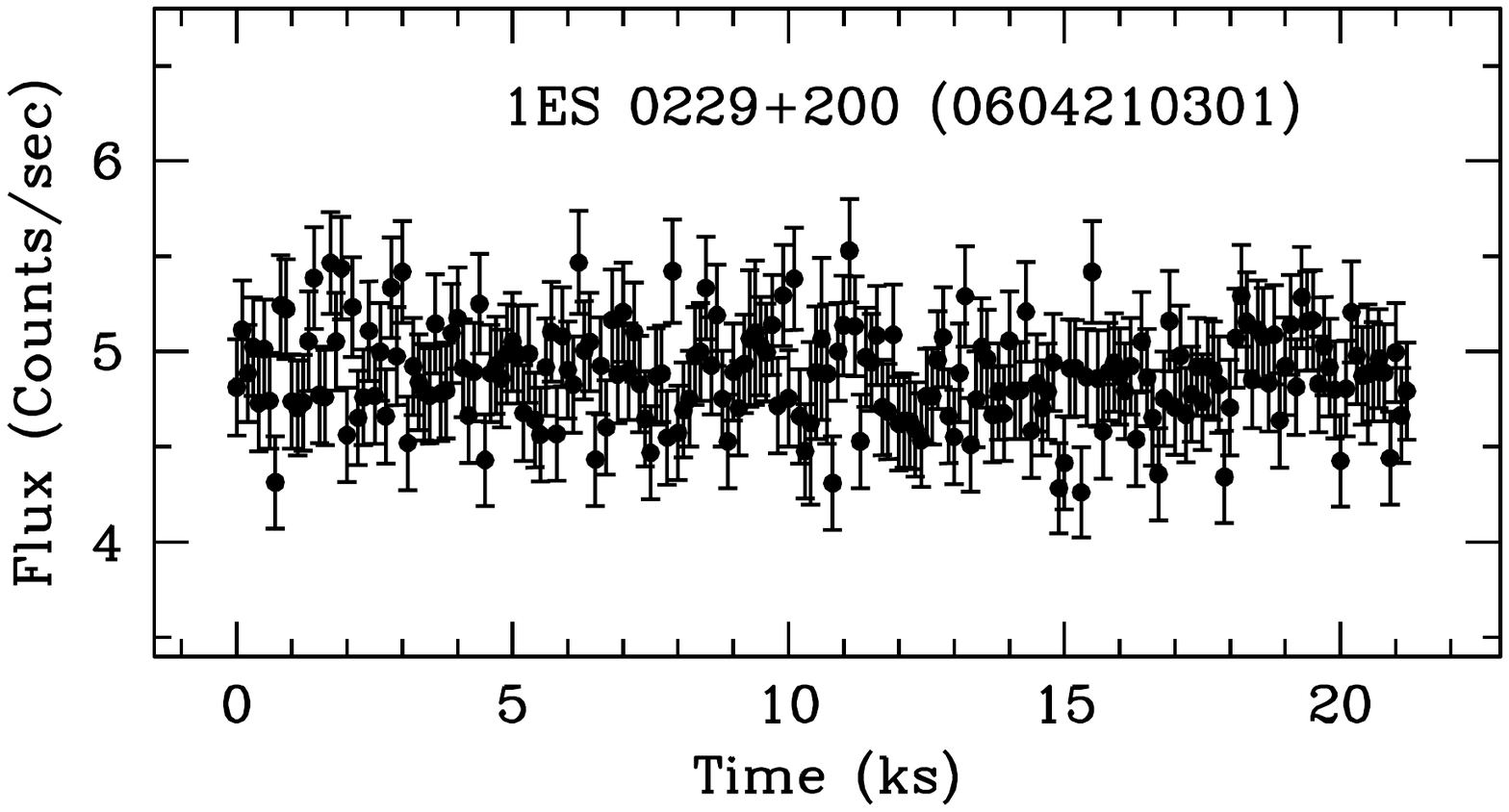}

\vspace*{-2.8in}
\includegraphics[scale=0.4]{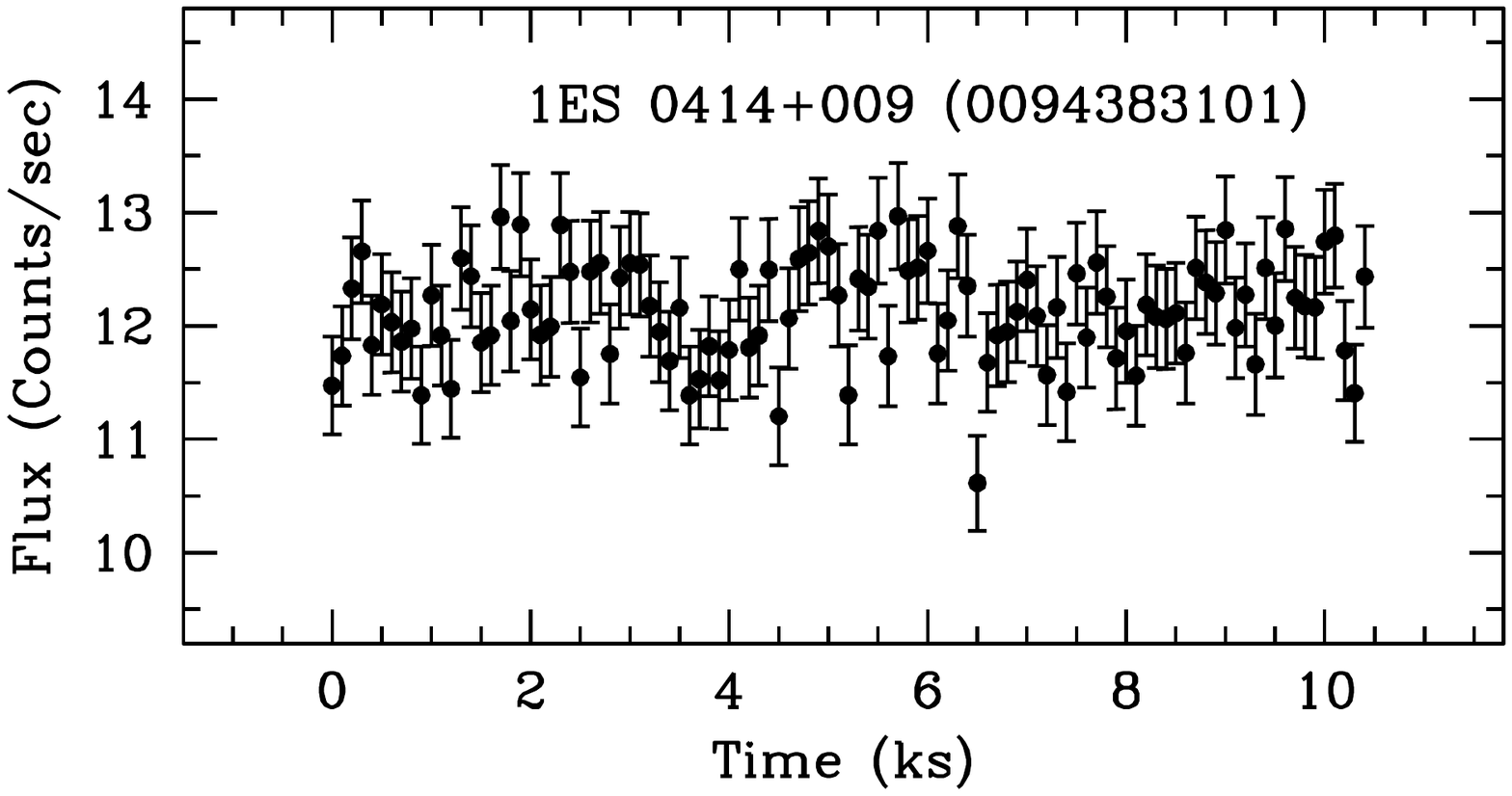}
\includegraphics[scale=0.4]{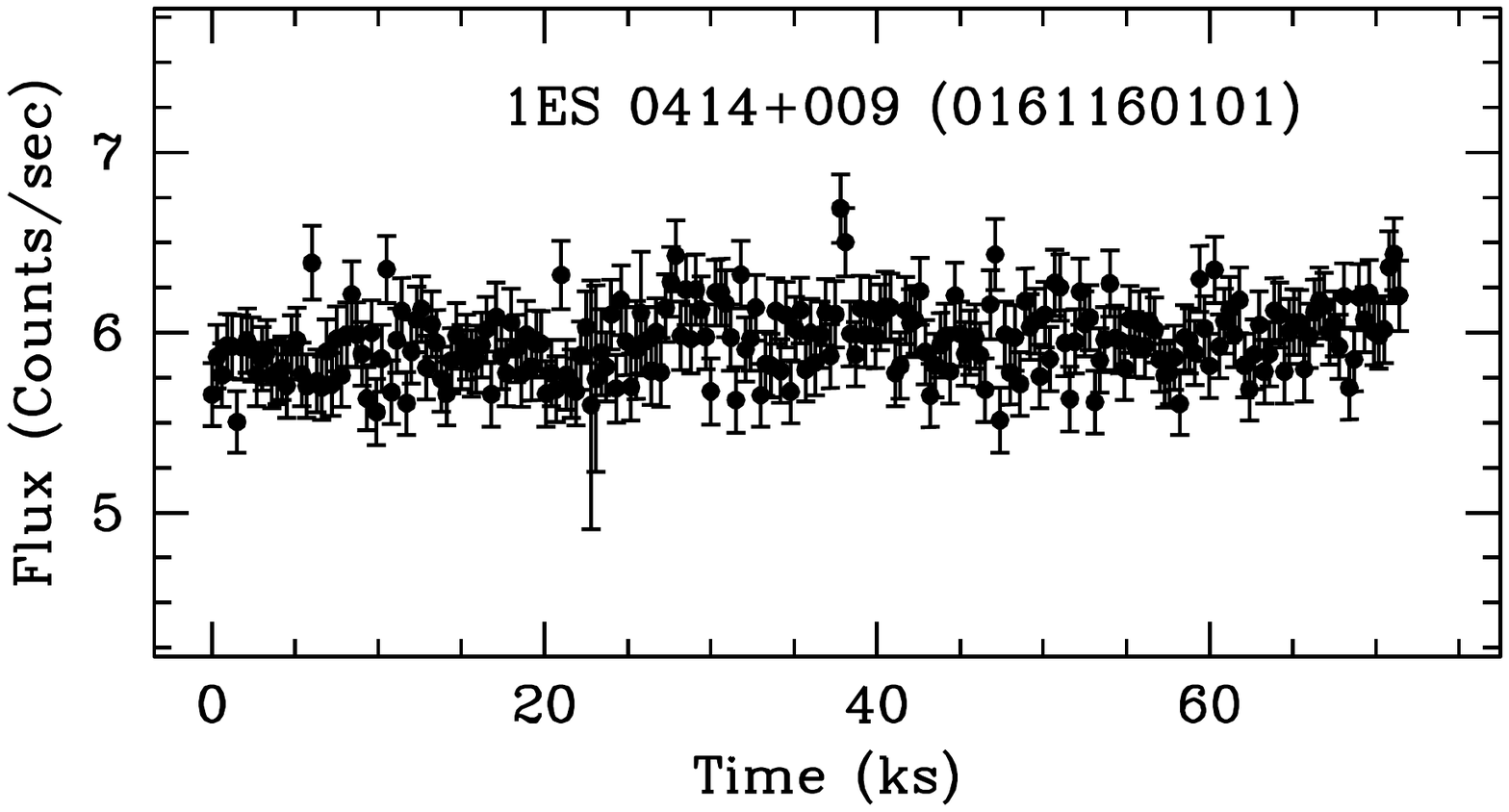}

\vspace*{-2.8in}
\includegraphics[scale=0.4]{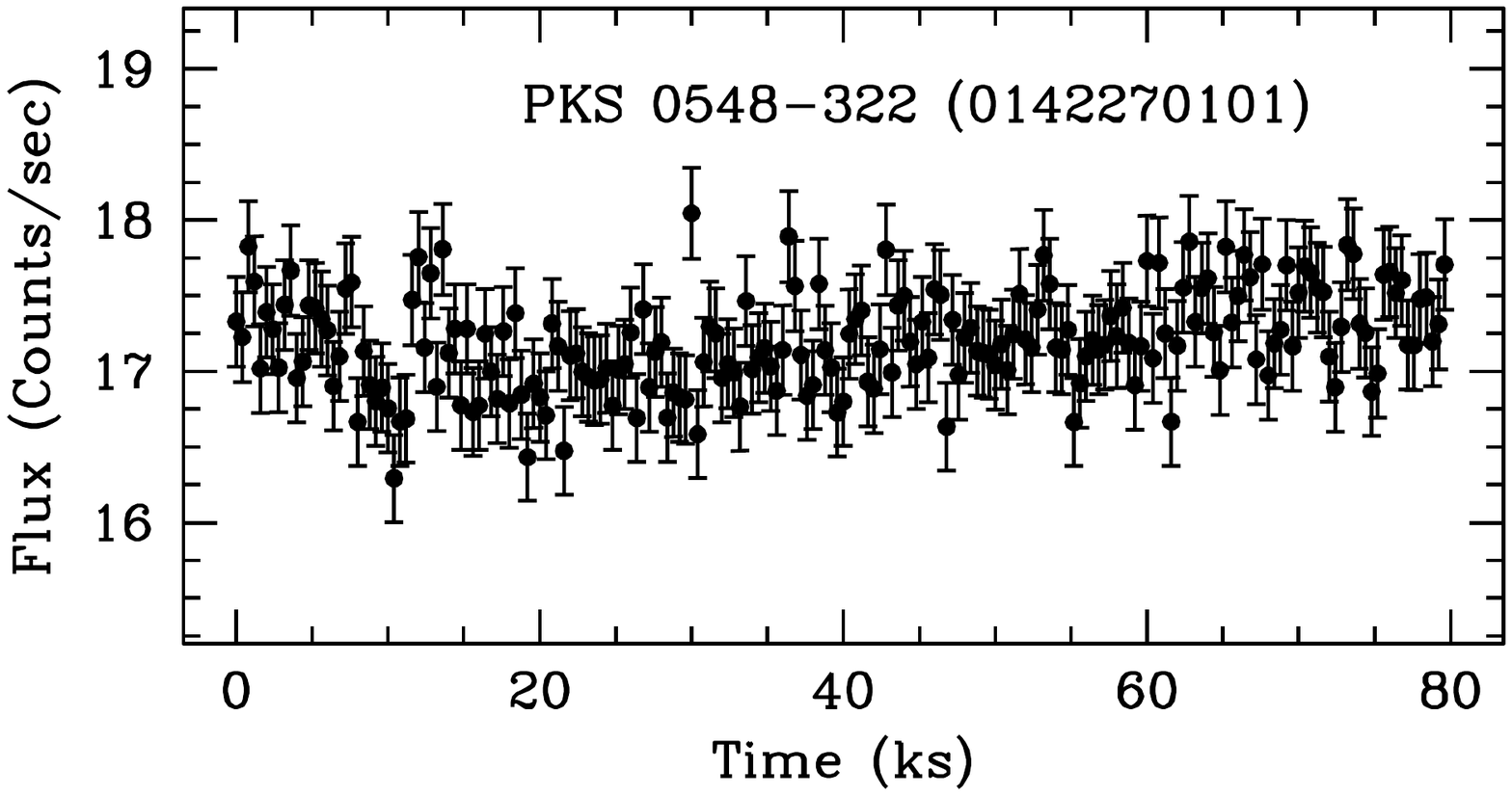}
\includegraphics[scale=0.4]{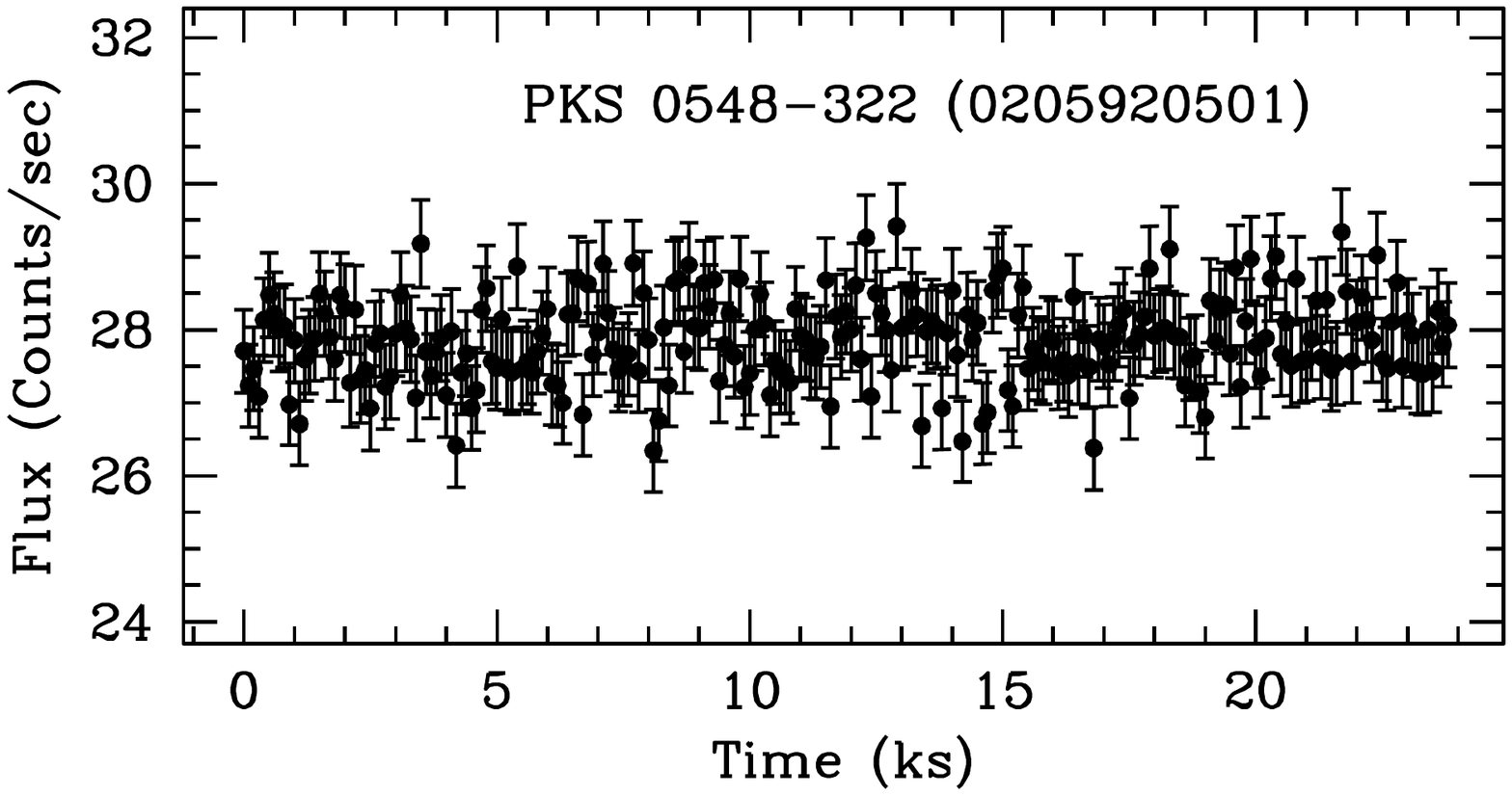}

\vspace*{-2.8in}
\includegraphics[scale=0.4]{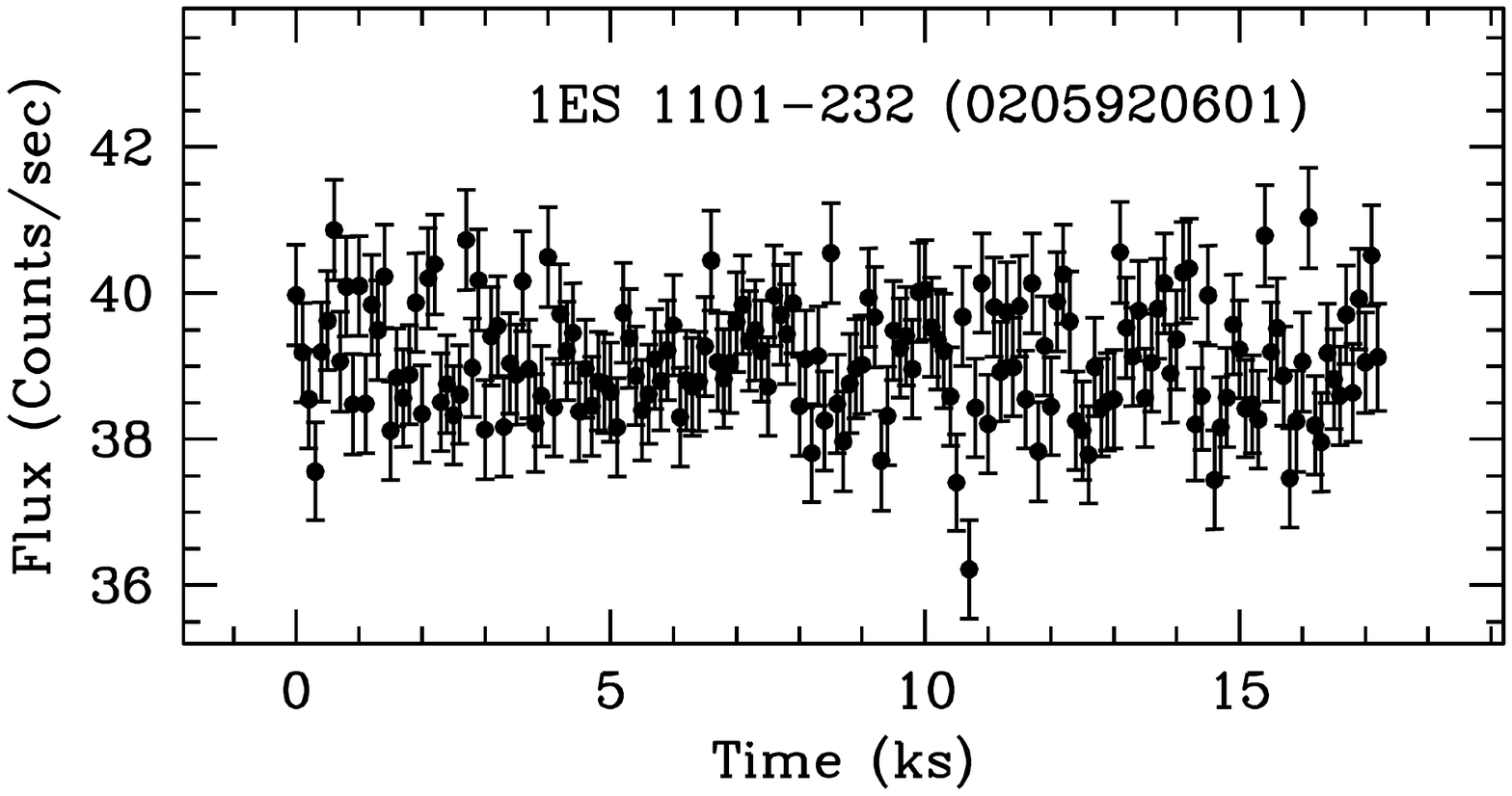}
\includegraphics[scale=0.4]{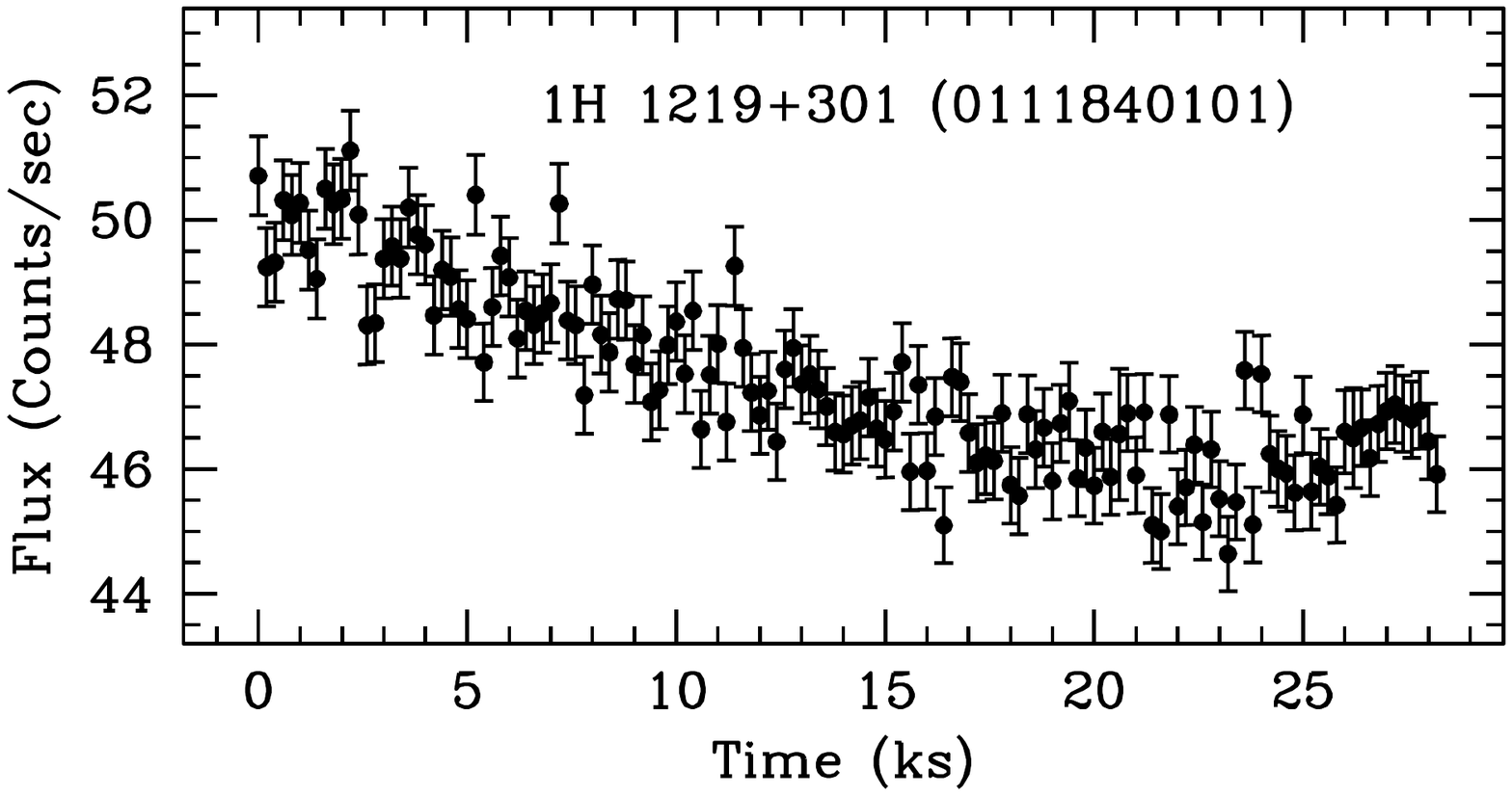}

\vspace{-0.7in}
\caption{Light curves (LCs) of 25 \emph{XMM-Newton} pointed observations of HBLs in total (0.3--10  keV) energy band. Source name and Observation ID are given in each plot.\label{A1}}

\end{figure*}

\clearpage
\setcounter{figure}{0}
\begin{figure*}
\centering

\vspace*{-1.4in}
\includegraphics[scale=0.4]{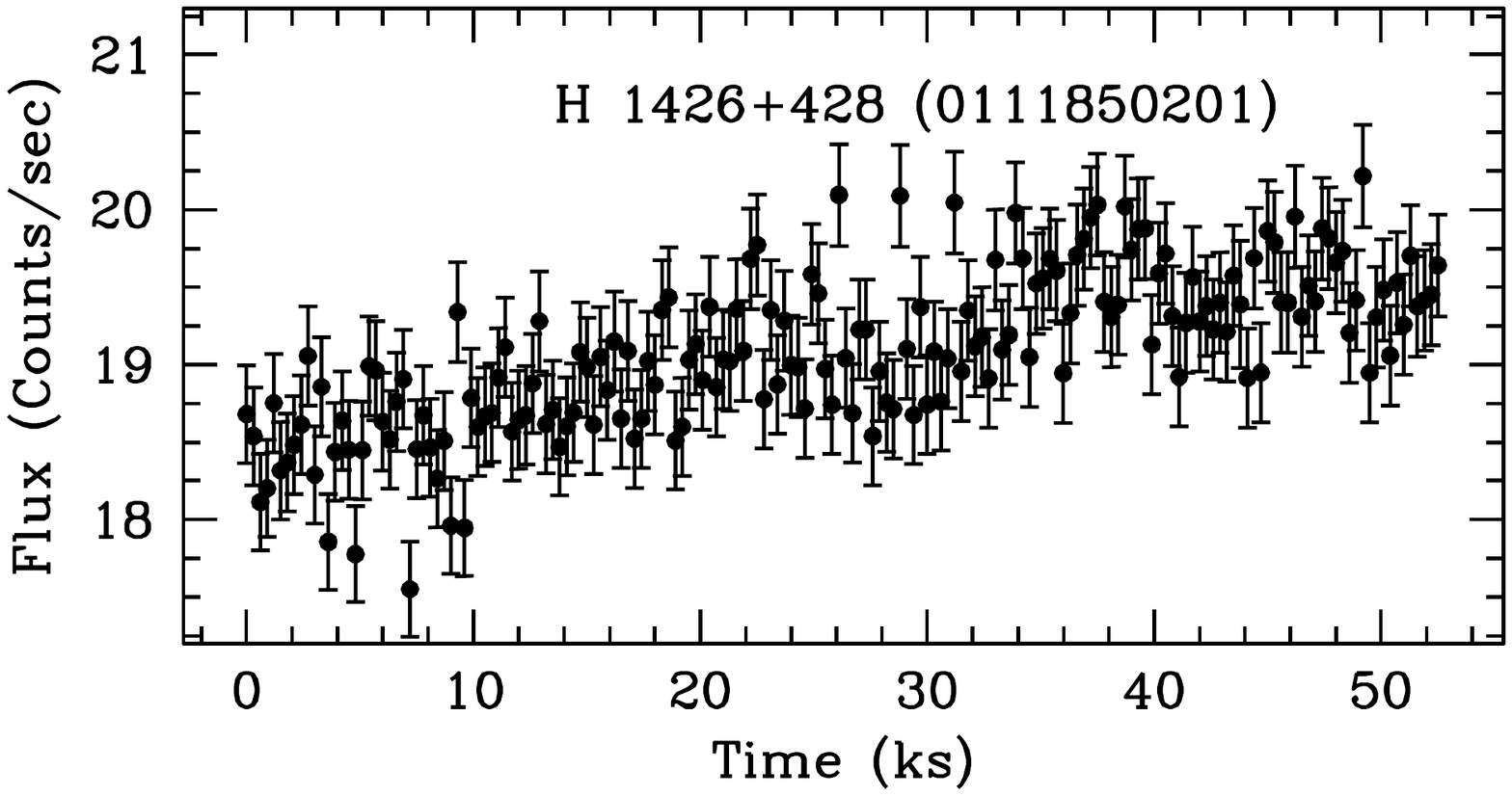}
\includegraphics[scale=0.4]{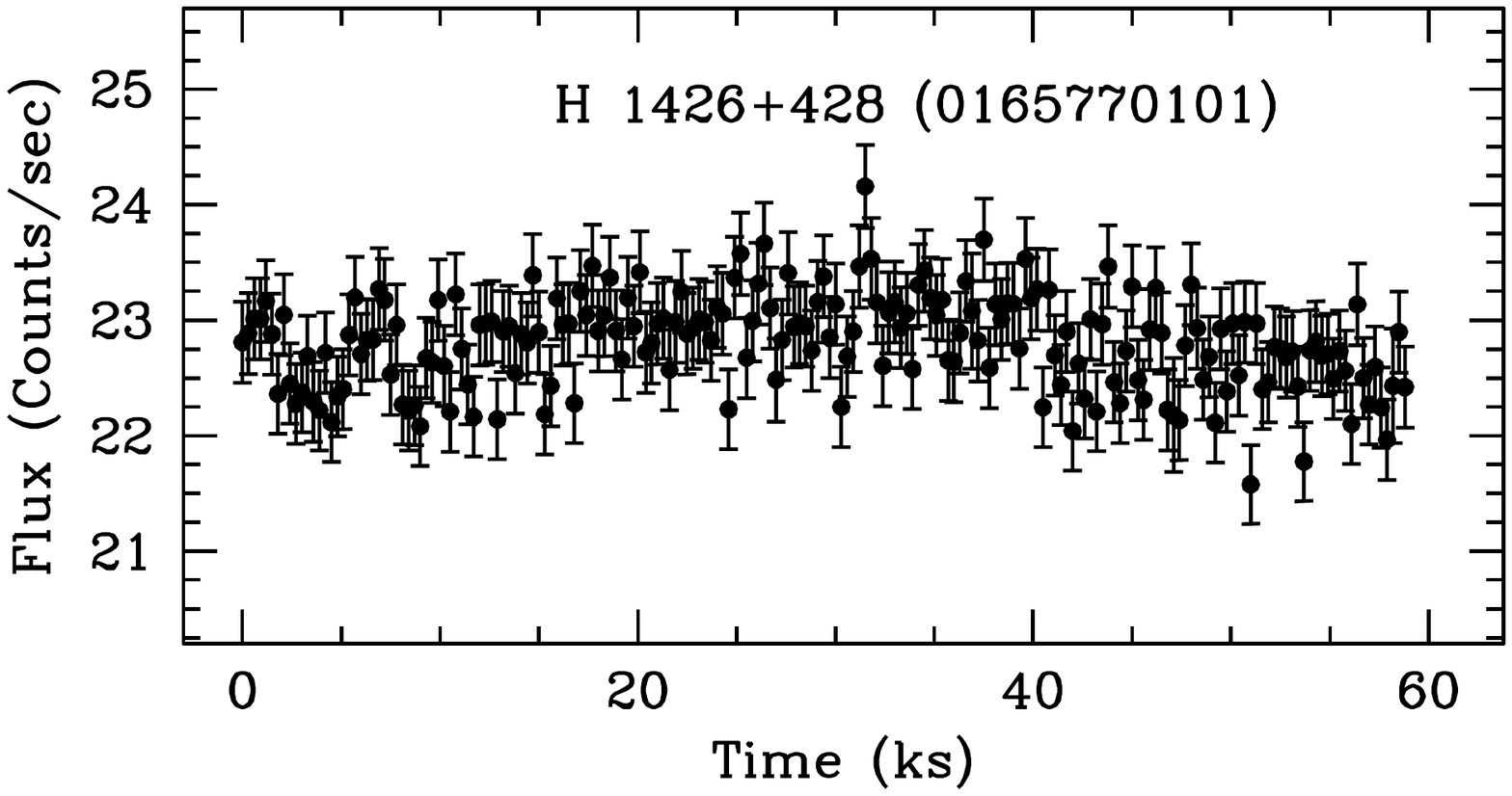}

\vspace*{-2.8in}
\includegraphics[scale=0.4]{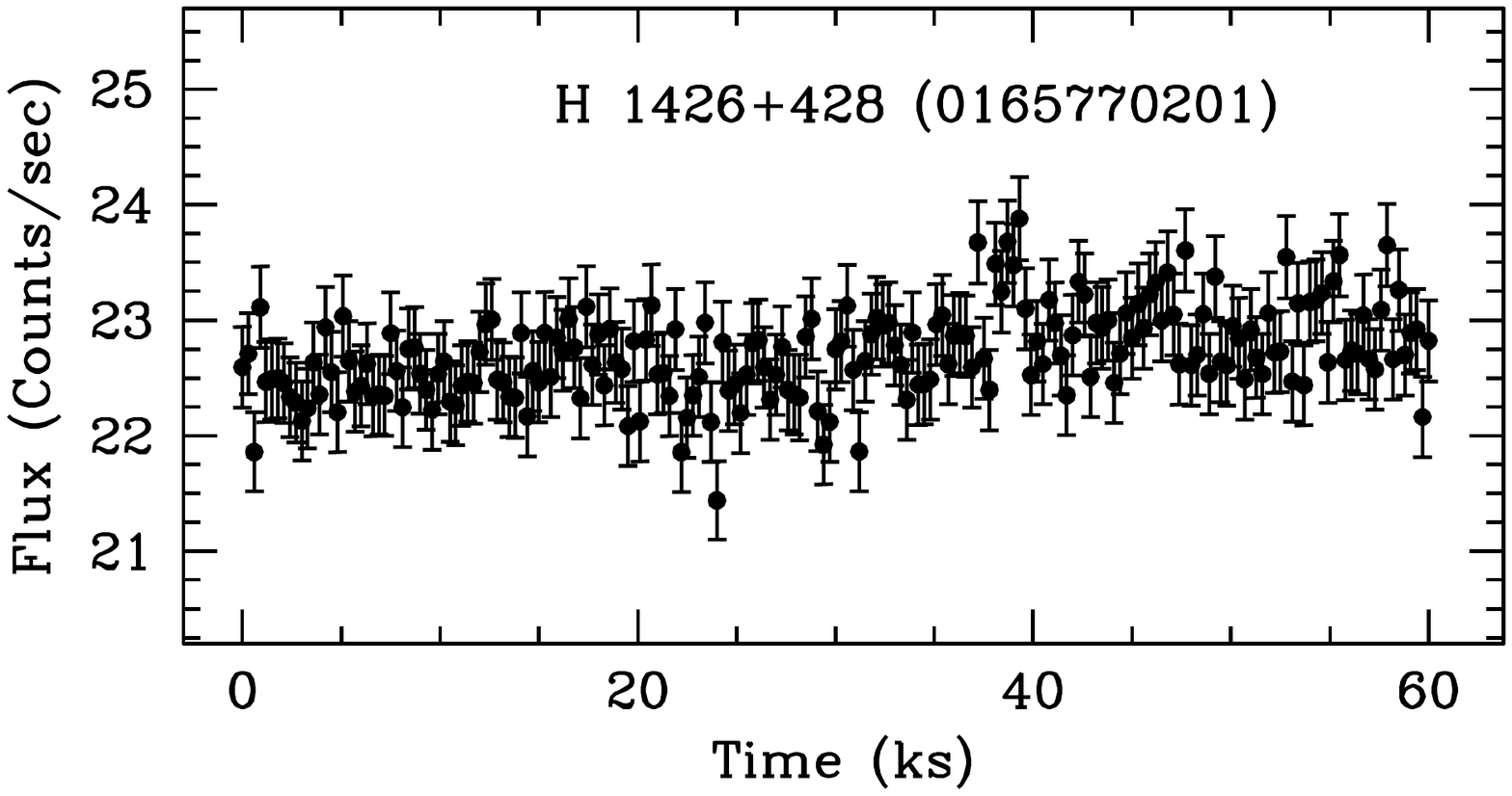}
\includegraphics[scale=0.4]{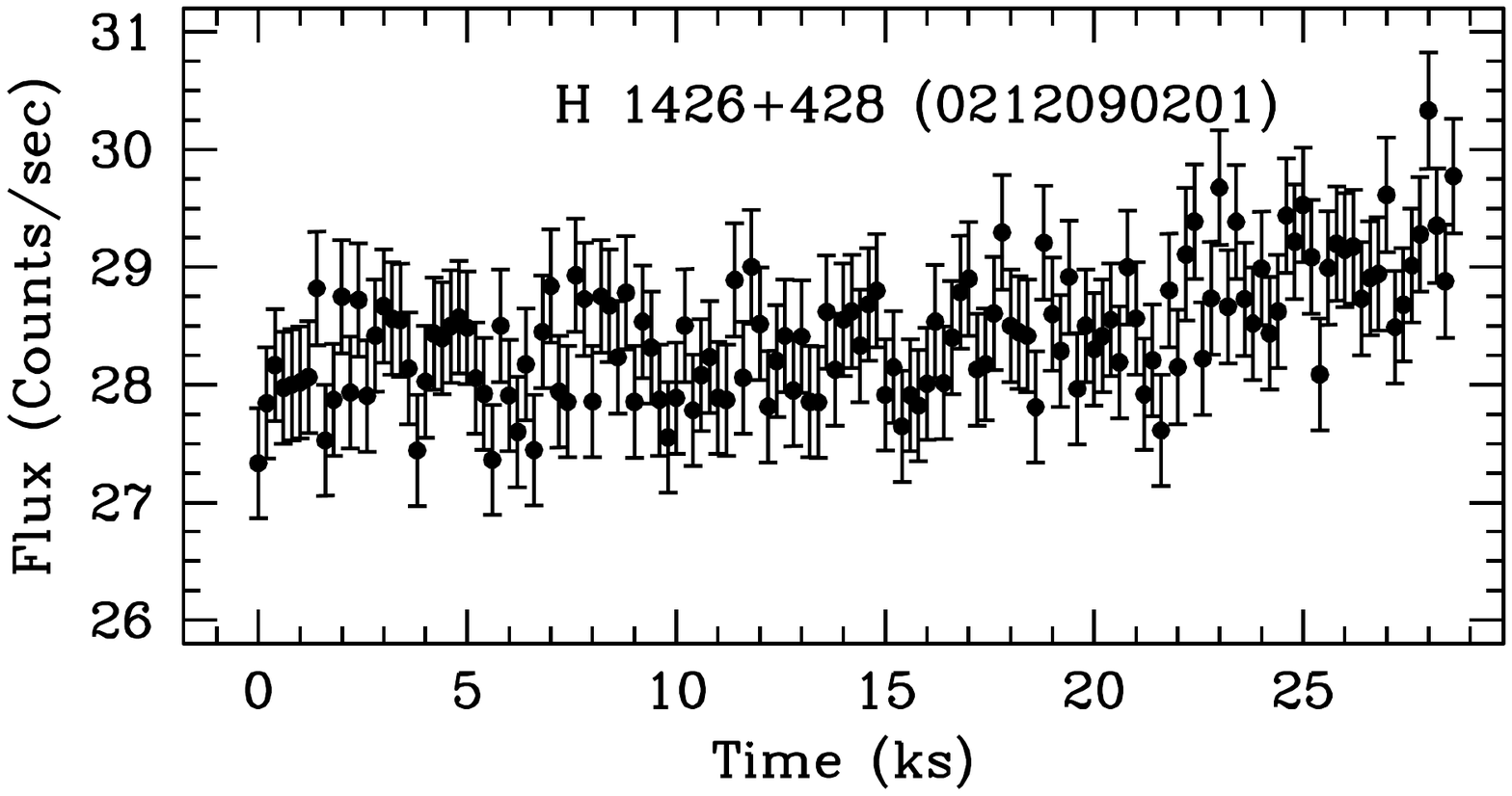}

\vspace*{-2.8in}
\includegraphics[scale=0.4]{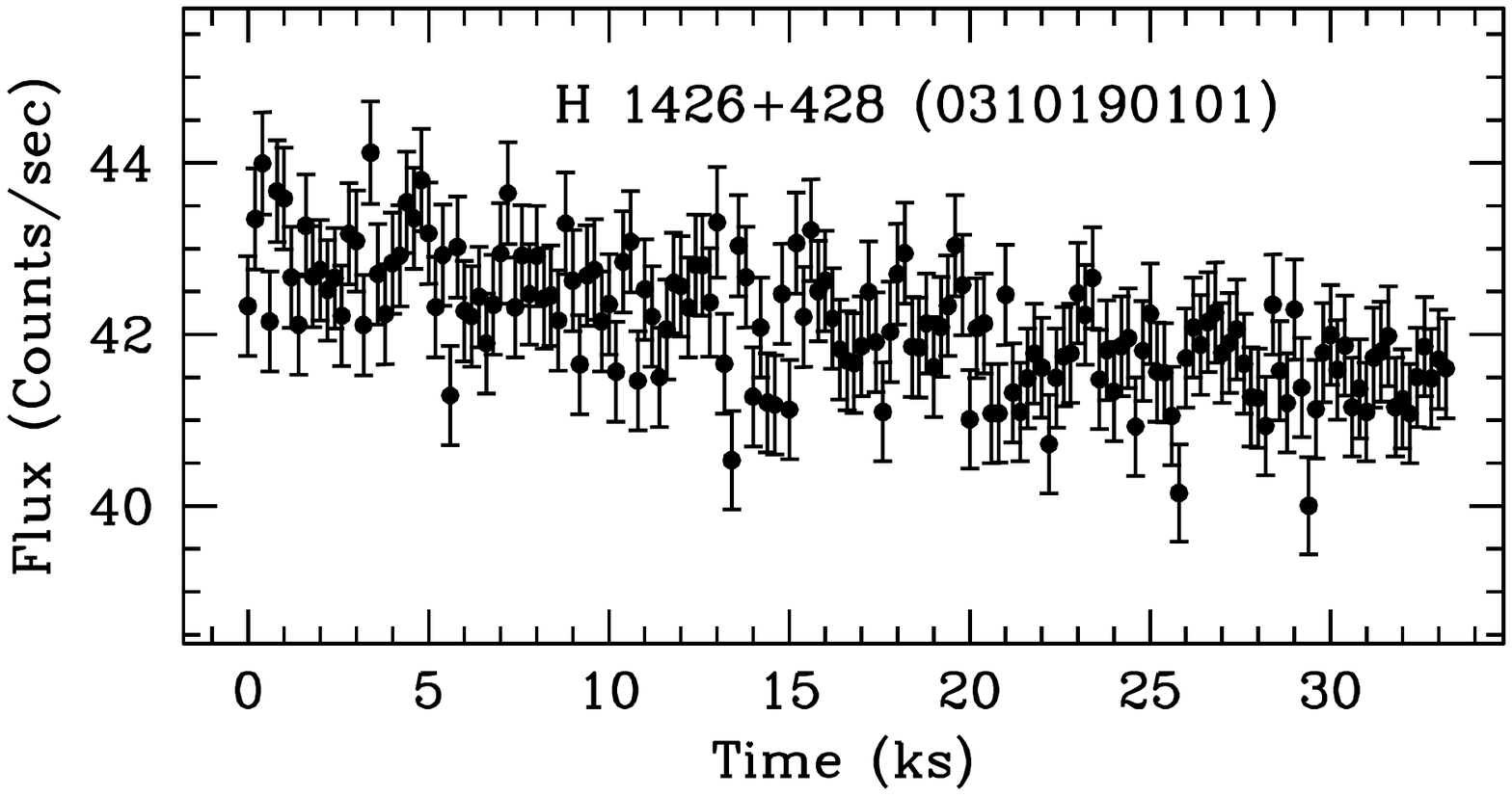}
\includegraphics[scale=0.4]{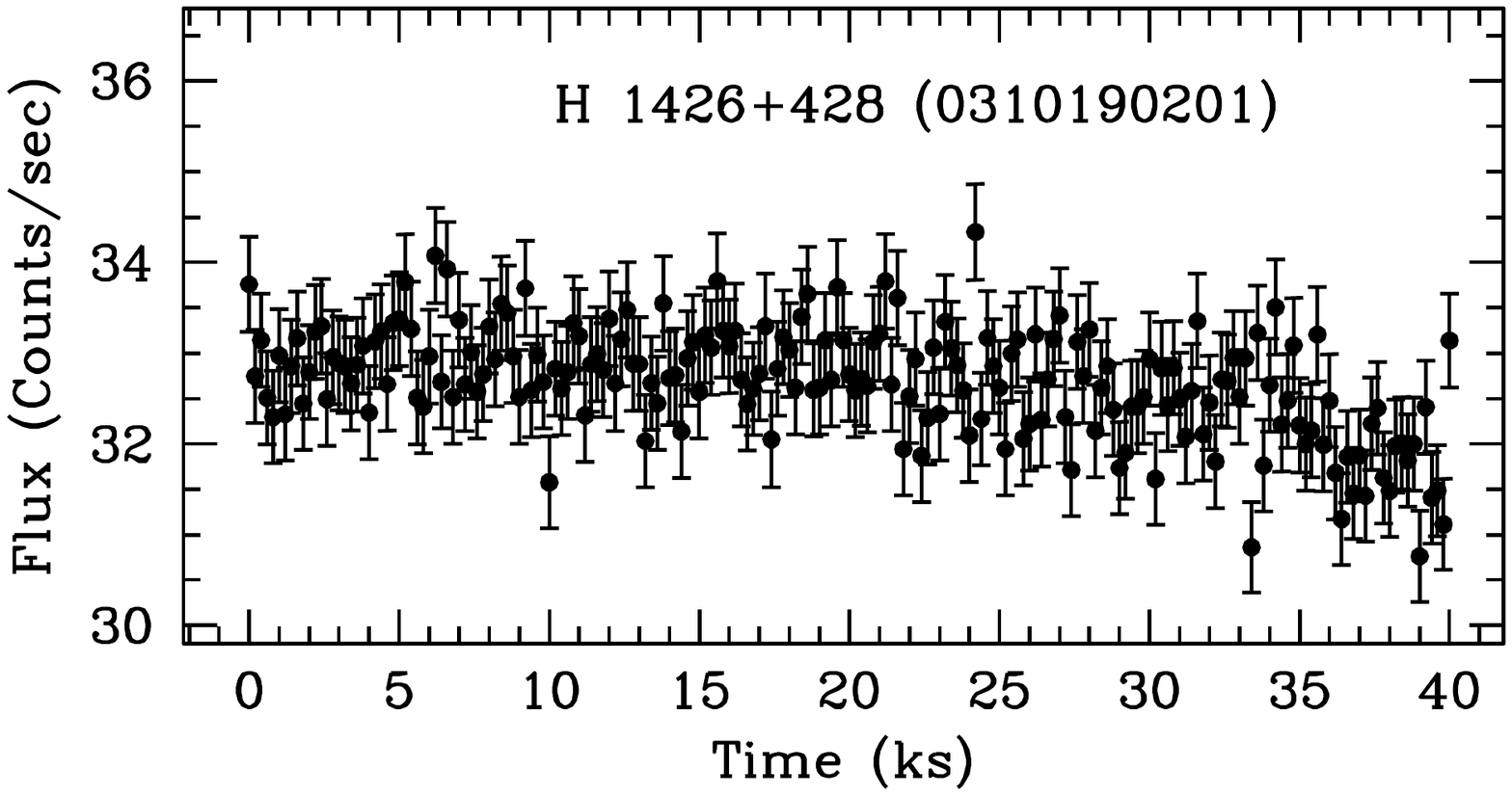}

\vspace*{-2.8in}
\includegraphics[scale=0.4]{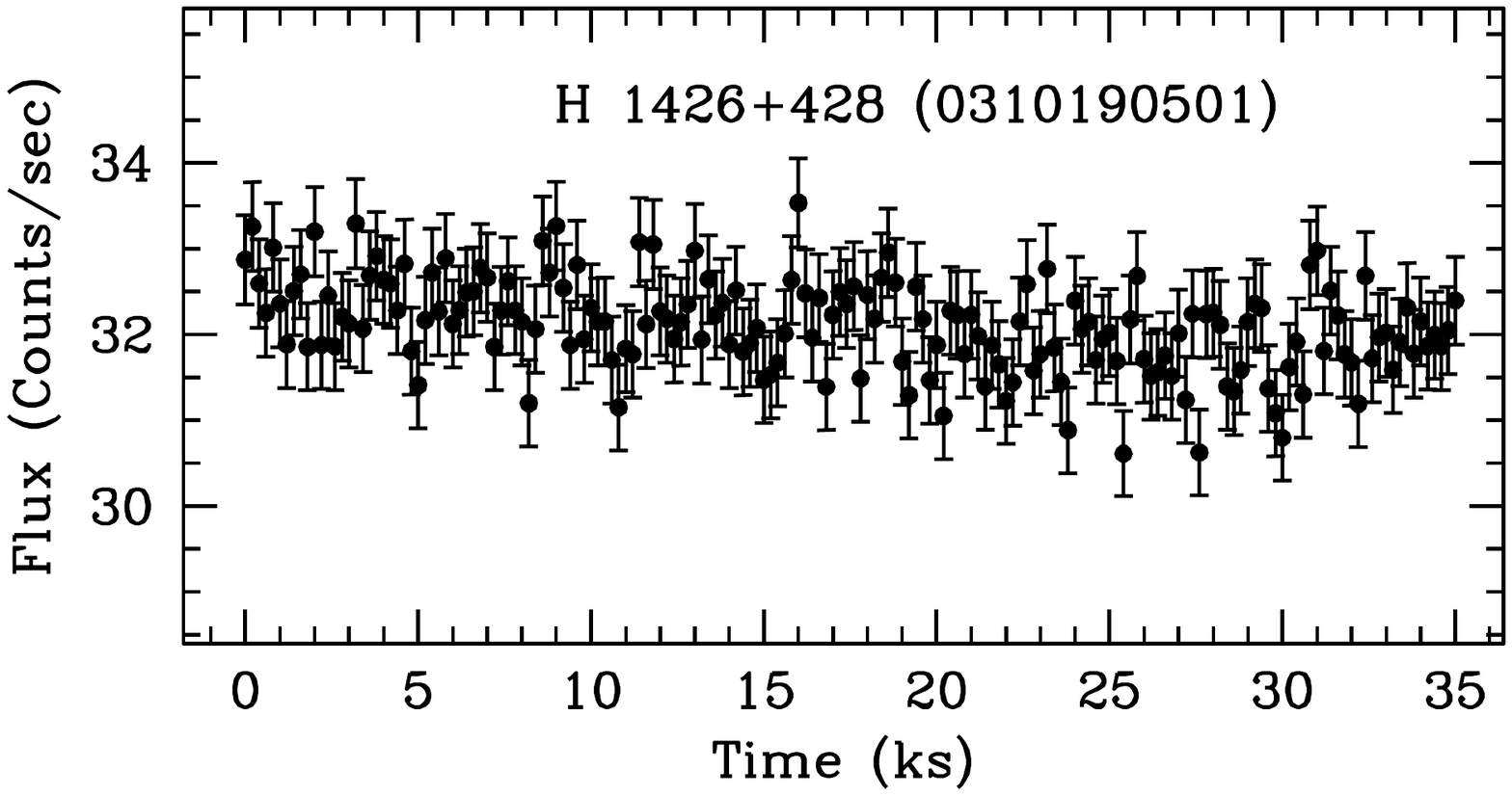}
\includegraphics[scale=0.4]{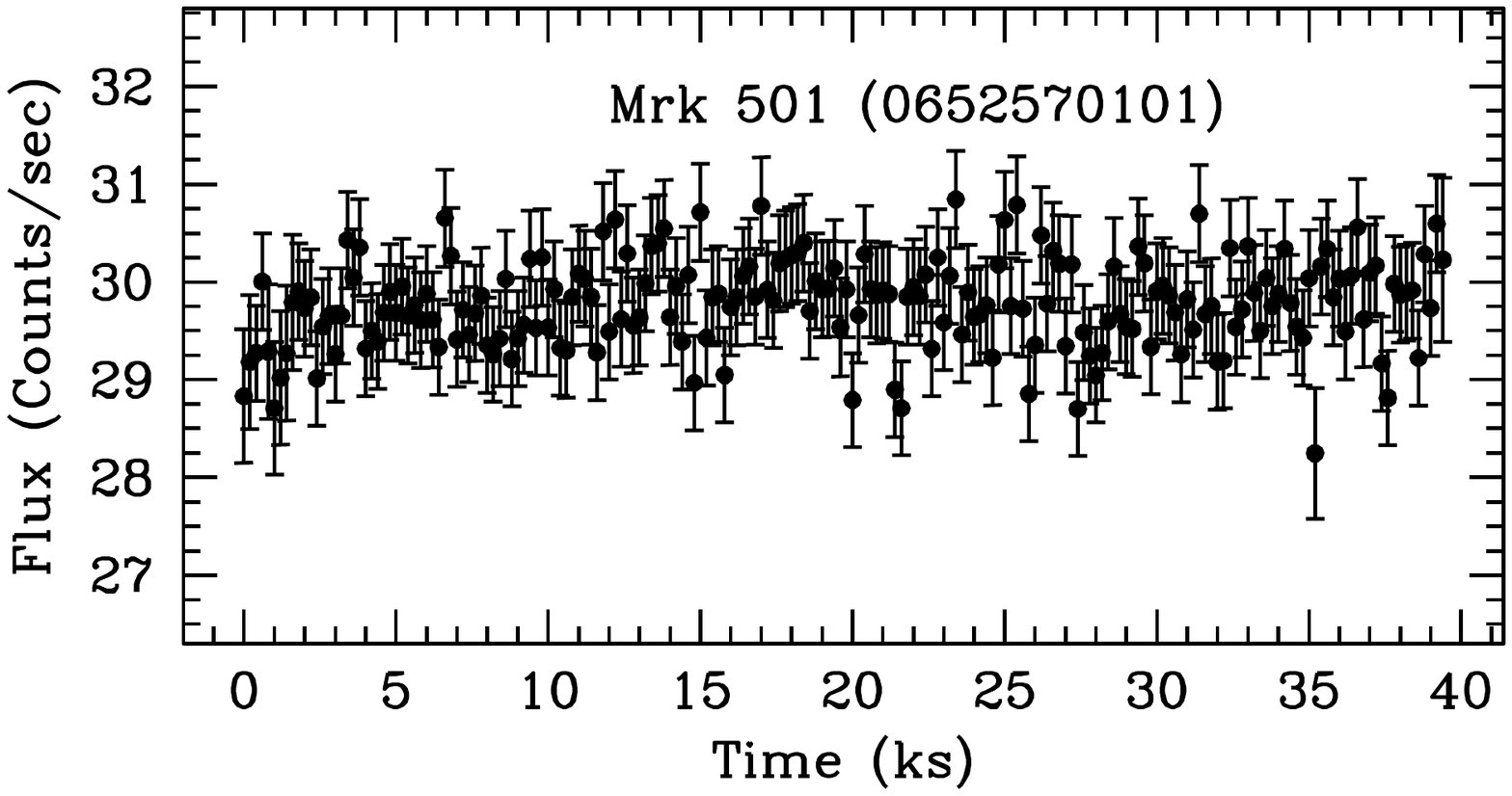}

\vspace{-0.7in}
\caption{Continued} 

\end{figure*}

\clearpage
\setcounter{figure}{0}
\begin{figure*}
\centering
\vspace*{-1.9in}

\includegraphics[scale=0.4]{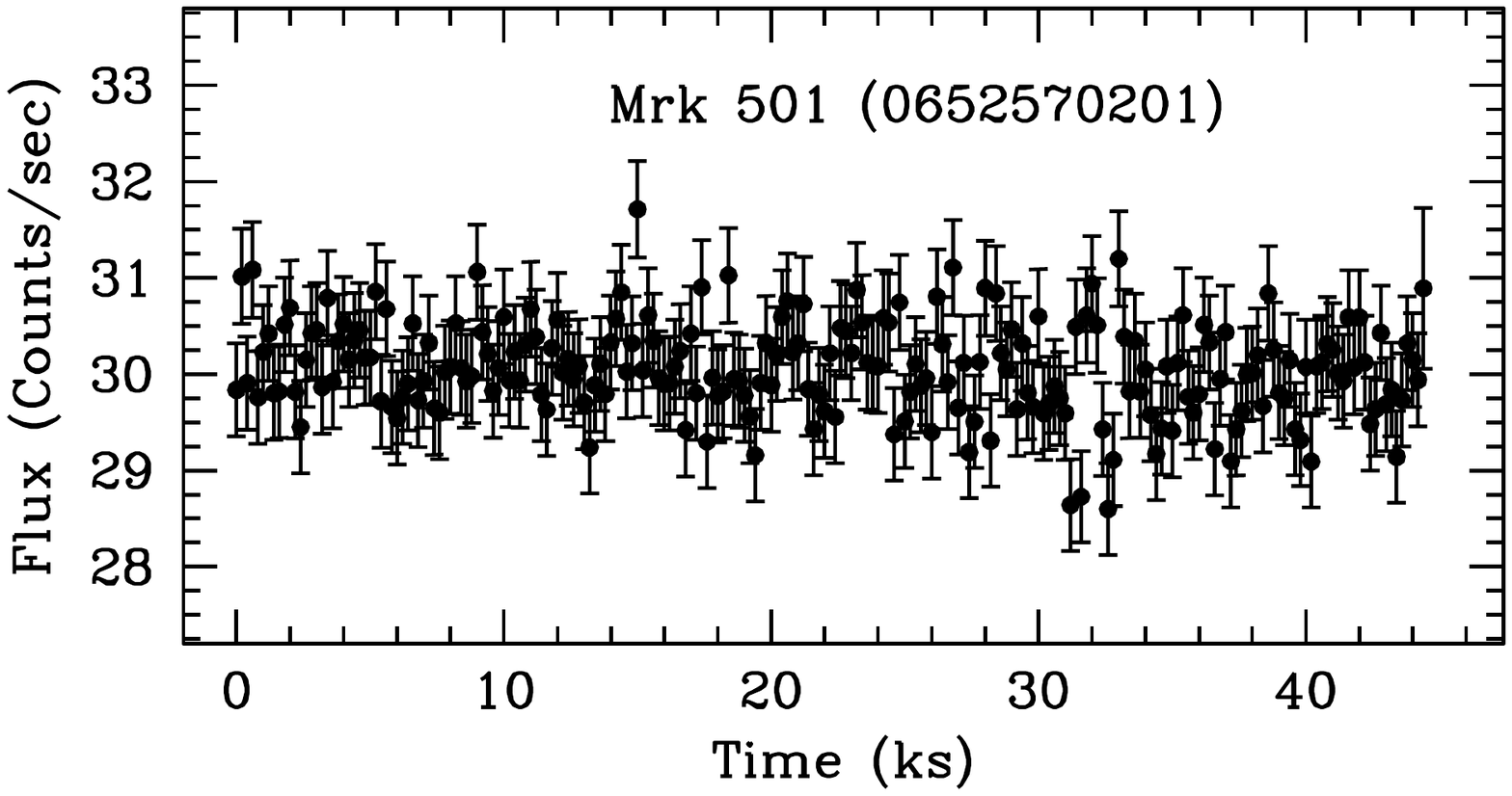}
\includegraphics[scale=0.4]{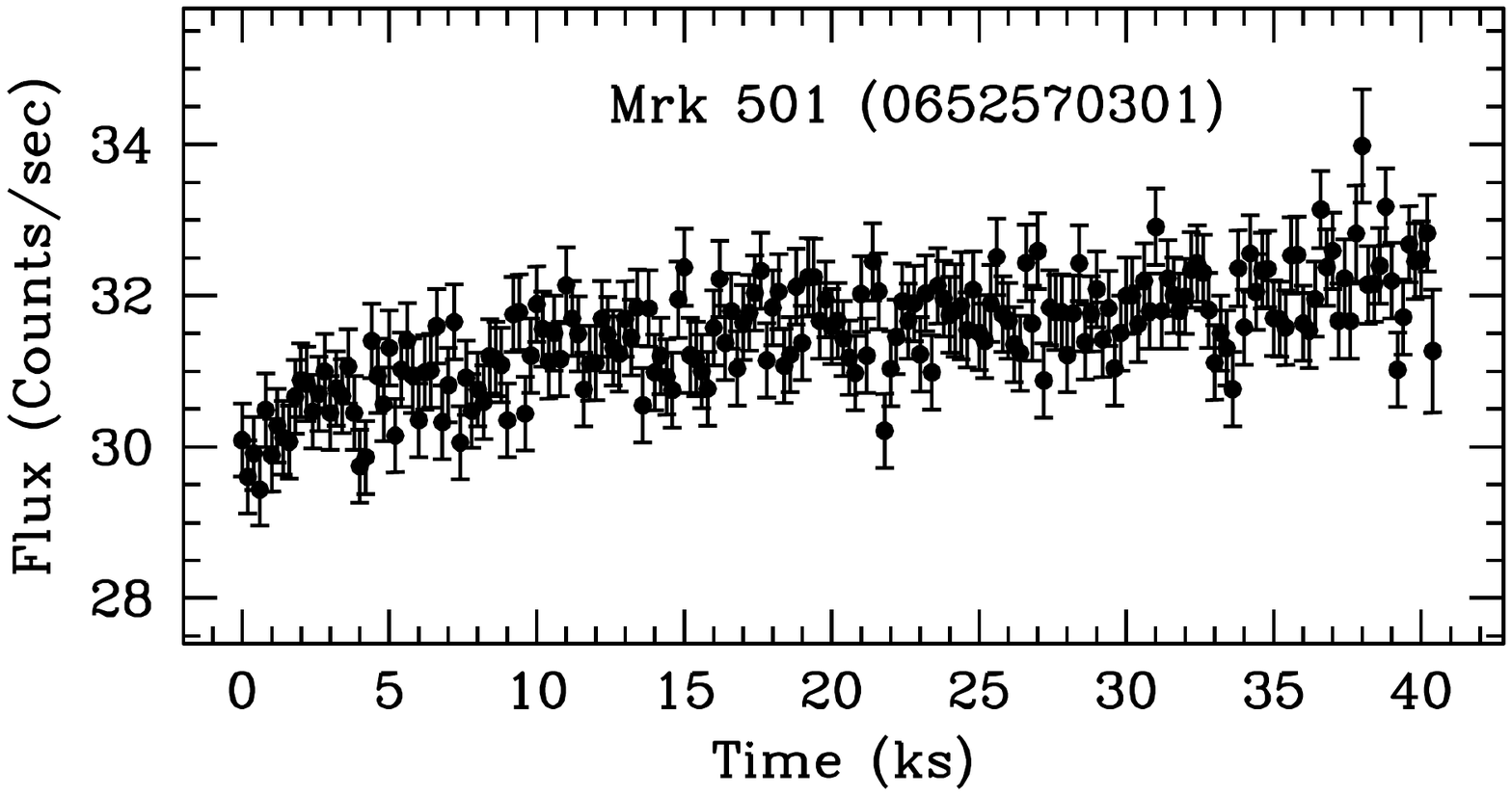}

\vspace*{-3.0in}
\includegraphics[scale=0.4]{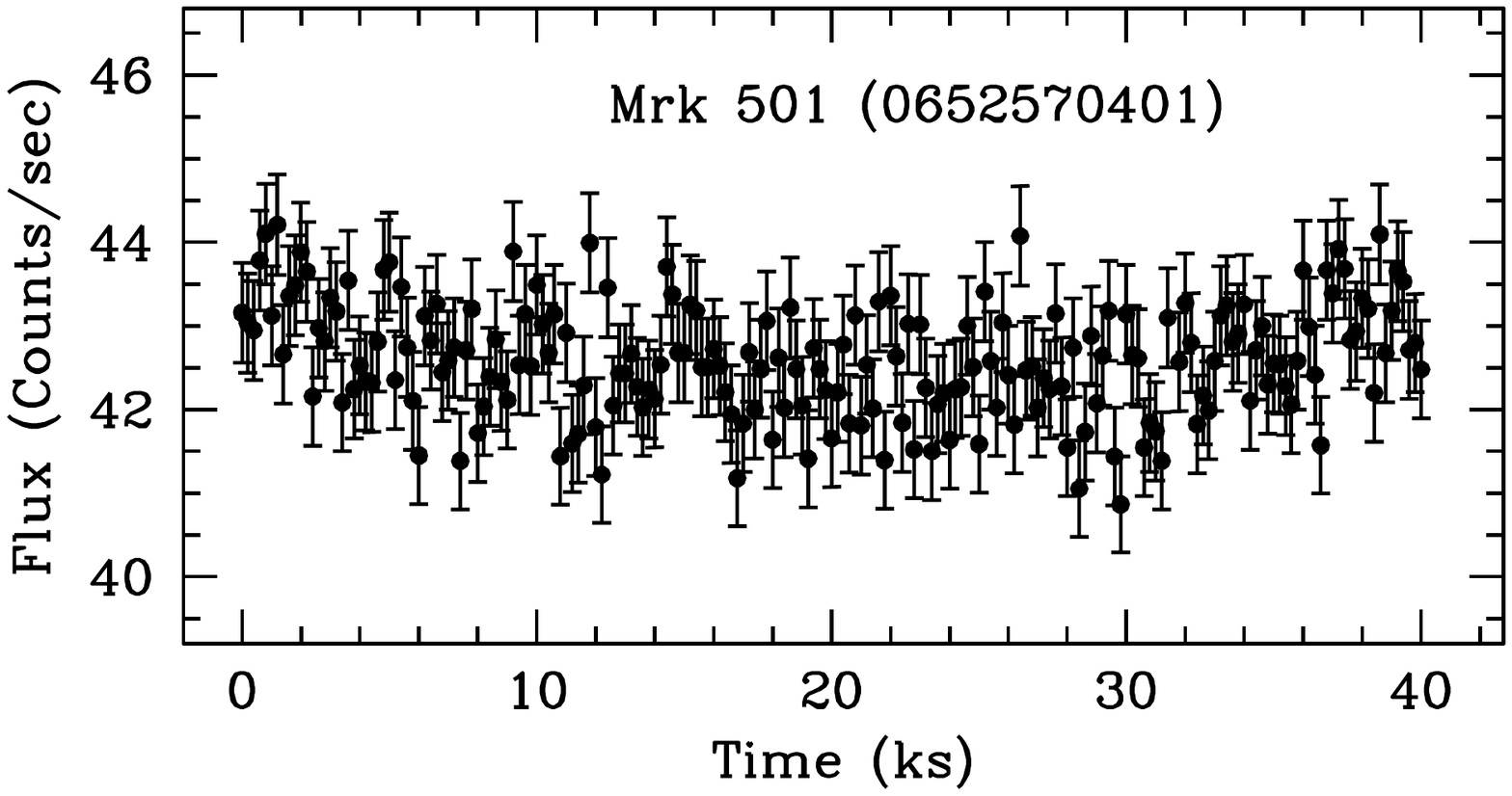}
\includegraphics[scale=0.4]{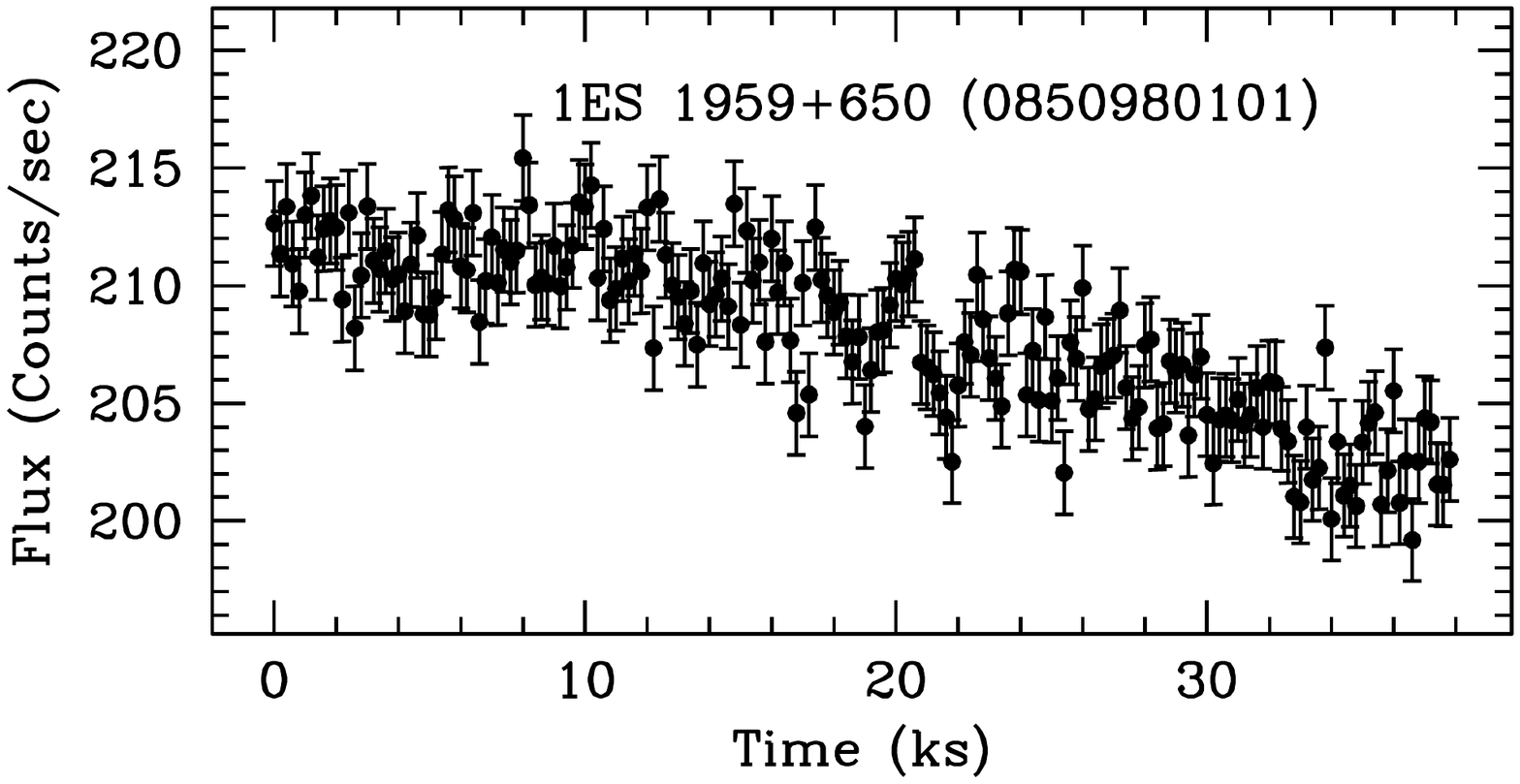}

\vspace*{-3.0in}
\includegraphics[scale=0.4]{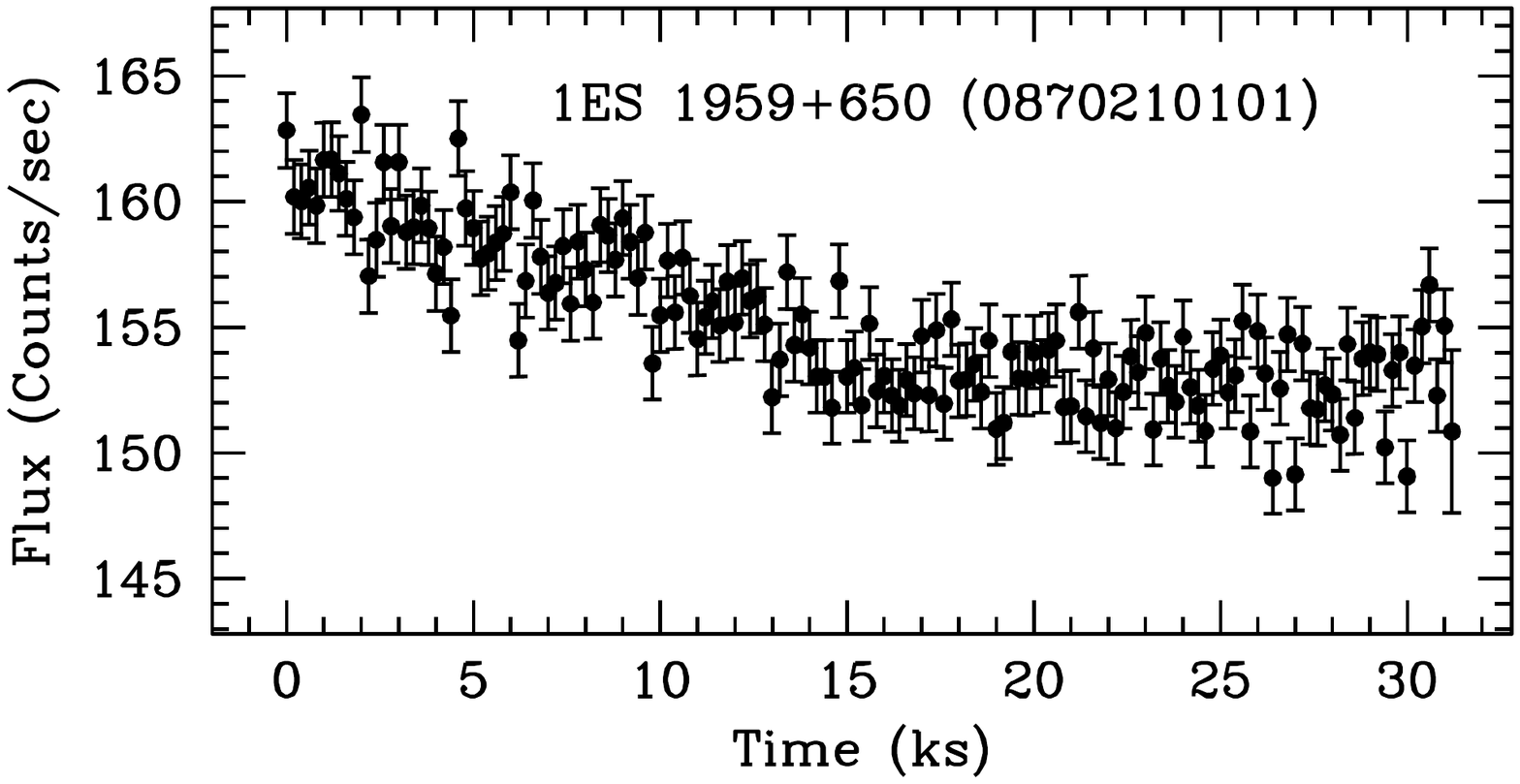}
\includegraphics[scale=0.4]{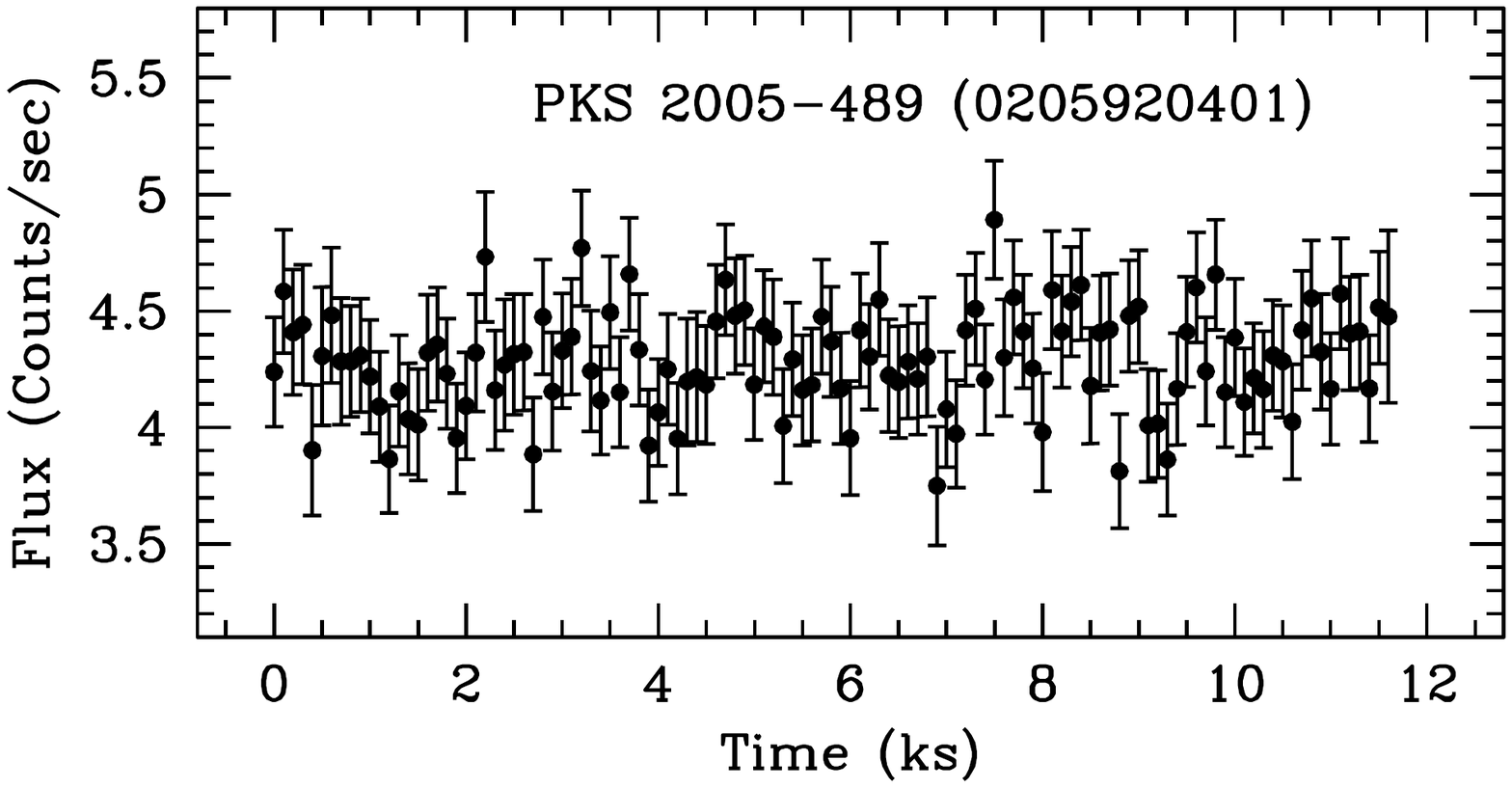}

\vspace*{-3.0in}
\includegraphics[scale=0.4]{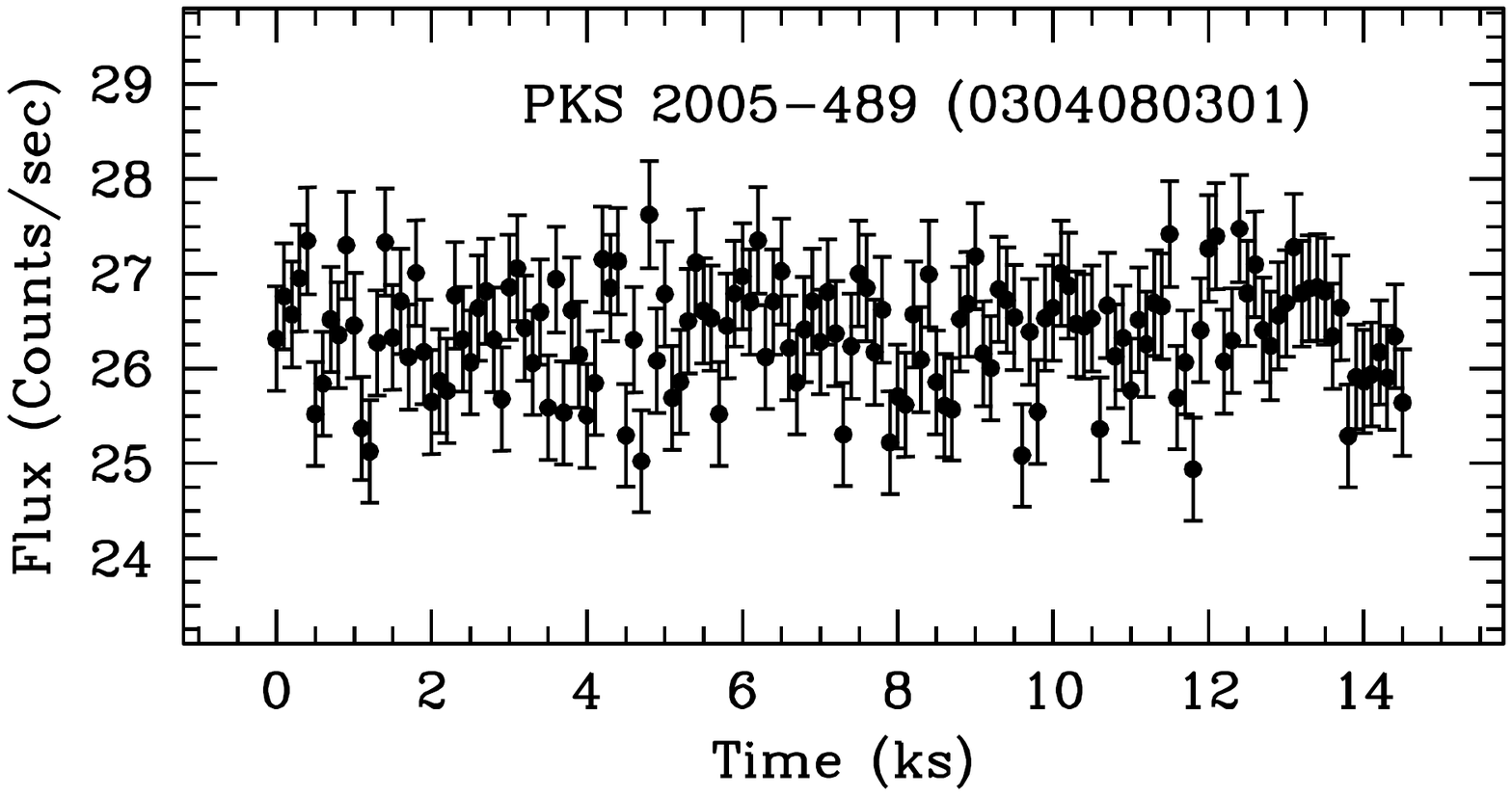}
\includegraphics[scale=0.4]{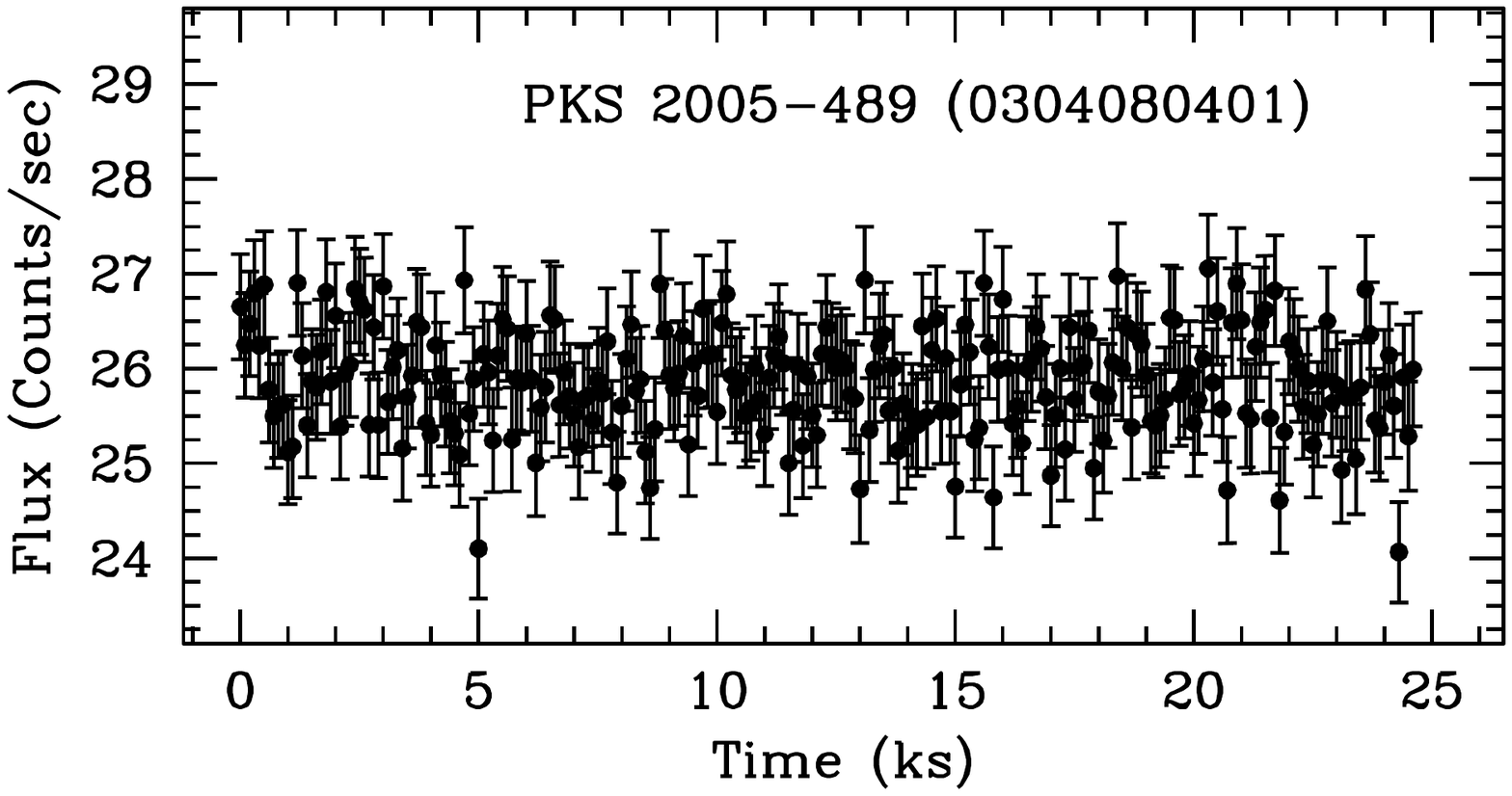}

\vspace*{-3.0in}
\includegraphics[scale=0.4]{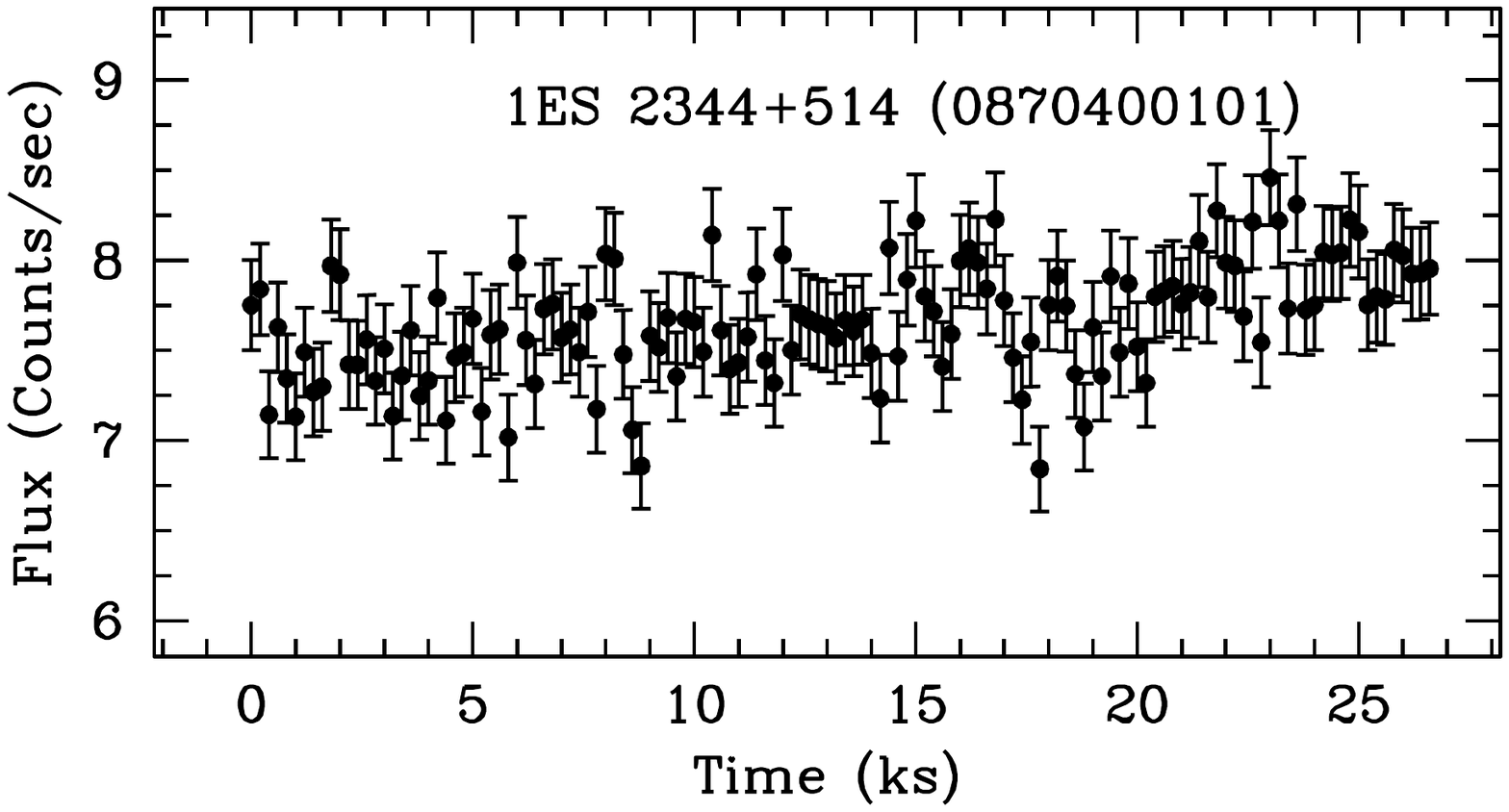}
\vspace{-0.9in}
\caption{Continued} 
\end{figure*}
\clearpage

\setcounter{figure}{1}
\begin{figure*}
\centering
\vspace*{-1.5in}
\includegraphics[scale=0.4]{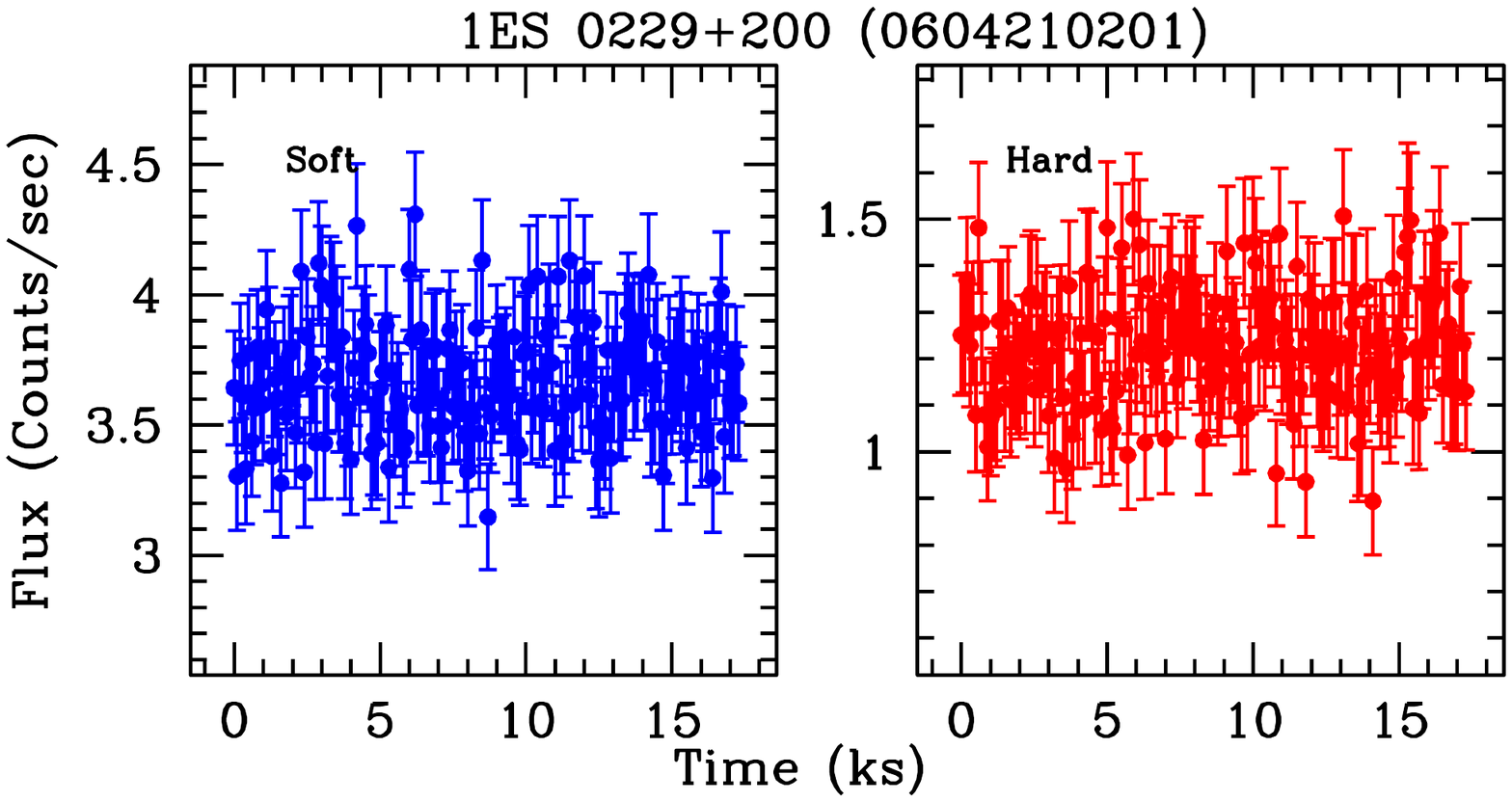}
\includegraphics[scale=0.4]{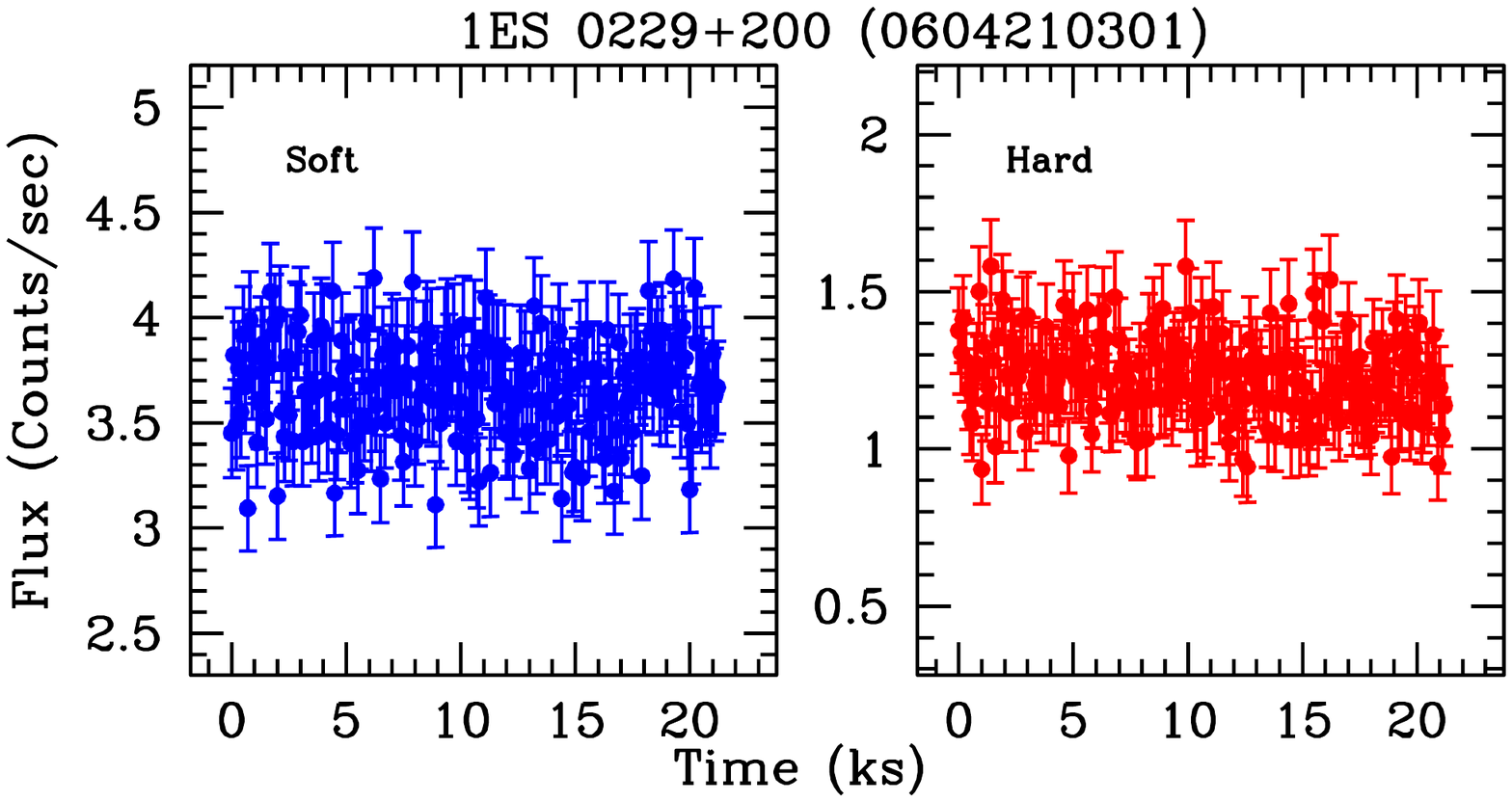}

\vspace*{-2.5in}
\includegraphics[scale=0.4]{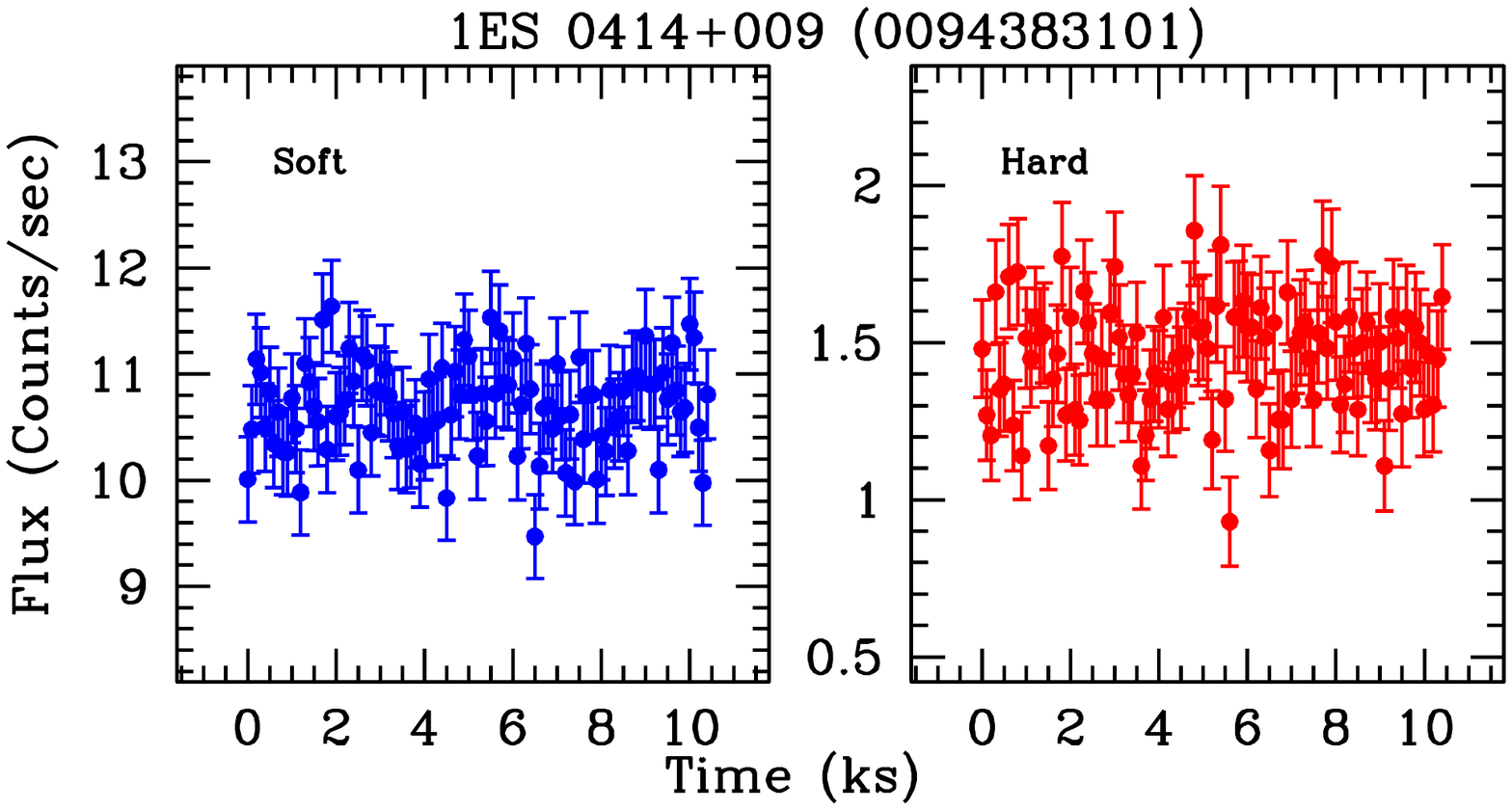}
\includegraphics[scale=0.4]{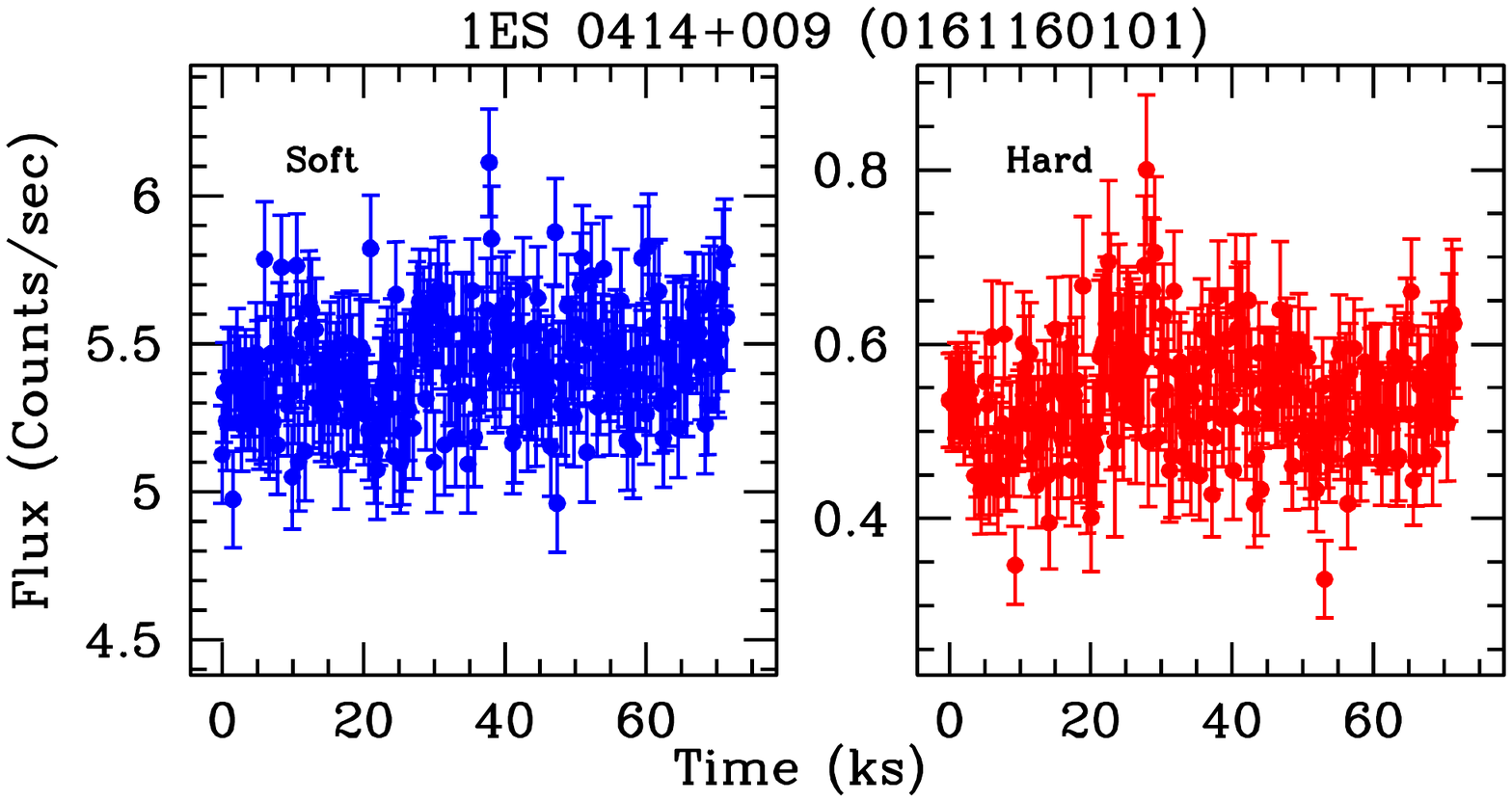}

\vspace*{-2.5in}
\includegraphics[scale=0.4]{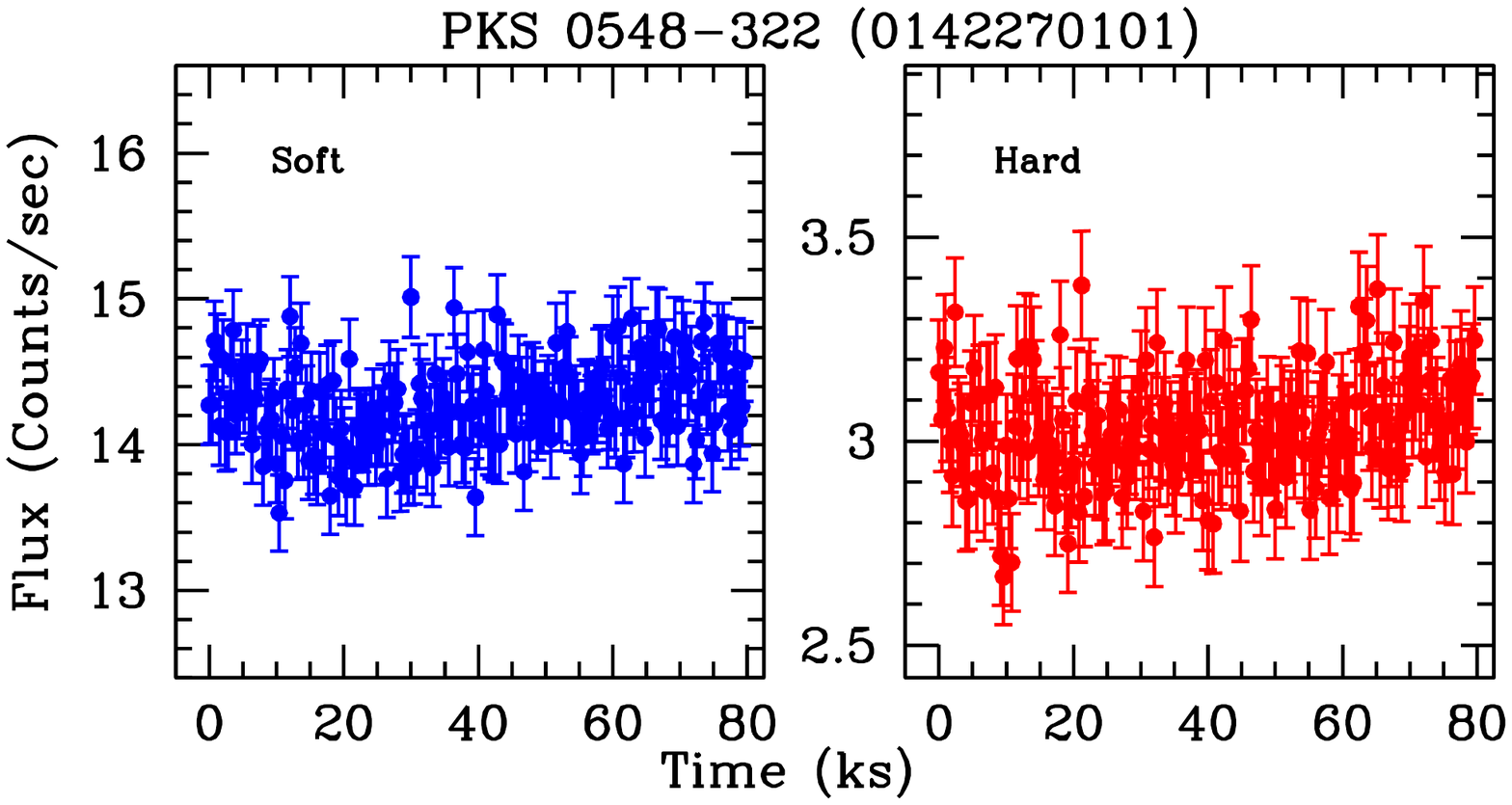}
\includegraphics[scale=0.4]{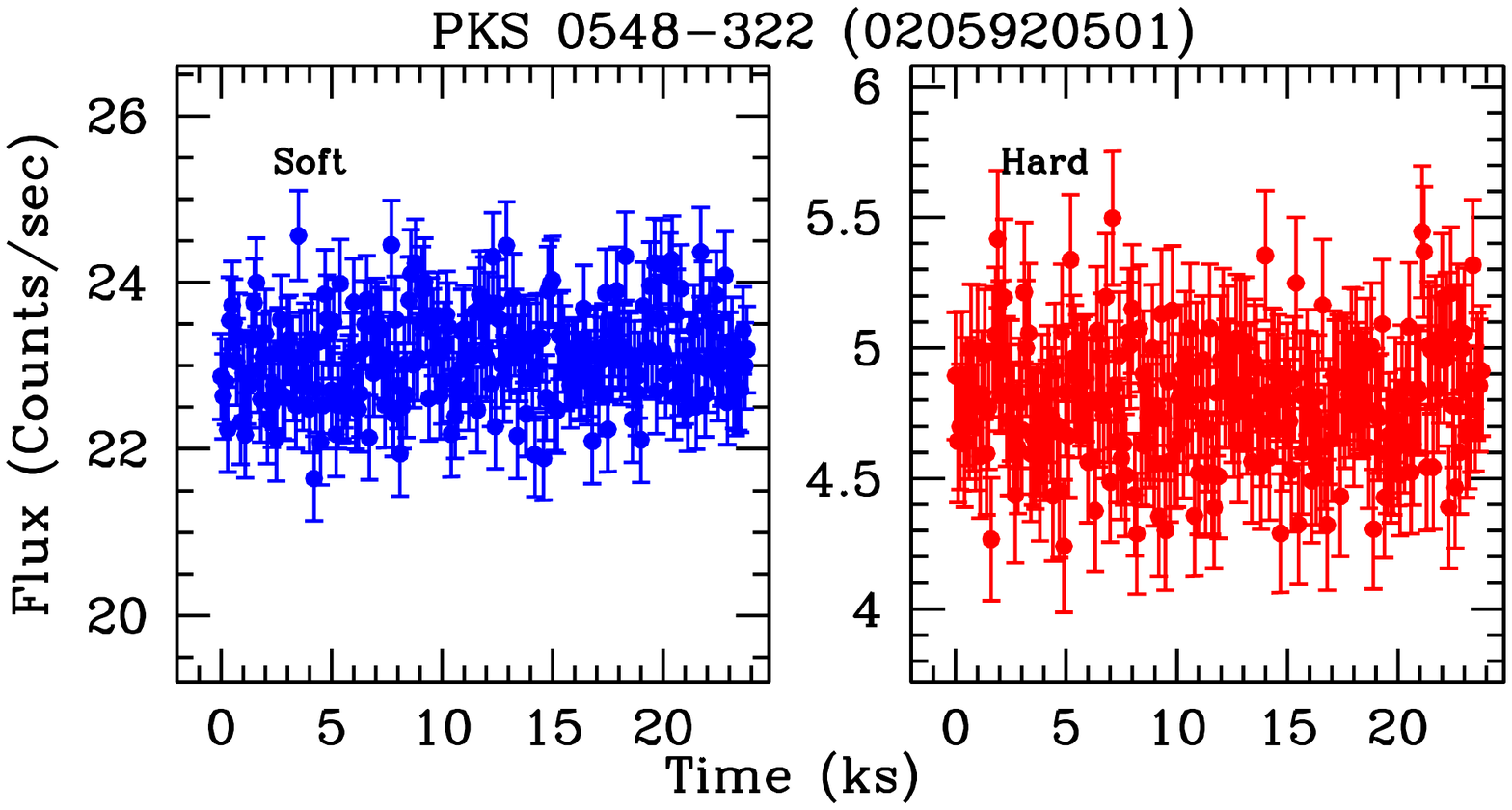}

\vspace*{-2.5in}
\includegraphics[scale=0.4]{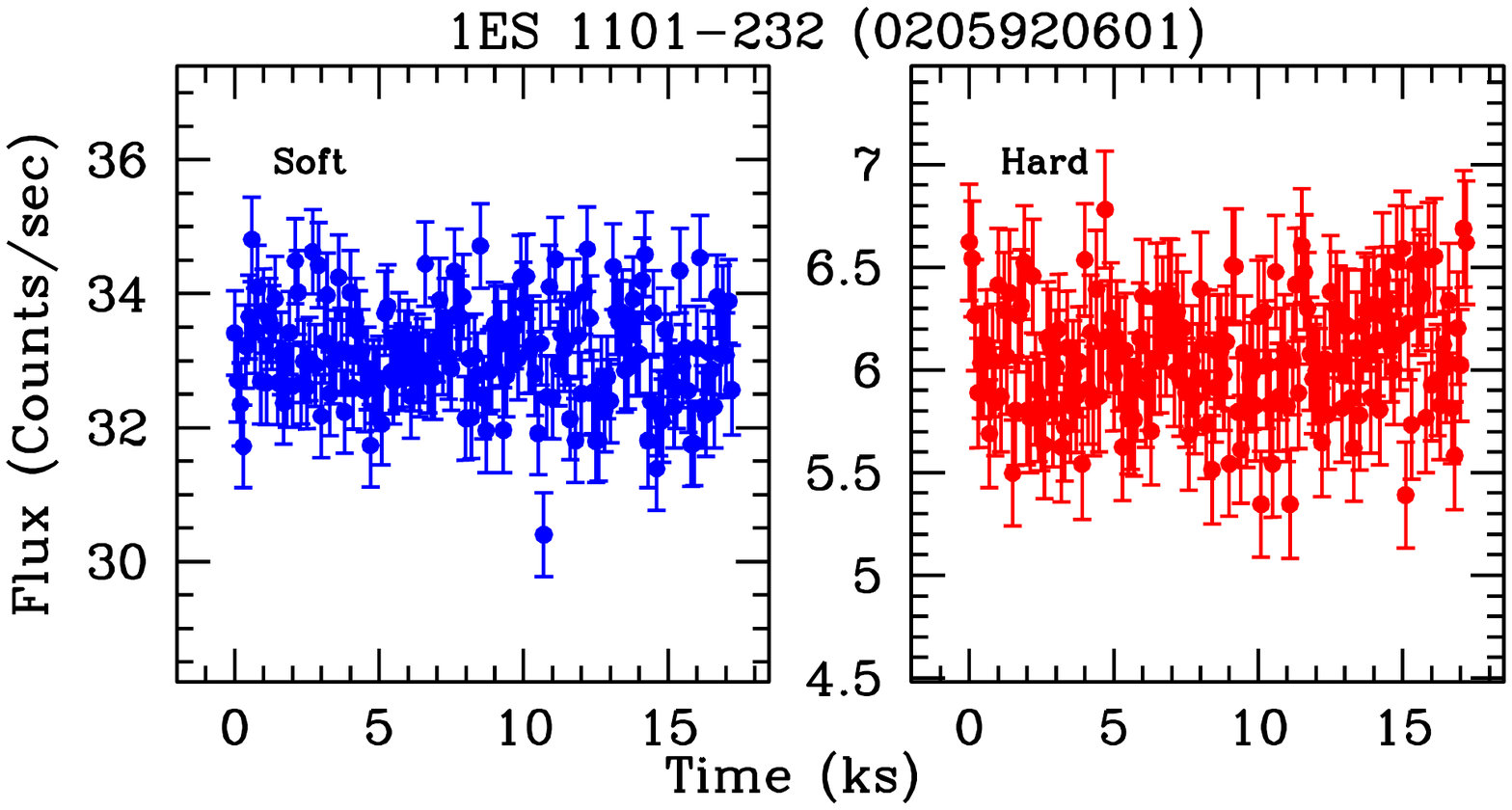}
\includegraphics[scale=0.4]{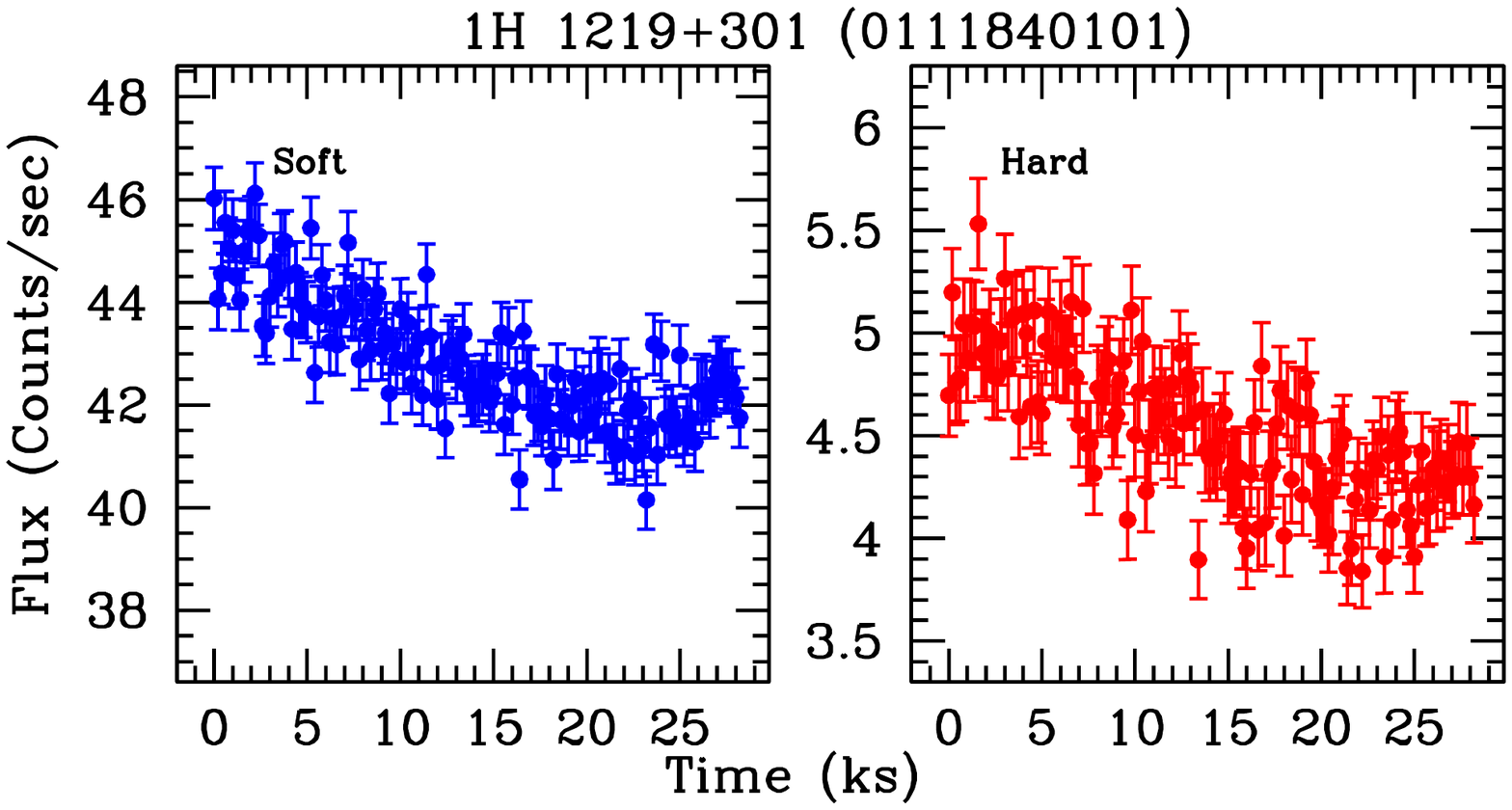}

\vspace{-0.7in}
\caption{LCs of 25 \emph{XMM-Newton} pointed observations in  the soft energy (0.3--2 keV; blue dots) and in the hard energy (2--10 keV; red dots). Source name and Observation ID are given above each plot.\label{A2}}

\end{figure*}

\clearpage
\setcounter{figure}{1}
\begin{figure*}
\centering

\vspace*{-1.5in}
\includegraphics[scale=0.4]{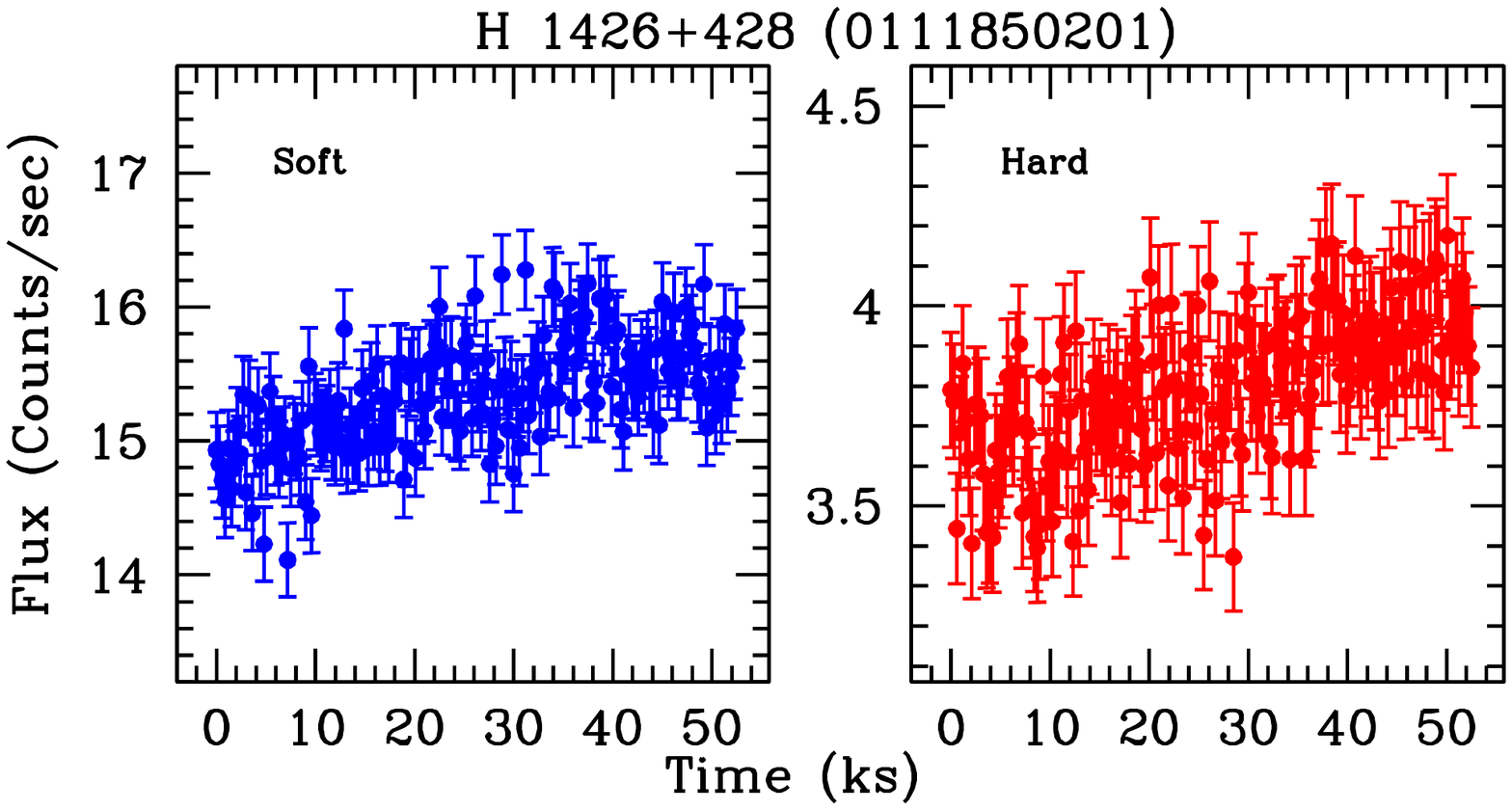}
\includegraphics[scale=0.4]{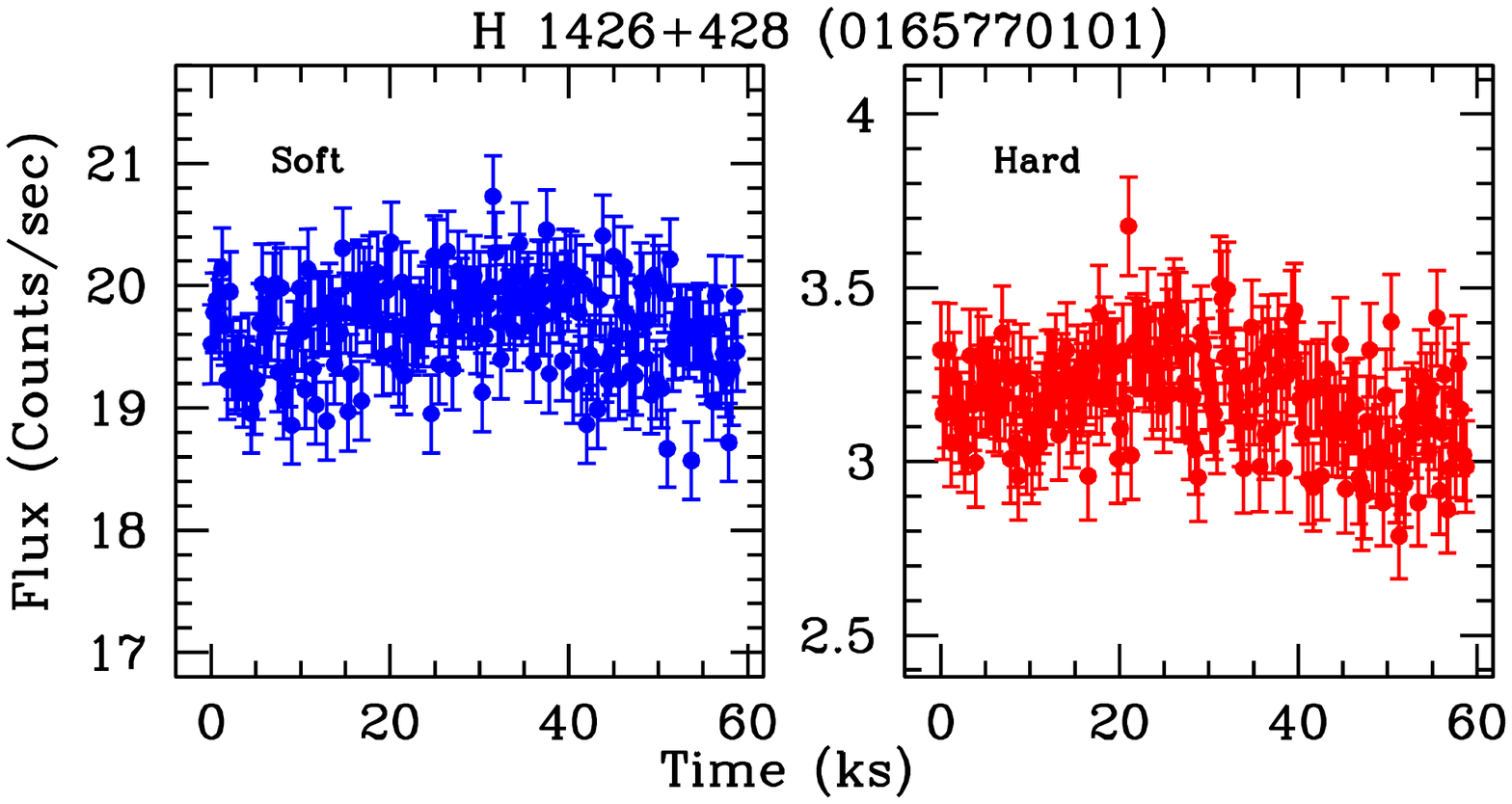}

\vspace*{-2.5in}
\includegraphics[scale=0.4]{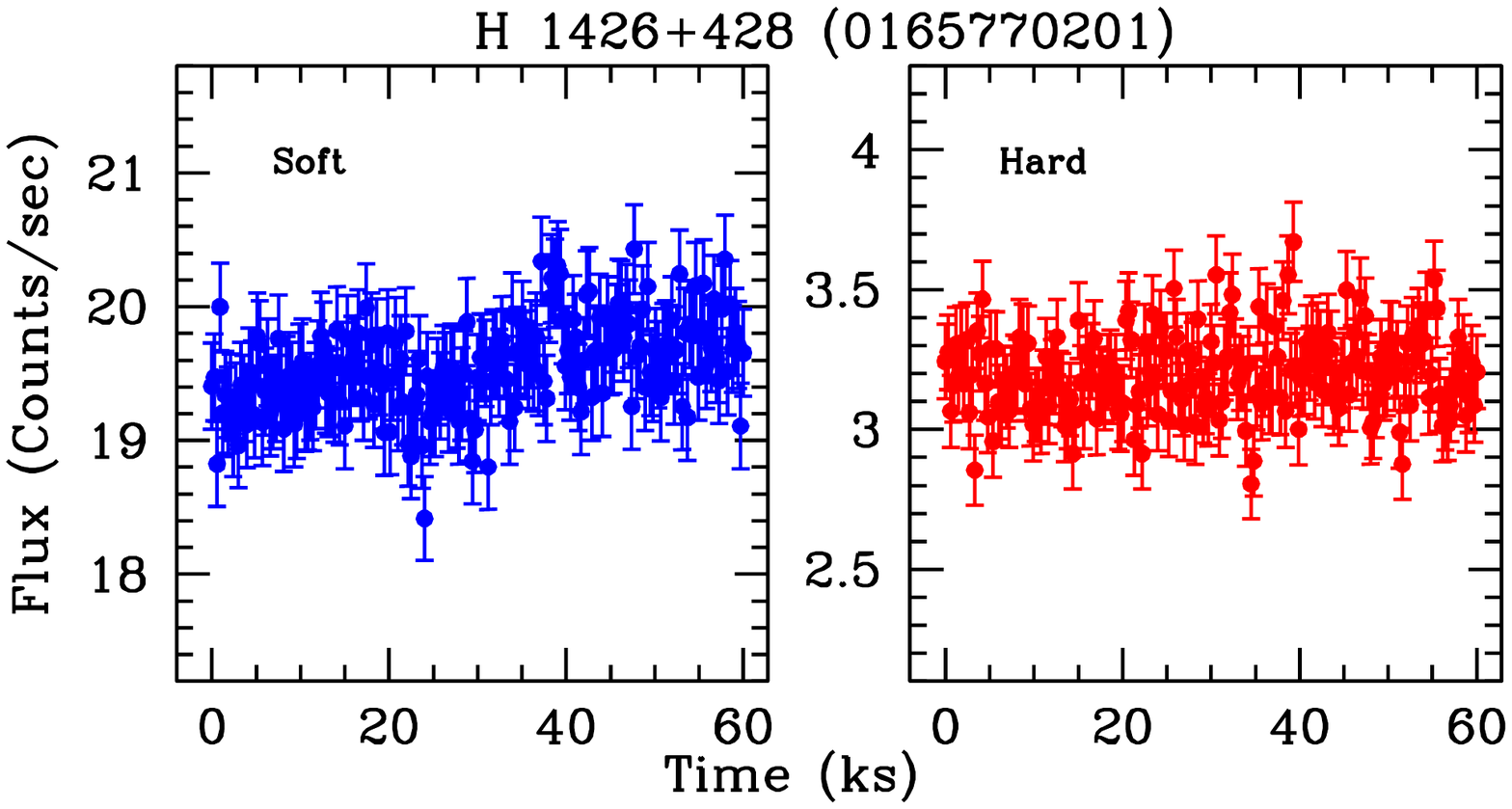}
\includegraphics[scale=0.4]{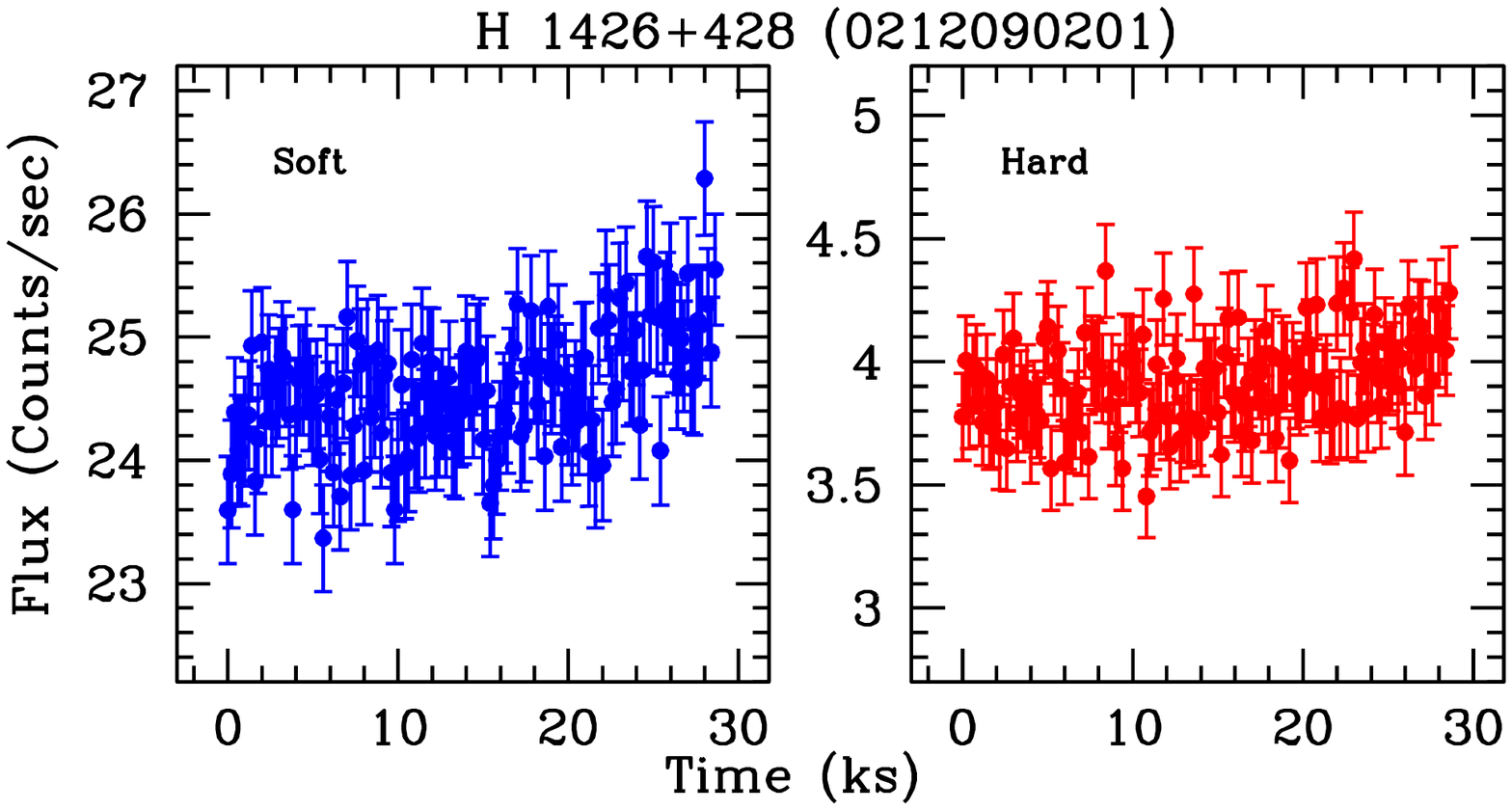}

\vspace*{-2.5in}
\includegraphics[scale=0.4]{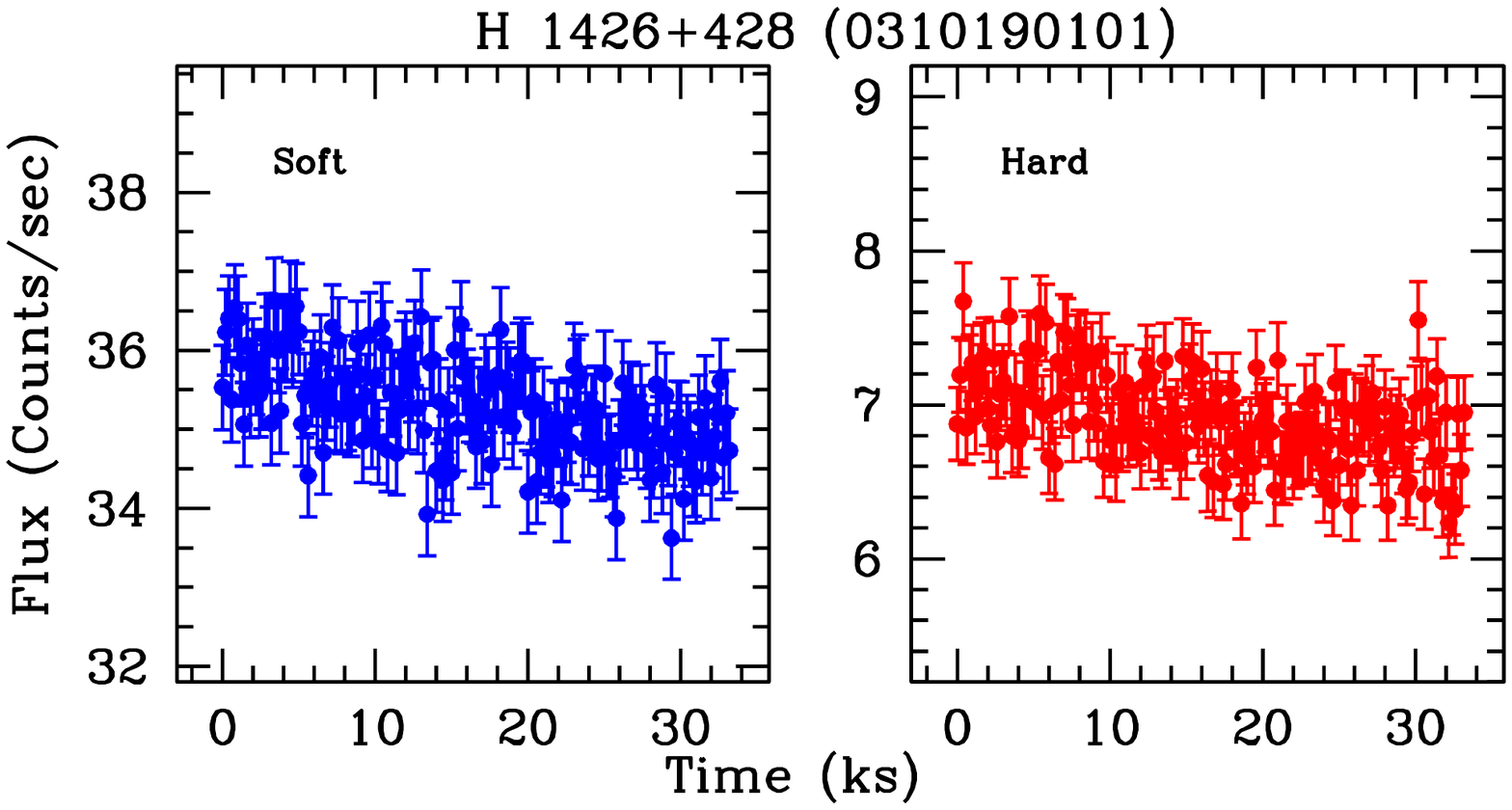}
\includegraphics[scale=0.4]{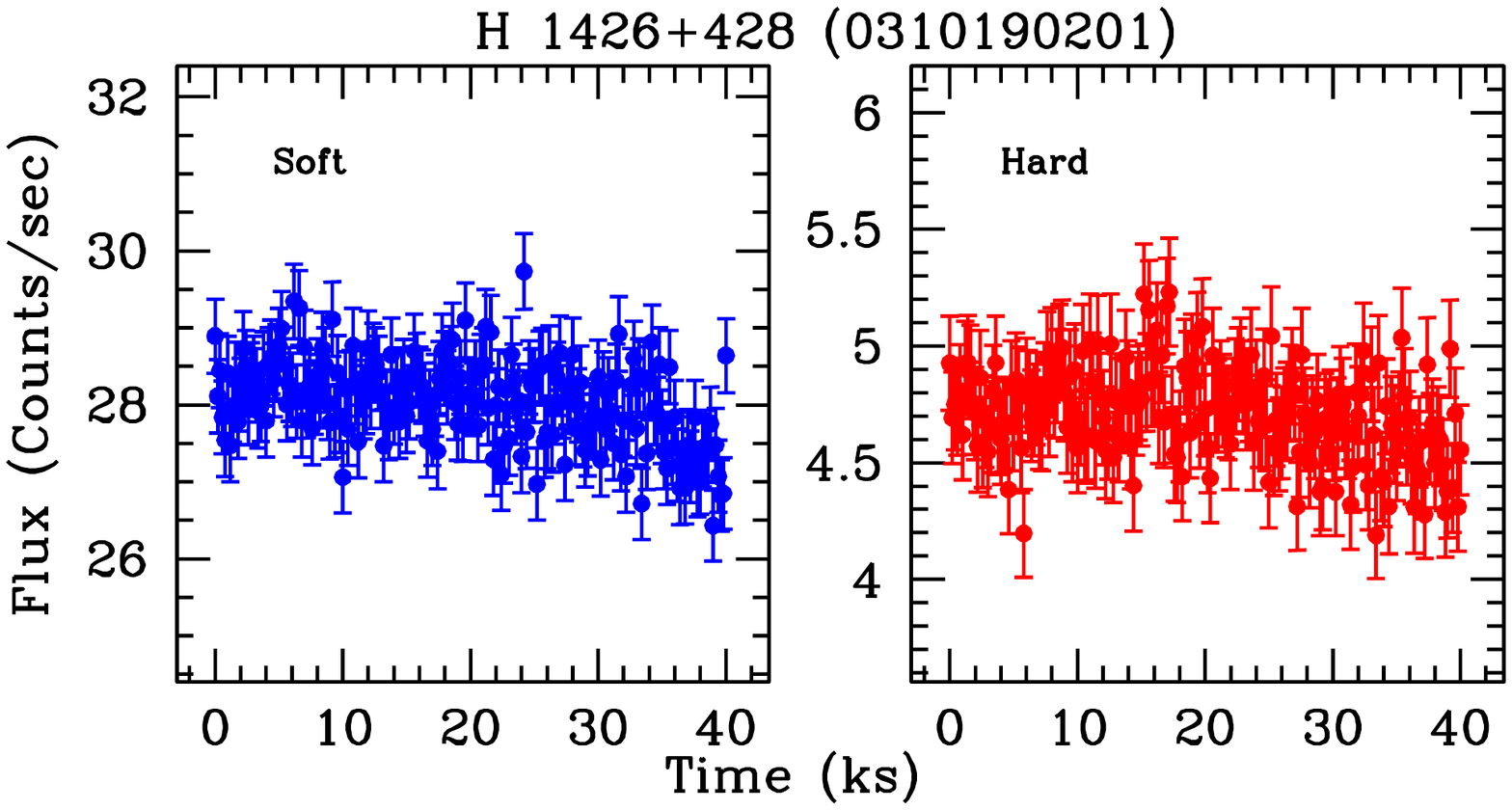}

\vspace*{-2.5in}
\includegraphics[scale=0.4]{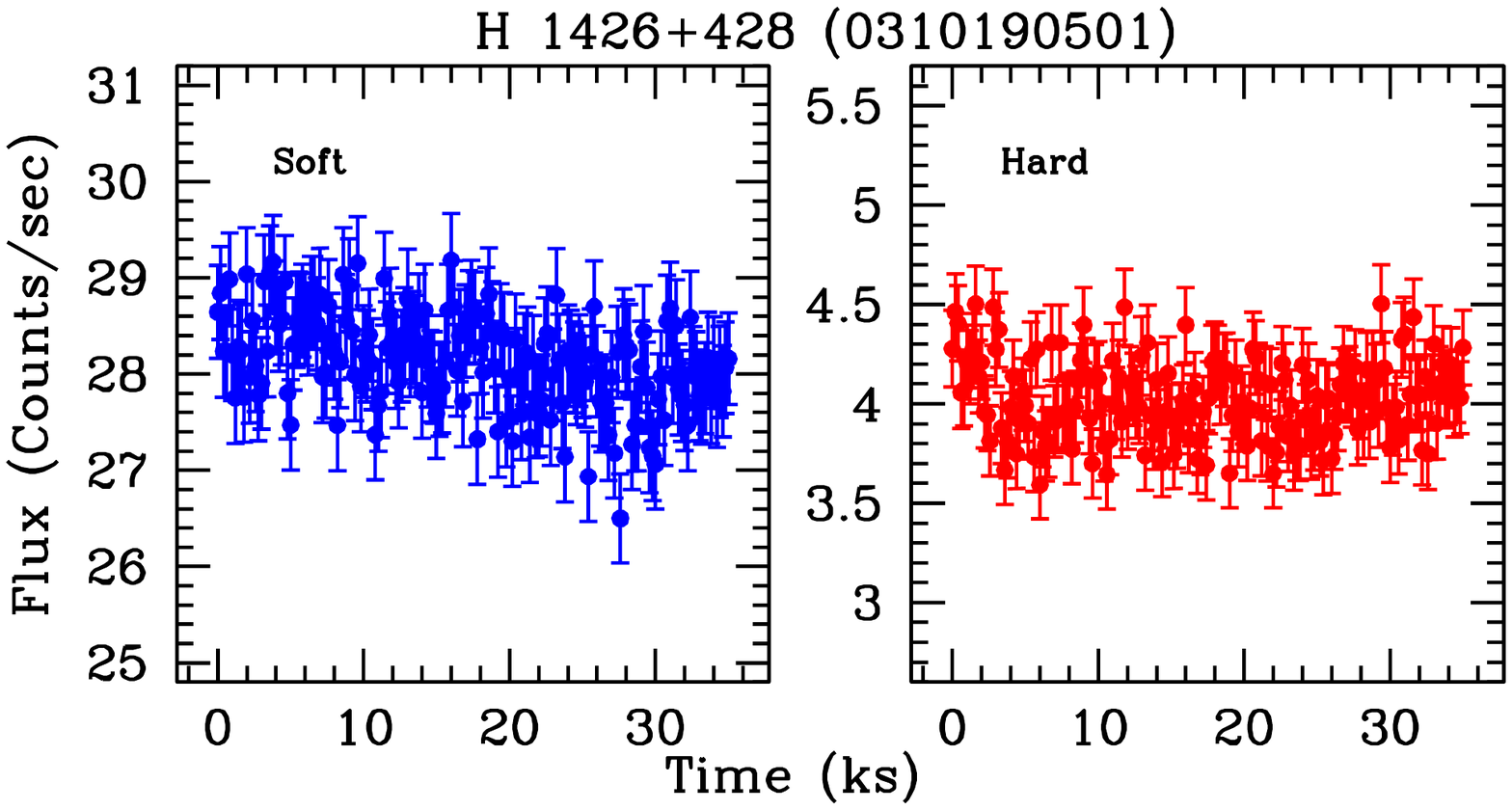}
\includegraphics[scale=0.4]{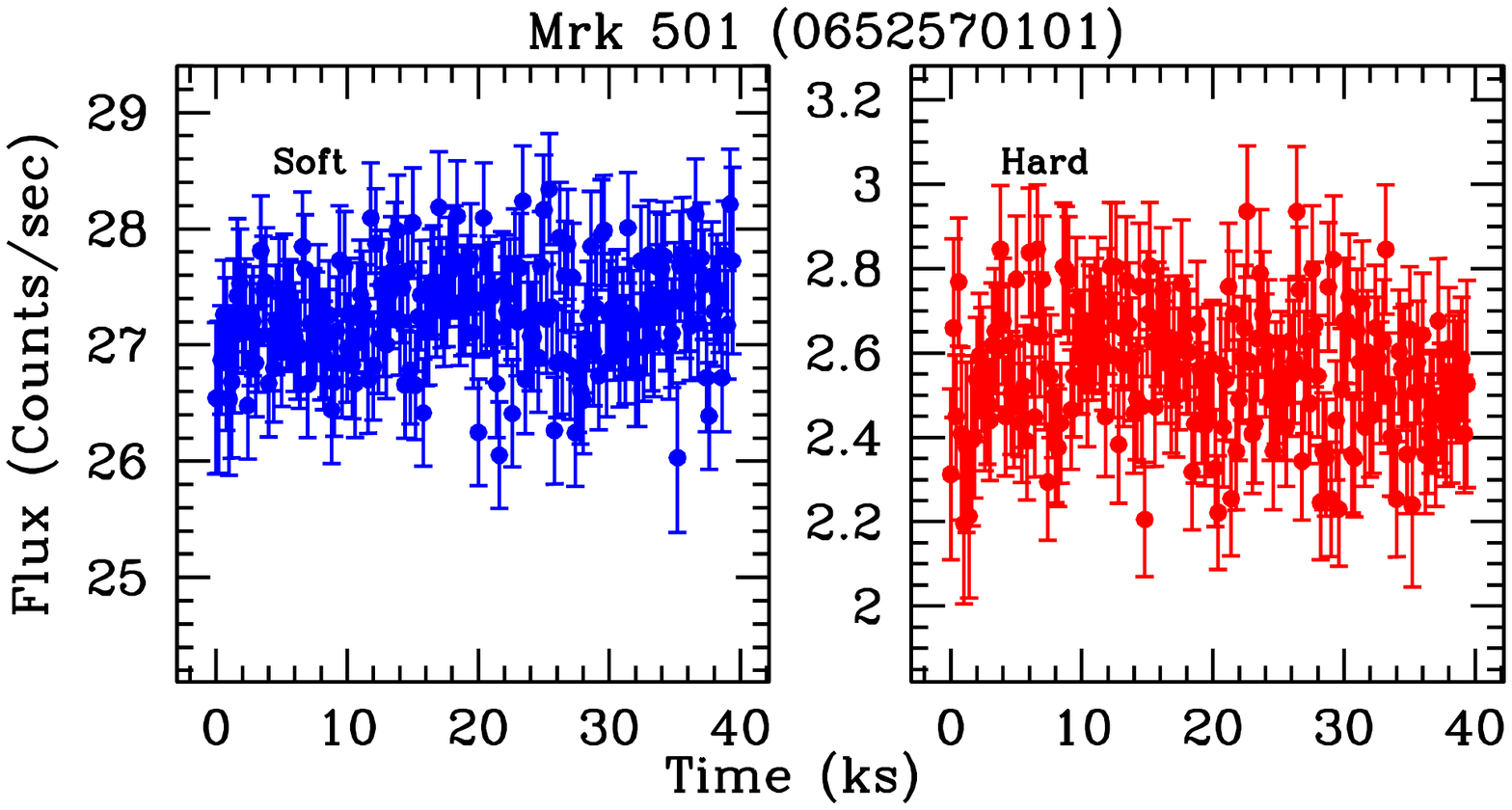}

\vspace{-0.7in}
\caption{Continued.}  

\end{figure*}

\clearpage
\setcounter{figure}{1}
\begin{figure*}
\centering

\vspace*{-1.5in}
\includegraphics[scale=0.4]{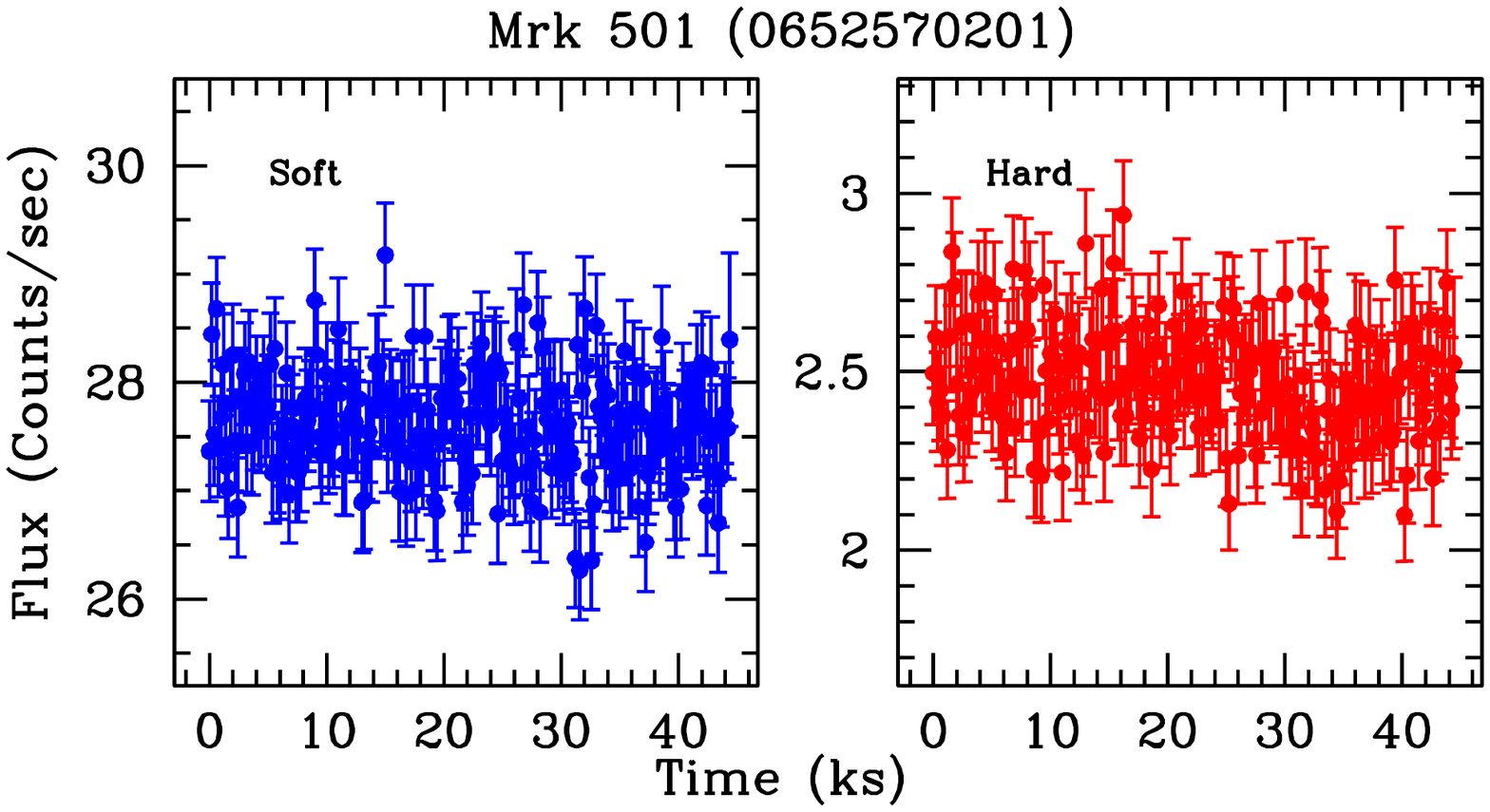}
\includegraphics[scale=0.4]{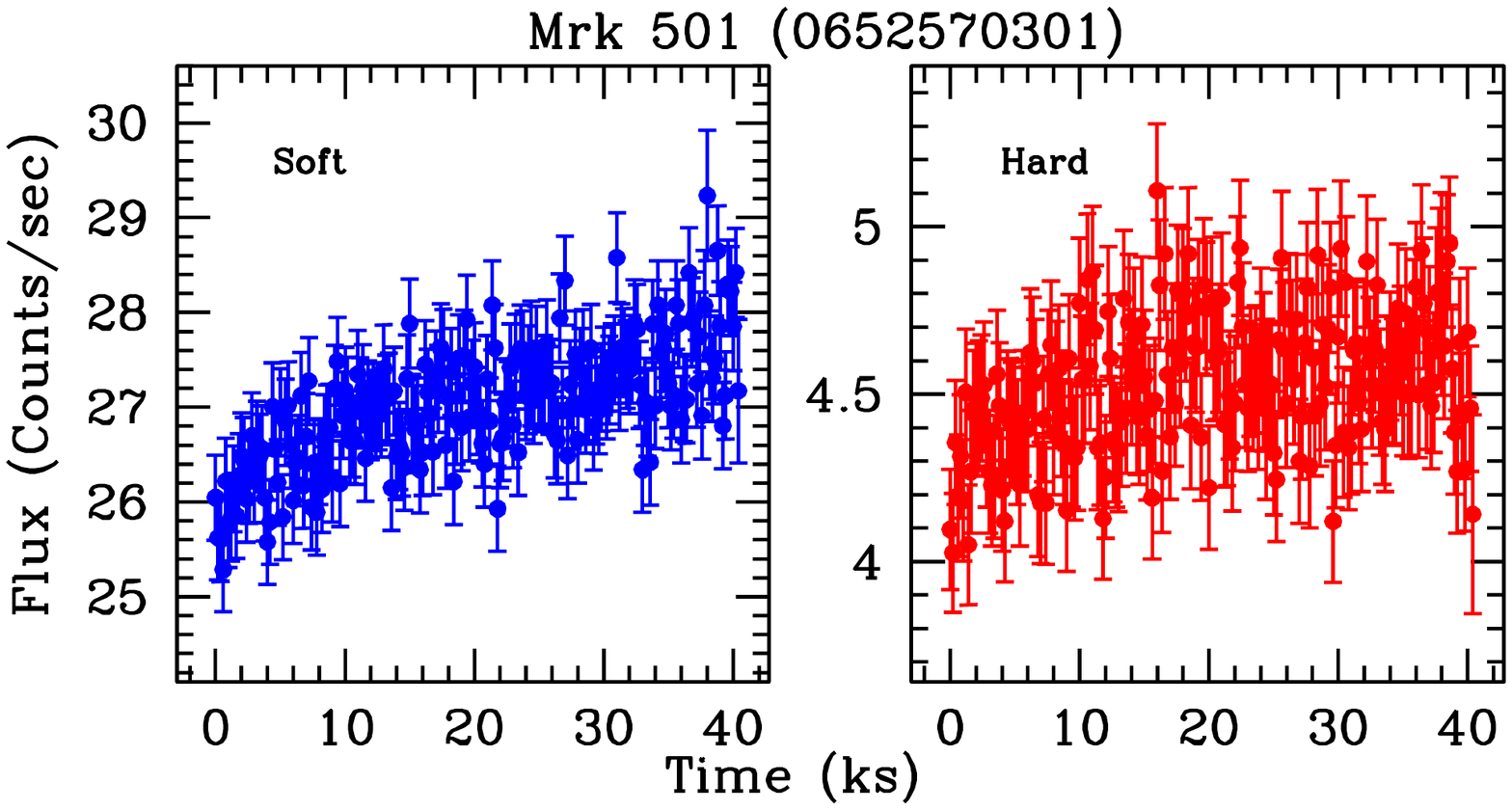}

\vspace*{-2.6in}
\includegraphics[scale=0.4]{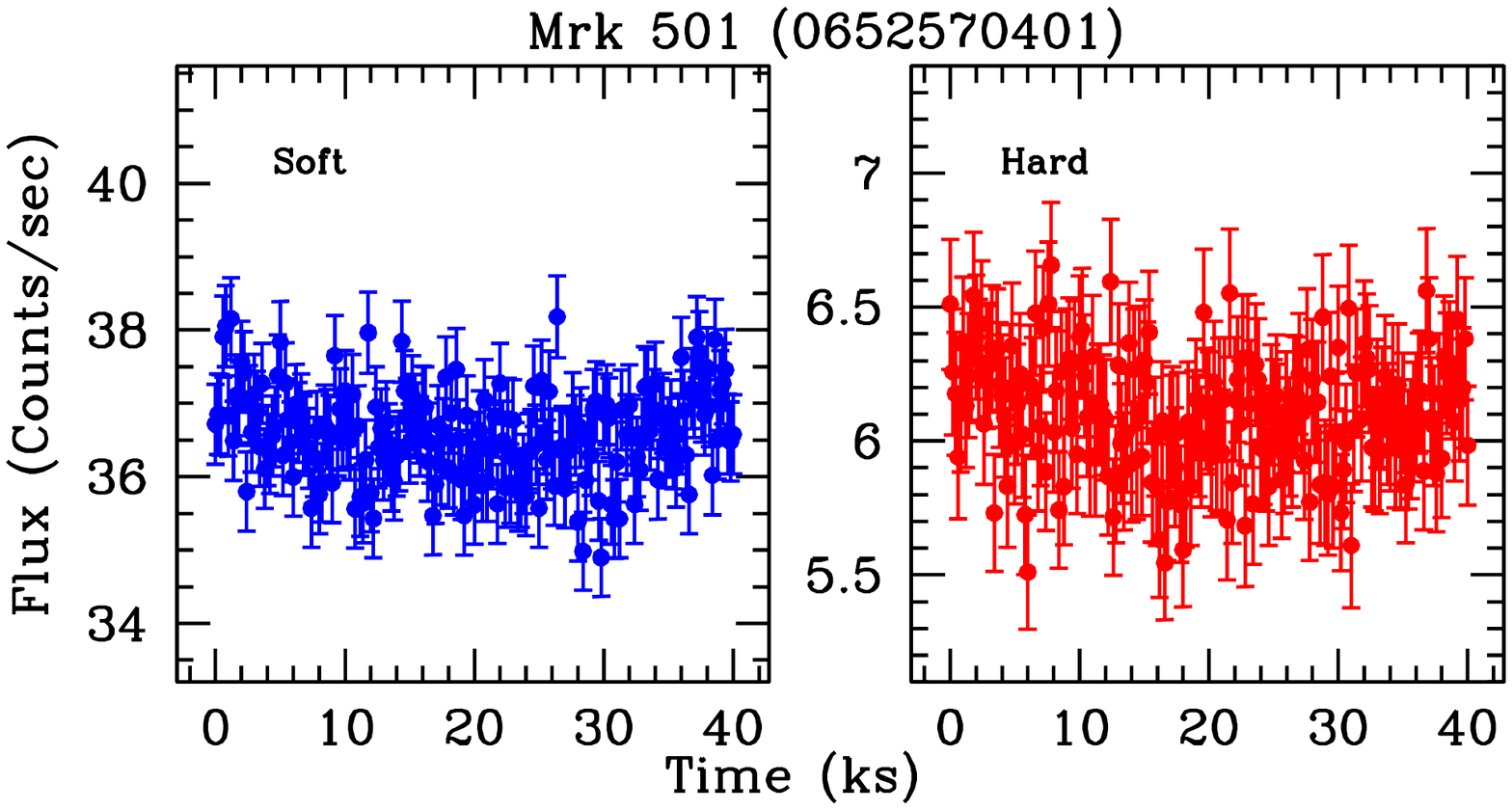}
\includegraphics[scale=0.4]{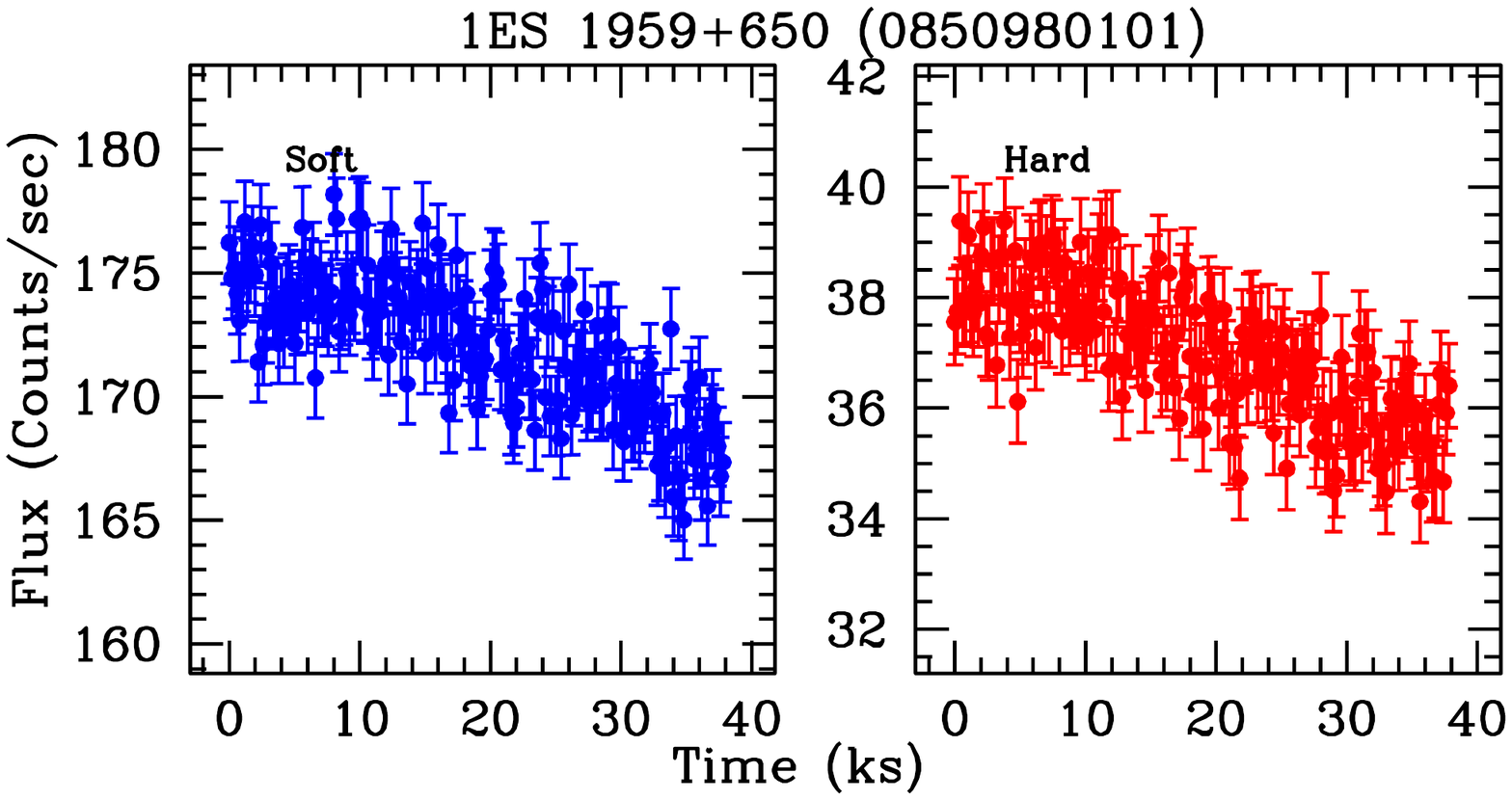}

\vspace*{-2.6in}
\includegraphics[scale=0.4]{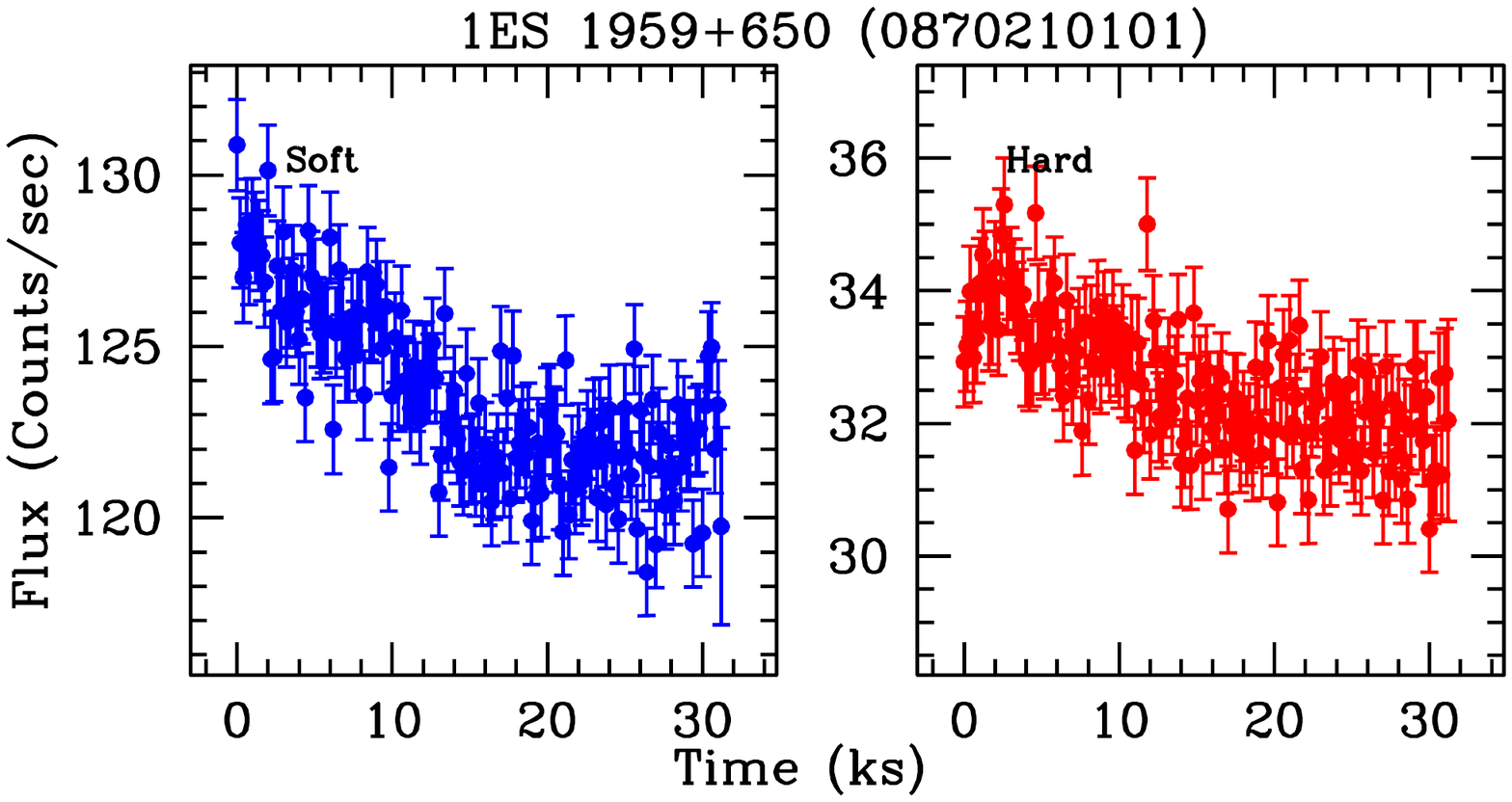}
\includegraphics[scale=0.4]{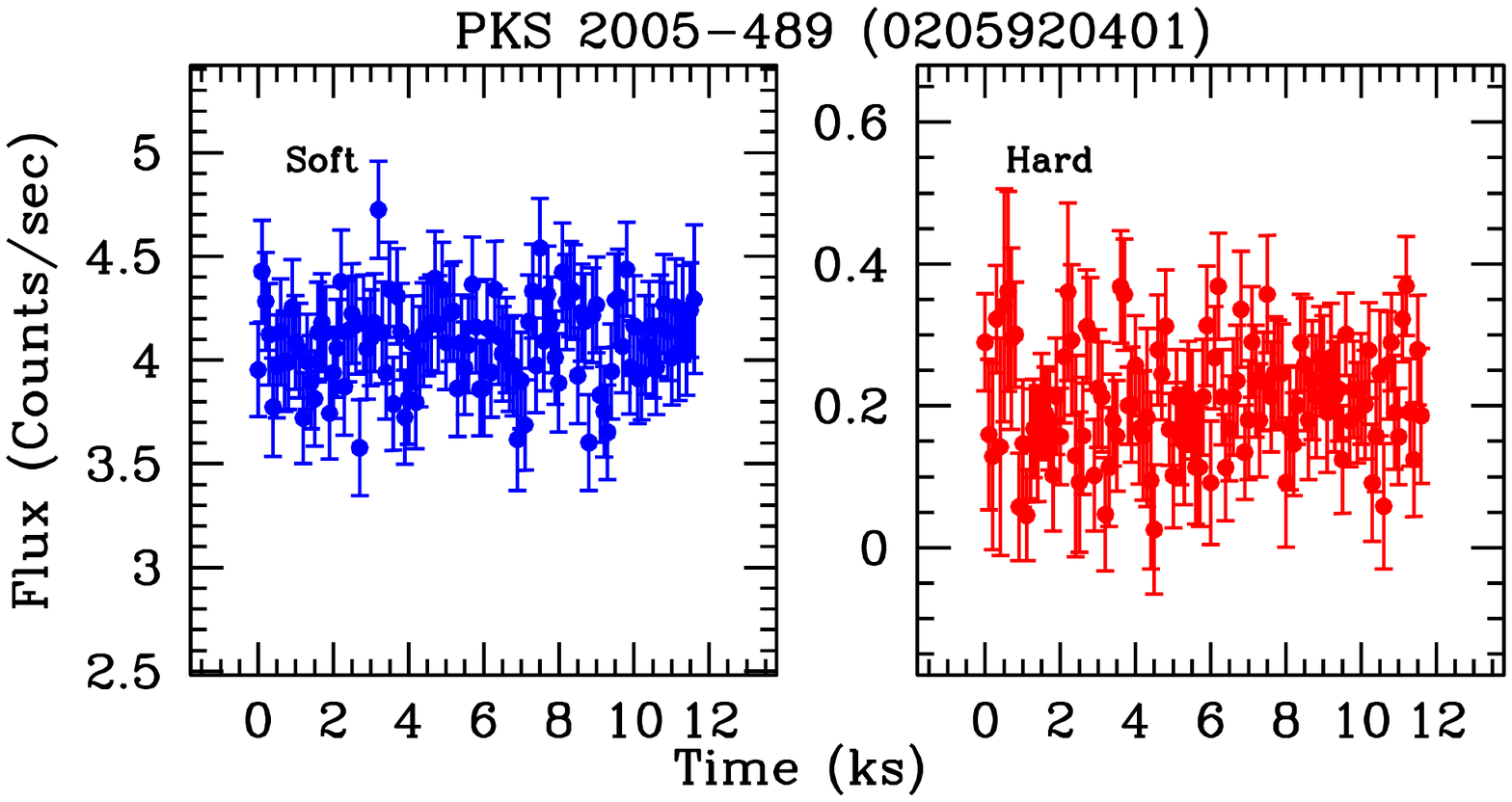}

\vspace*{-2.6in}
\includegraphics[scale=0.4]{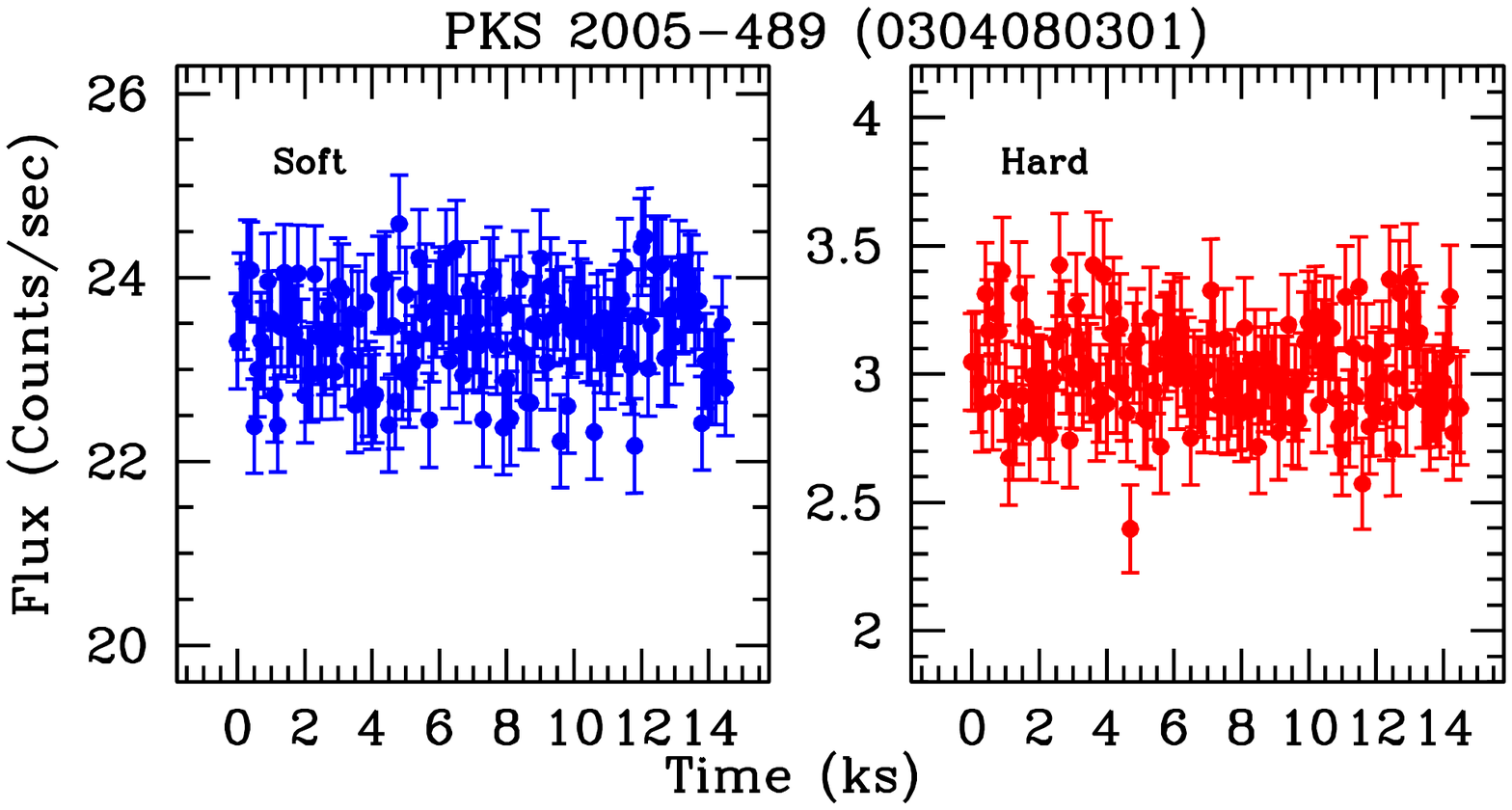}
\includegraphics[scale=0.4]{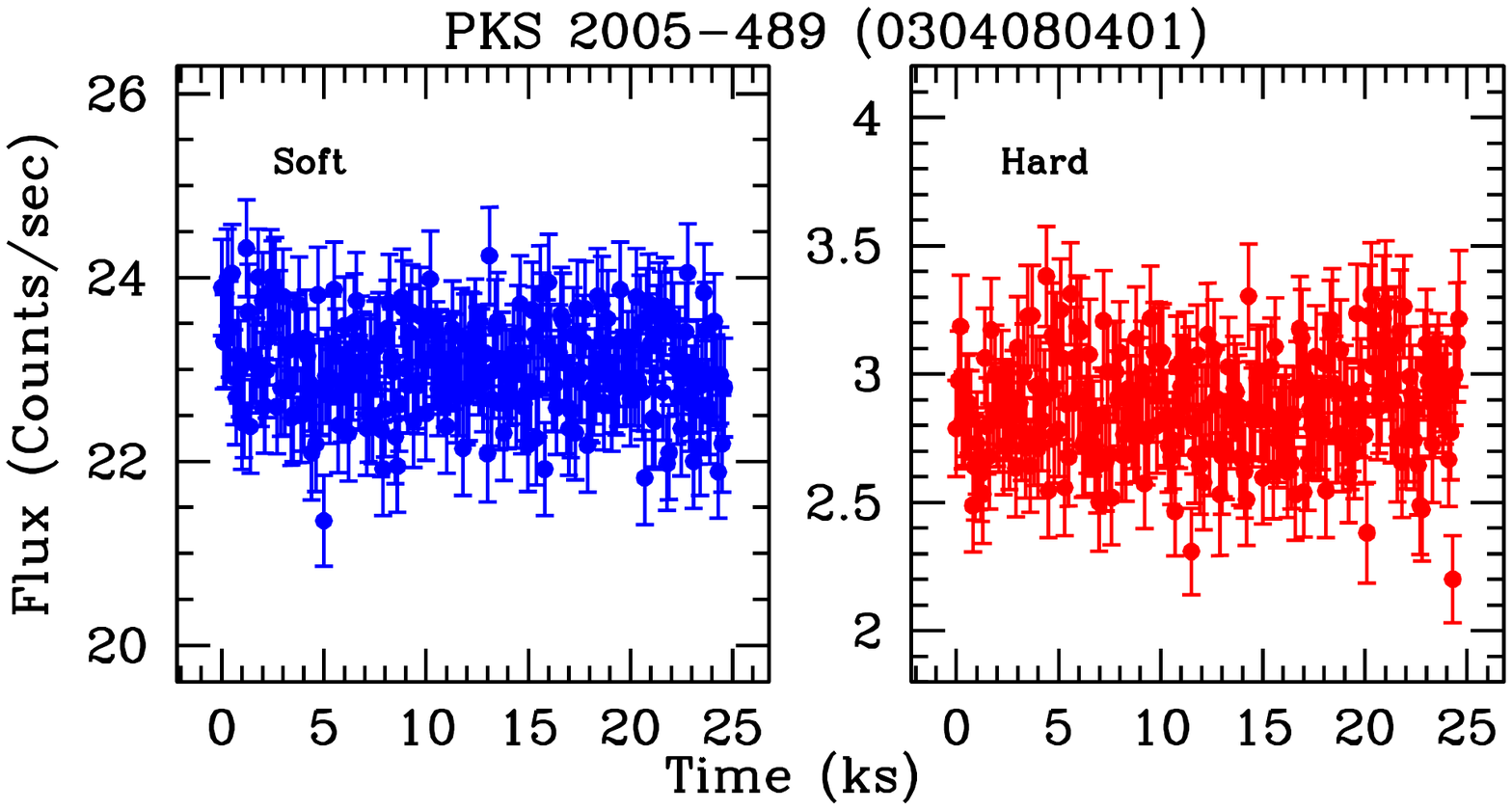}

\vspace*{-2.6in}
\includegraphics[scale=0.4]{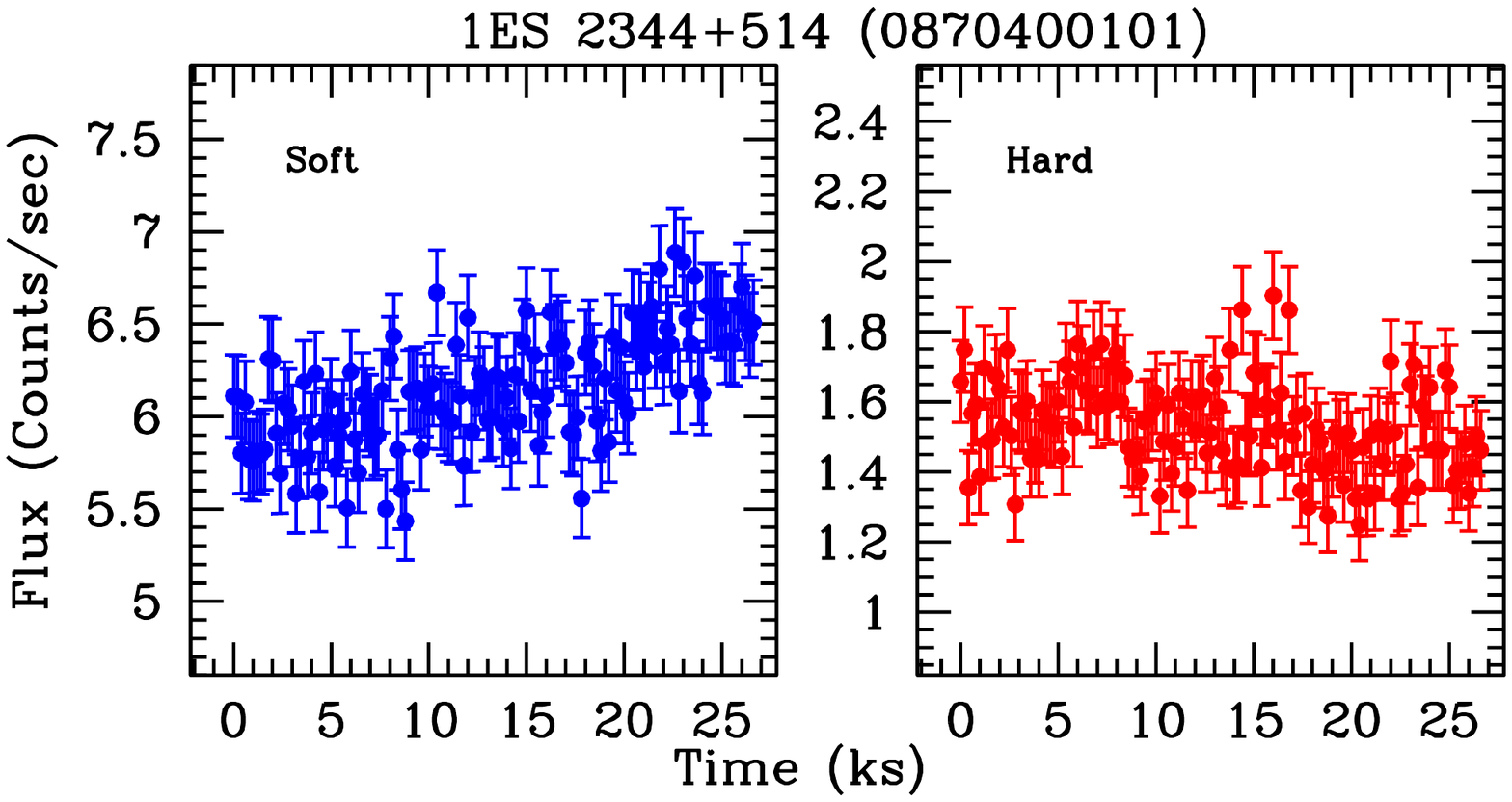}

\vspace{-0.7in}
\caption{Continued.}

\end{figure*}


\setcounter{figure}{2}

\begin{figure*}
\centering
\vspace*{-1.5in}
\includegraphics[scale=0.4]{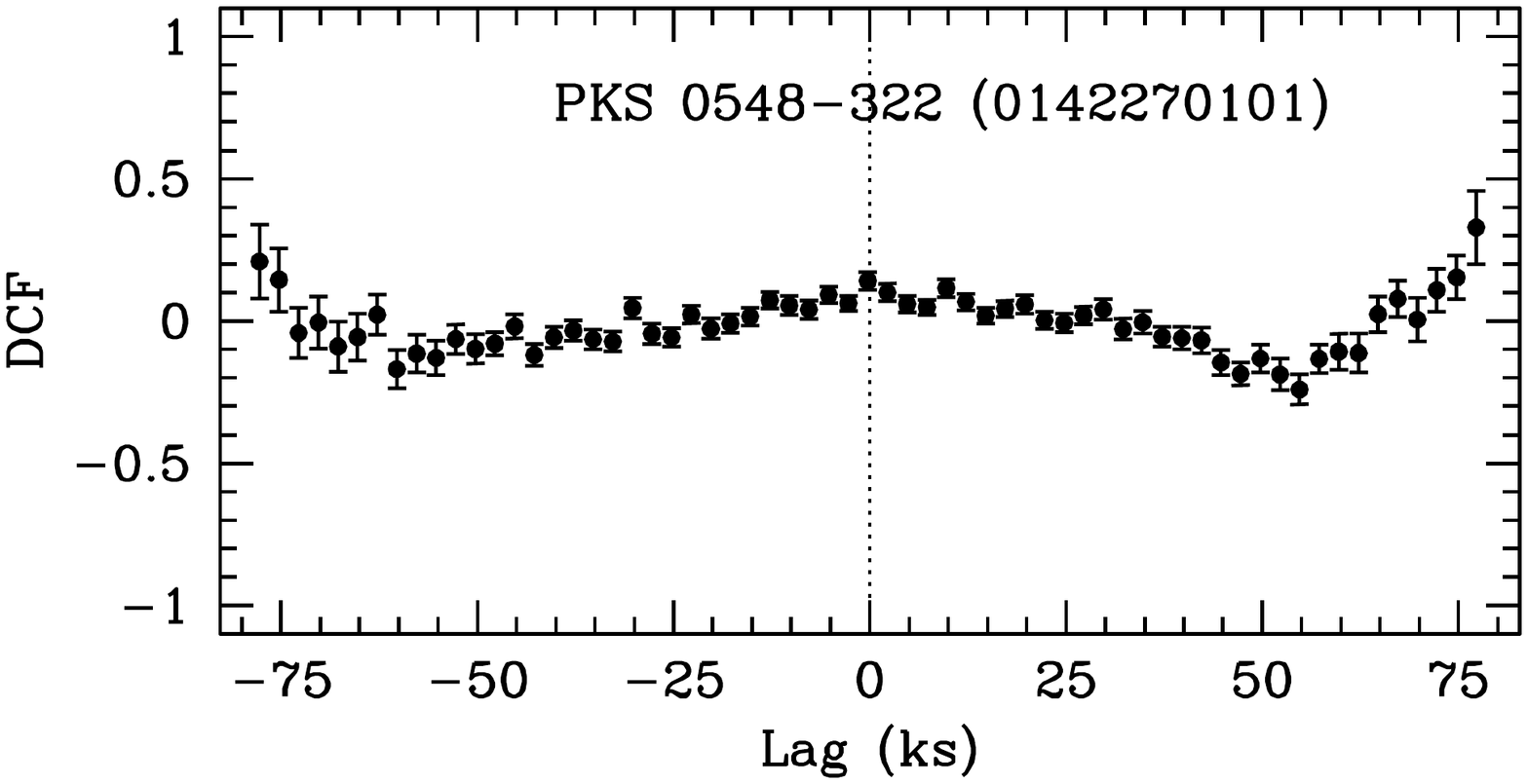}
\includegraphics[scale=0.4]{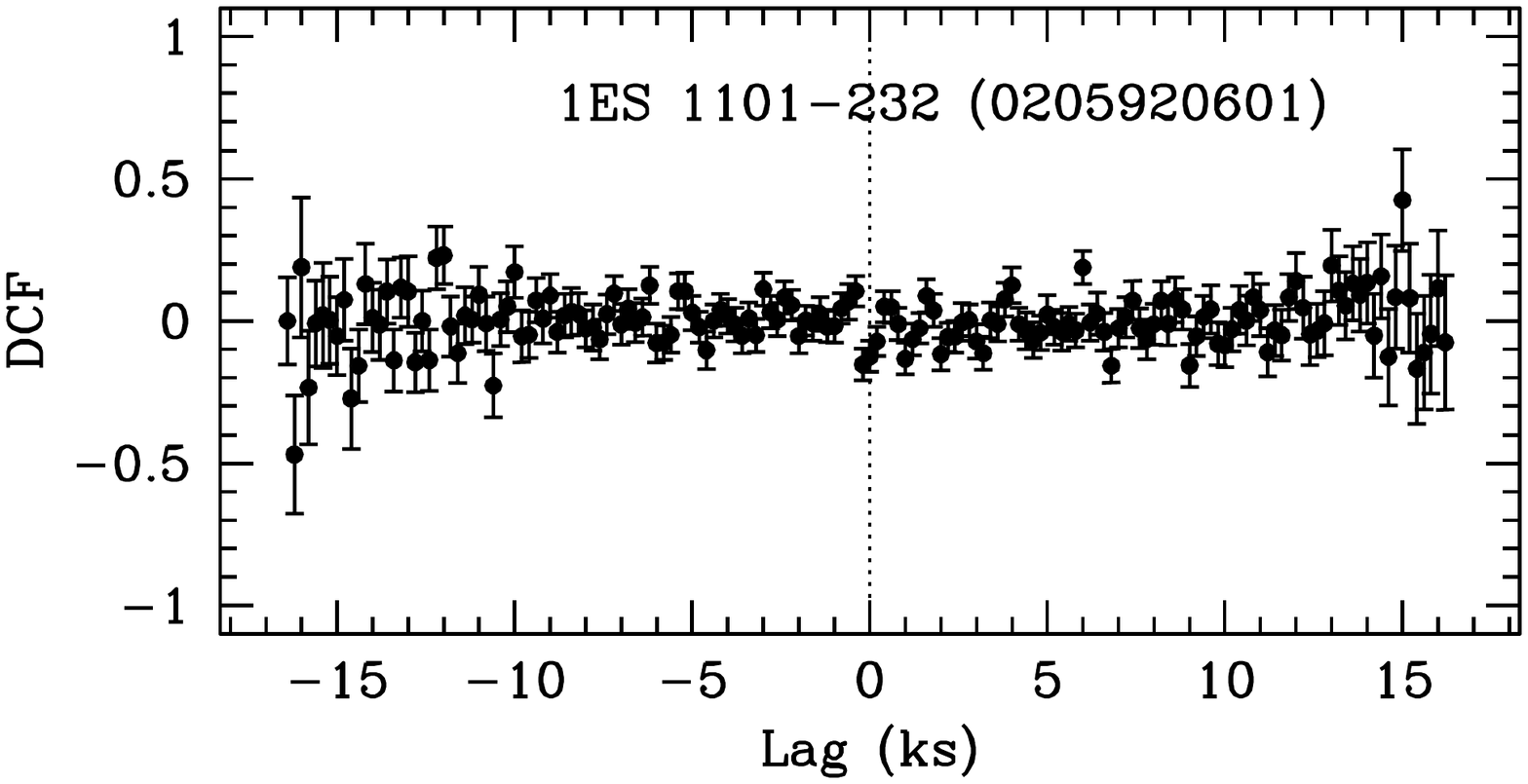}

\vspace*{-2.8in}
\includegraphics[scale=0.4]{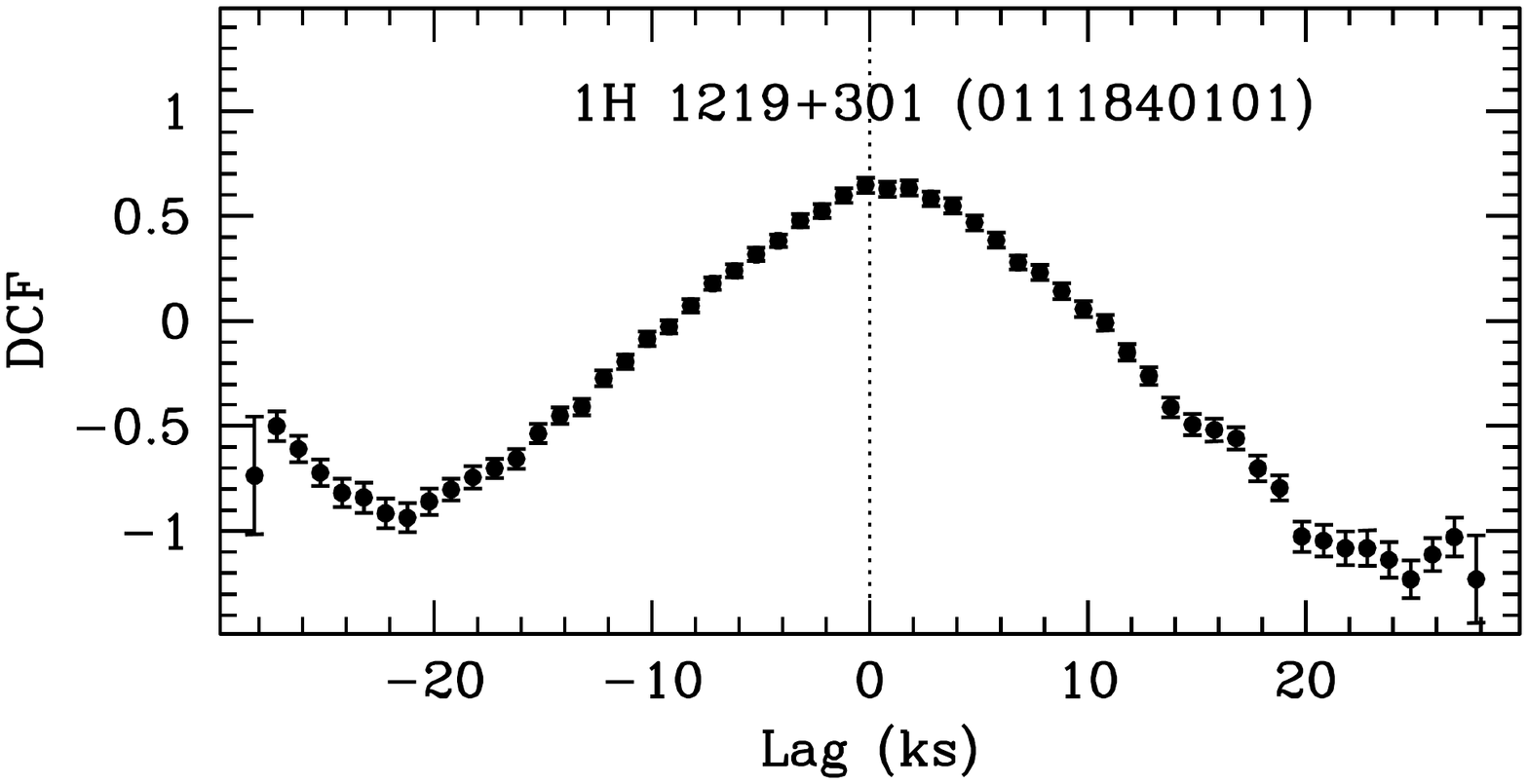}
\includegraphics[scale=0.4]{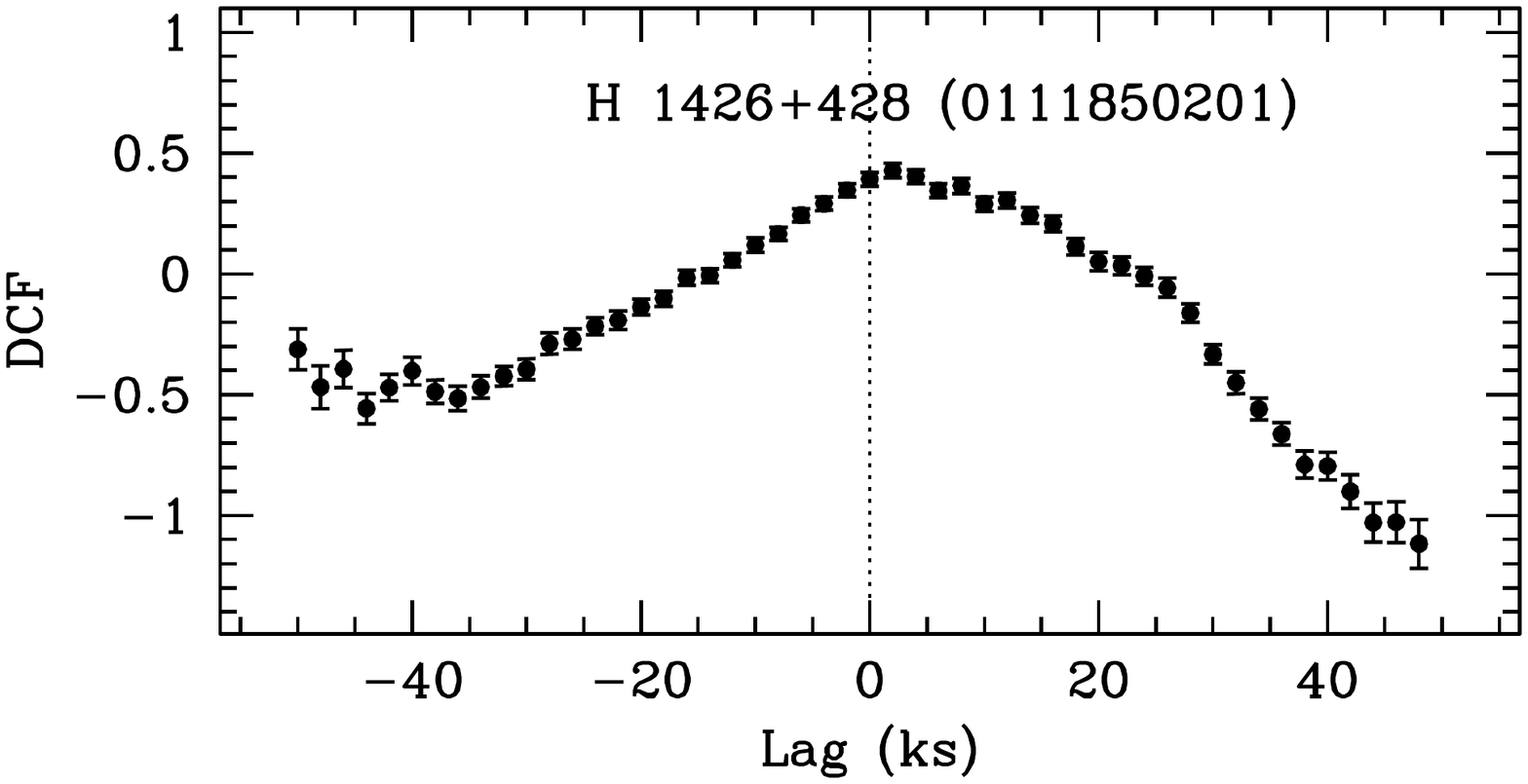}

\vspace*{-2.8in}
\includegraphics[scale=0.4]{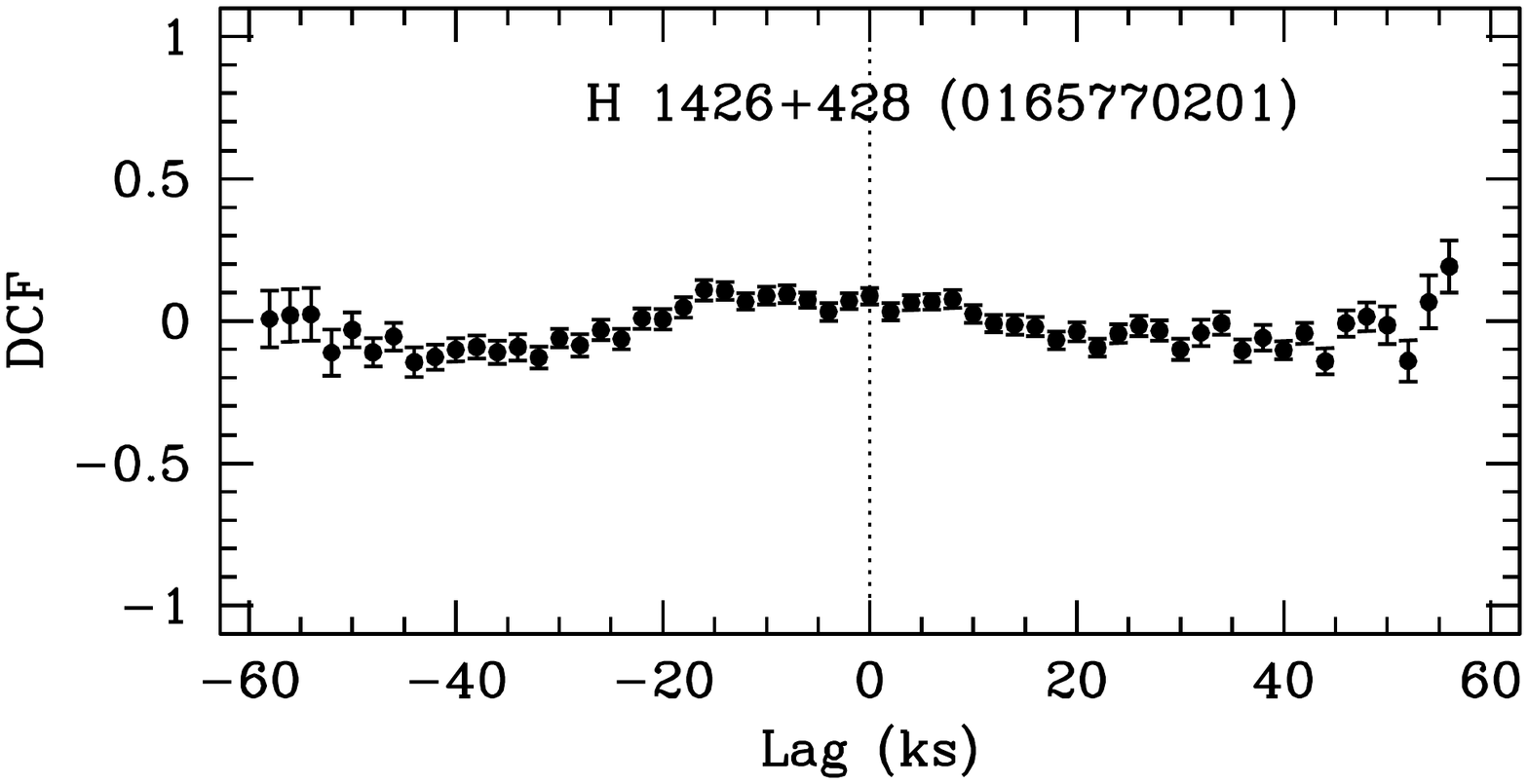}
\includegraphics[scale=0.4]{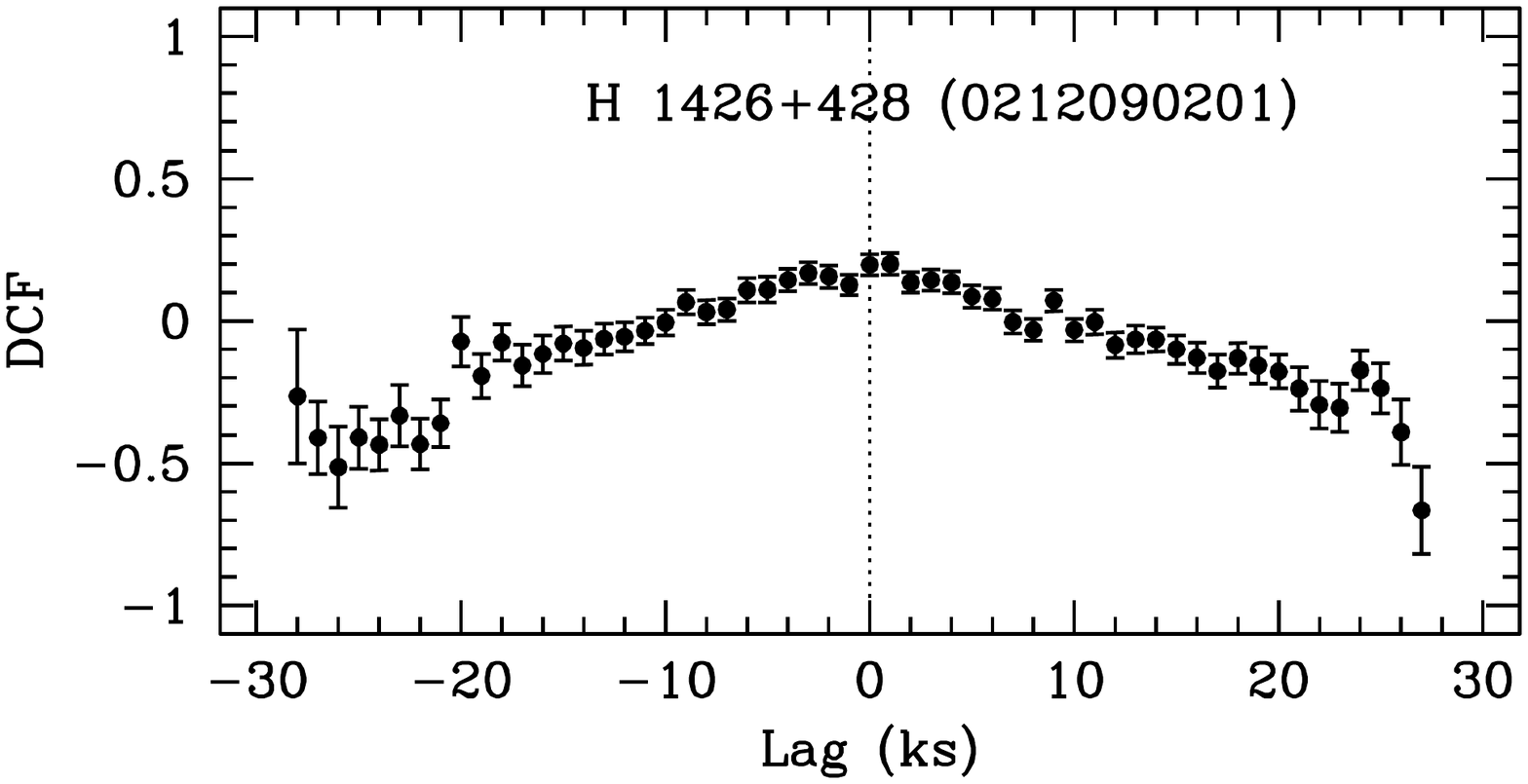}

\vspace*{-2.8in}
\includegraphics[scale=0.4]{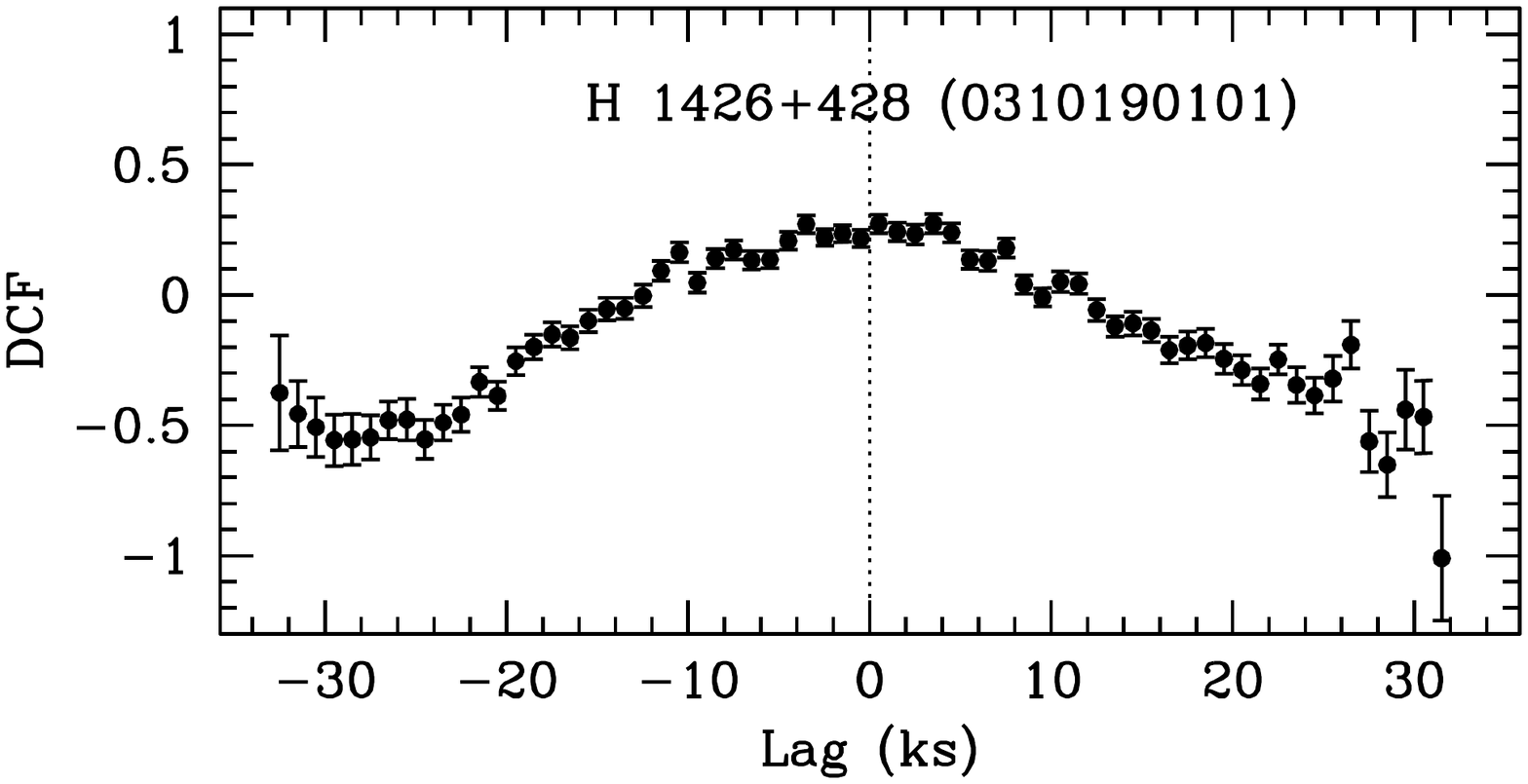}
\includegraphics[scale=0.4]{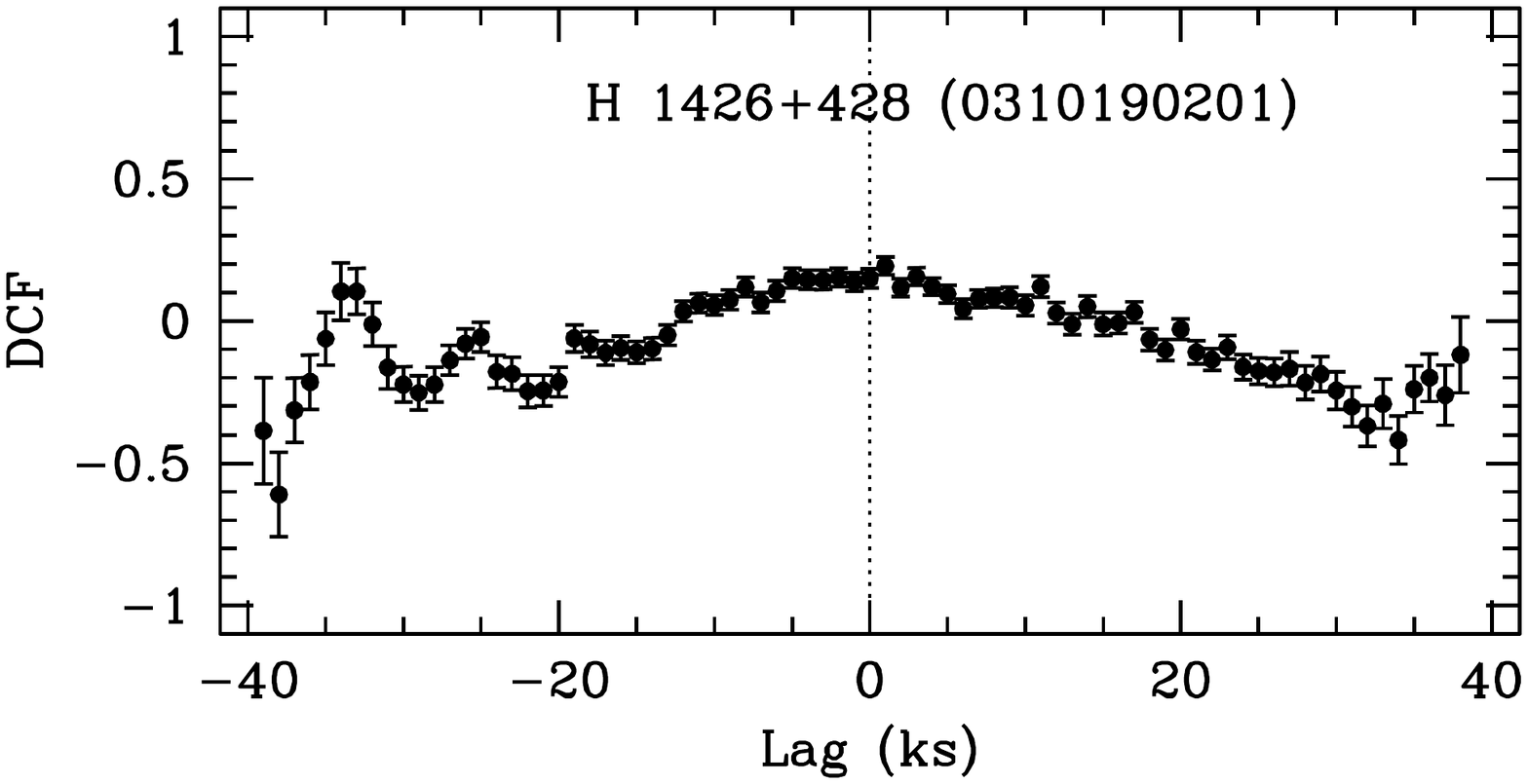}

\vspace{-0.7in}
\caption{Discrete Correlation Function (DCF) plots for variable light  curves labeled with source names and Observation IDs.\label{A3}
}

\end{figure*}
\clearpage


\setcounter{figure}{2}

\begin{figure*}
\centering

\vspace*{-1.5in}
\includegraphics[scale=0.4]{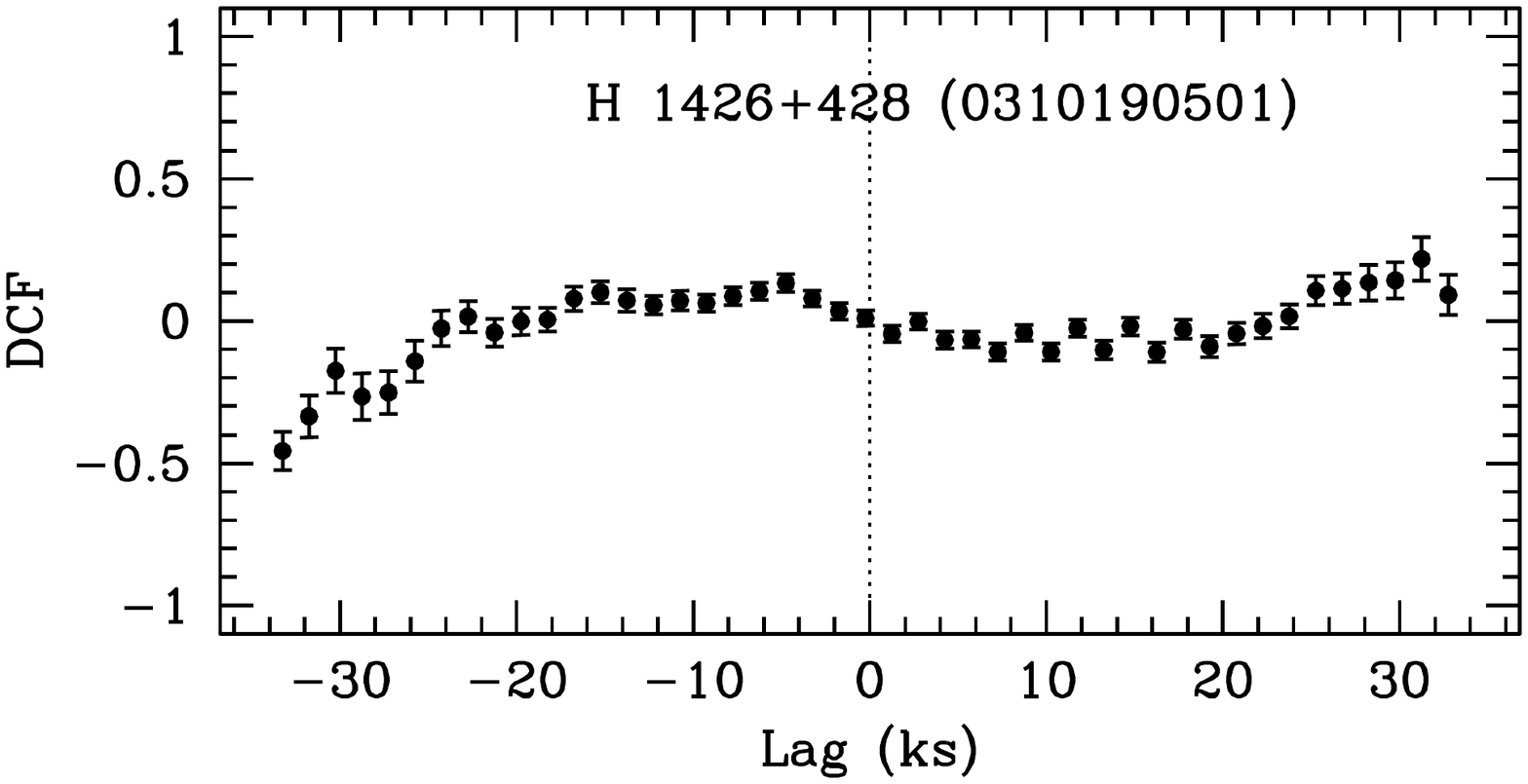}
\includegraphics[scale=0.4]{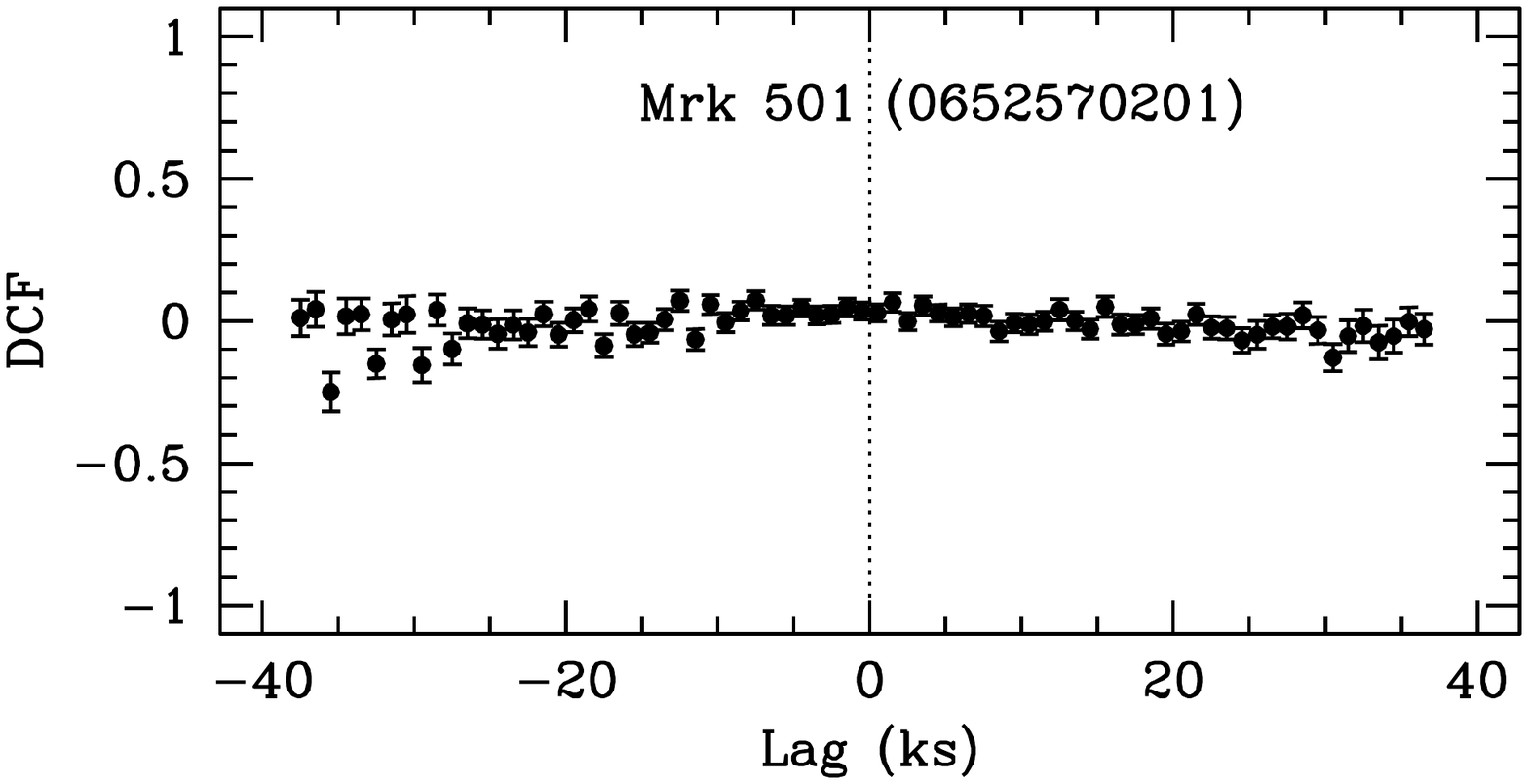}

\vspace*{-2.8in}
\includegraphics[scale=0.4]{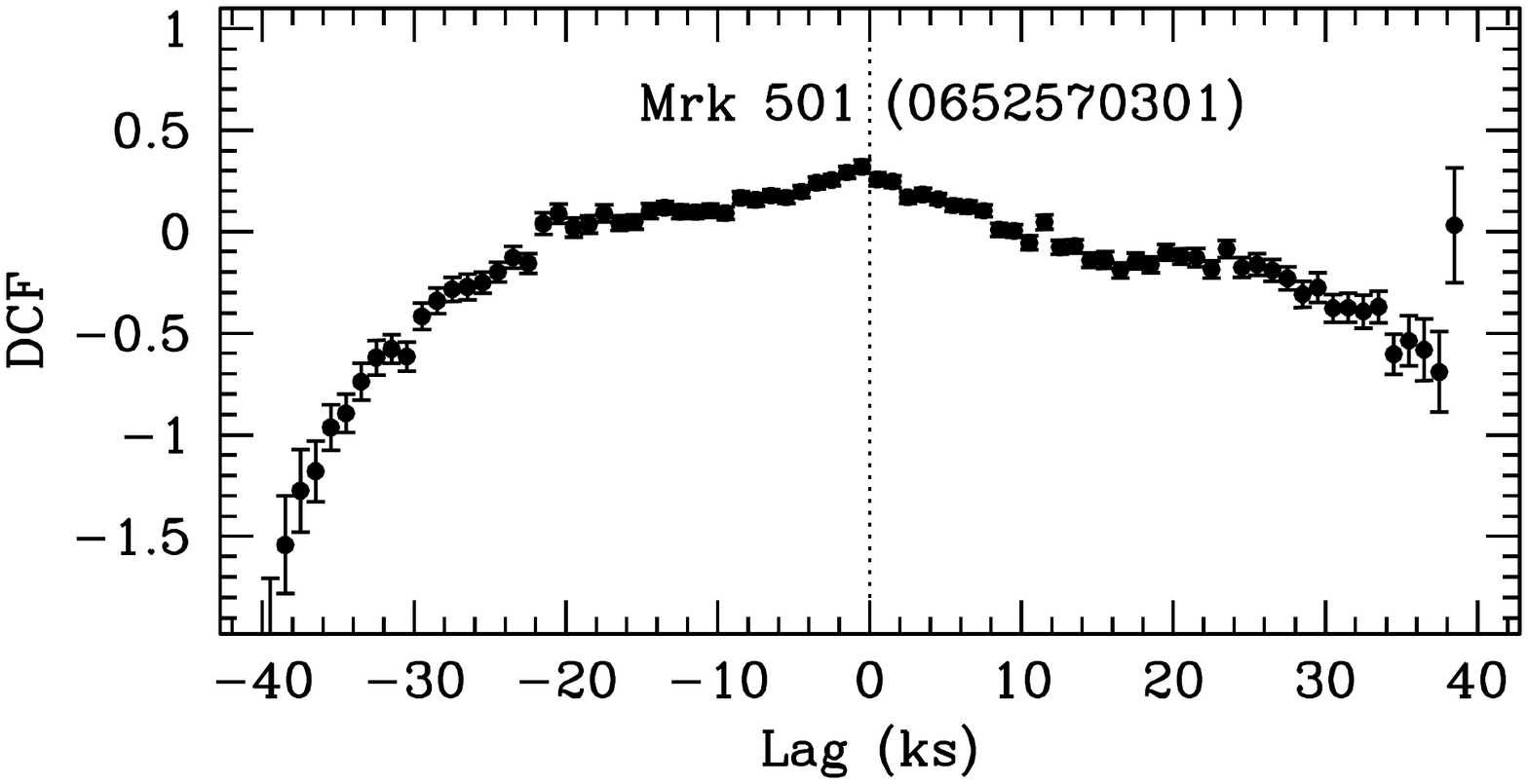}
\includegraphics[scale=0.4]{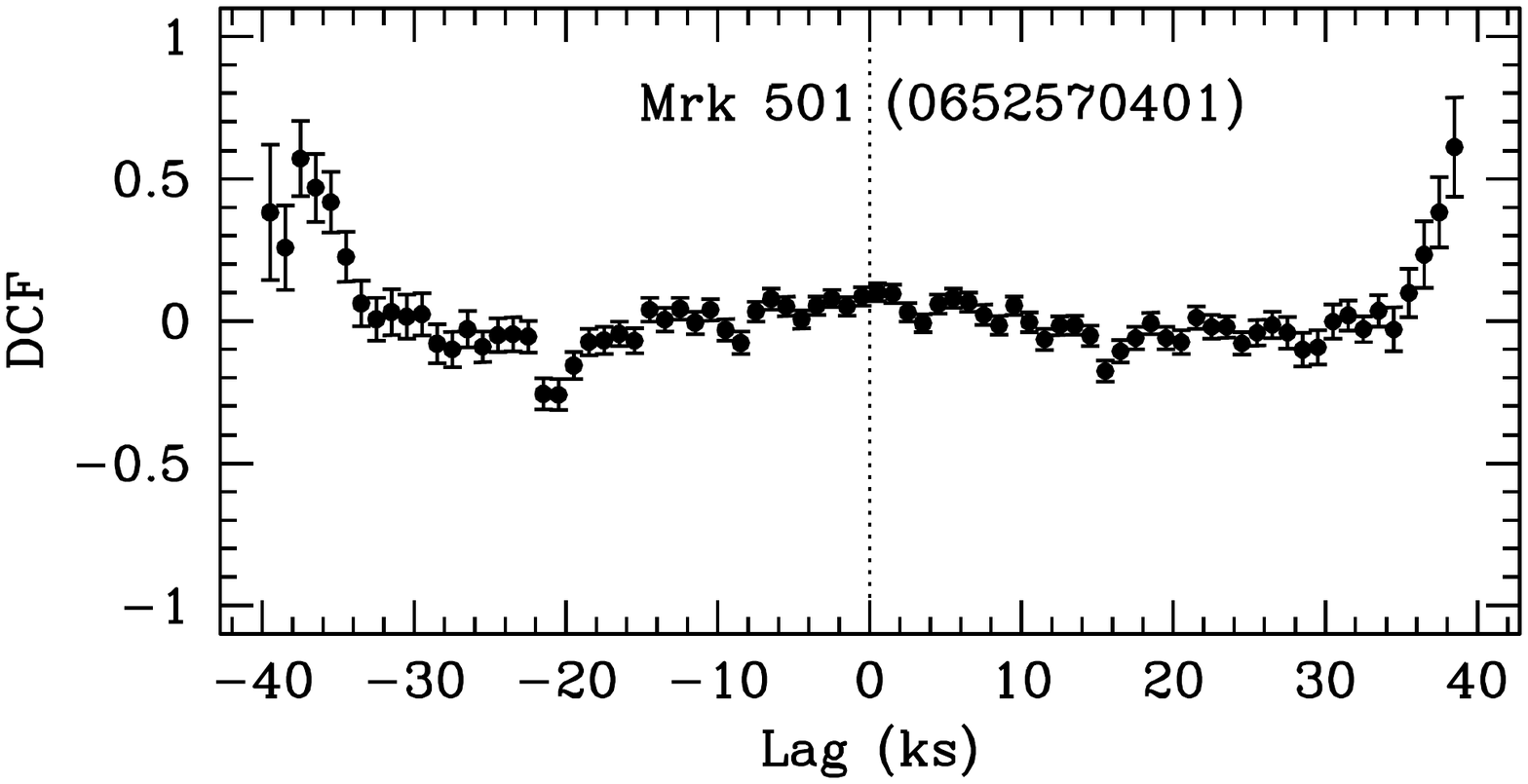}

\vspace*{-2.8in}
\includegraphics[scale=0.4]{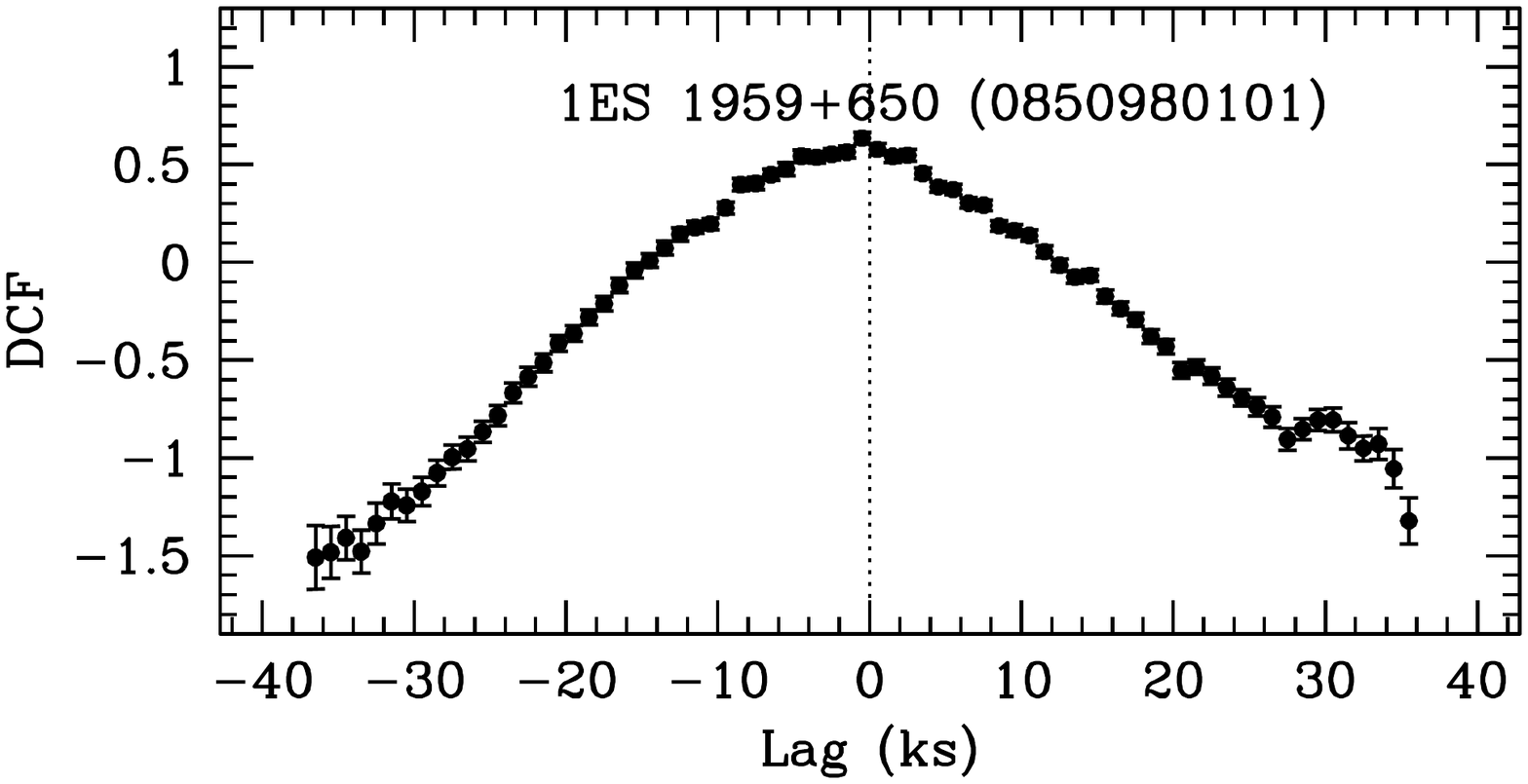}
\includegraphics[scale=0.4]{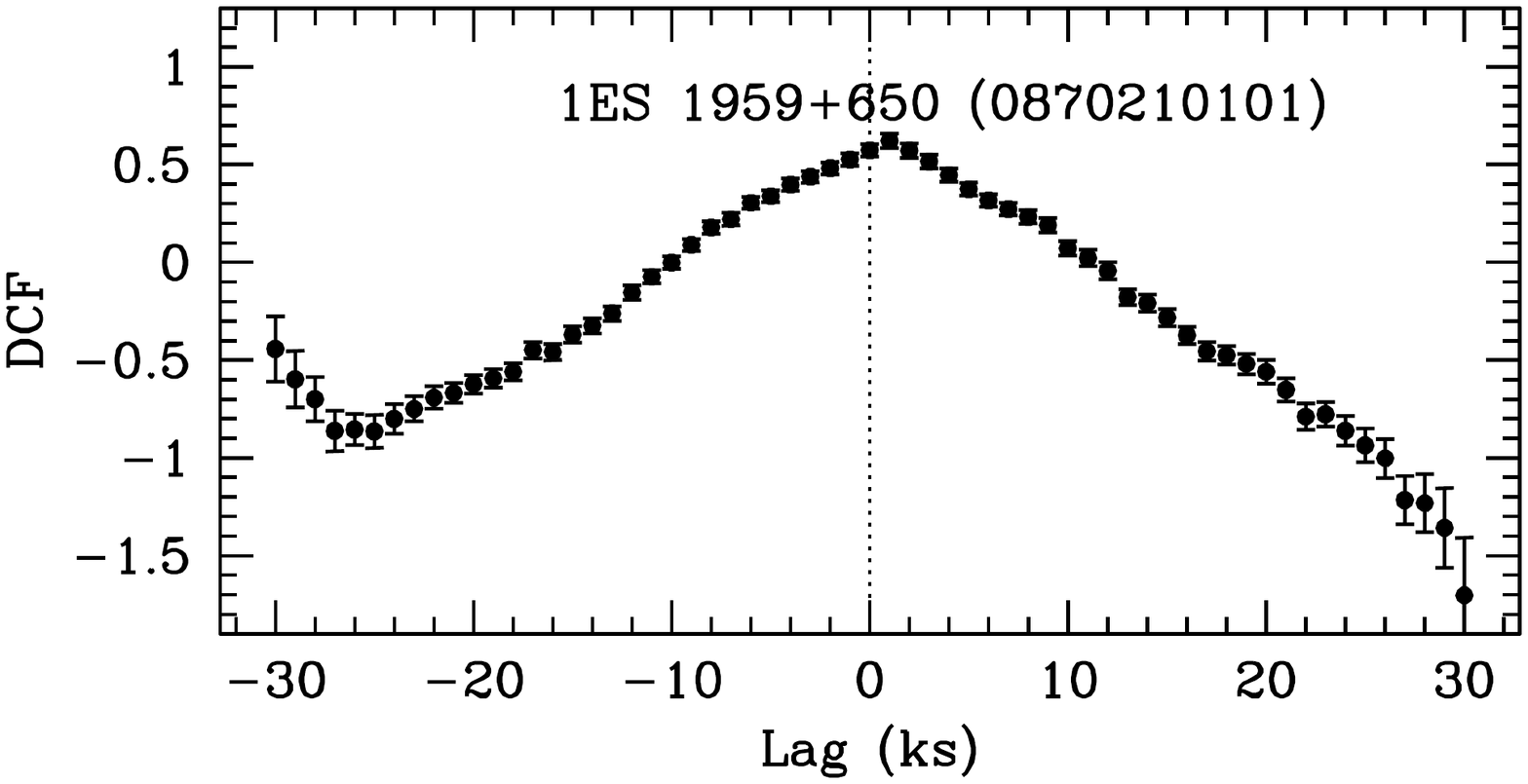}

\vspace*{-2.8in}
\includegraphics[scale=0.4]{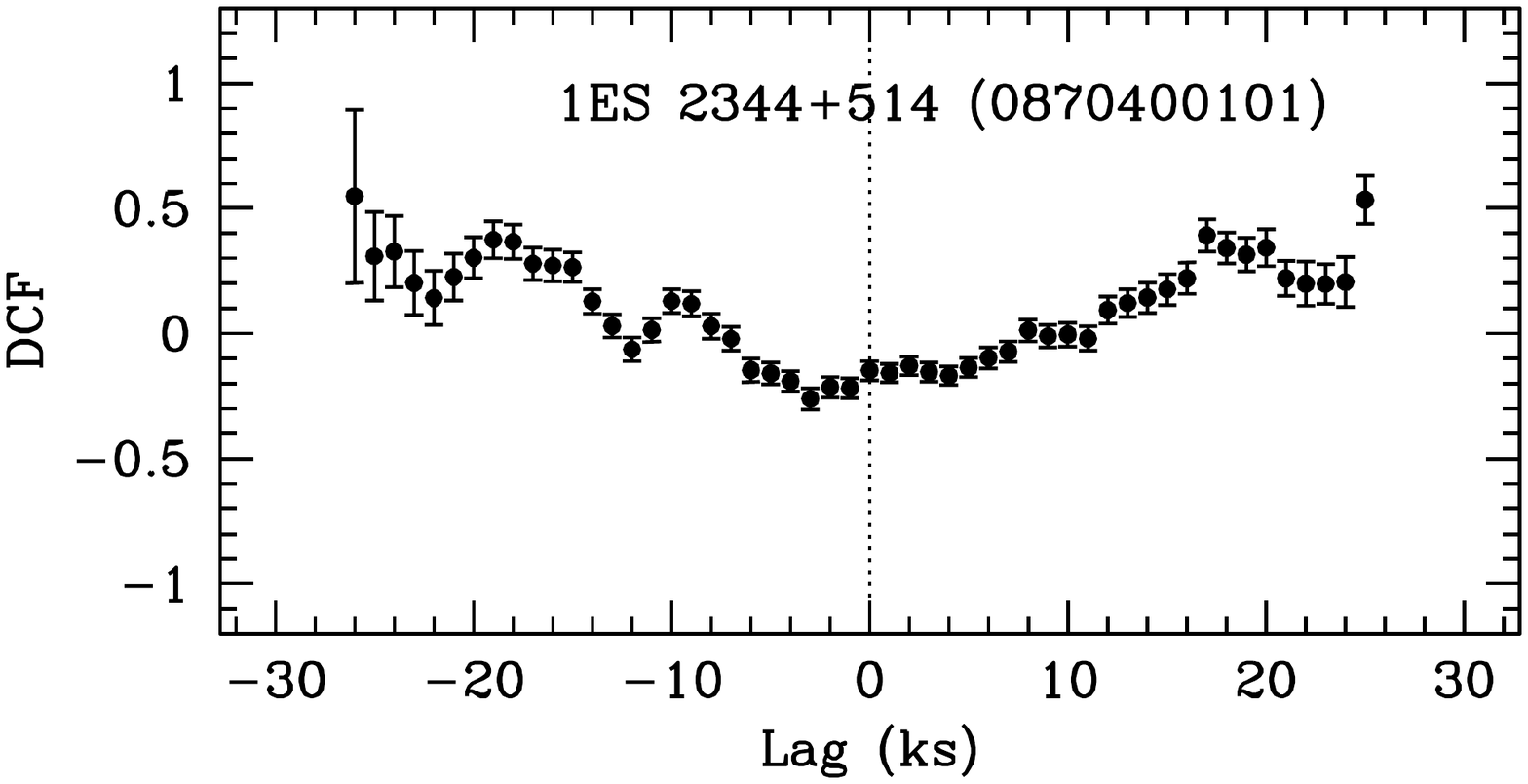}

\vspace{-0.7in}
\caption{Continued.}  

\end{figure*}

\setcounter{figure}{3}

\begin{figure*}
\centering
\vspace*{-1.5in}
\includegraphics[scale=0.4]{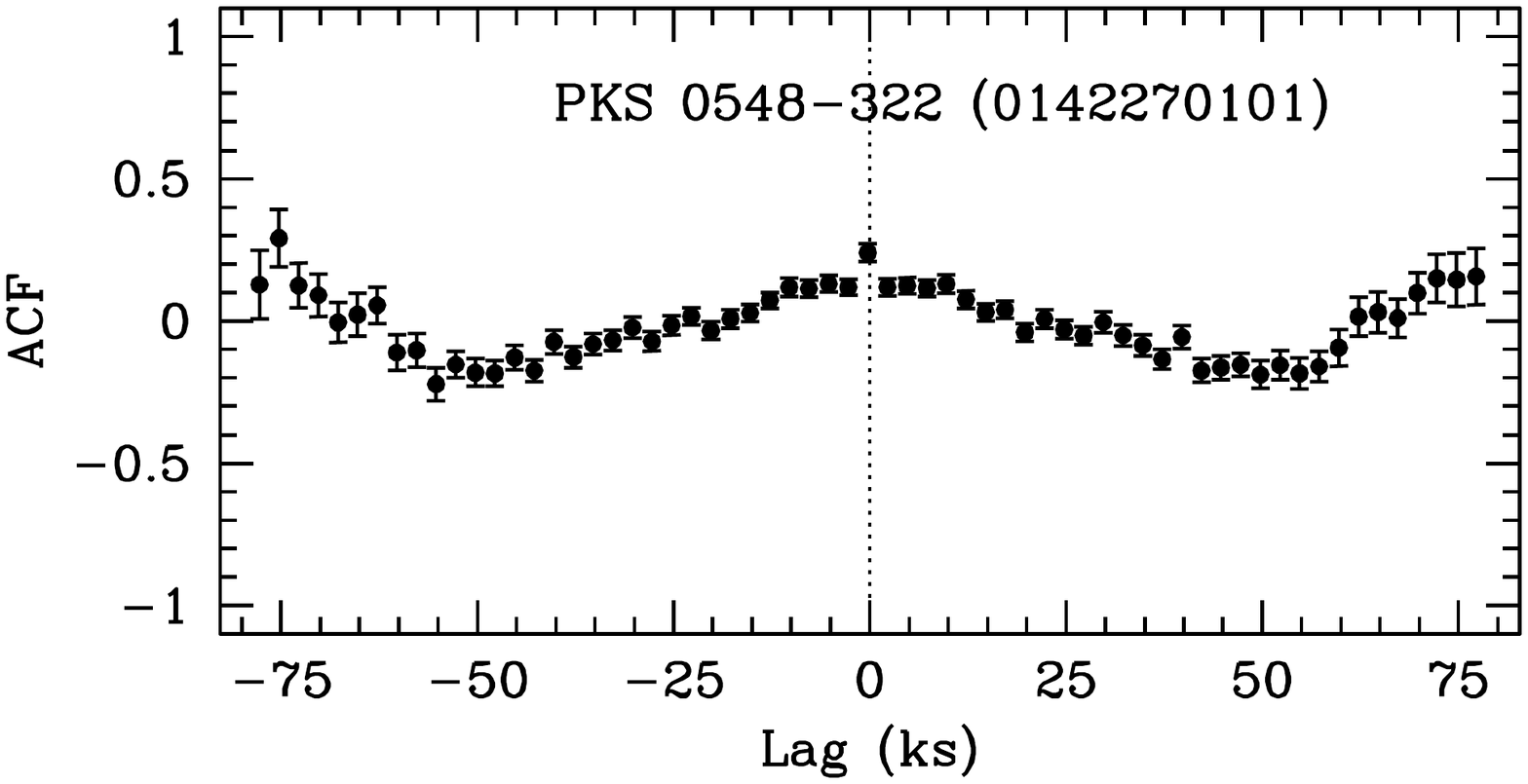}
\includegraphics[scale=0.4]{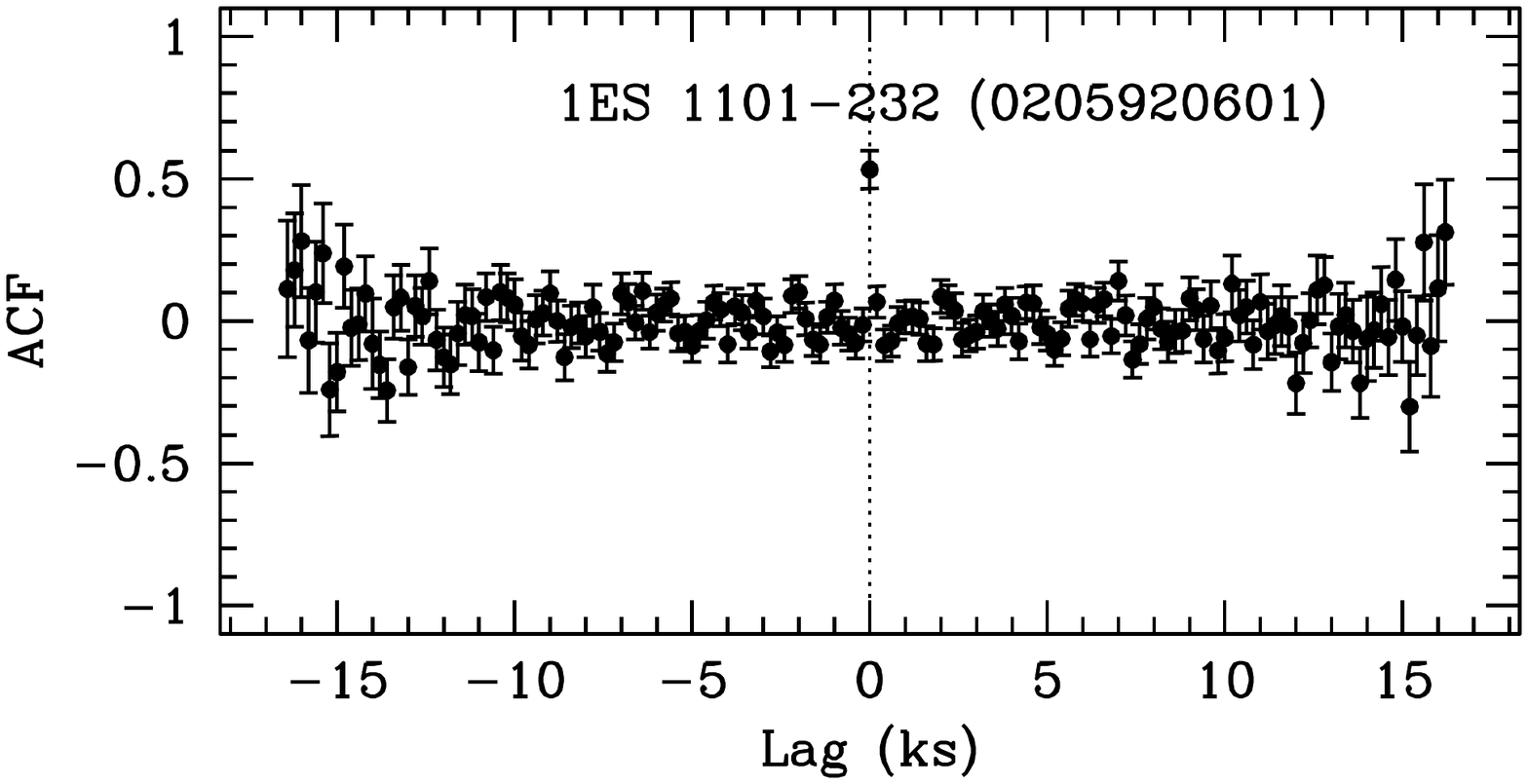}

\vspace*{-2.8in}
\includegraphics[scale=0.4]{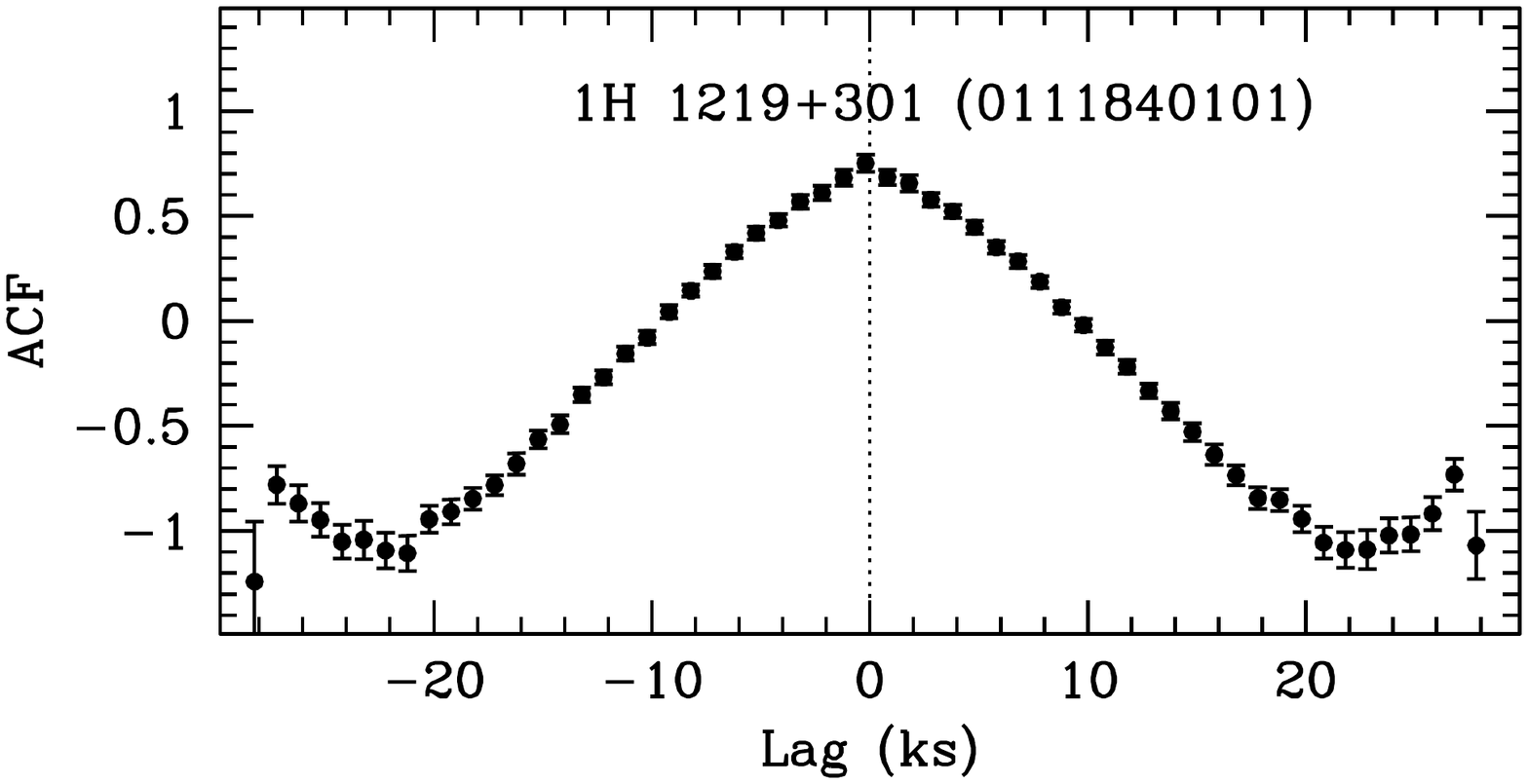}
\includegraphics[scale=0.4]{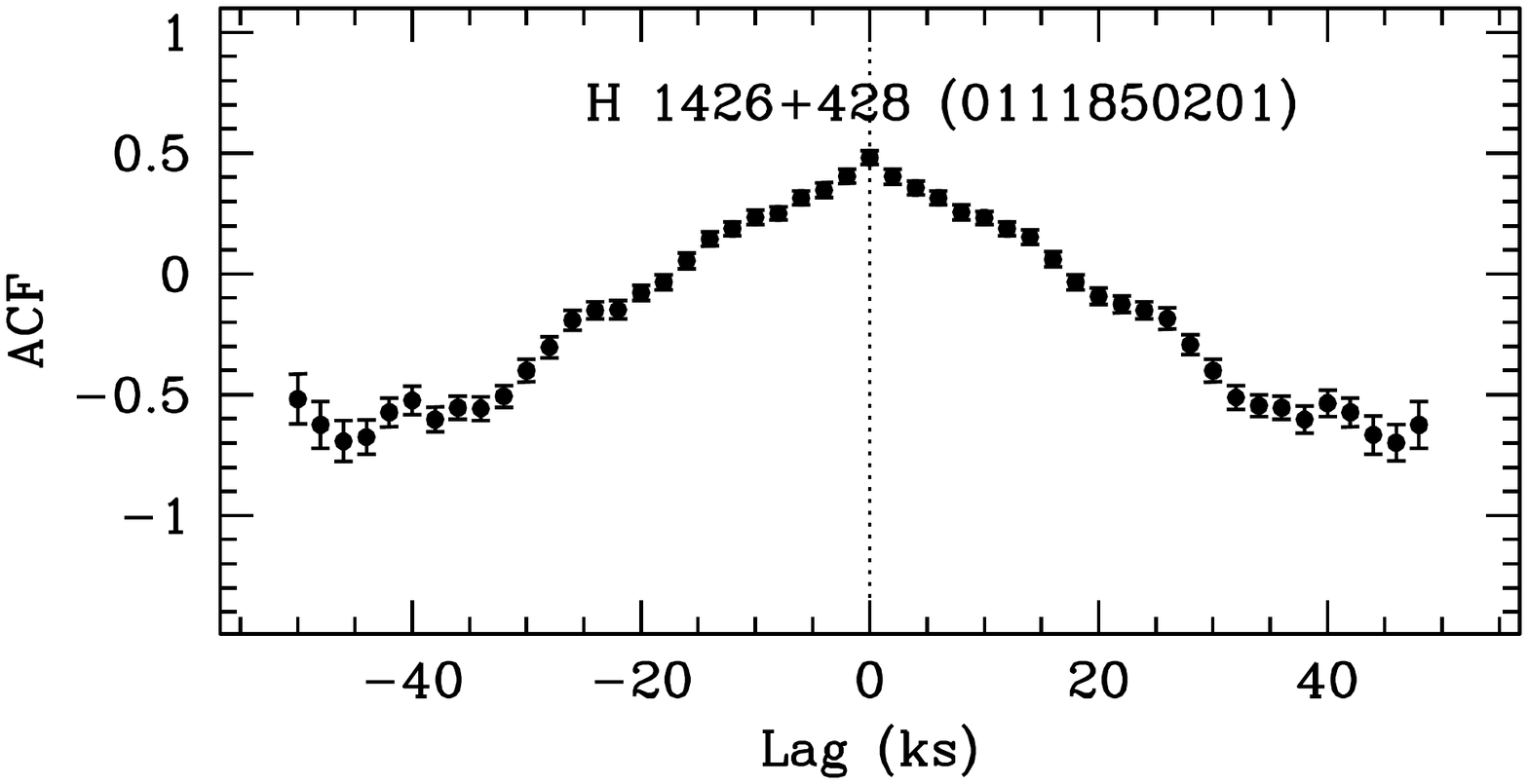}

\vspace*{-2.8in}
\includegraphics[scale=0.4]{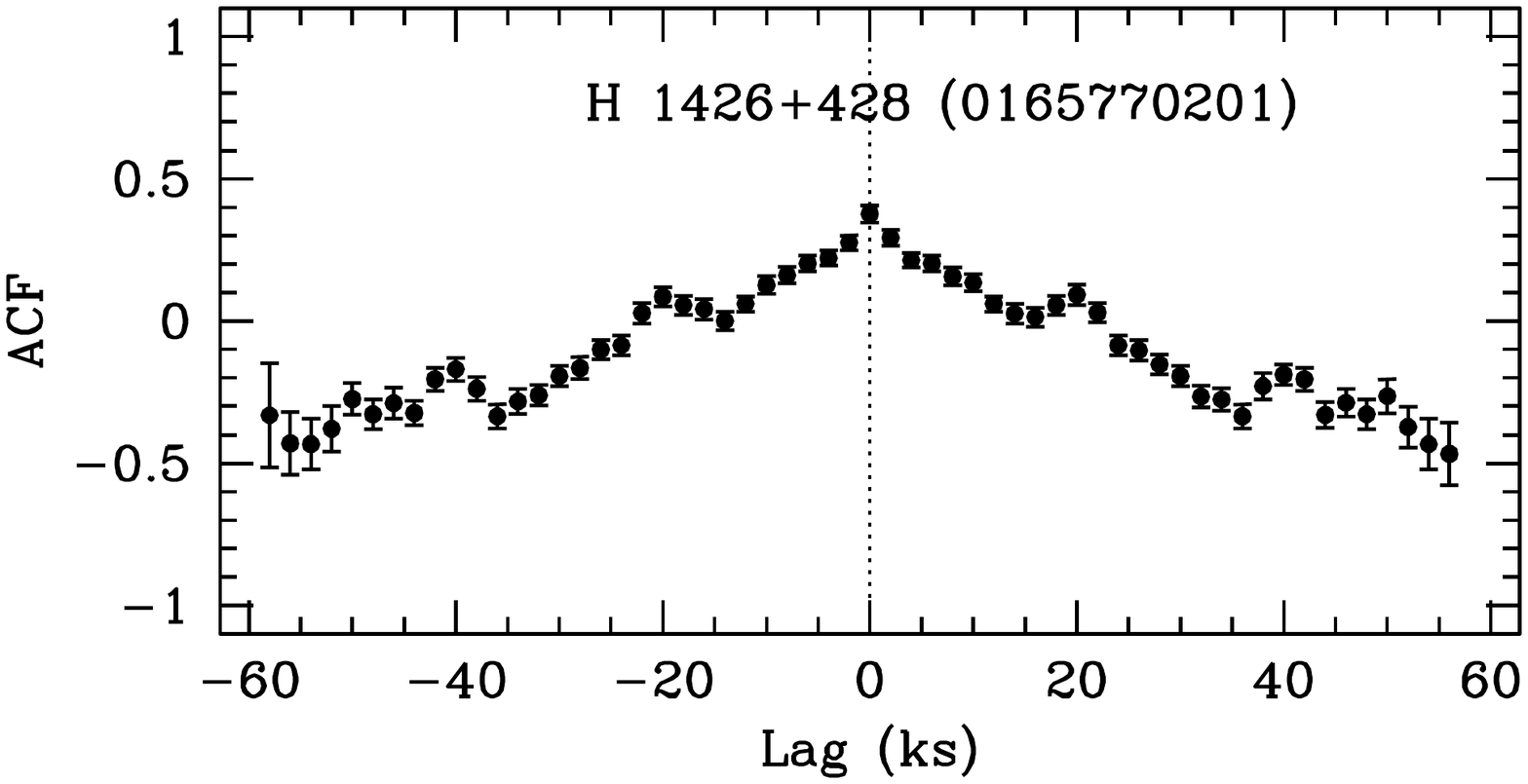}
\includegraphics[scale=0.4]{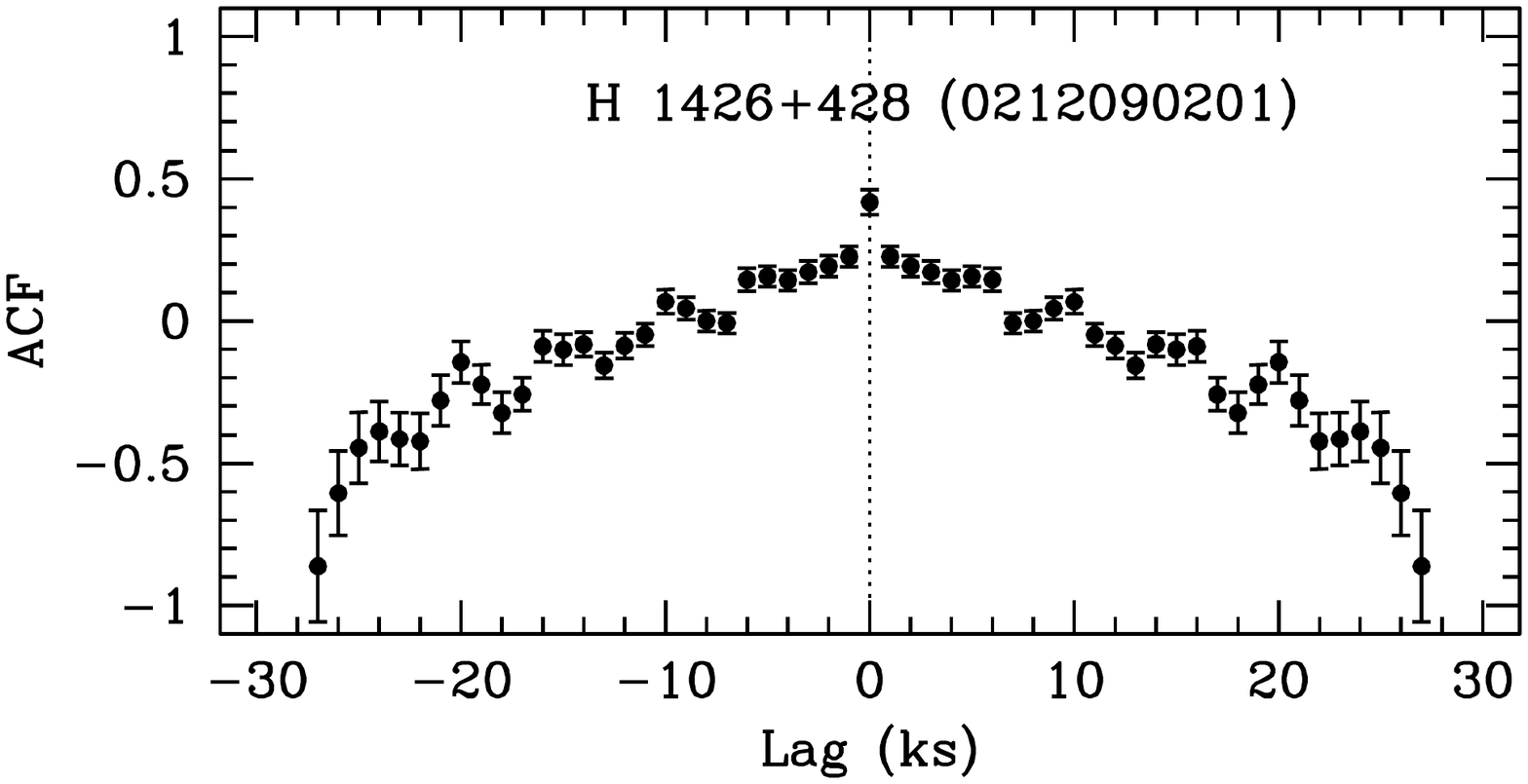}

\vspace*{-2.8in}
\includegraphics[scale=0.4]{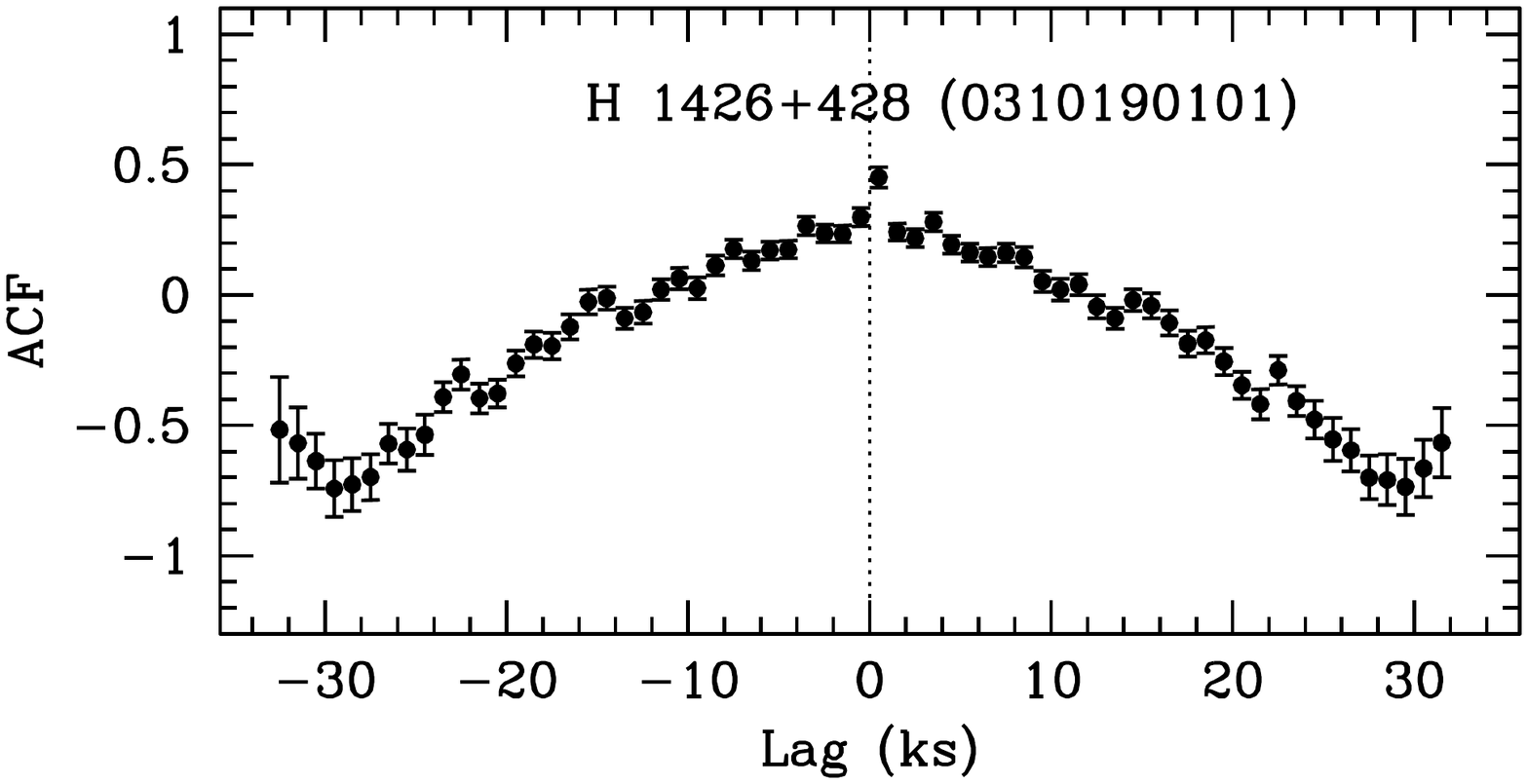}
\includegraphics[scale=0.4]{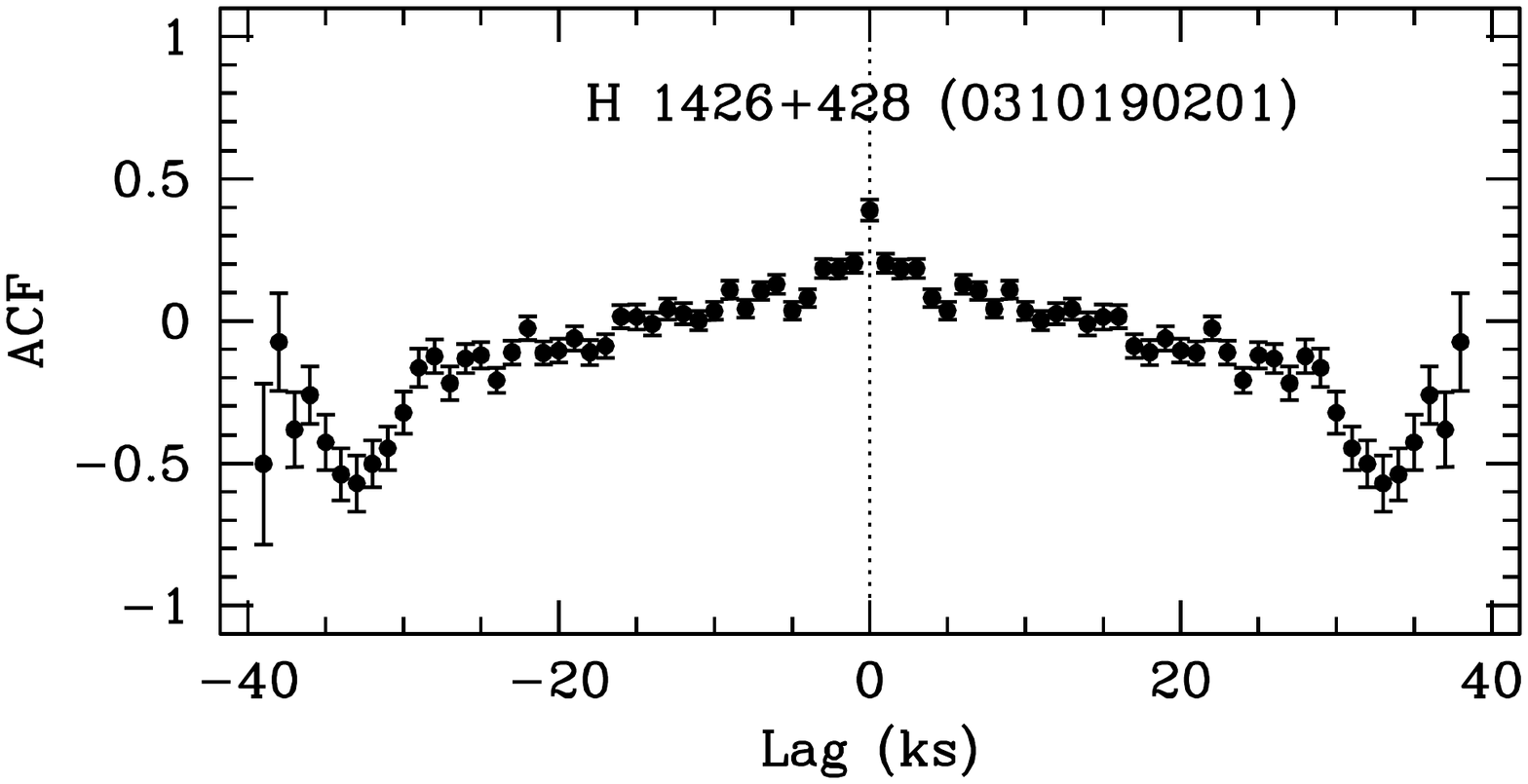}

\vspace{-0.7in}
\caption{Auto Correlation Function (ACF) plots for variable light  curves labeled with source names and Observation IDs.\label{A4}}

\end{figure*}
\clearpage


\setcounter{figure}{3}

\begin{figure*}
\centering

\vspace*{-1.5in}
\includegraphics[scale=0.4]{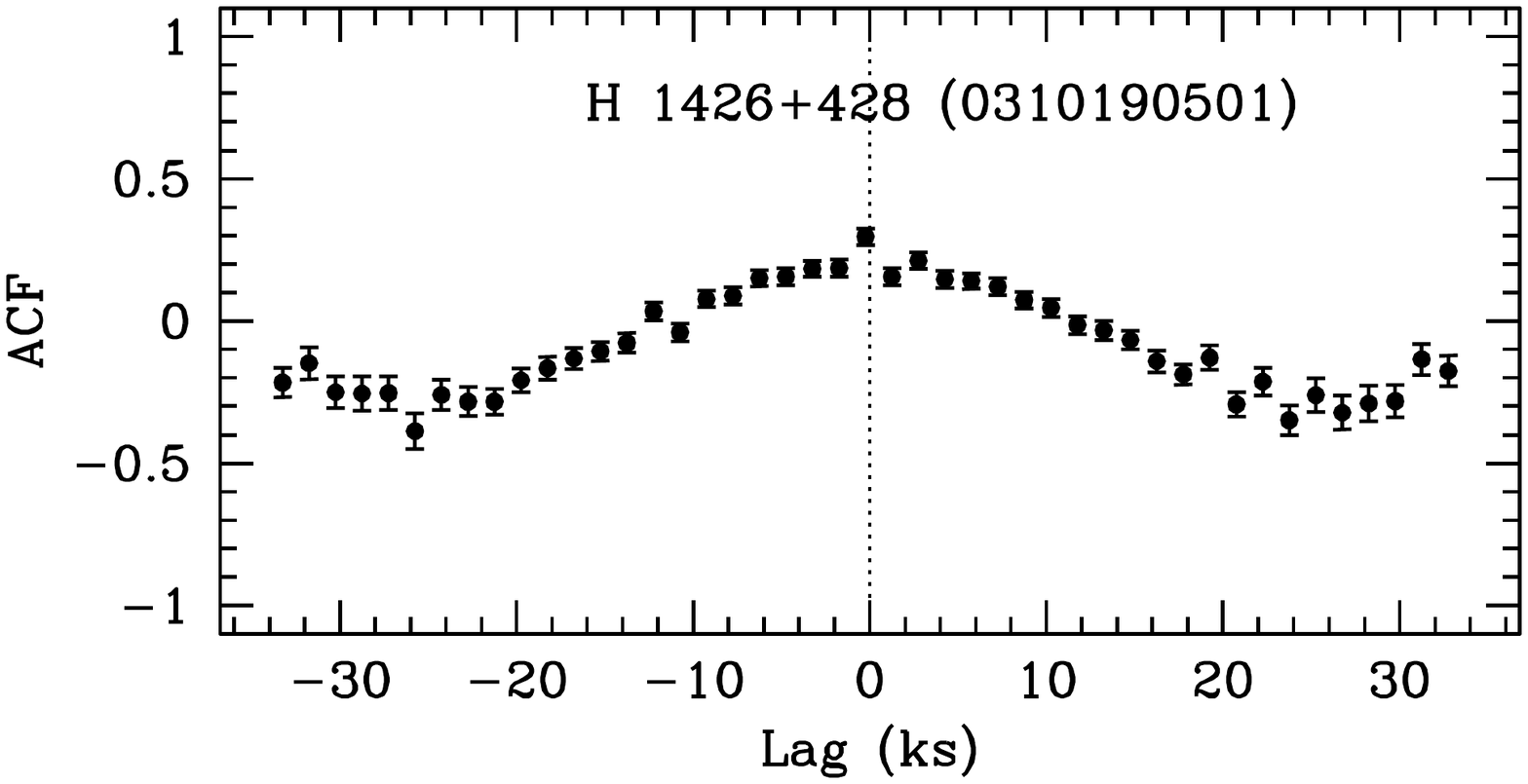}
\includegraphics[scale=0.4]{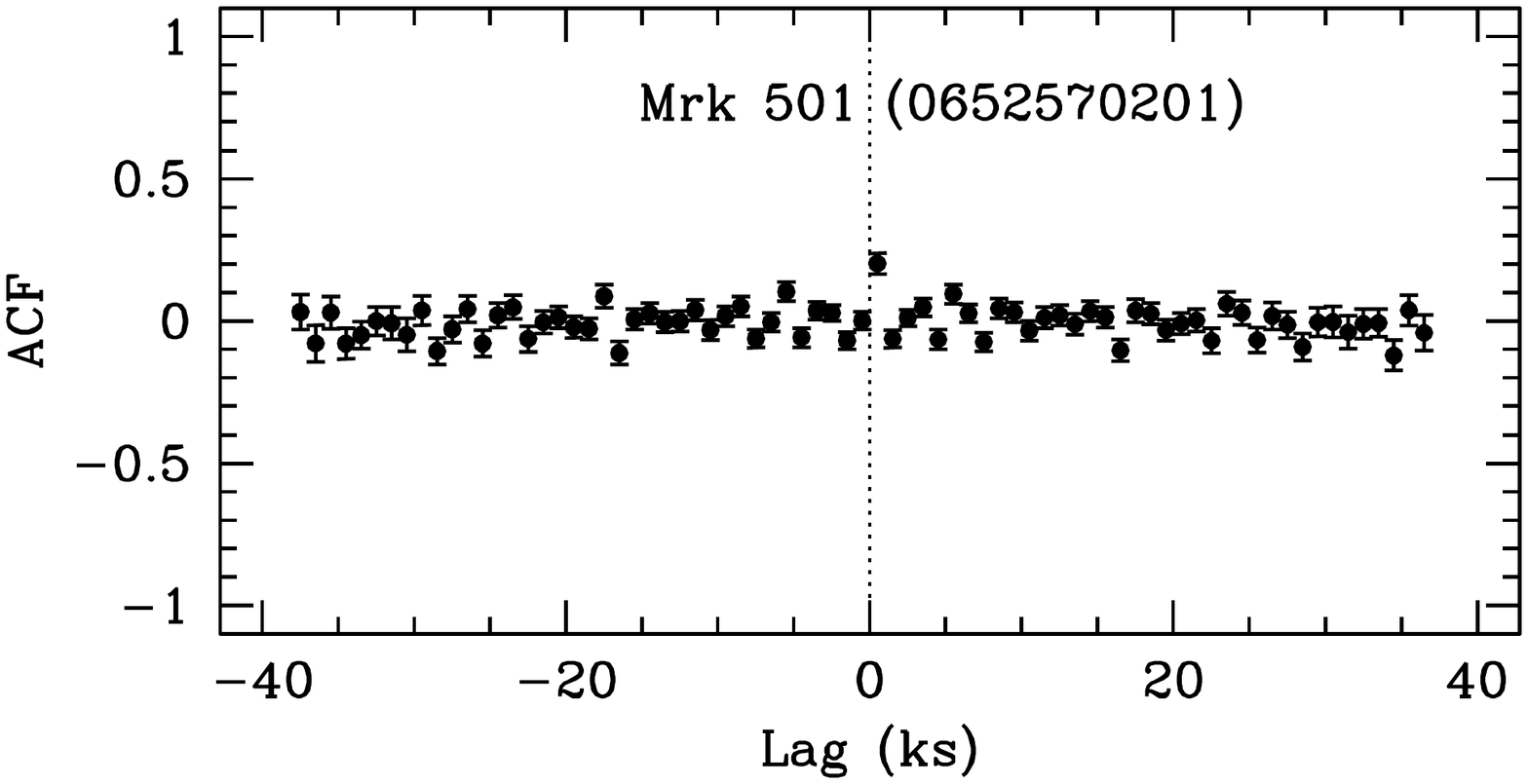}

\vspace*{-2.8in}
\includegraphics[scale=0.4]{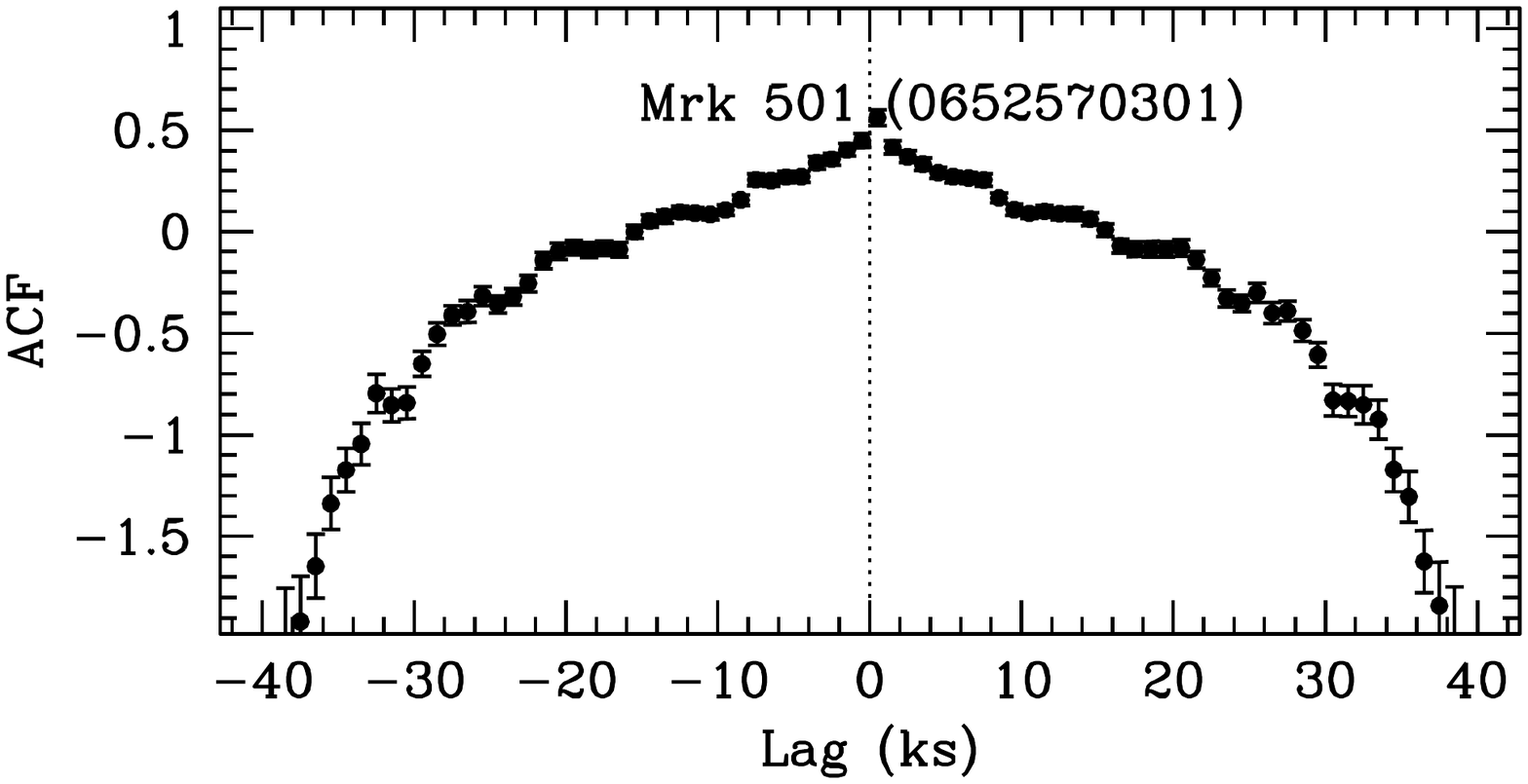}
\includegraphics[scale=0.4]{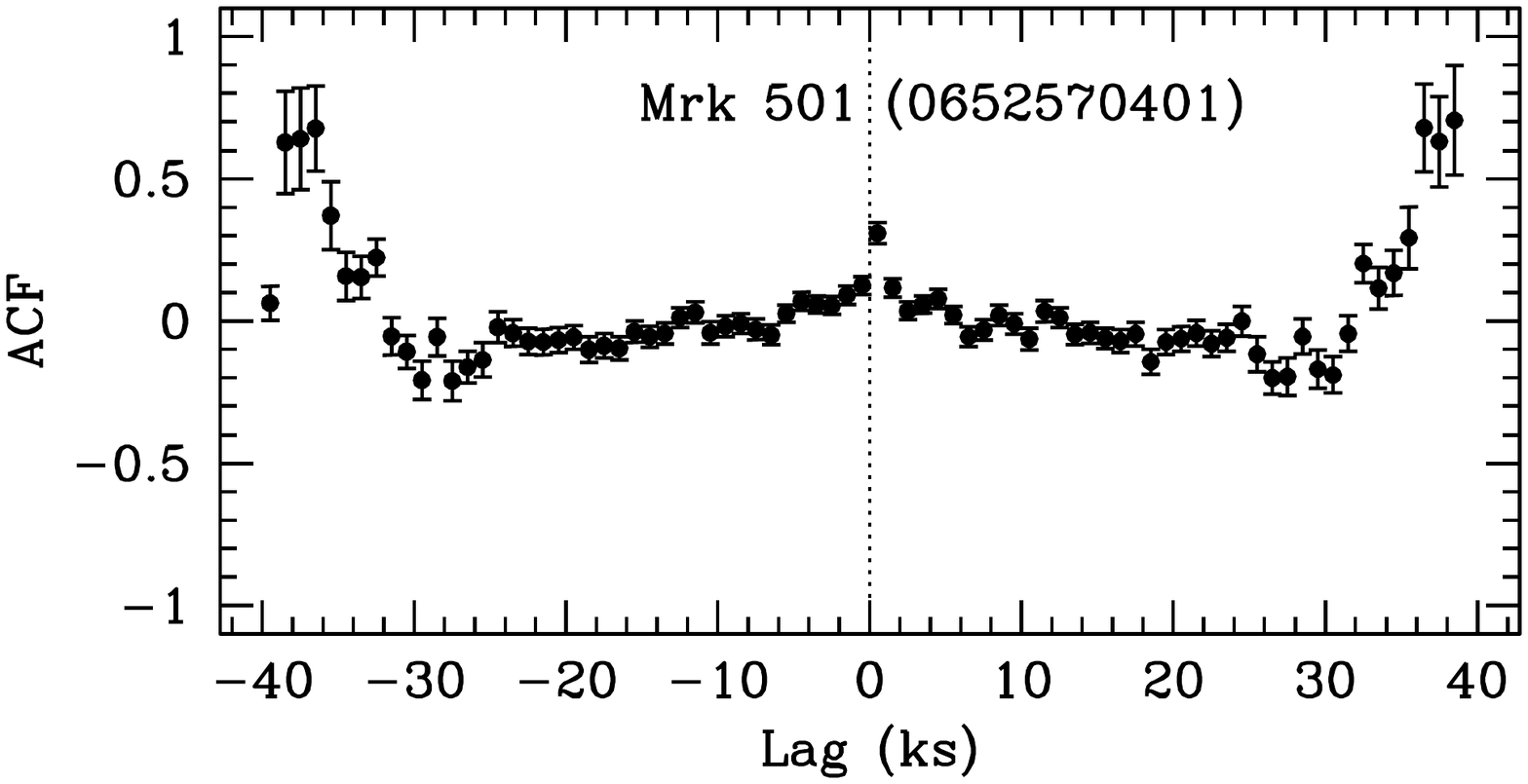}

\vspace*{-2.8in}
\includegraphics[scale=0.4]{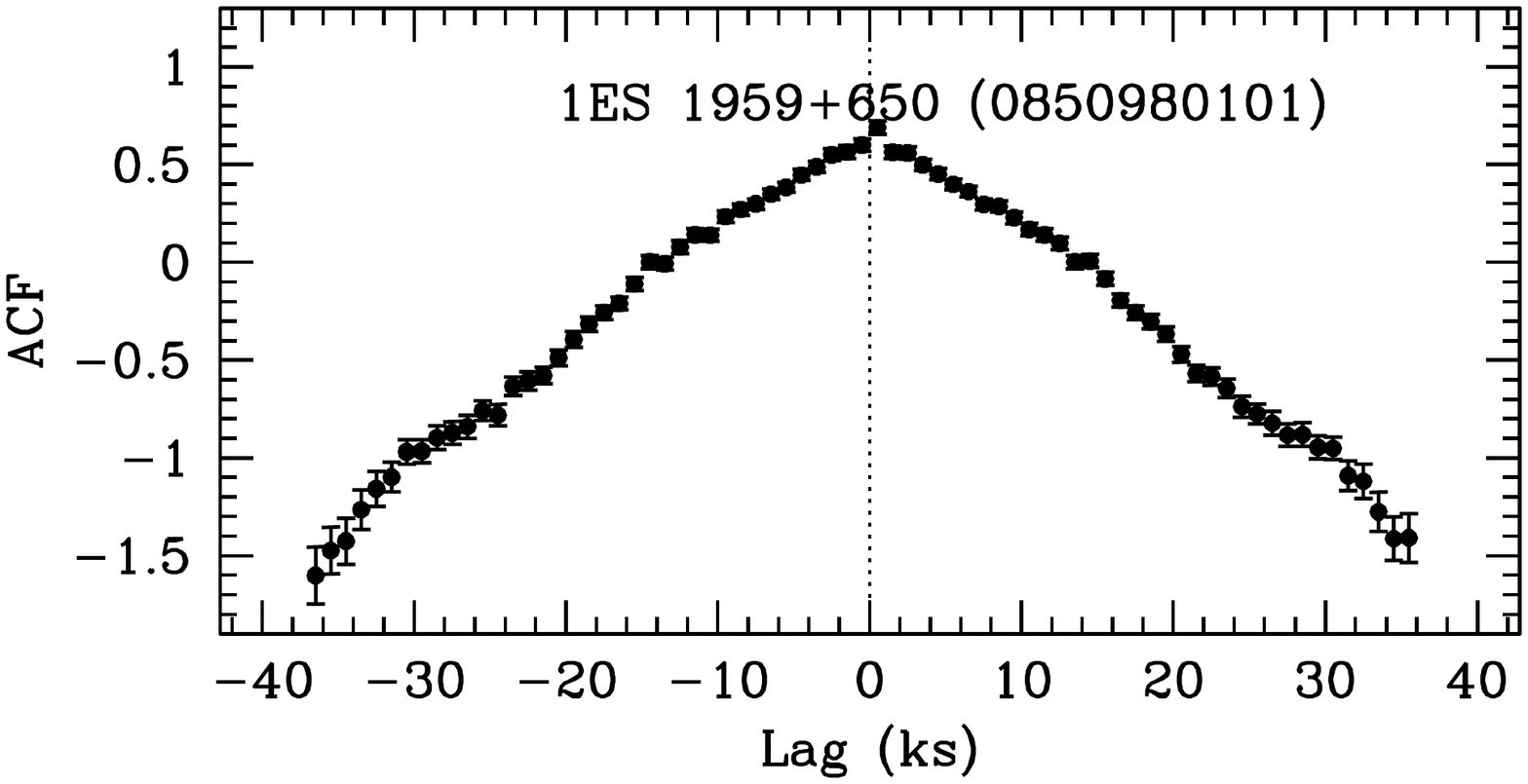}
\includegraphics[scale=0.4]{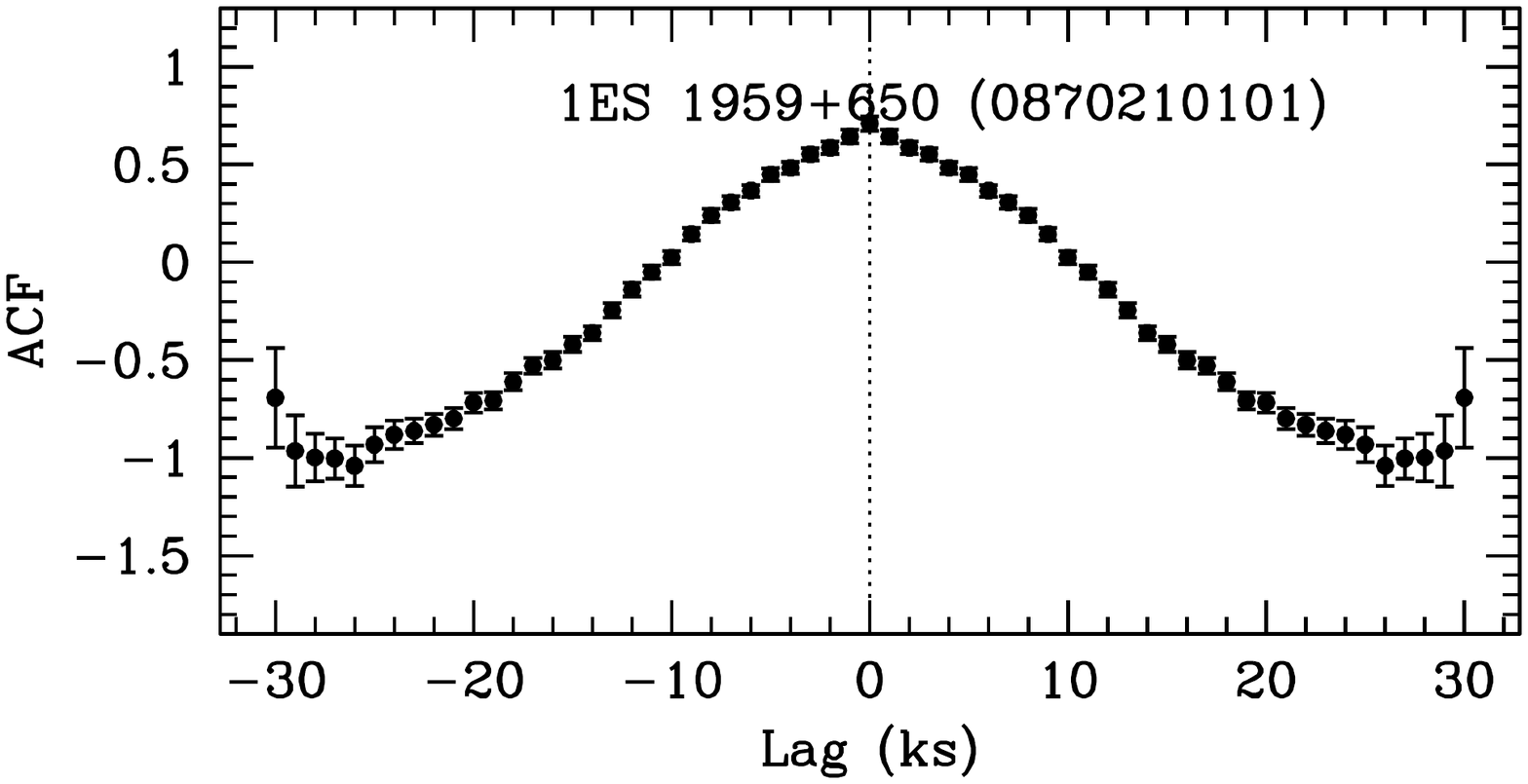}

\vspace*{-2.8in}
\includegraphics[scale=0.4]{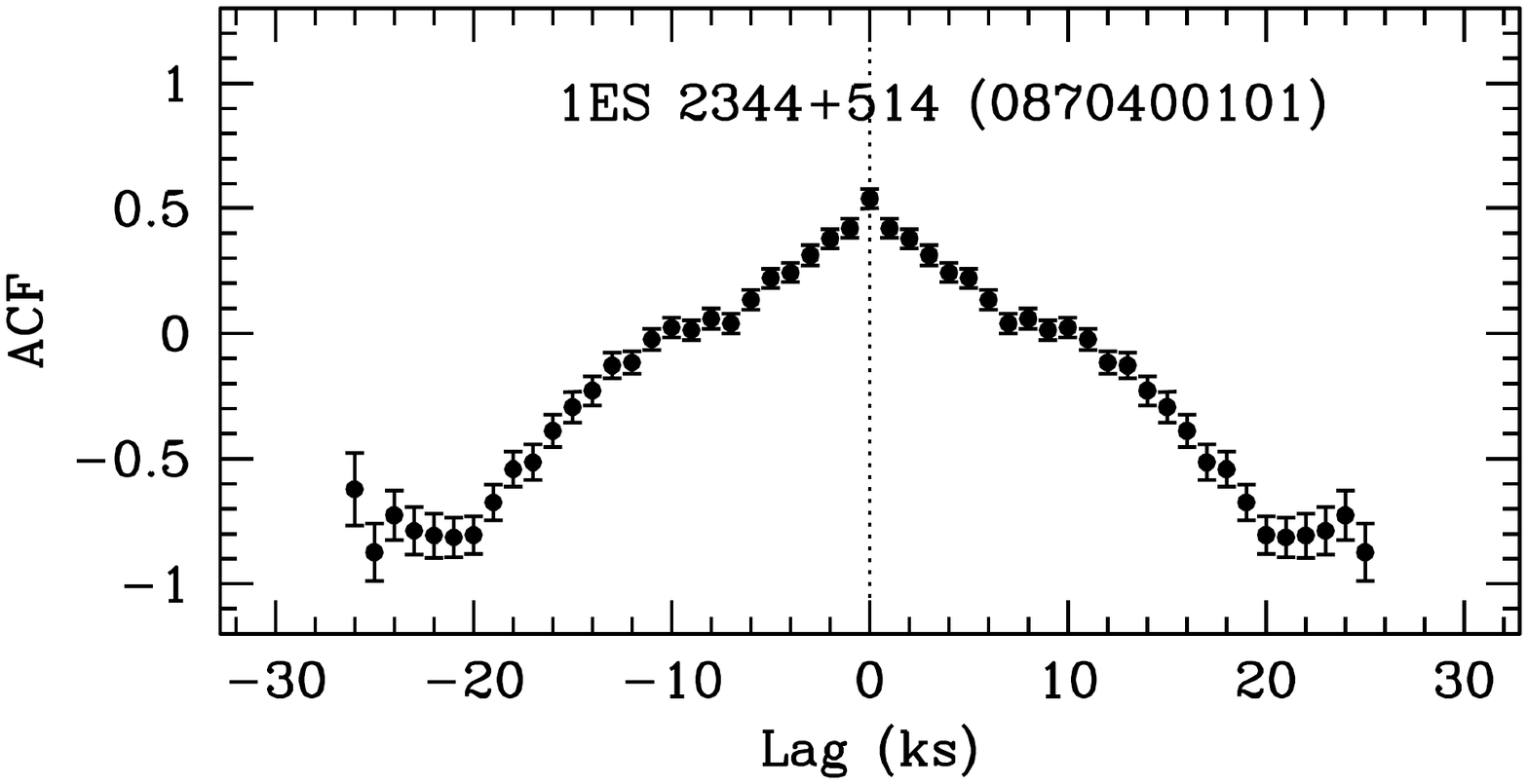}

\vspace{-0.7in}
\caption{Continued.}  

\end{figure*}

\setcounter{figure}{4}

\begin{figure*}
\centering
\vspace*{-1.5in}
\includegraphics[scale=0.4]{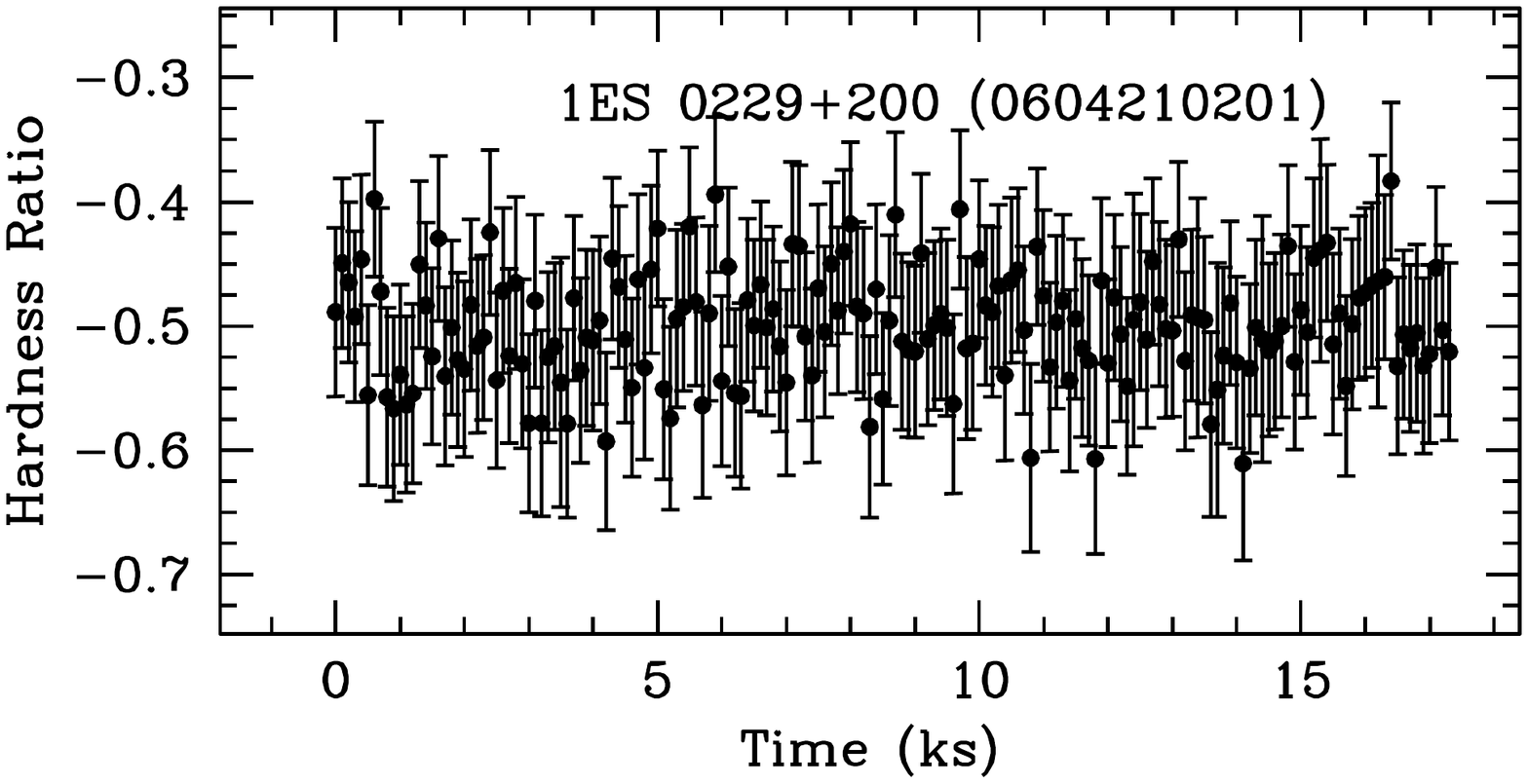}
\includegraphics[scale=0.4]{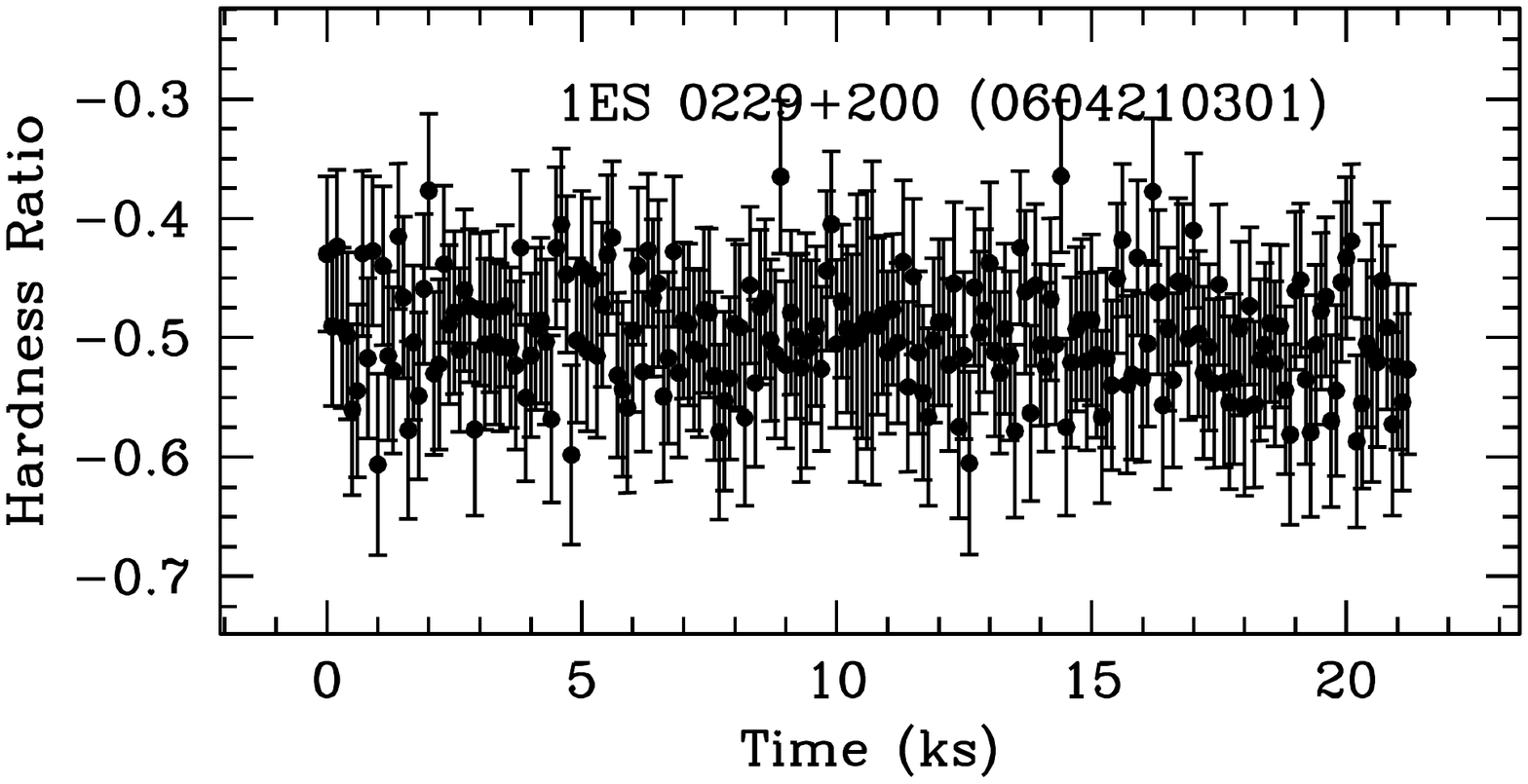}

\vspace*{-2.8in}
\includegraphics[scale=0.4]{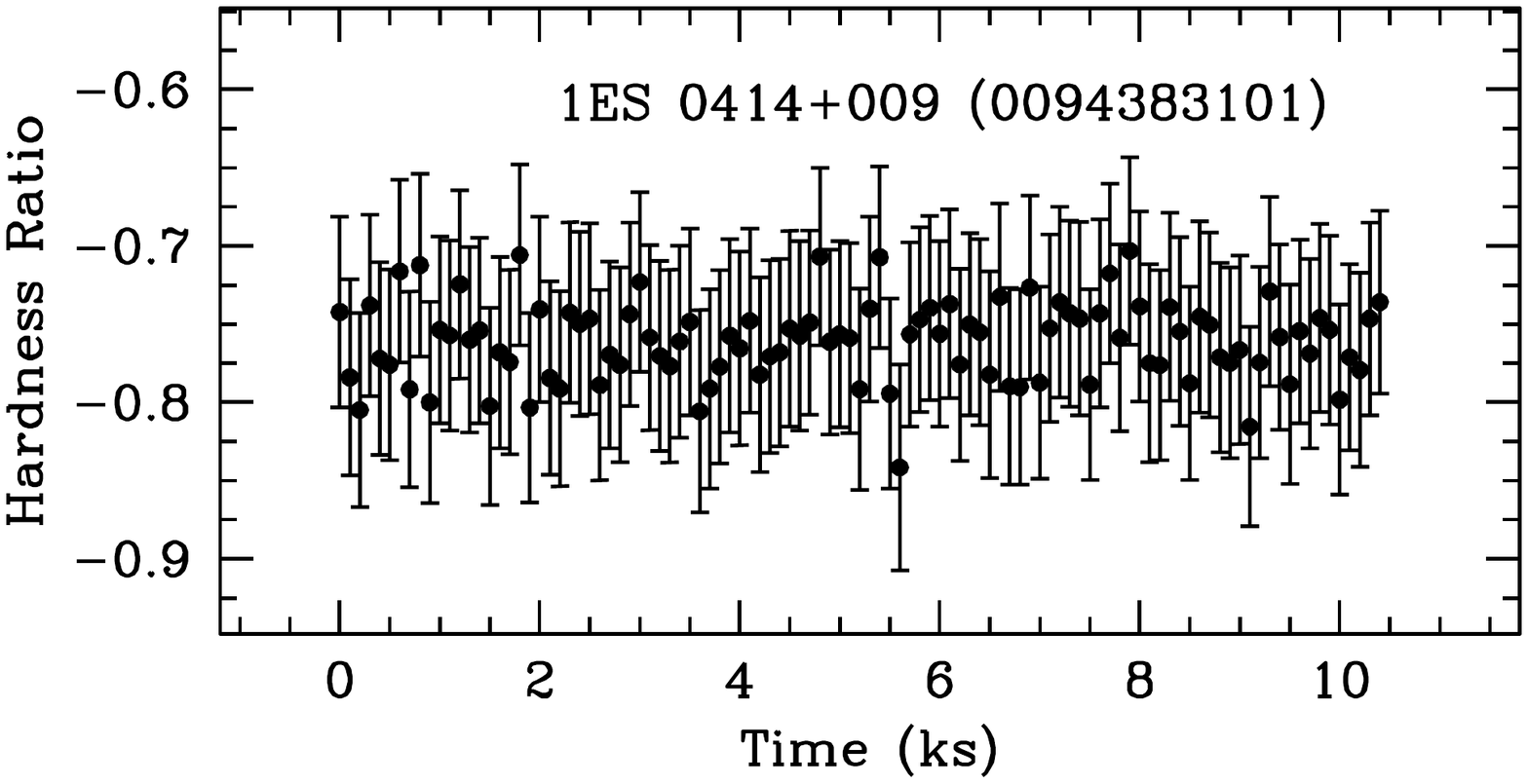}
\includegraphics[scale=0.4]{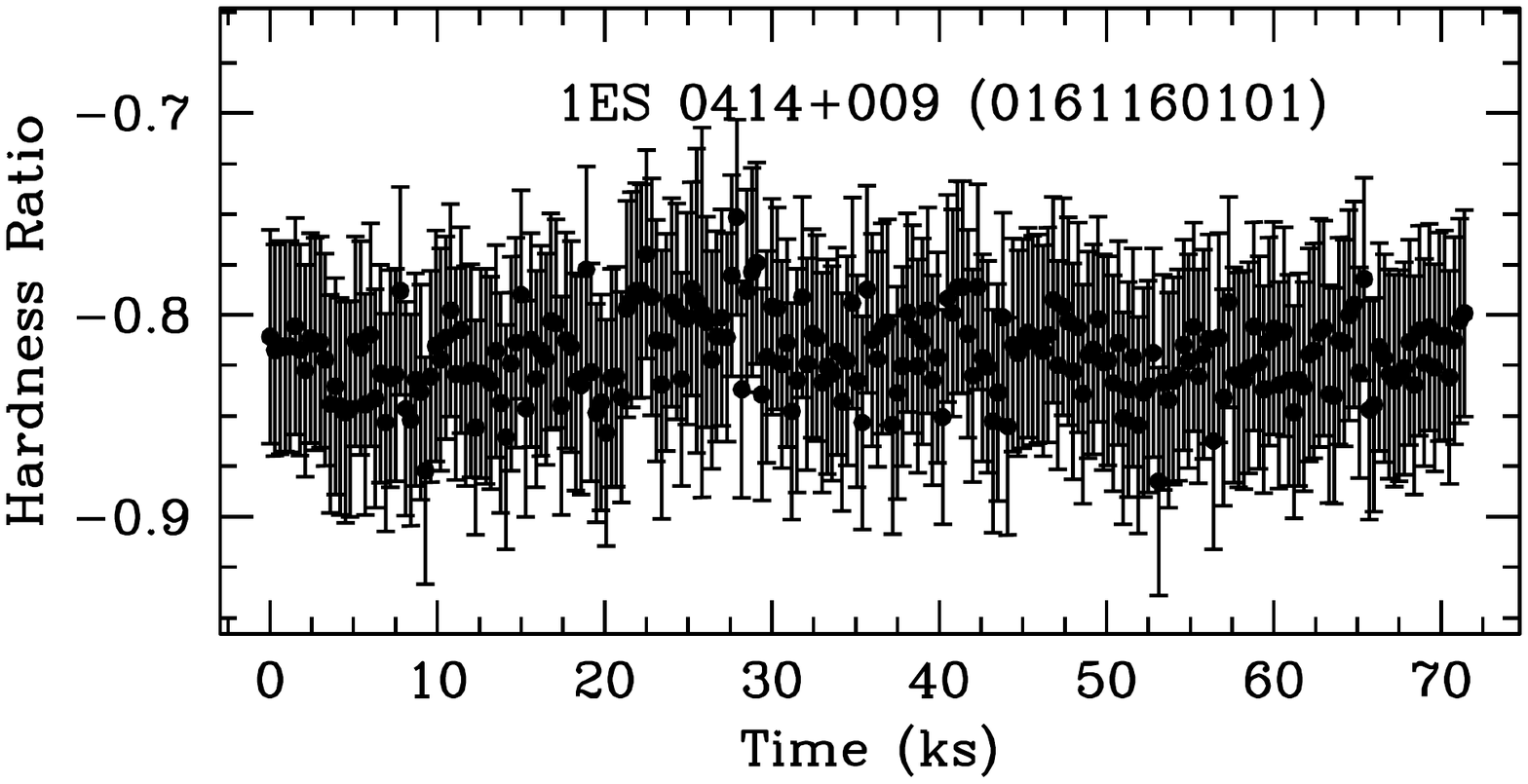}

\vspace*{-2.8in}
\includegraphics[scale=0.4]{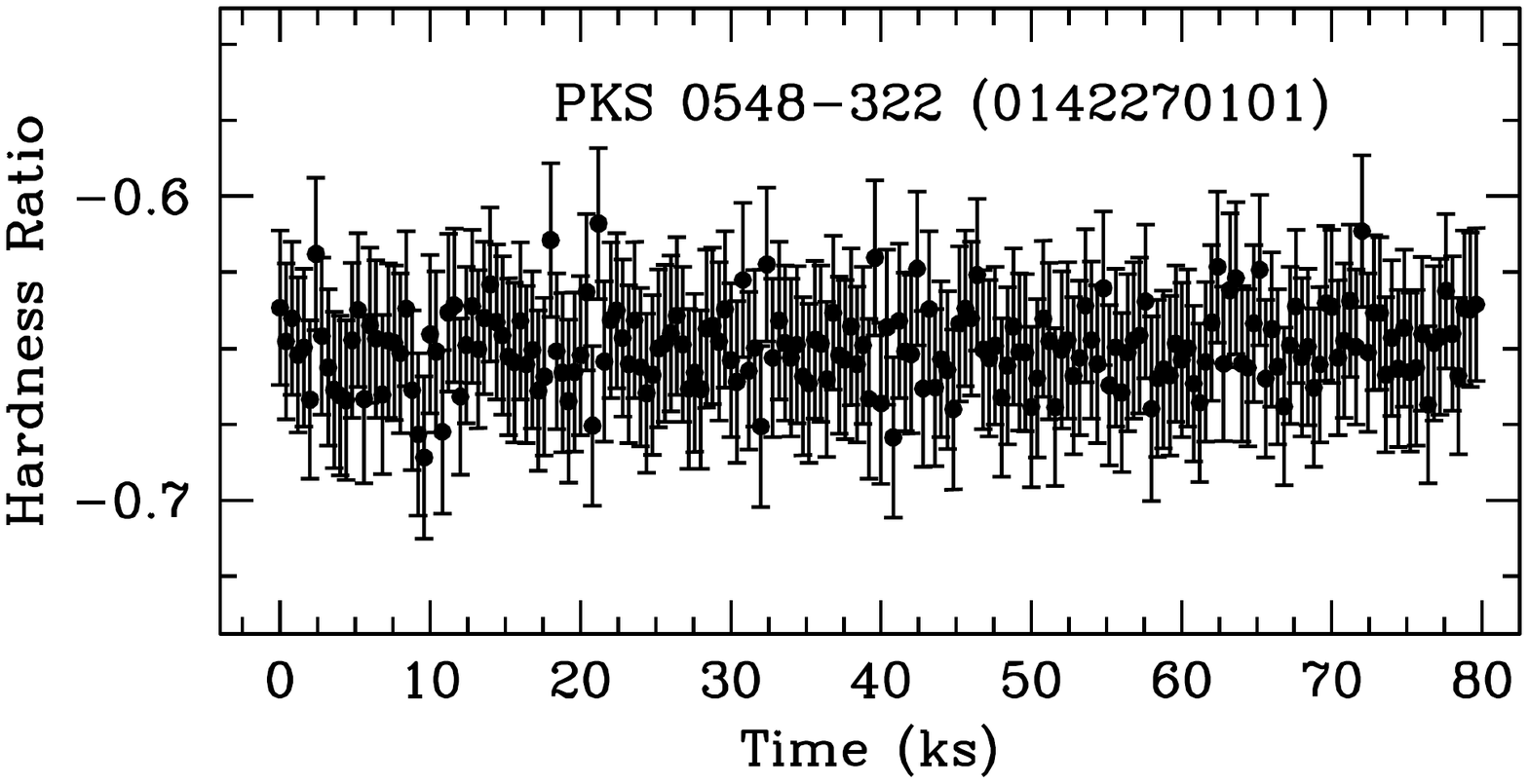}
\includegraphics[scale=0.4]{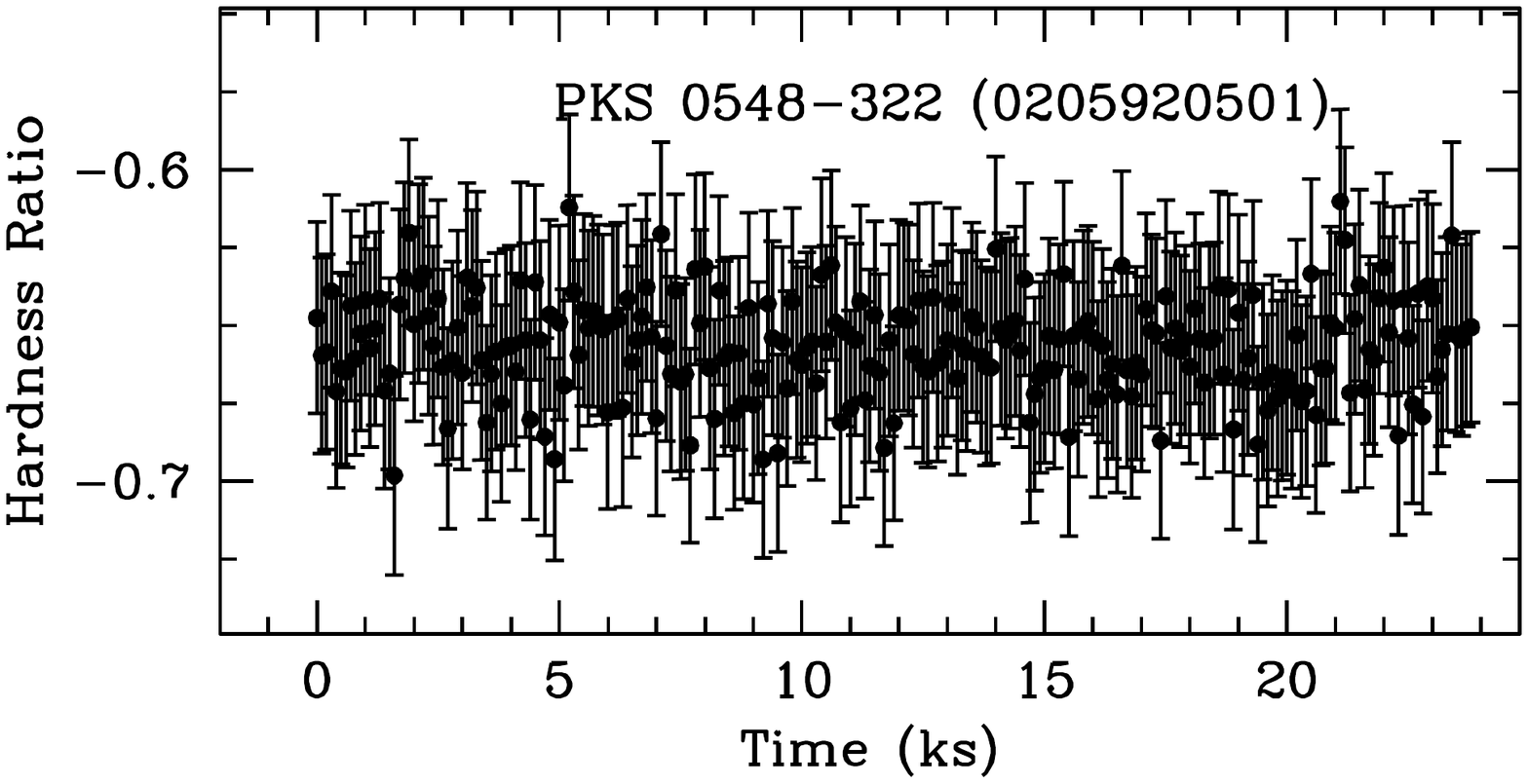}

\vspace*{-2.8in}
\includegraphics[scale=0.4]{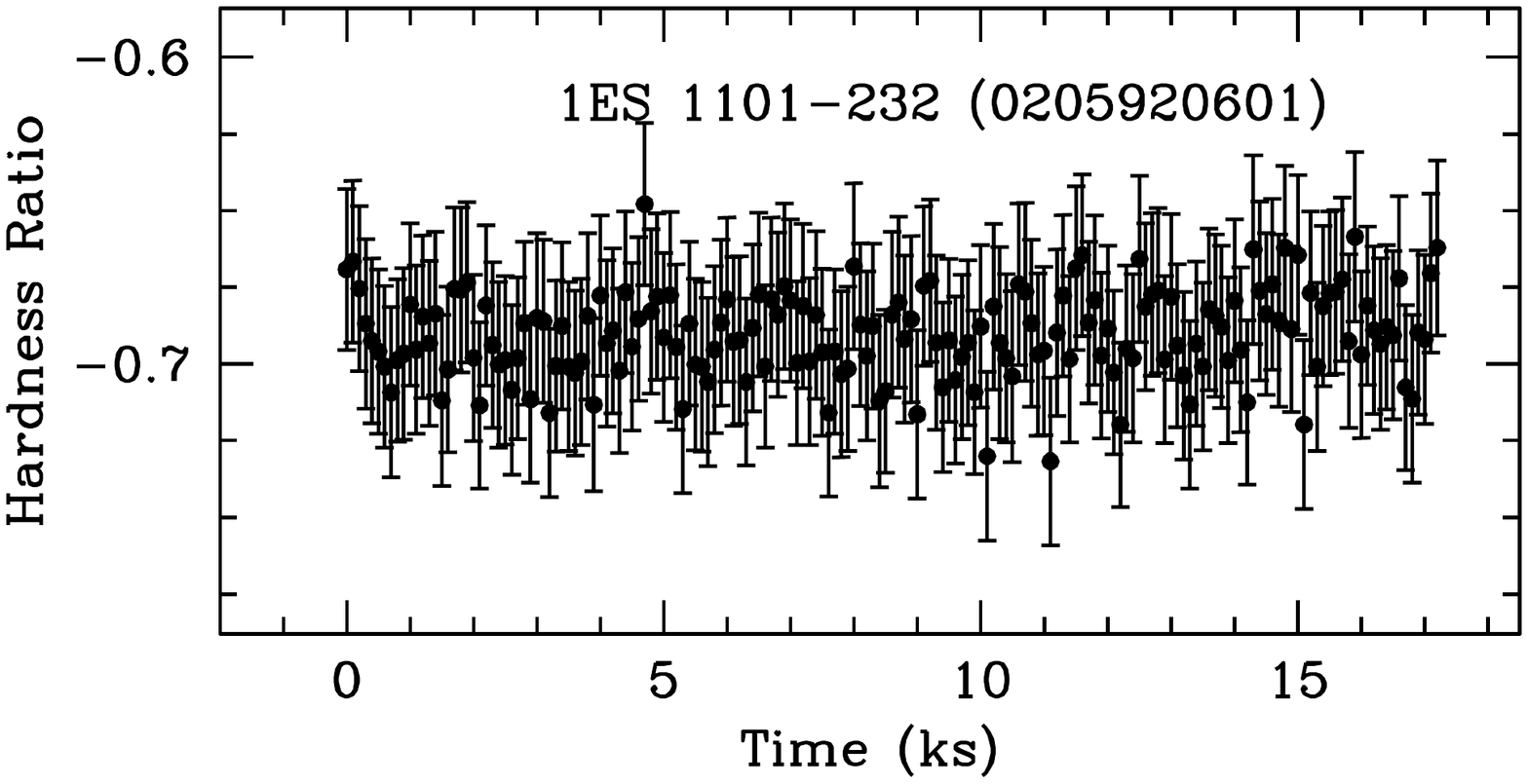}
\includegraphics[scale=0.4]{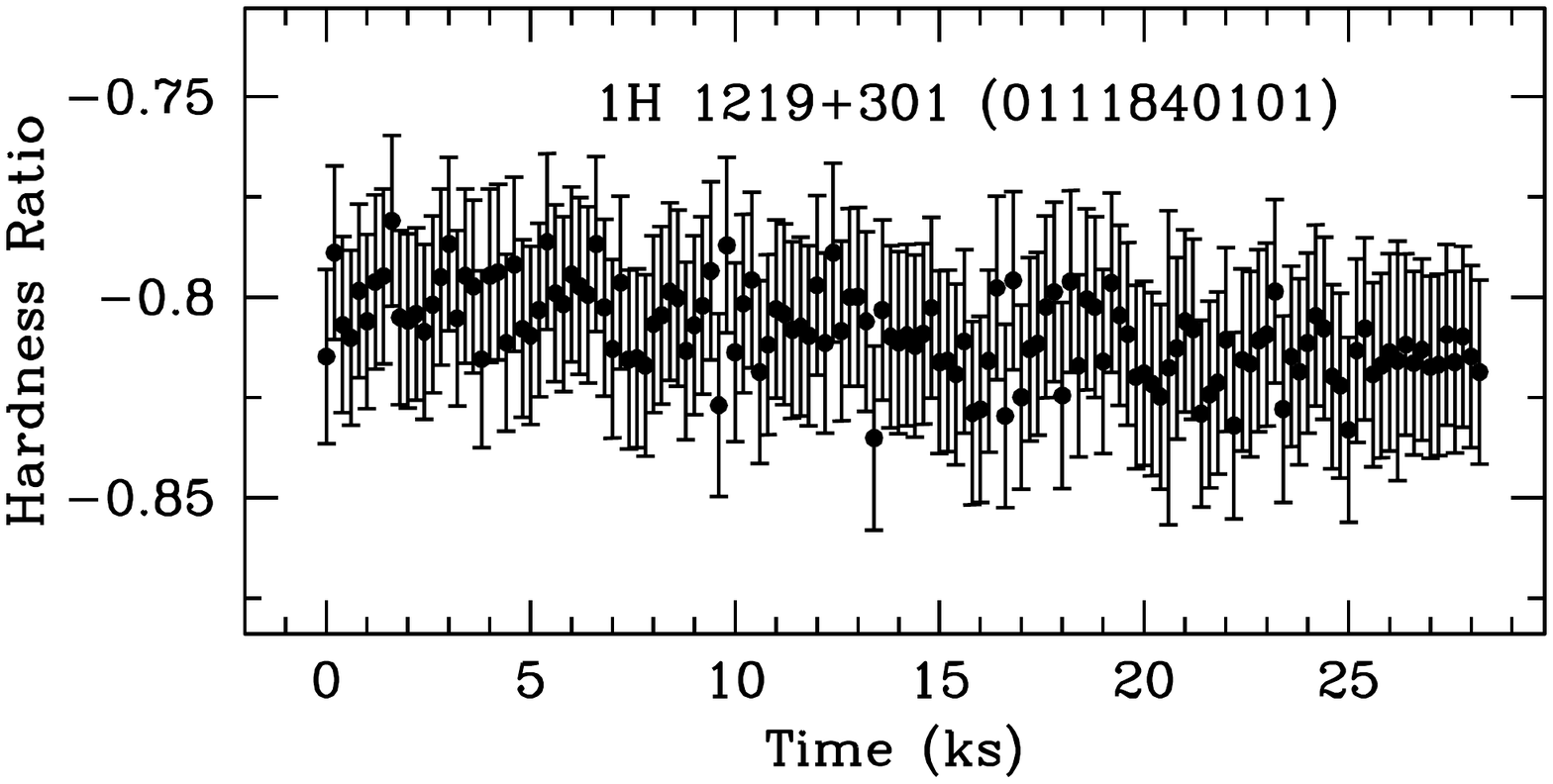}

\vspace{-0.7in}
\caption{Hardness Ratio (HR) of 25 \emph{XMM-Newton} pointed observations labeled with source name and Observation ID.\label{A5}}

\end{figure*}
\clearpage

\setcounter{figure}{4}

\begin{figure*}
\centering

\vspace*{-1.5in}
\includegraphics[scale=0.4]{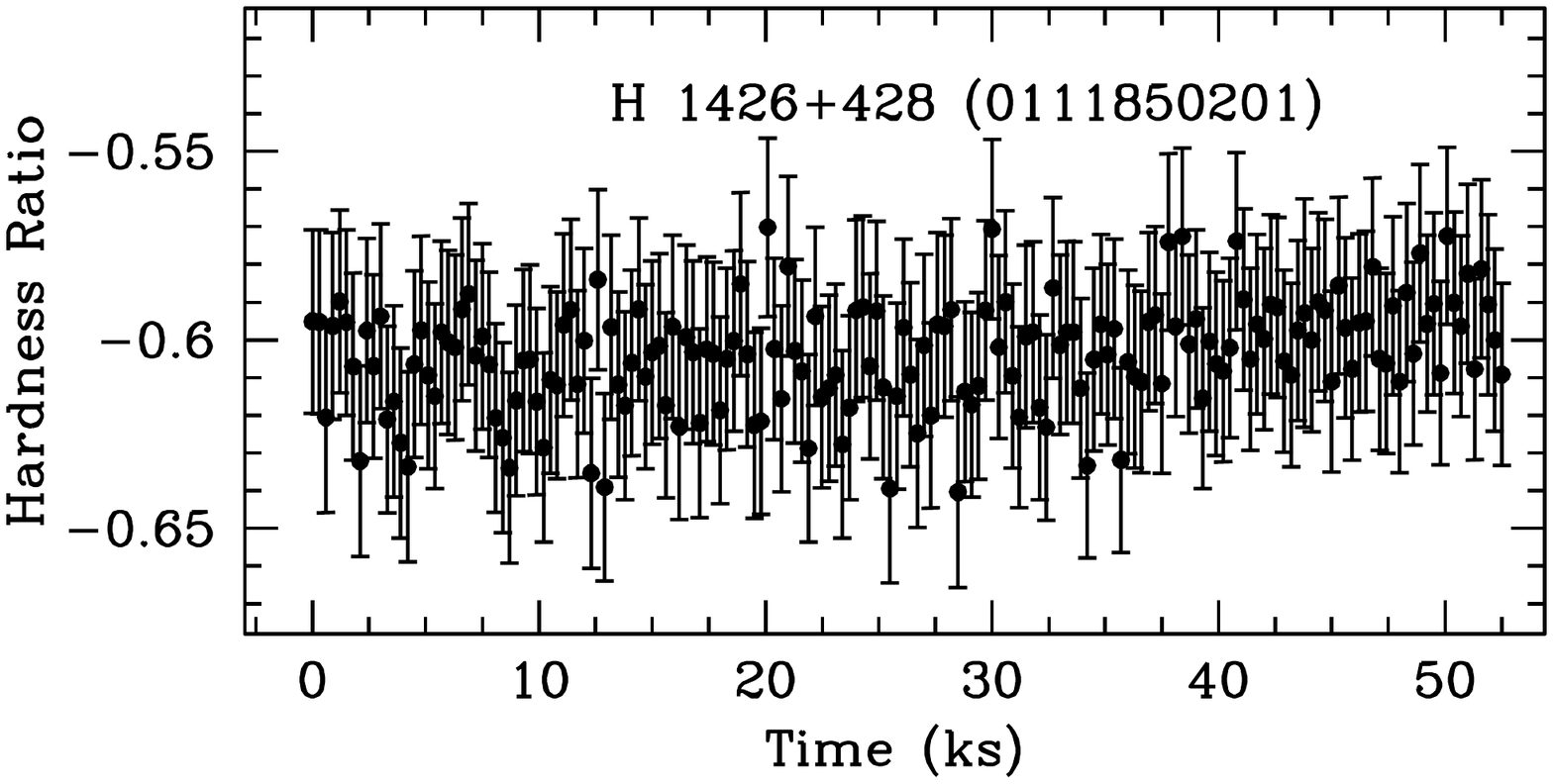}
\includegraphics[scale=0.4]{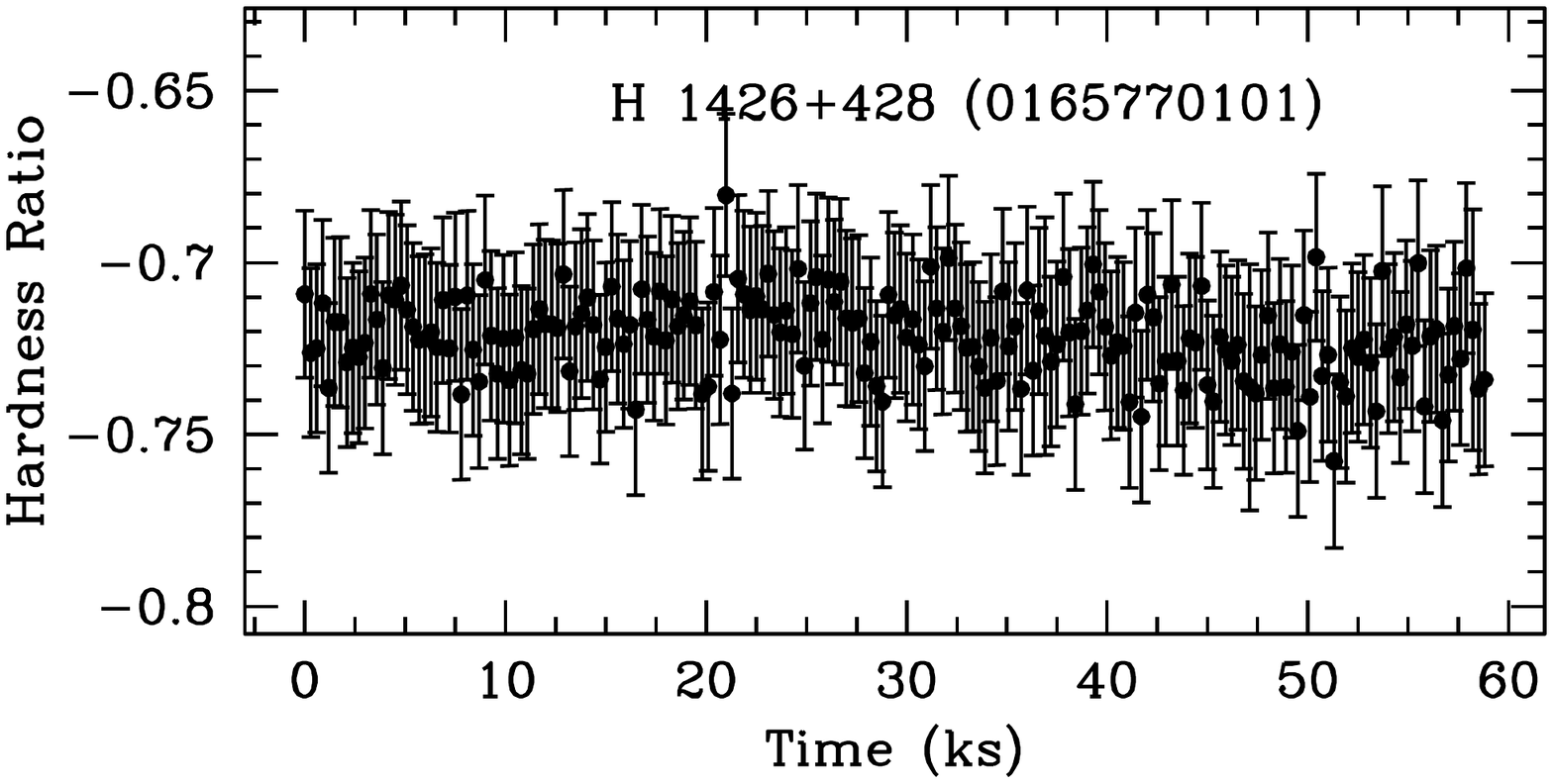}

\vspace*{-2.8in}
\includegraphics[scale=0.4]{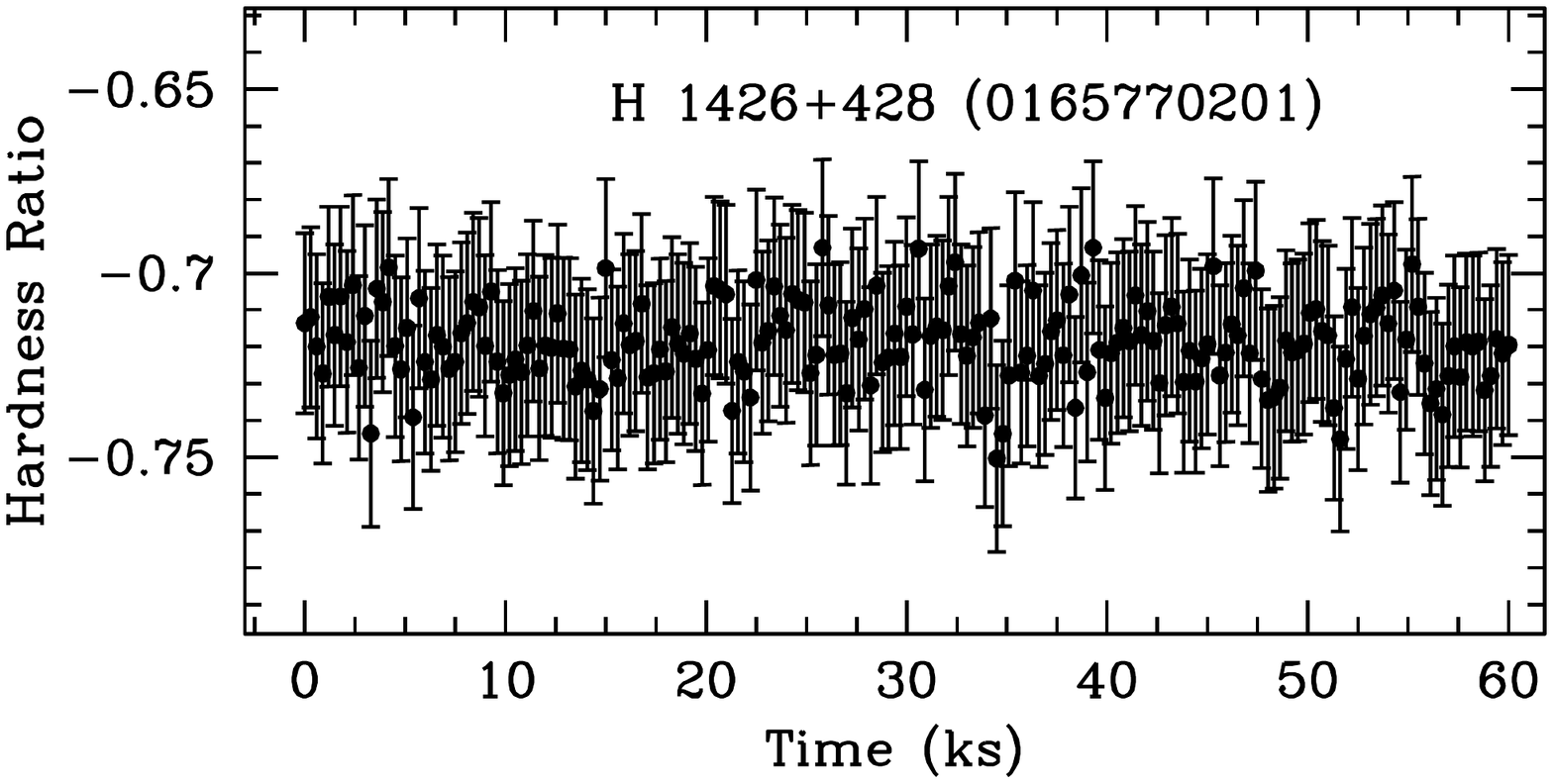}
\includegraphics[scale=0.4]{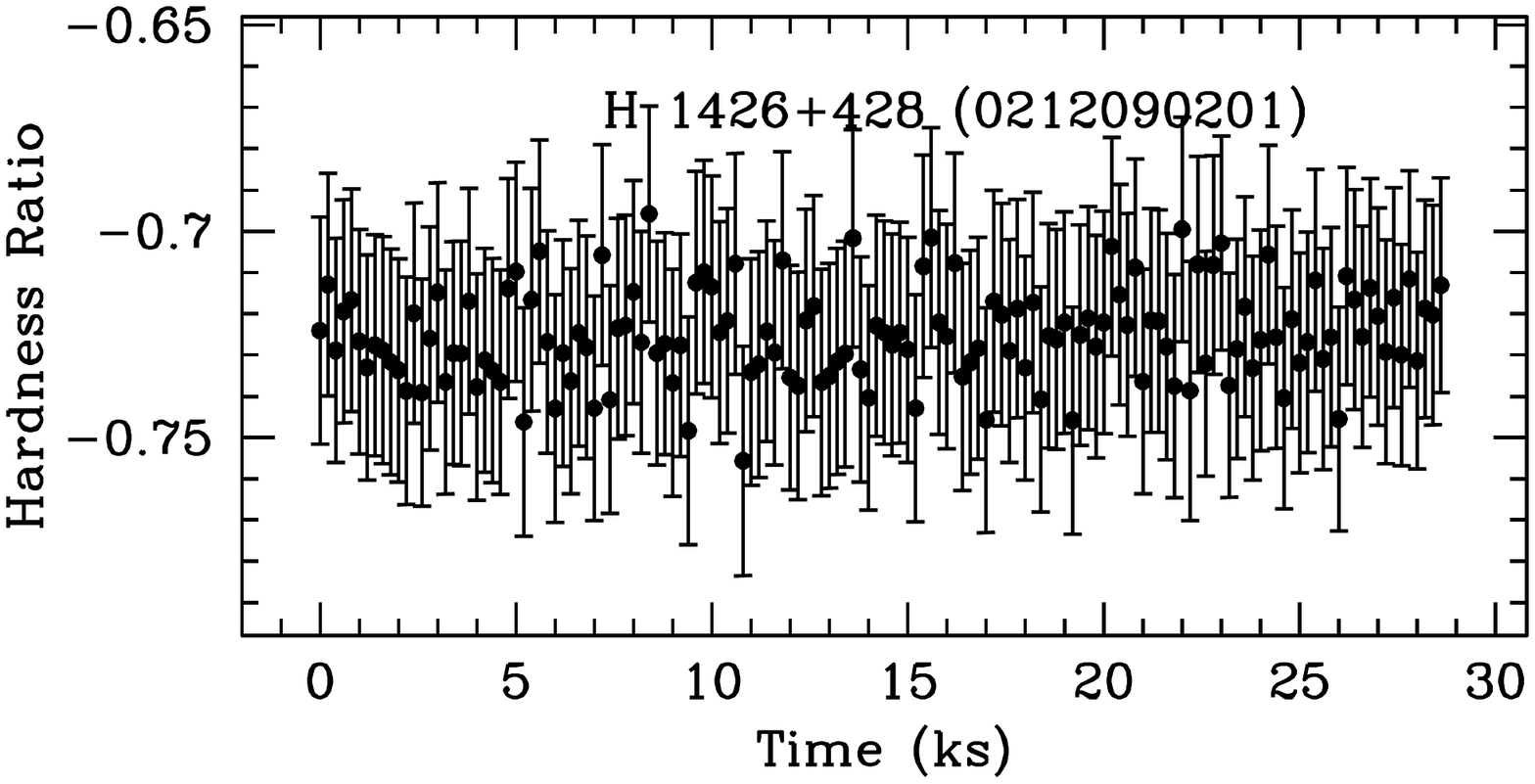}

\vspace*{-2.8in}
\includegraphics[scale=0.4]{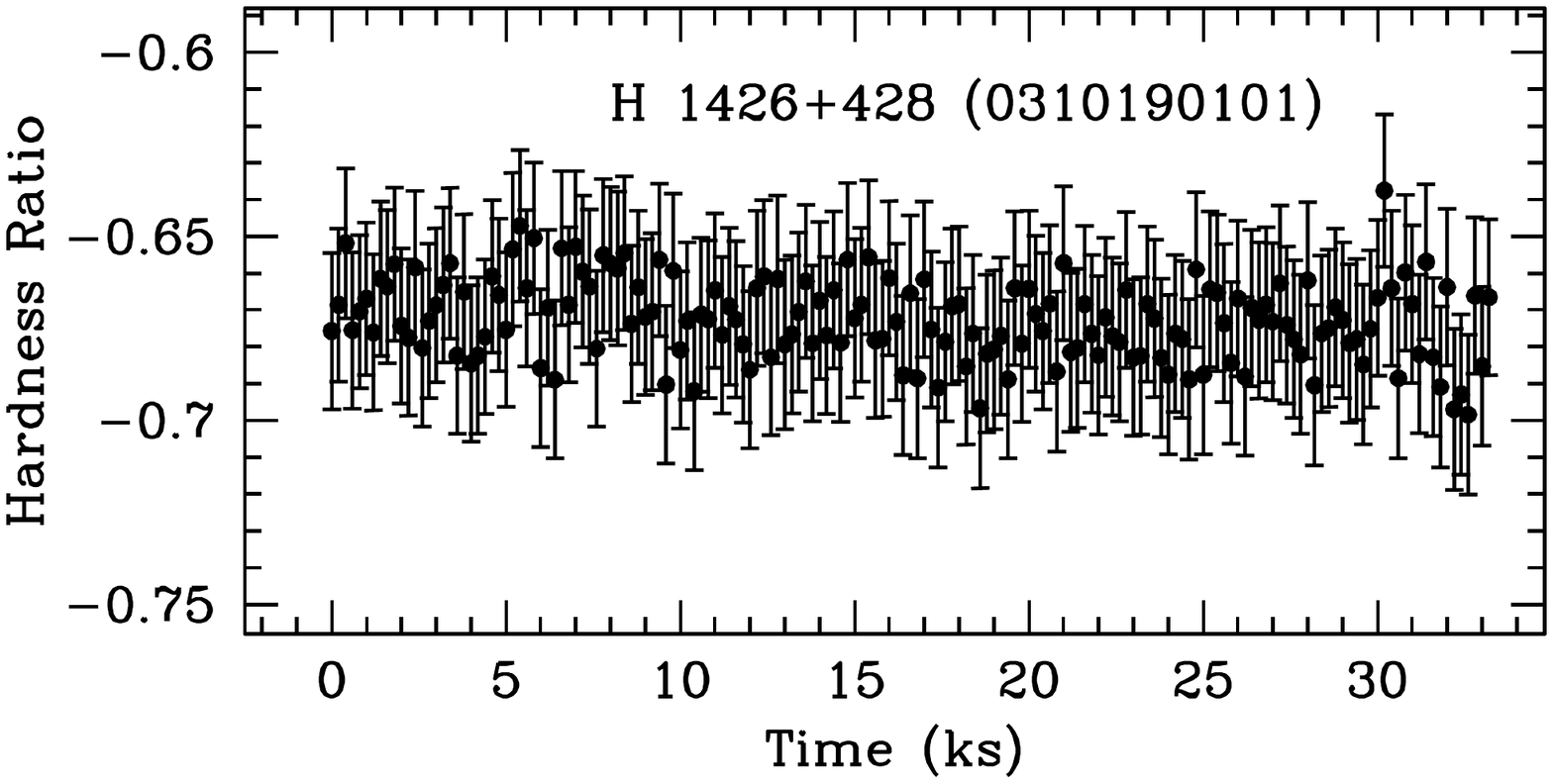}
\includegraphics[scale=0.4]{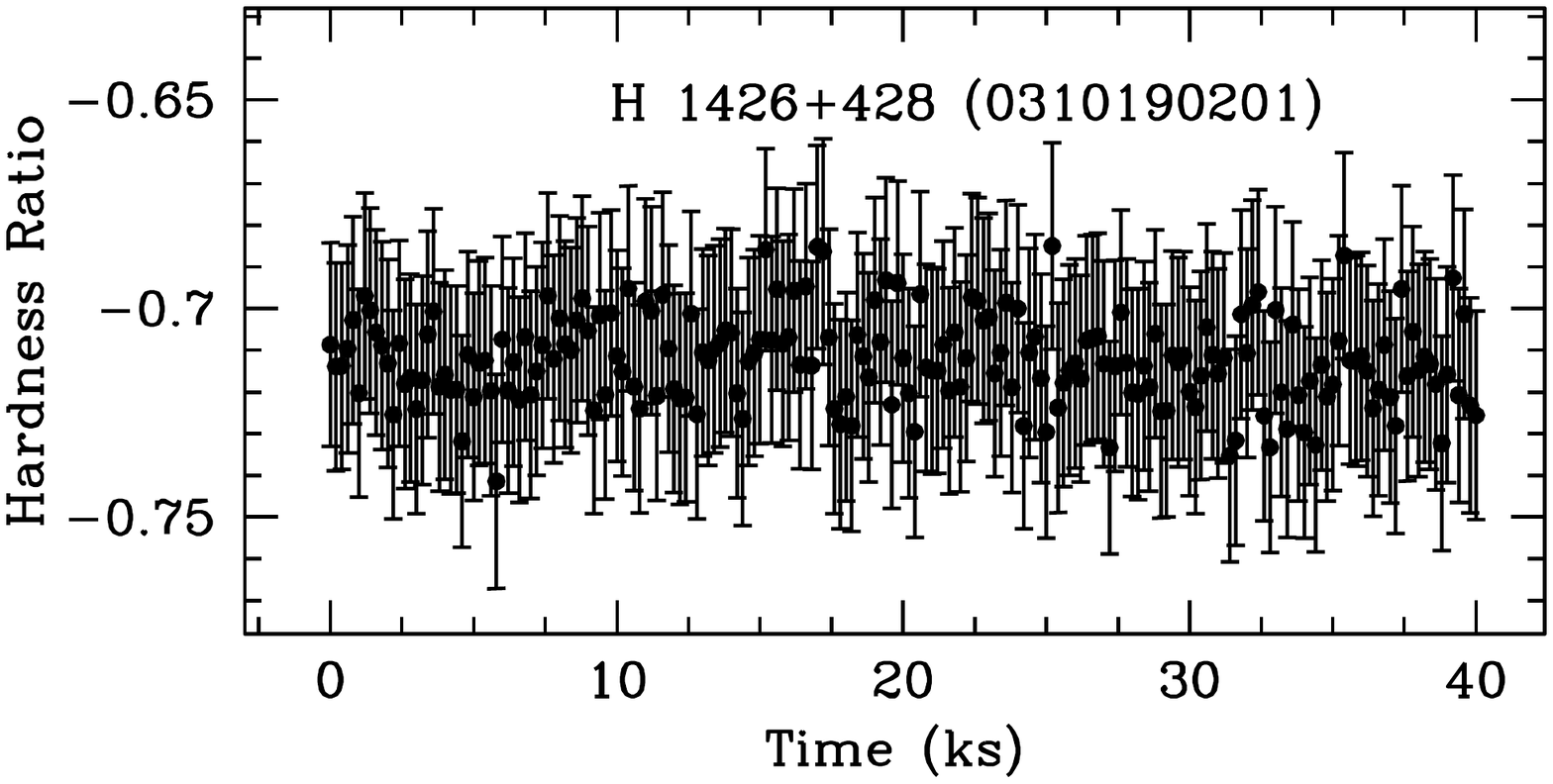}

\vspace*{-2.8in}
\includegraphics[scale=0.4]{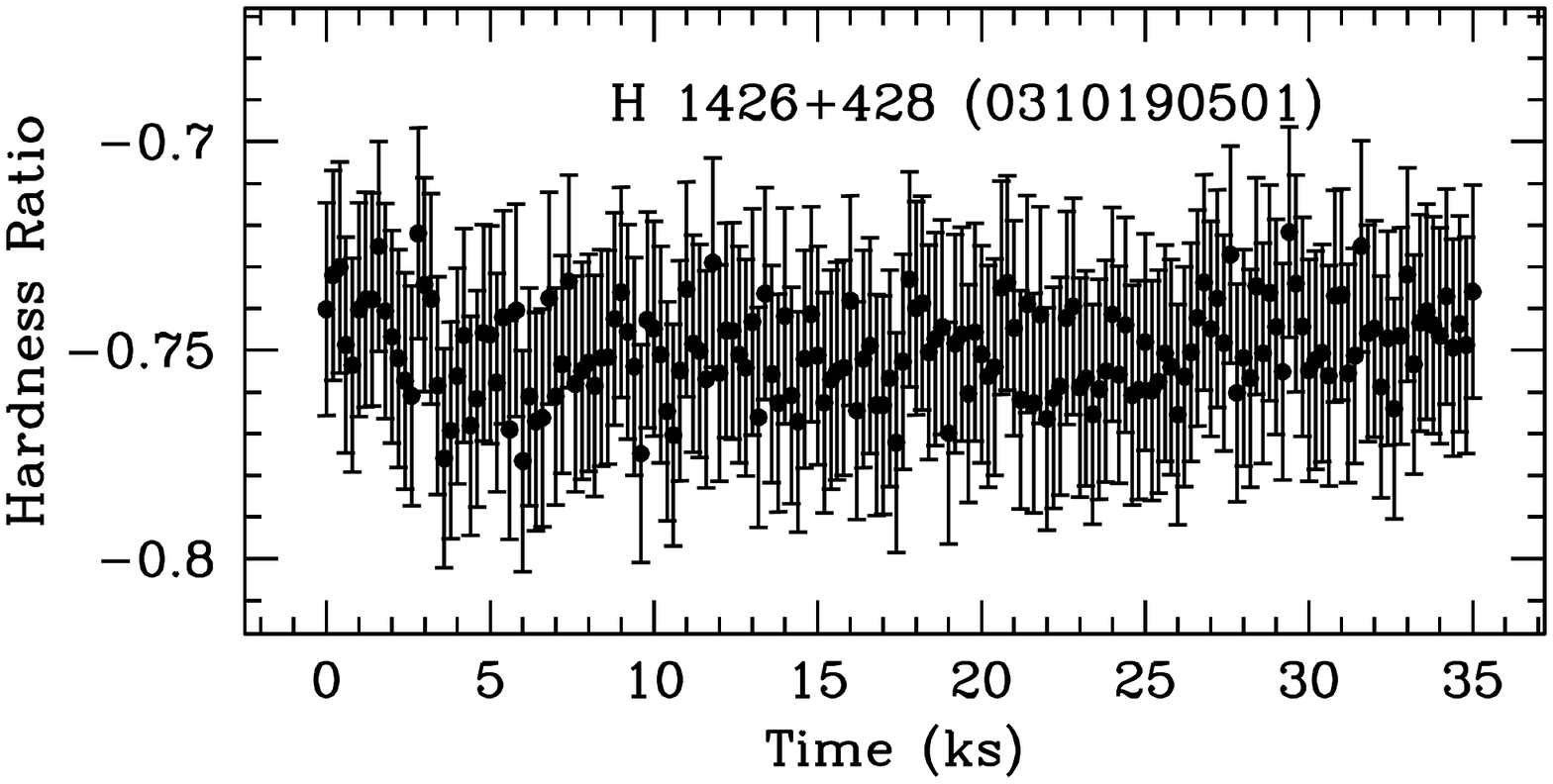}
\includegraphics[scale=0.4]{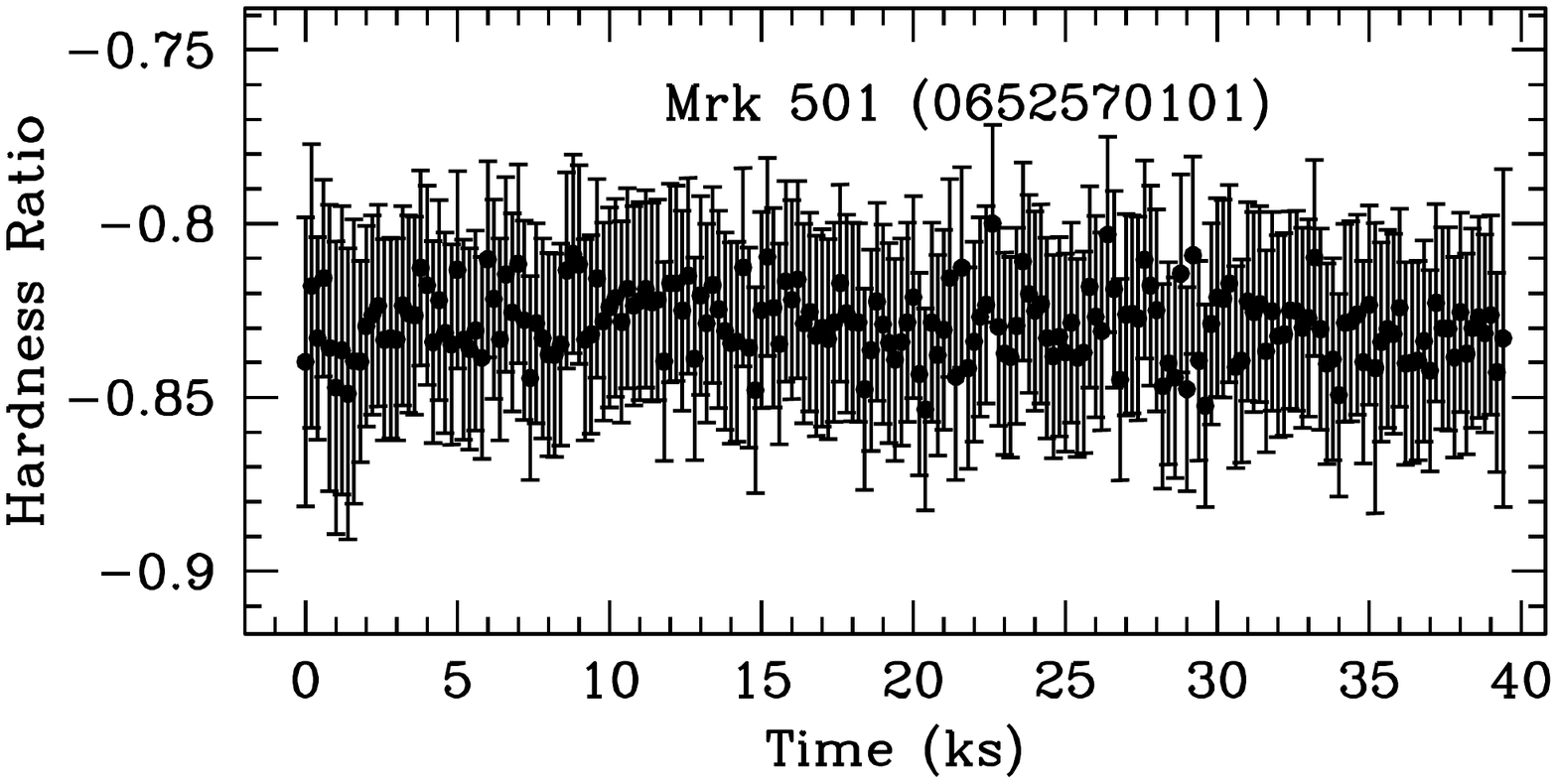}

\vspace{-0.7in}
\caption{Continued.}  

\end{figure*}

\setcounter{figure}{4}
\begin{figure*}
\centering

\vspace*{-1.5in}
\includegraphics[scale=0.4]{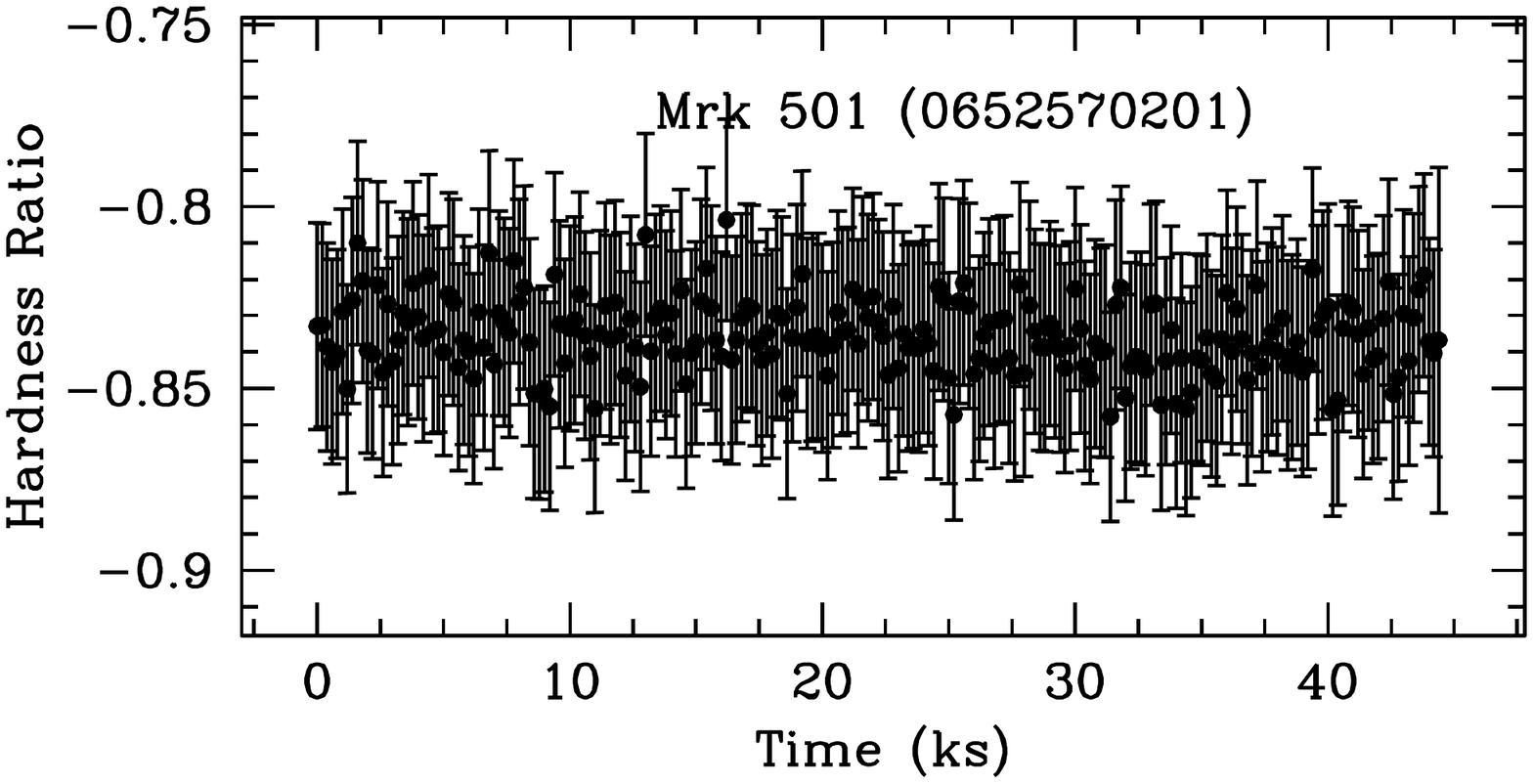}
\includegraphics[scale=0.4]{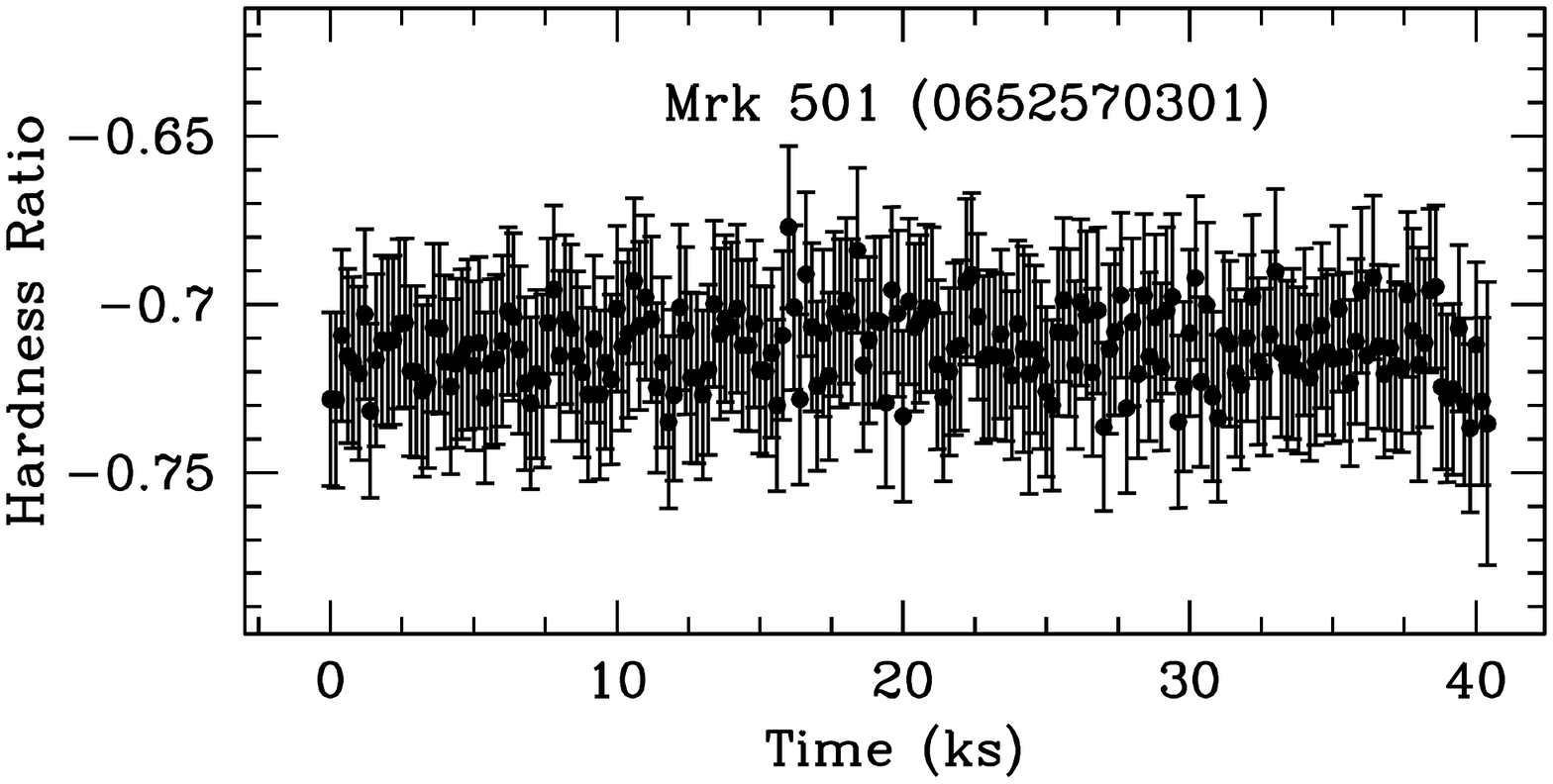}

\vspace*{-3.0in}
\includegraphics[scale=0.4]{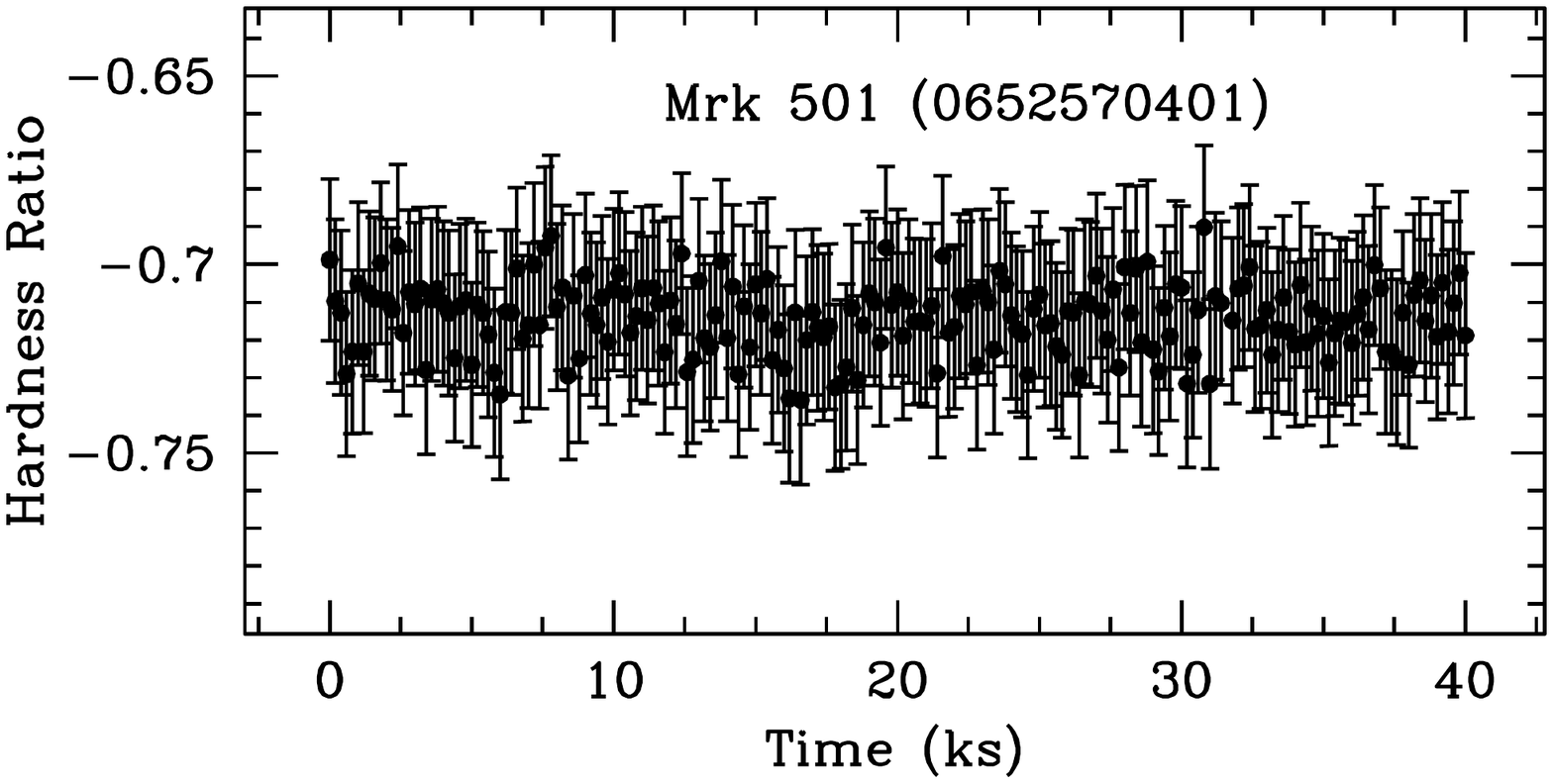}
\includegraphics[scale=0.4]{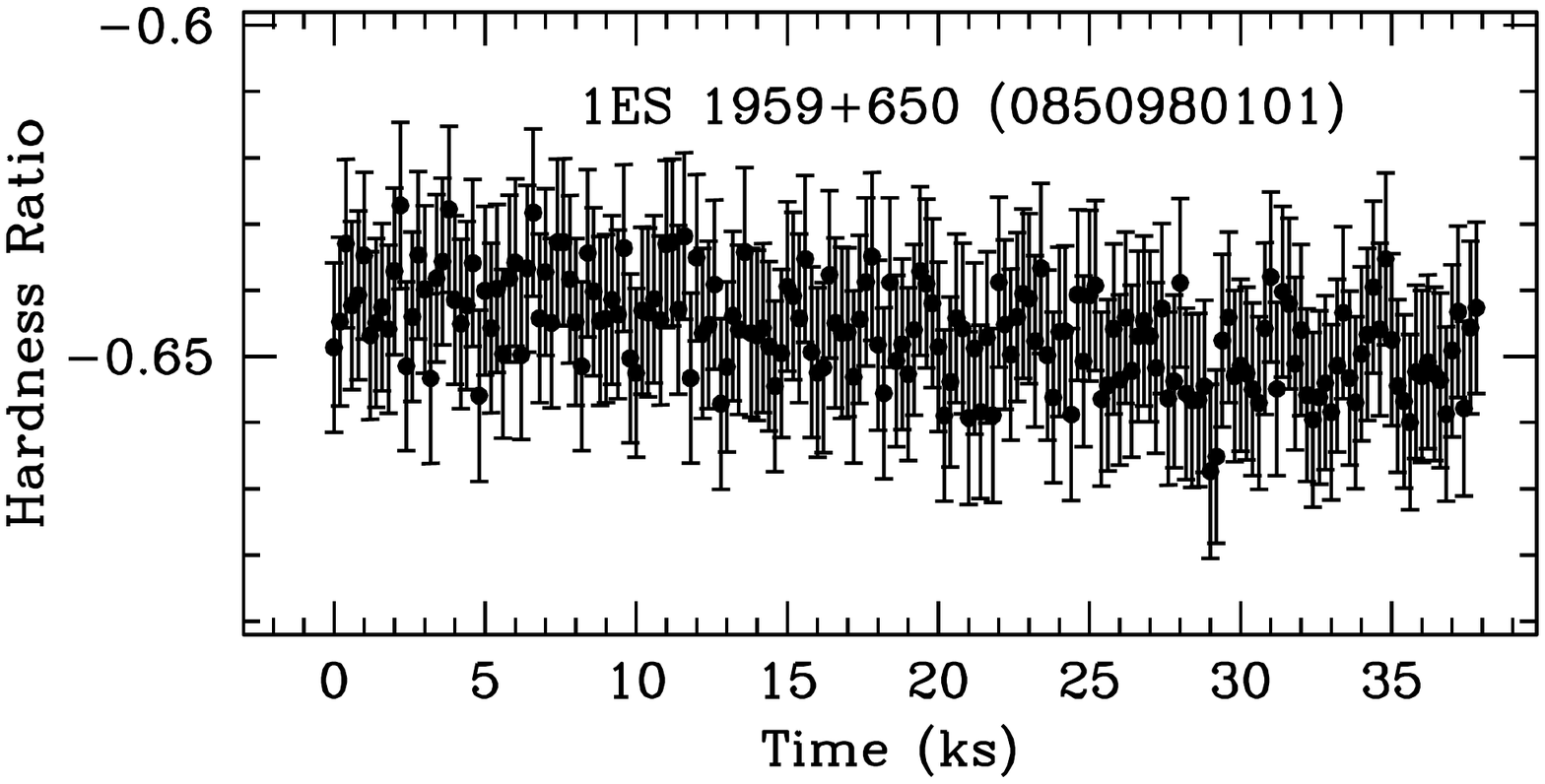}

\vspace*{-3.0in}
\includegraphics[scale=0.4]{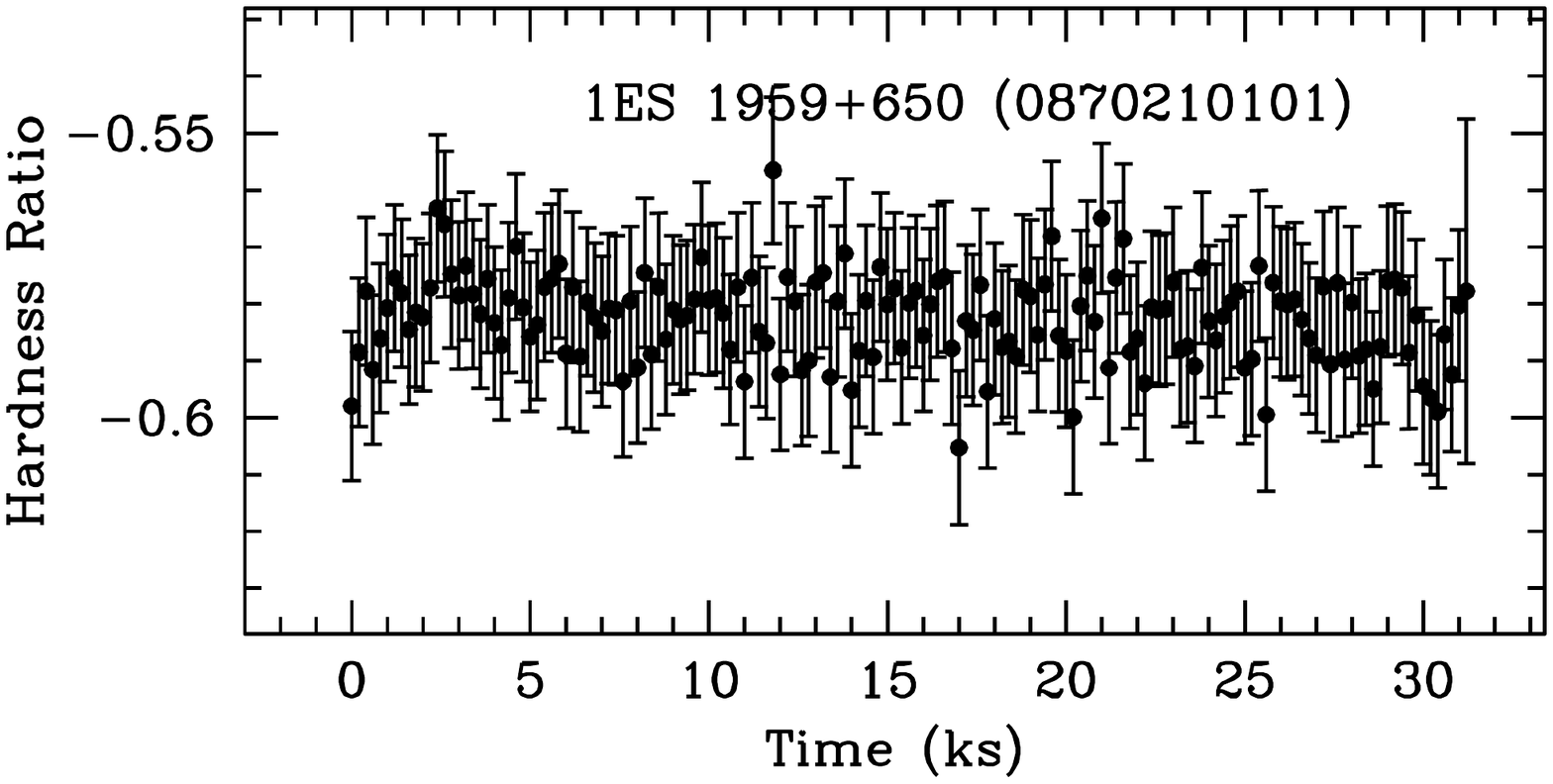}
\includegraphics[scale=0.4]{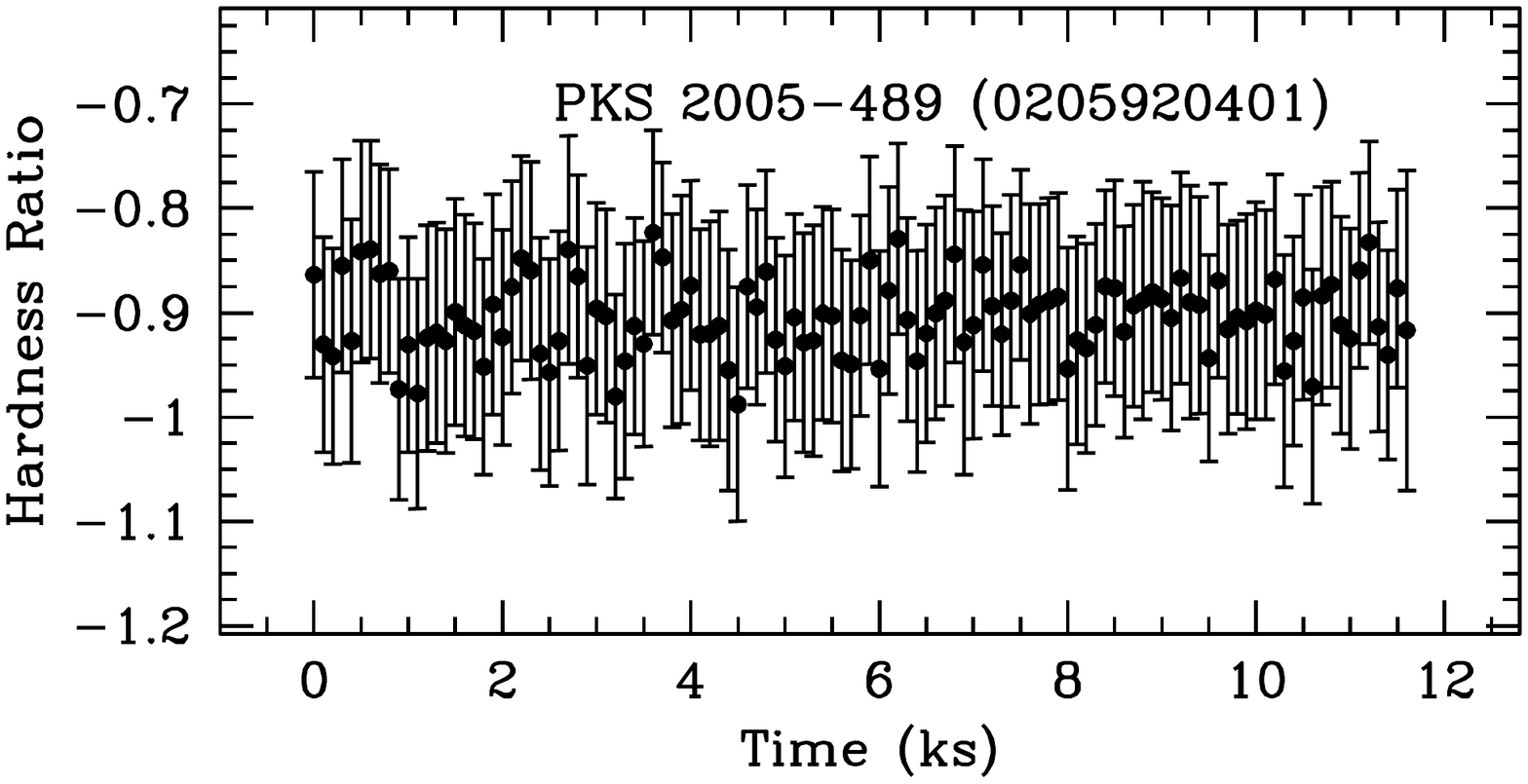}

\vspace*{-3.0in}
\includegraphics[scale=0.4]{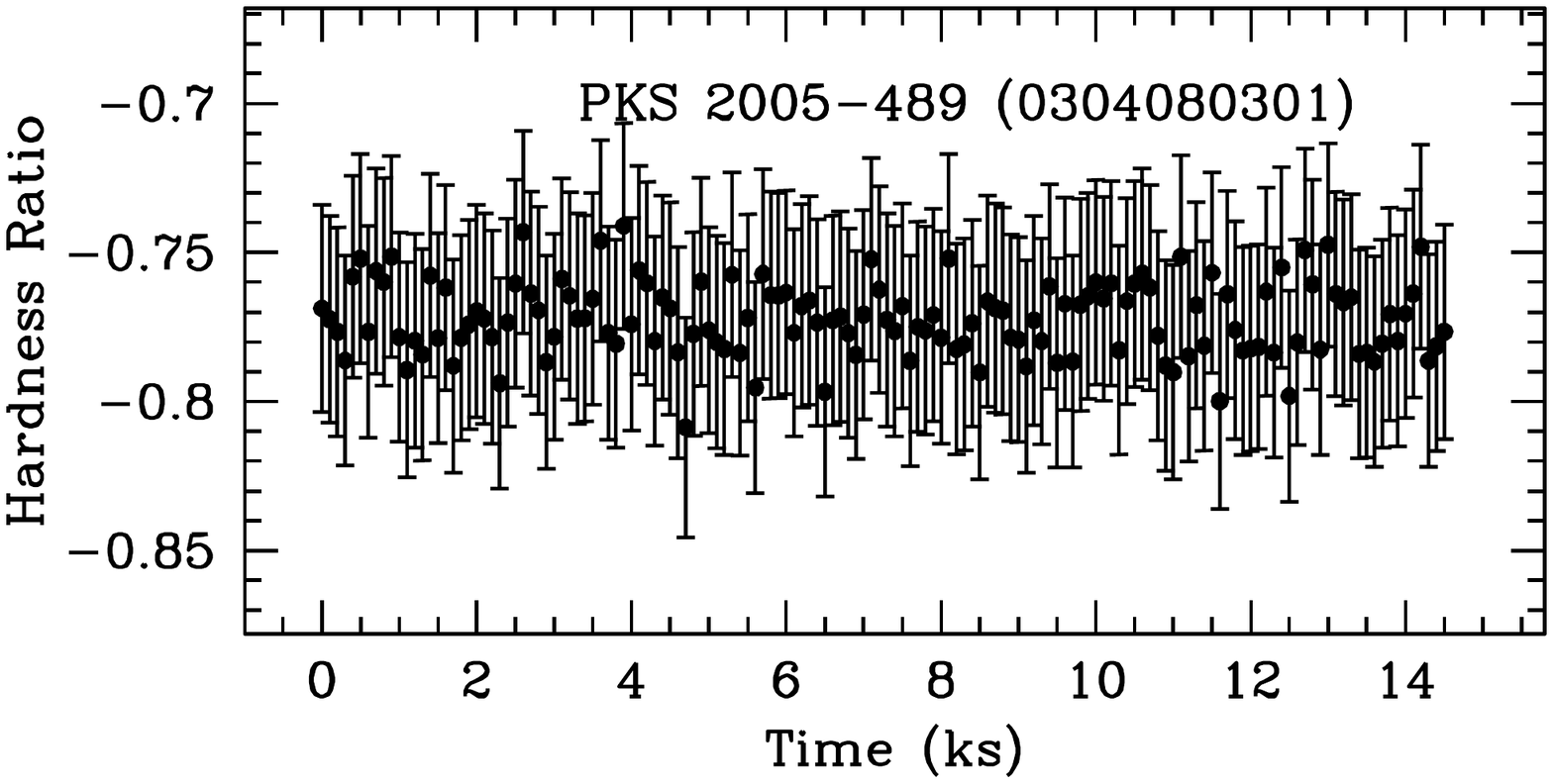}
\includegraphics[scale=0.4]{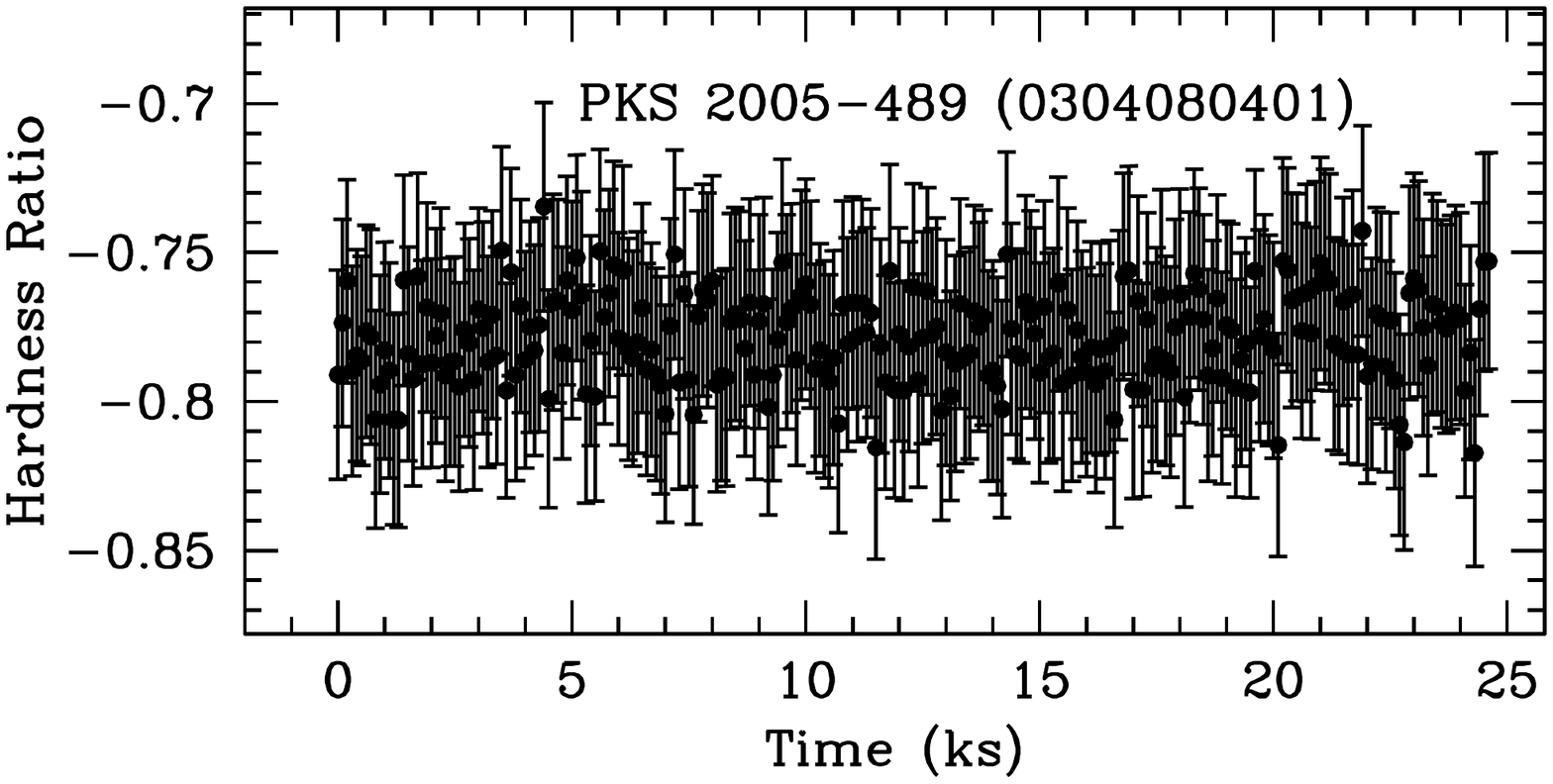}

\vspace*{-3.0in}
\includegraphics[scale=0.4]{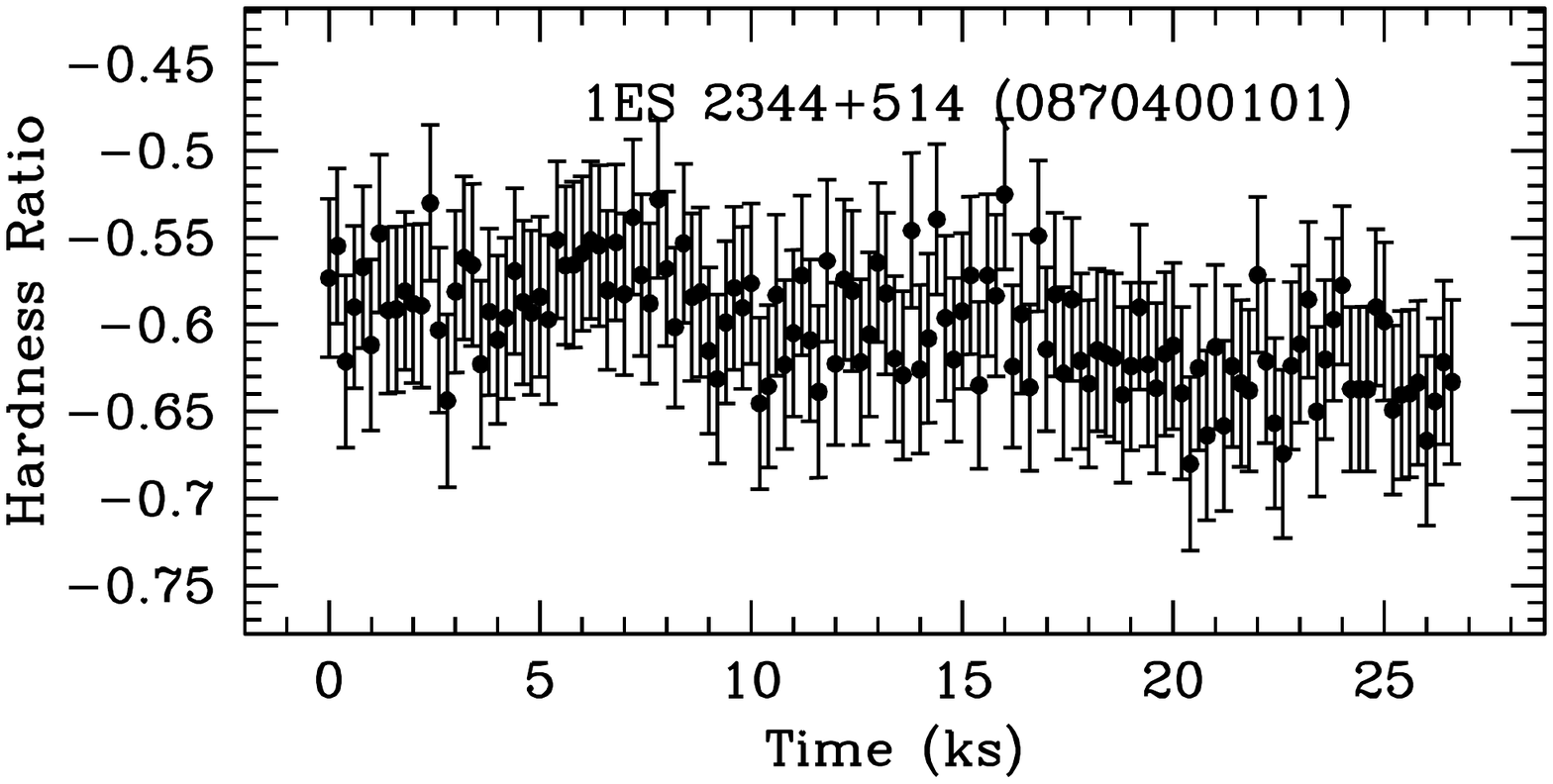}

\vspace{-0.7in}
\caption{Continued.}  

\end{figure*}

\clearpage

\setcounter{figure}{5}
\begin{figure*}
\centering

\vspace*{-0.1in}
\includegraphics[scale=0.2]{Fig6a.pdf}
\includegraphics[scale=0.2]{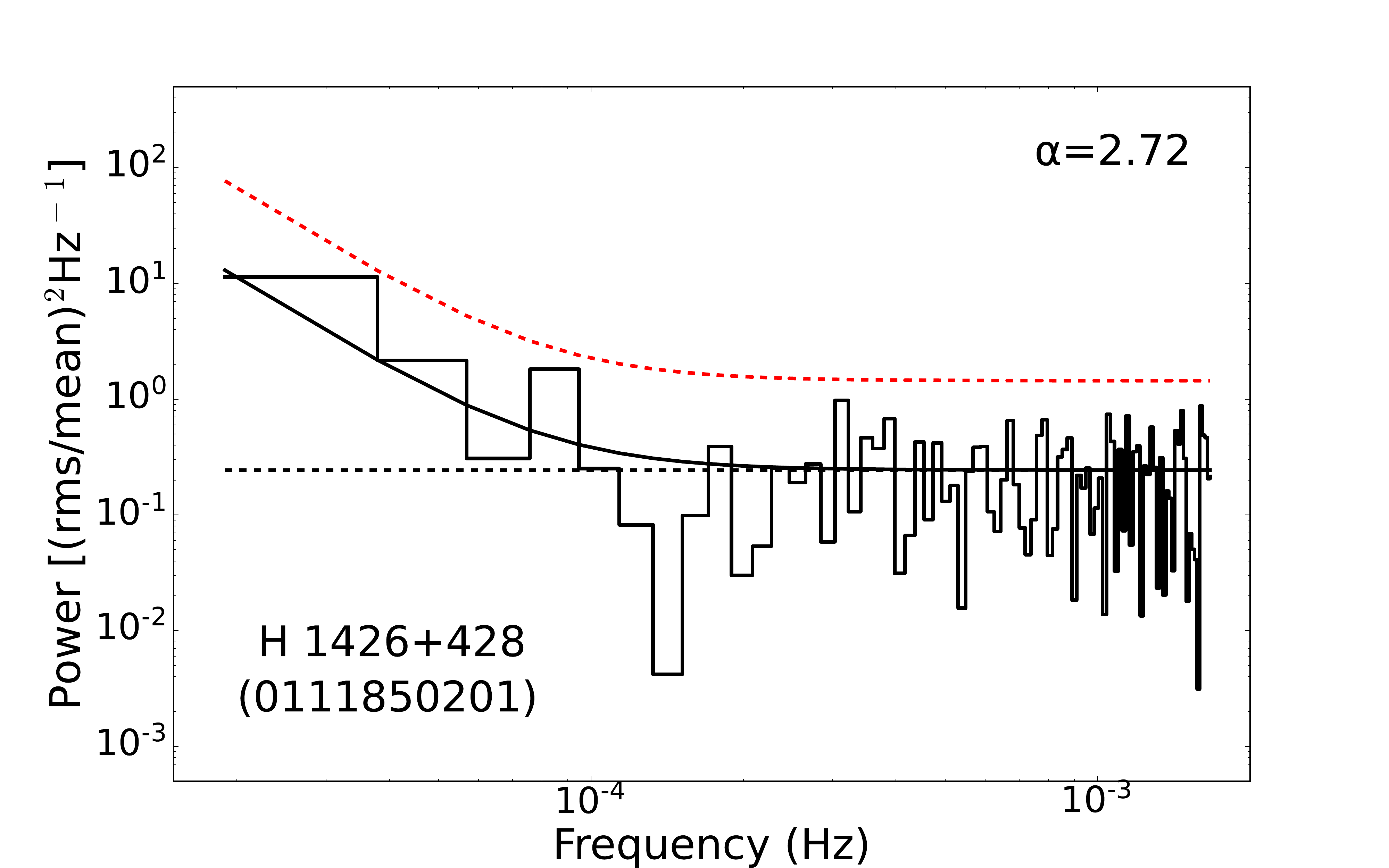}

\vspace*{0.01in}
\includegraphics[scale=0.2]{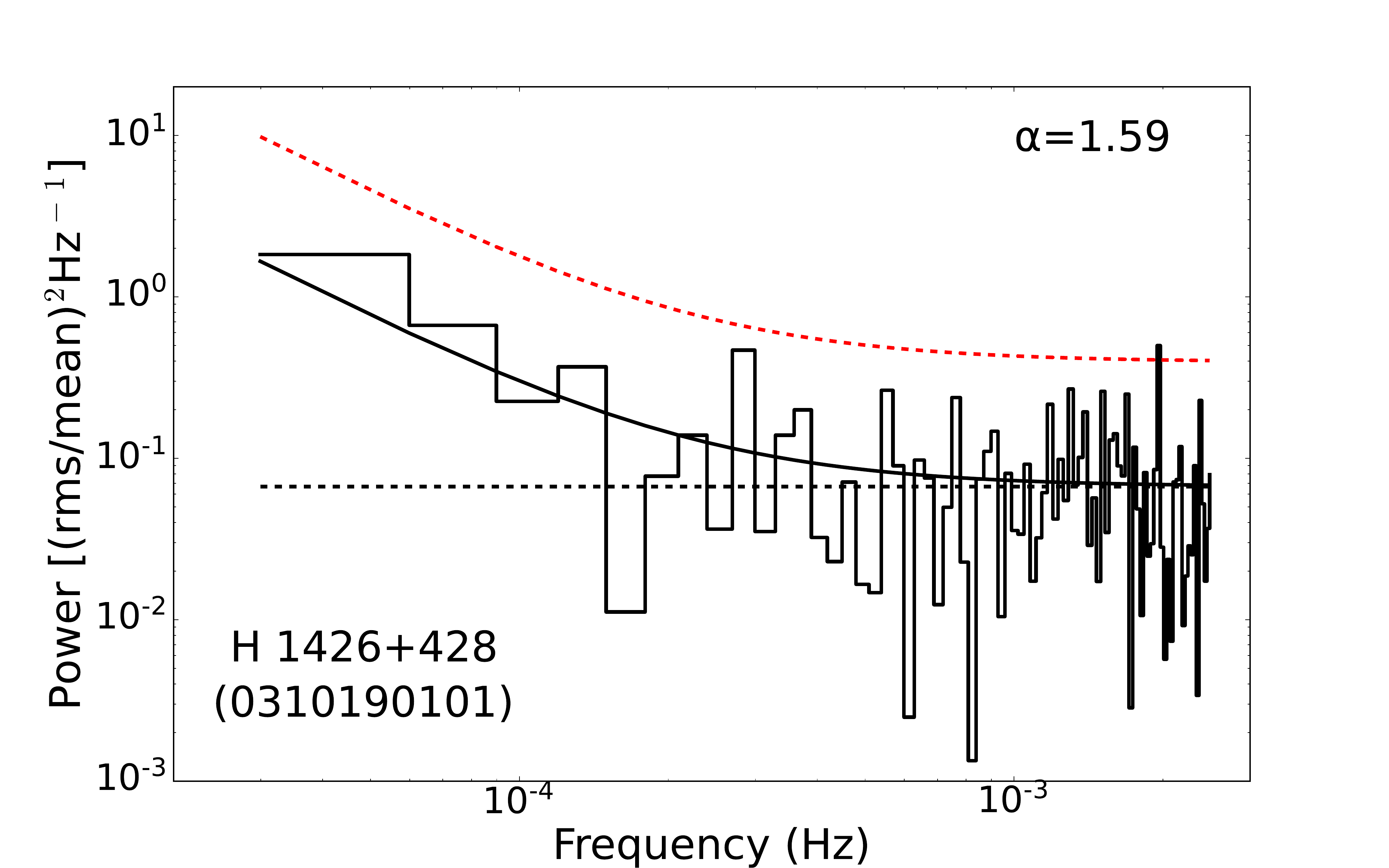}
\includegraphics[scale=0.2]{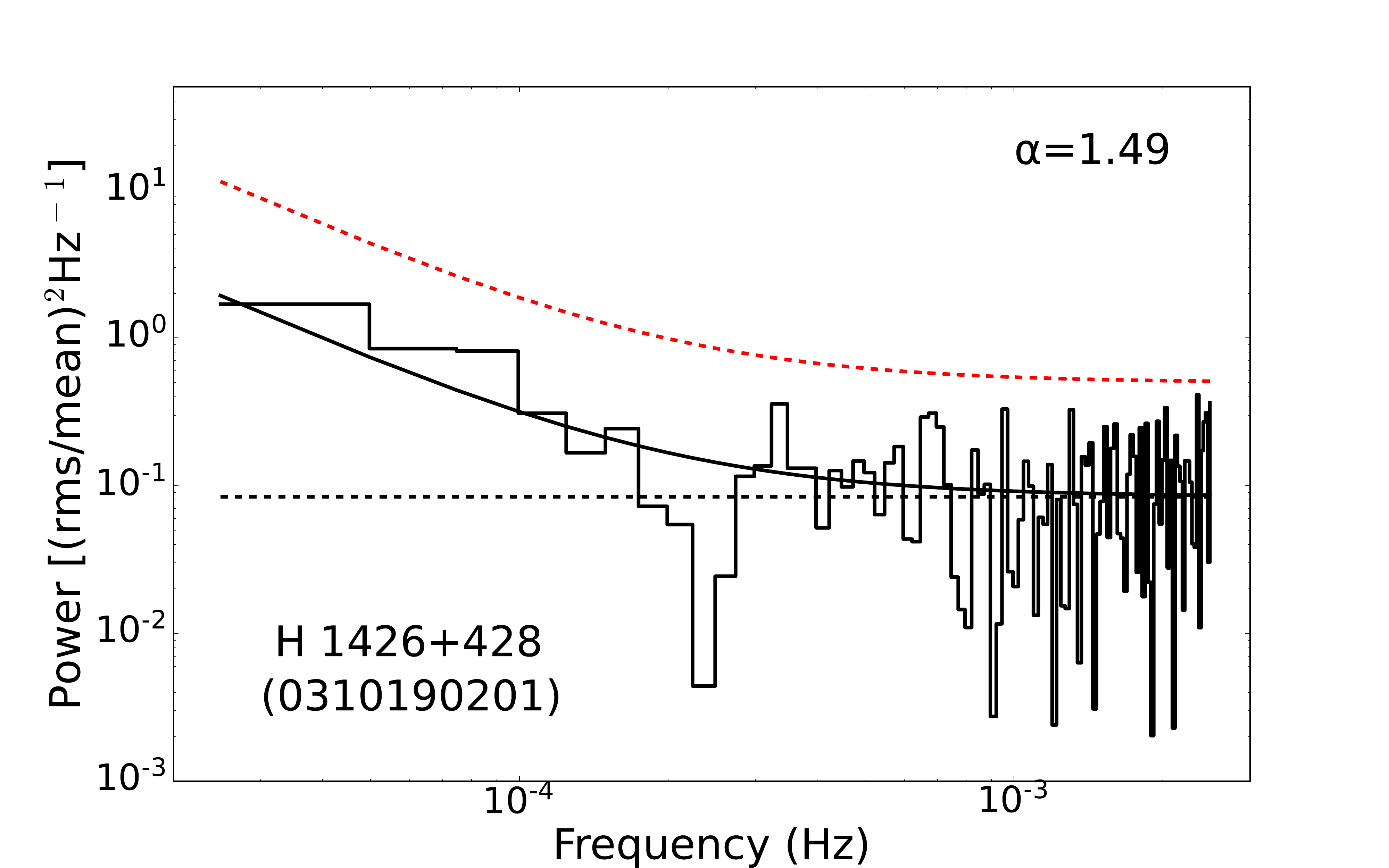}

\vspace*{0.01in}
\includegraphics[scale=0.2]{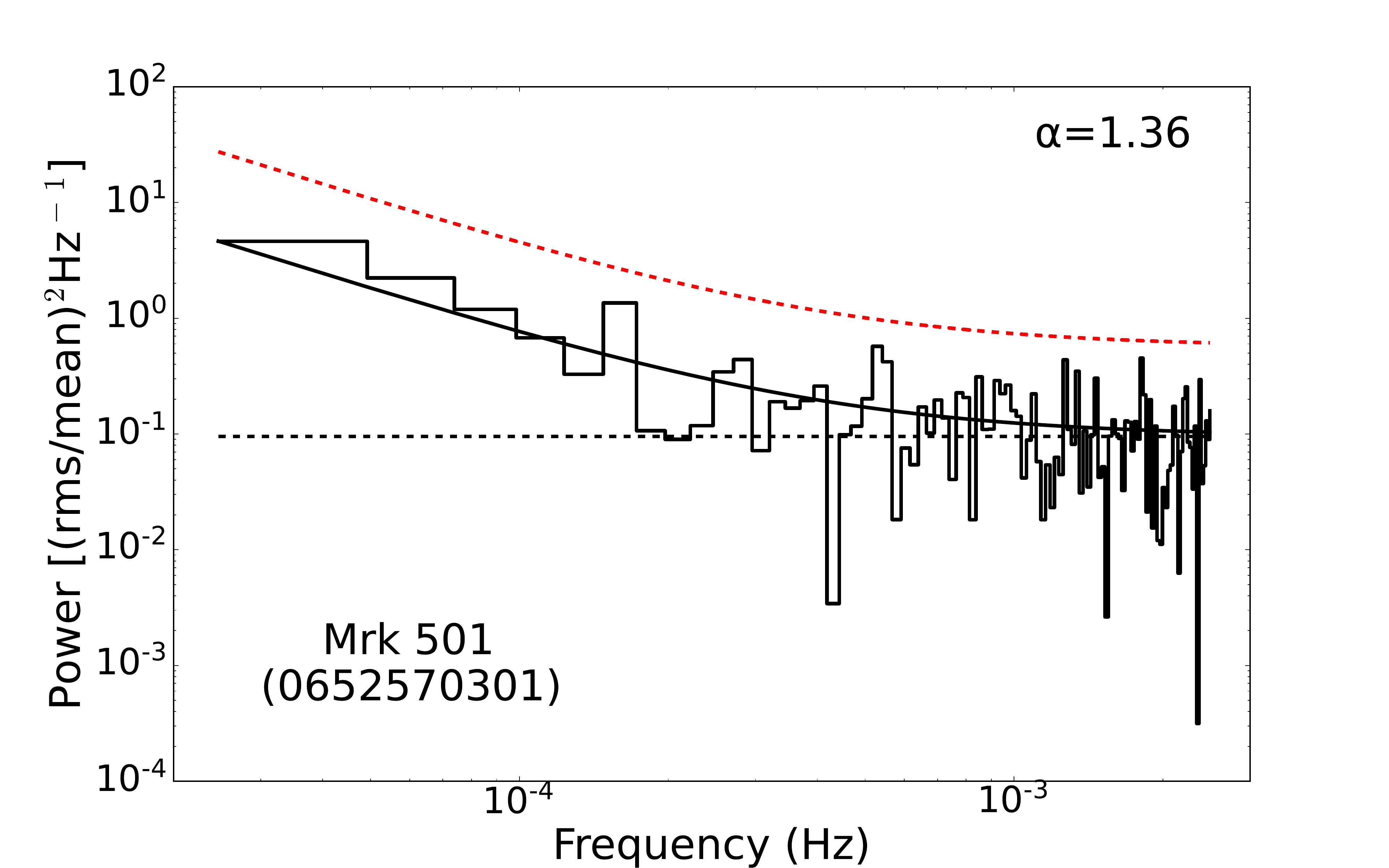}
\includegraphics[scale=0.2]{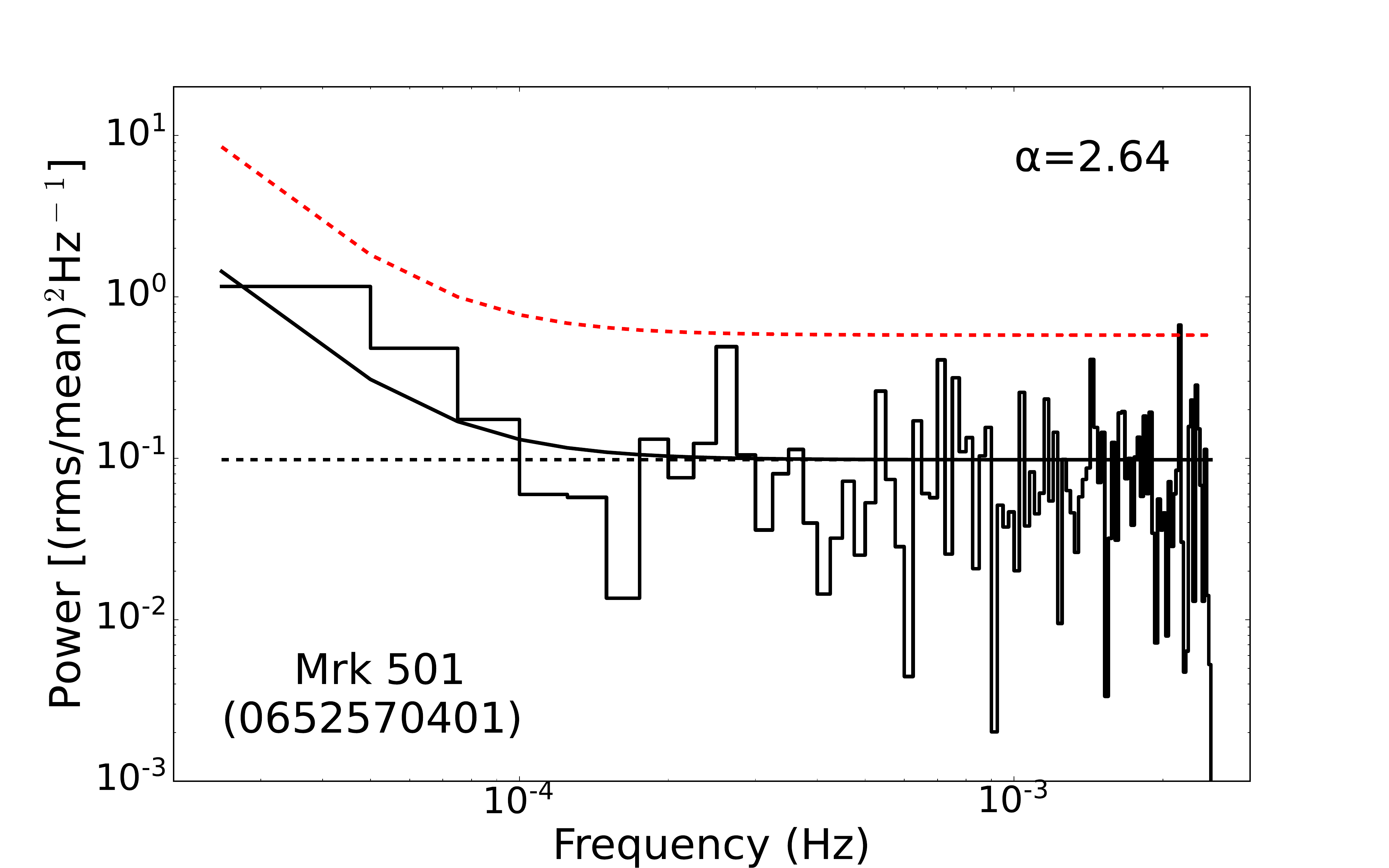}

\vspace*{0.01in}
\includegraphics[scale=0.2]{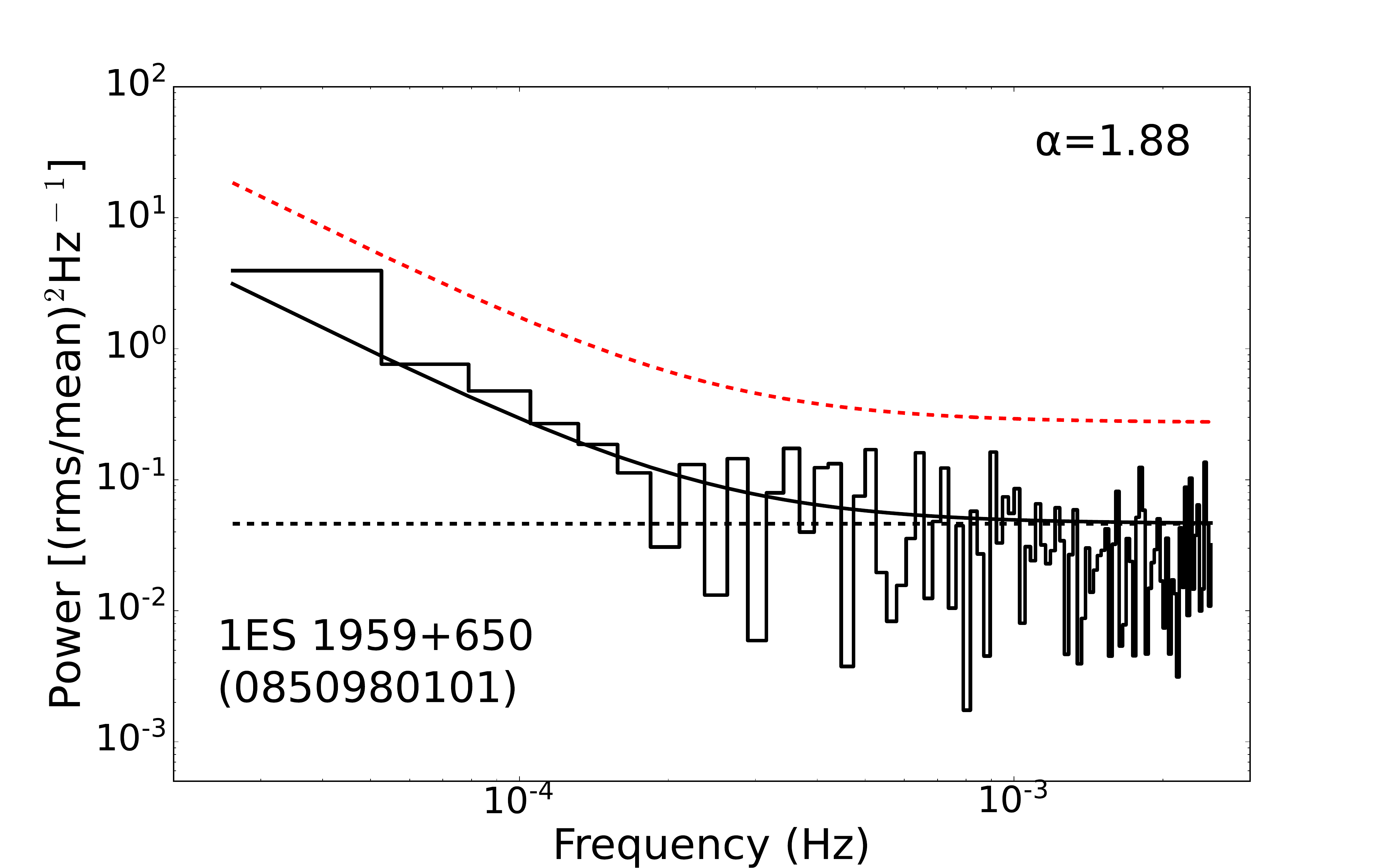}
\includegraphics[scale=0.2]{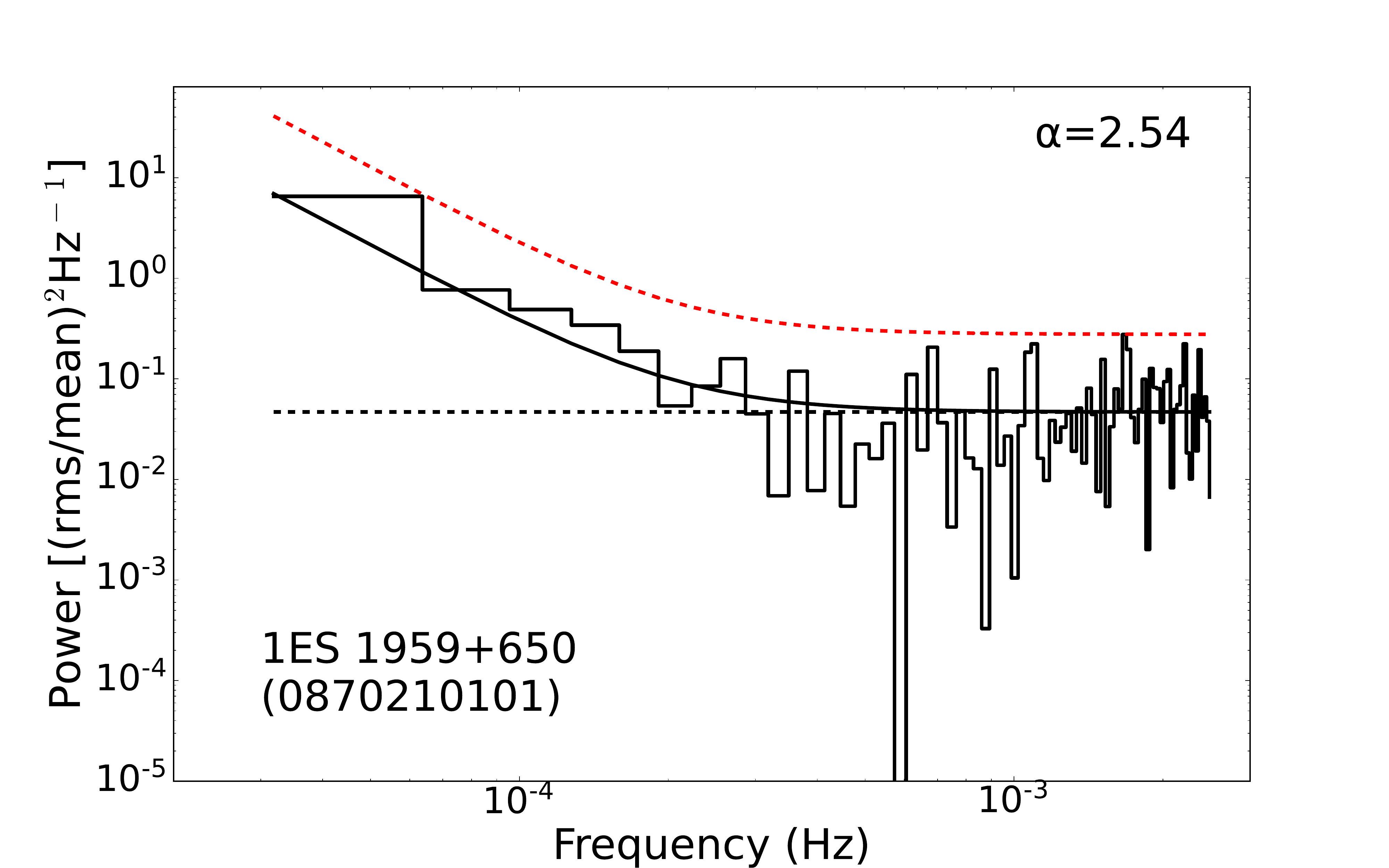}

\vspace{0.3in}
\caption{Power Spectral Density (PSD) plots for   soft  energy band light curves.  Also shown are fits to the red noise (black curve), the white noise level (dotted line) and 3$\sigma$ level above the noise (red dotted curve). The source and observation ID as well as the PSD index  are given in each plot. \label{A6}}
\end{figure*}


\clearpage
\begin{figure*}
\centering
\vspace*{-0.1in}
\includegraphics[scale=0.2]{Fig7a.pdf}
\includegraphics[scale=0.2]{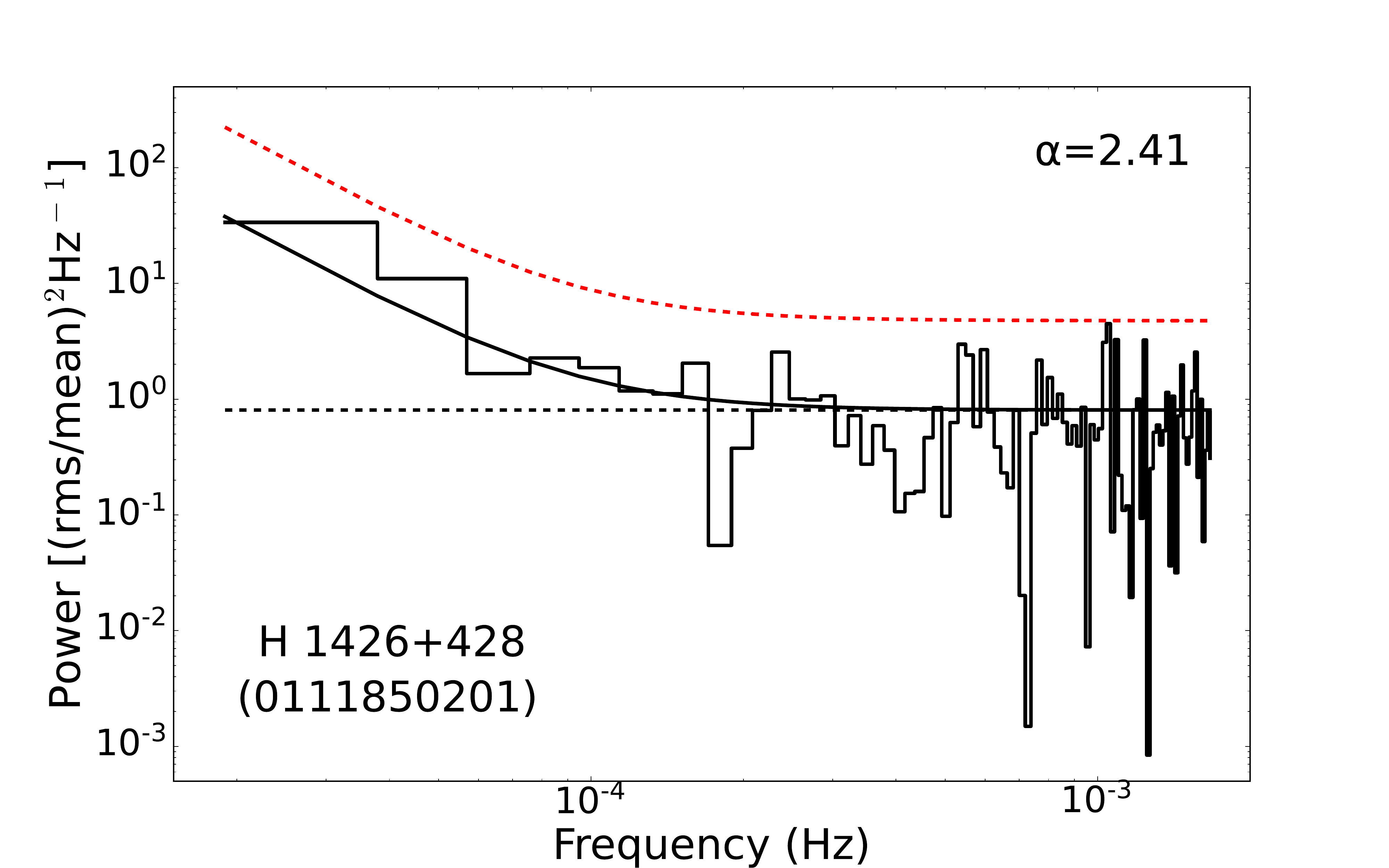}

\vspace*{0.01in}
\includegraphics[scale=0.2]{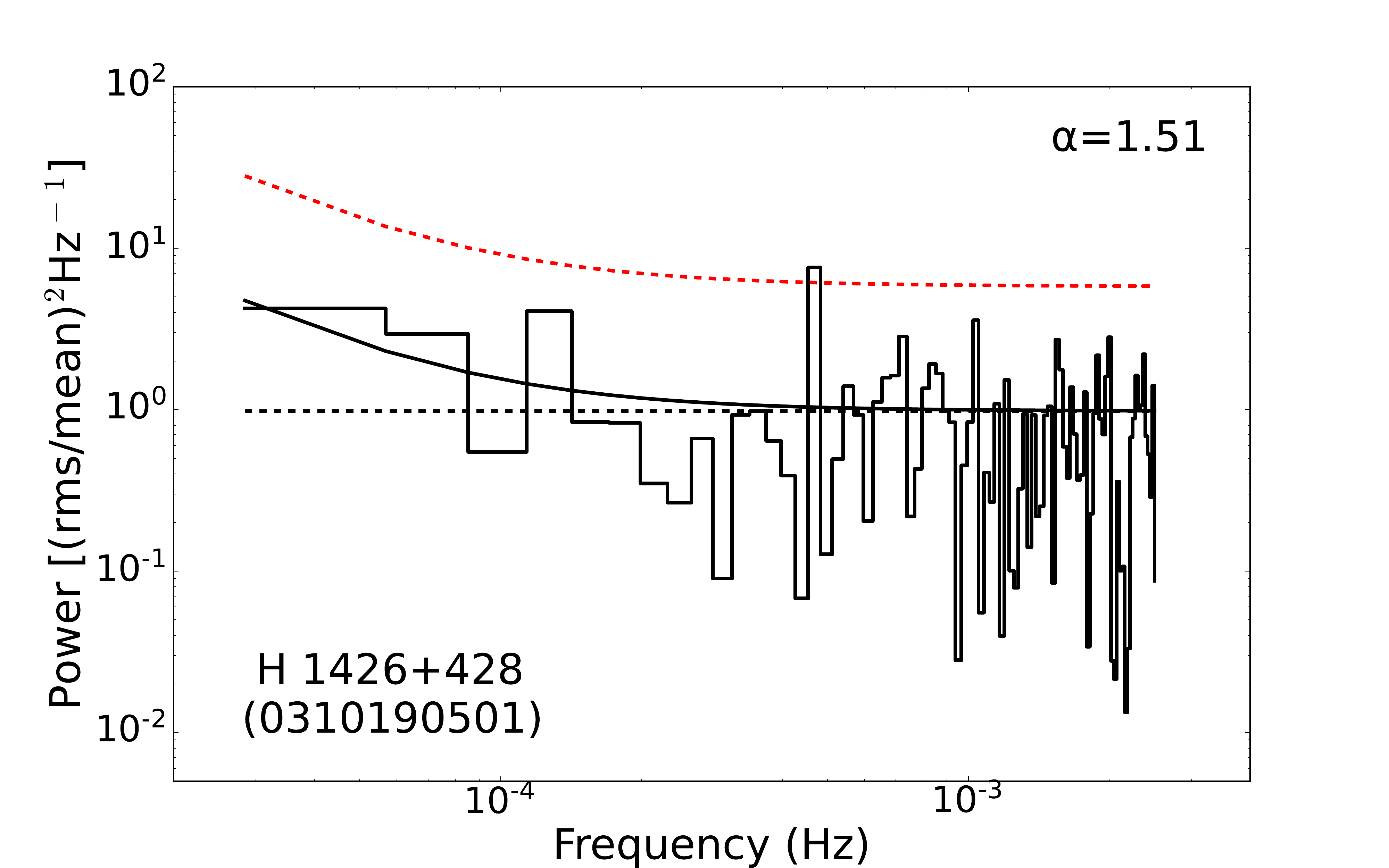}
\includegraphics[scale=0.2]{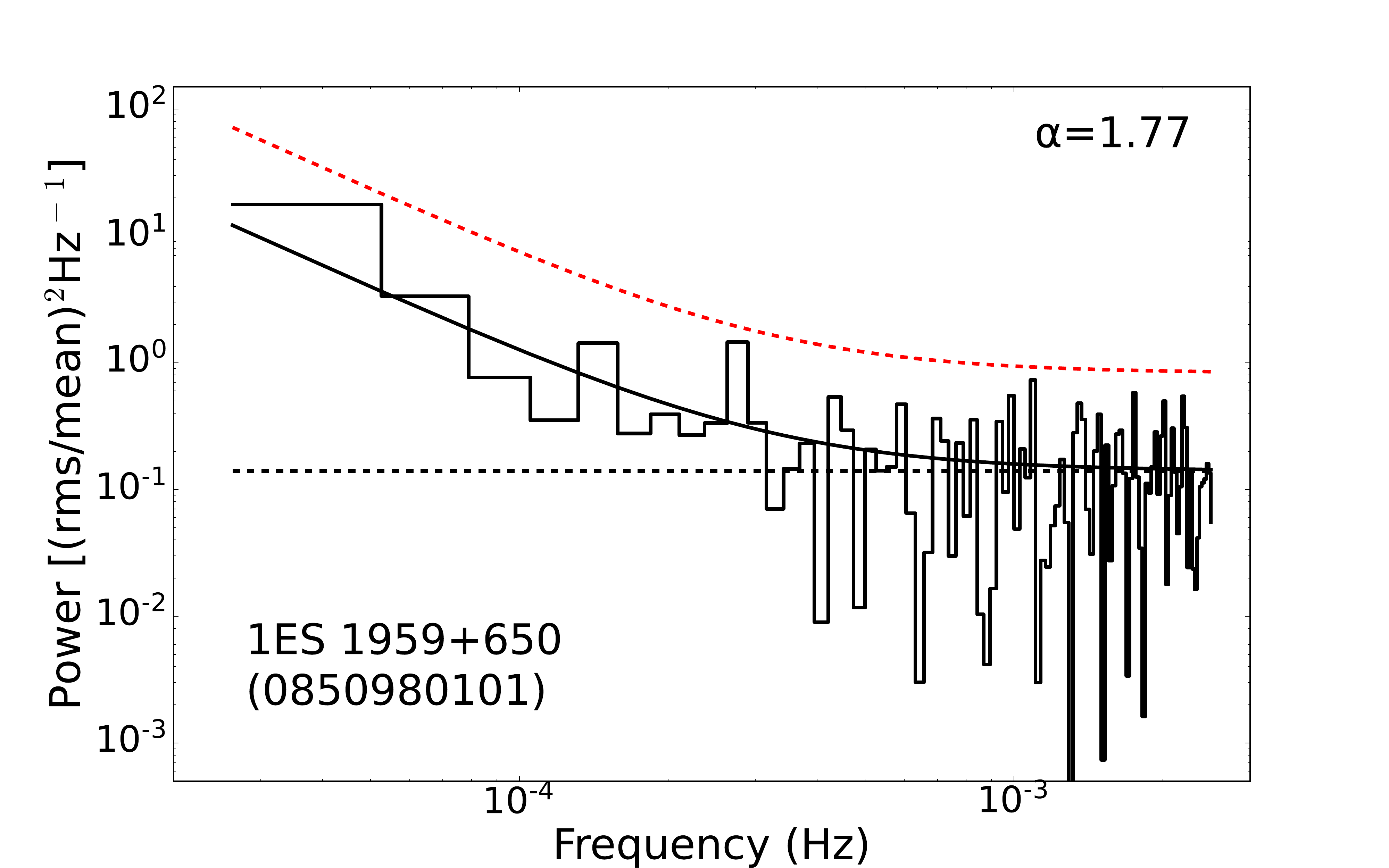}

\vspace*{0.01in}
\includegraphics[scale=0.2]{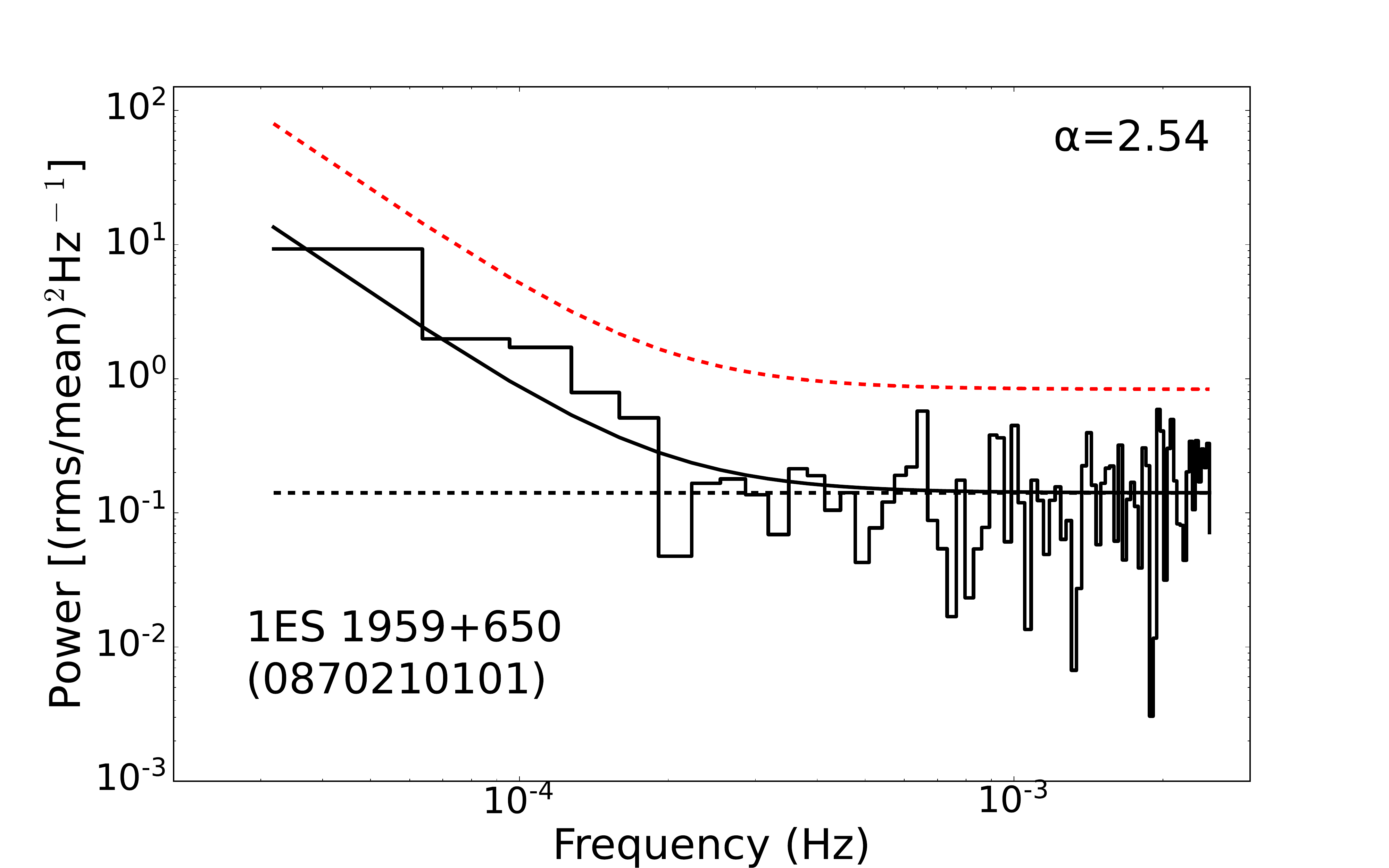}
\includegraphics[scale=0.2]{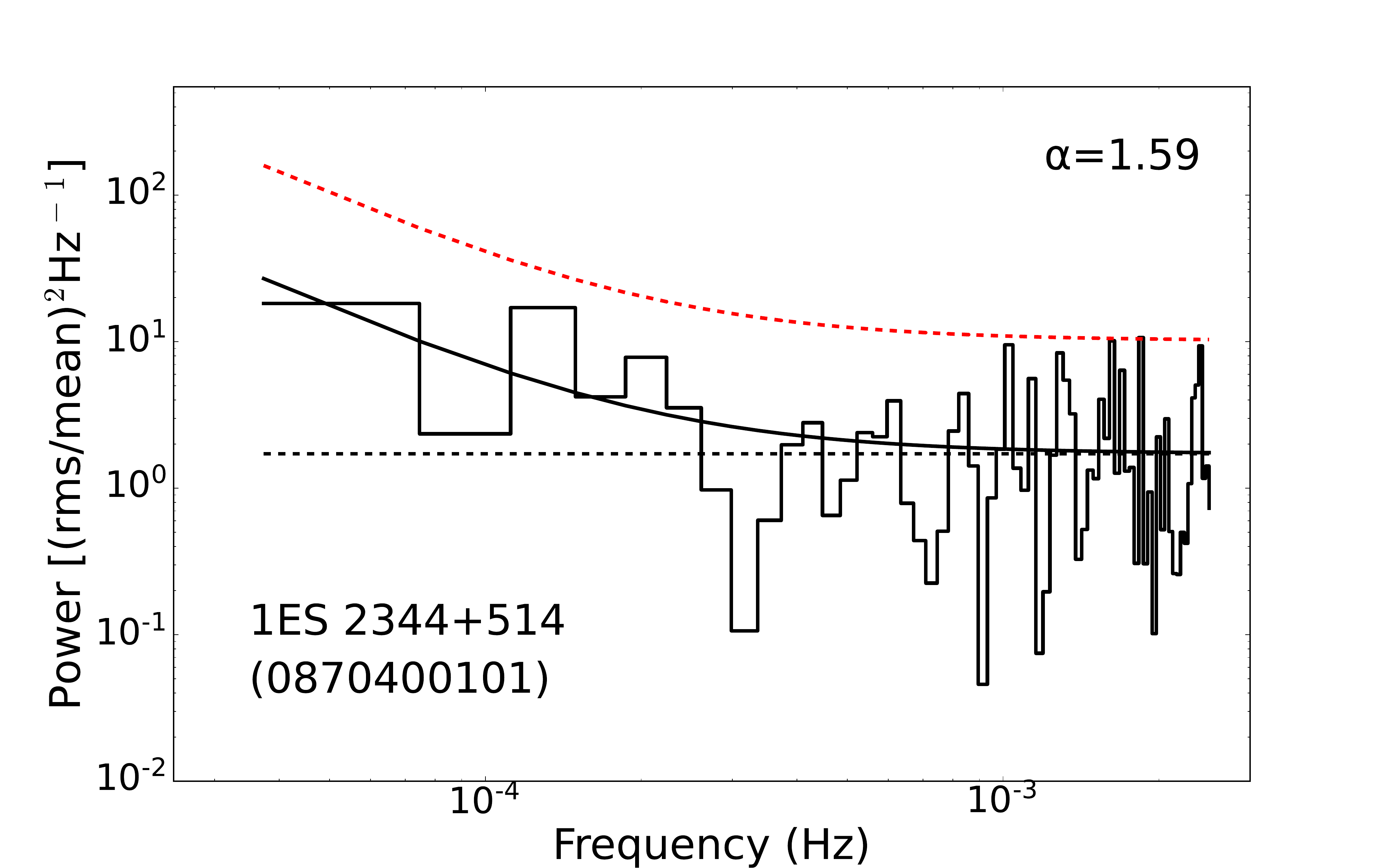}

\vspace*{0.01in}

\caption{PSD for the hard energy band. The labelling is same as that of Figure \ref{A6}\label{A7}} 
\end{figure*}

\clearpage

\setcounter{figure}{7}
\begin{figure*}
\centering

\vspace*{-0.1in}
\includegraphics[scale=0.2]{Fig8a.pdf}
\includegraphics[scale=0.2]{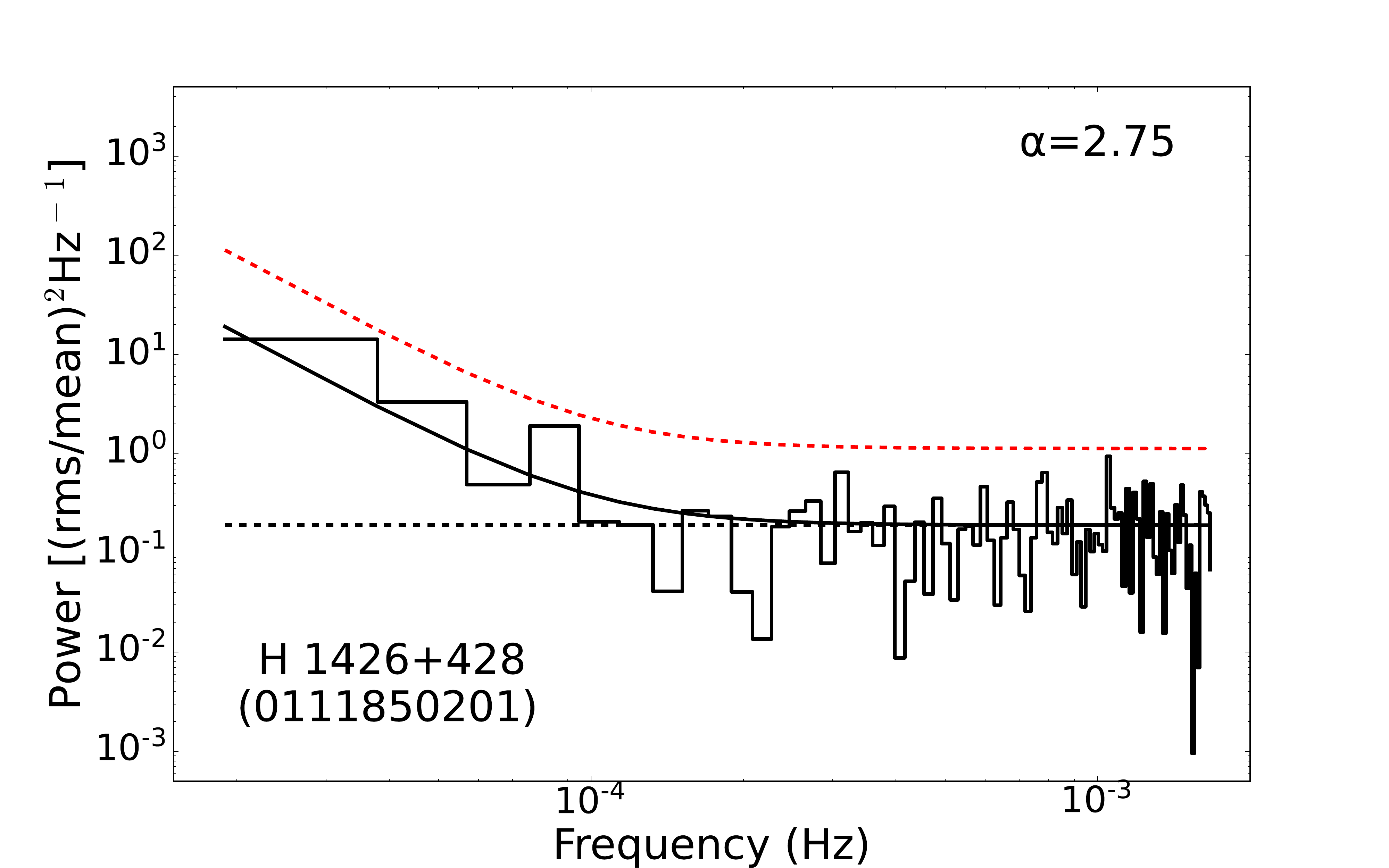}

\vspace*{0.005in}
\includegraphics[scale=0.2]{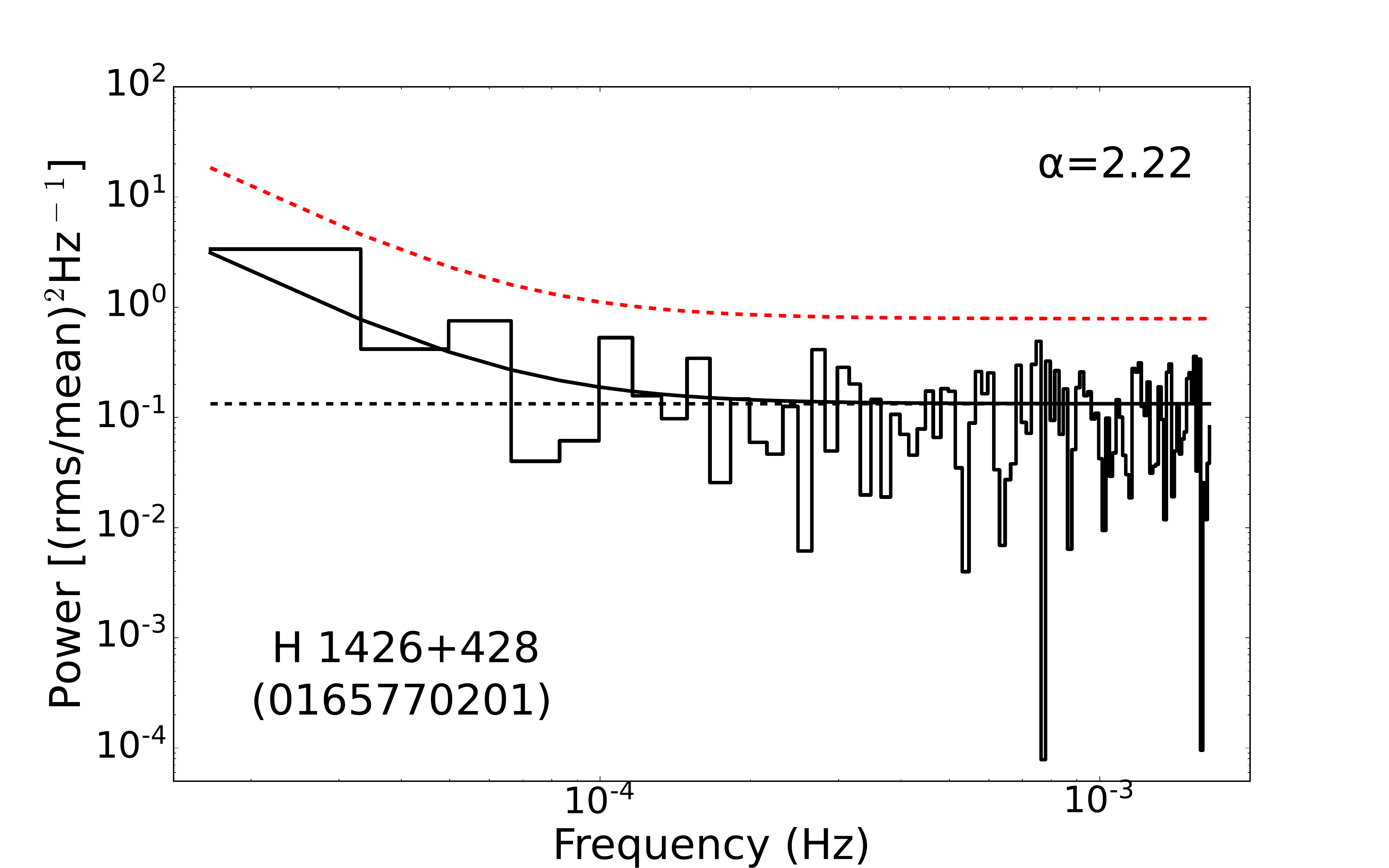}
\includegraphics[scale=0.2]{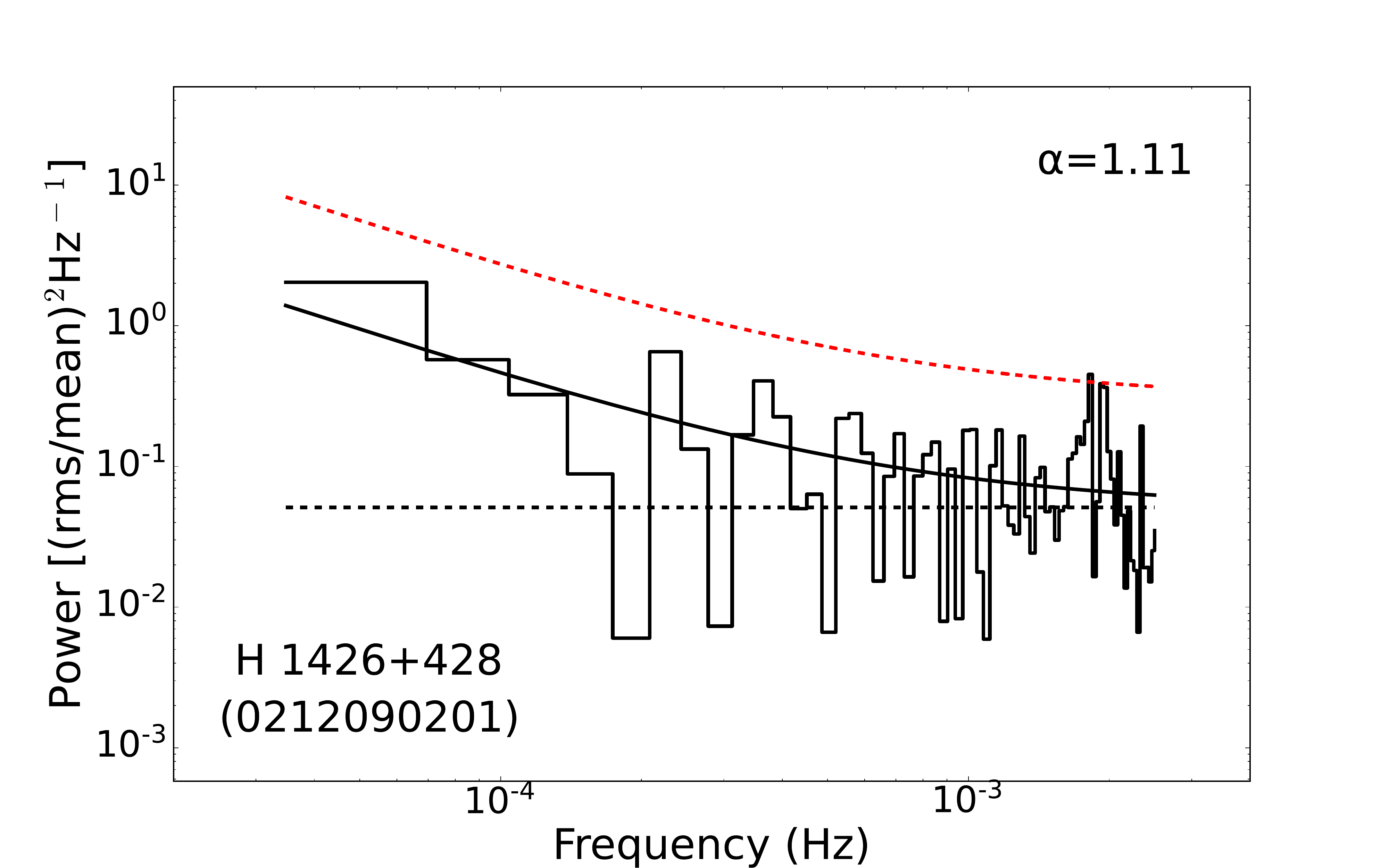}

\vspace*{0.005in}
\includegraphics[scale=0.2]{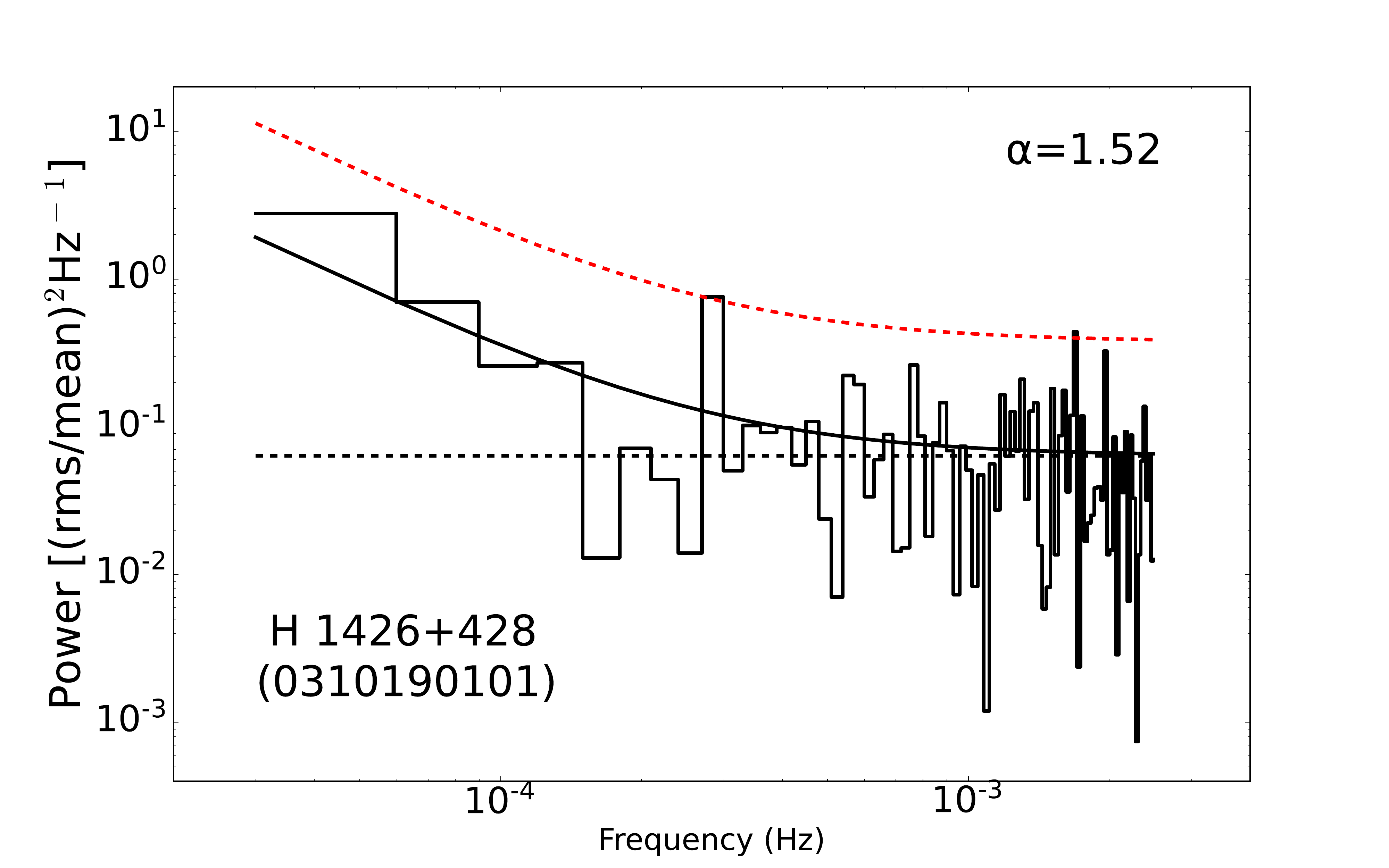}
\includegraphics[scale=0.2]{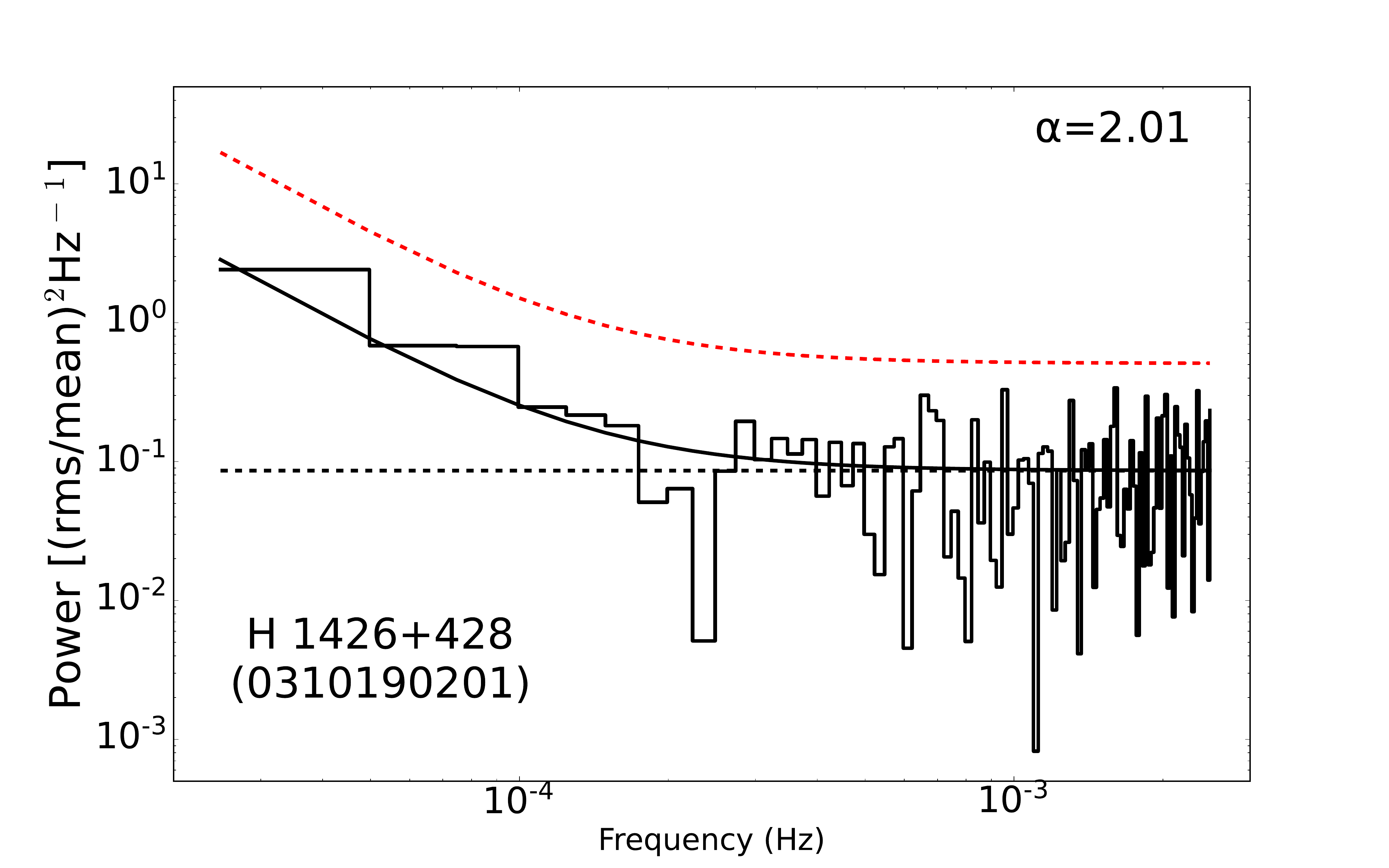}

\vspace*{0.005in}
\includegraphics[scale=0.2]{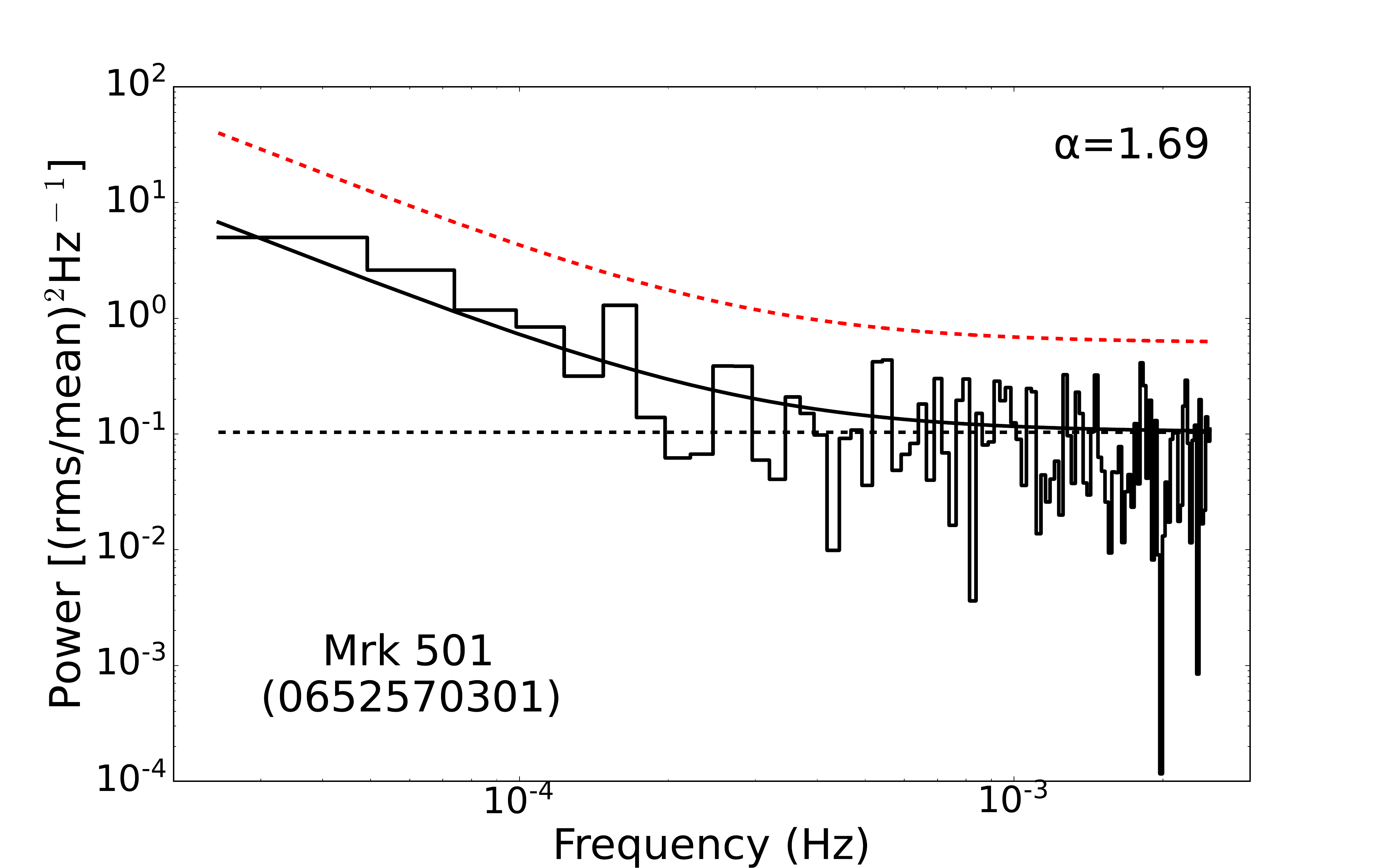}
\includegraphics[scale=0.2]{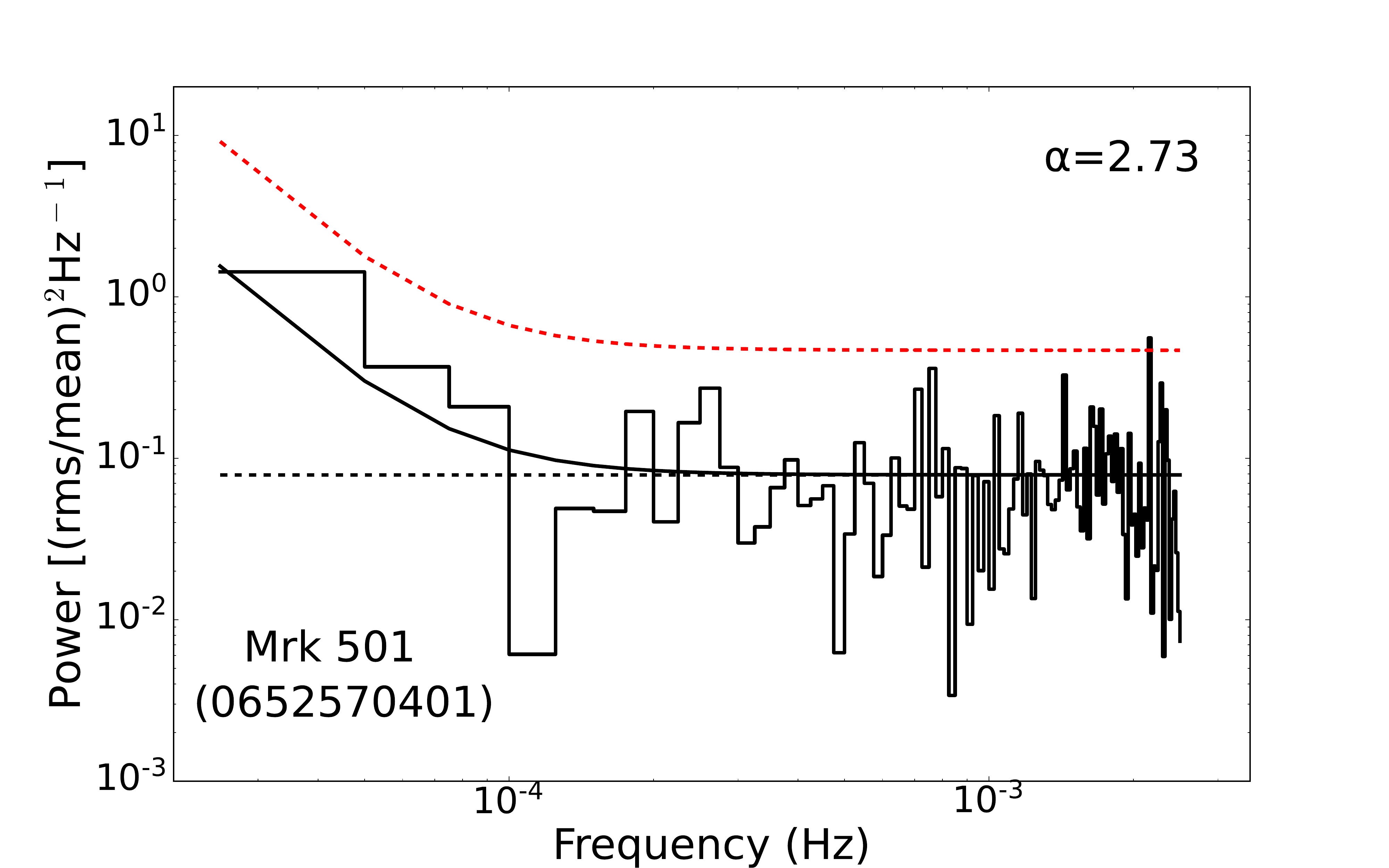}

\vspace{0.1in}
\caption{PSD for the total energy band. The labelling is same as that of Figure \ref{A6}\label{A8}}
\end{figure*}
\setcounter{figure}{7}
\begin{figure*}
\centering
\vspace*{-0.1in}
\includegraphics[scale=0.2]{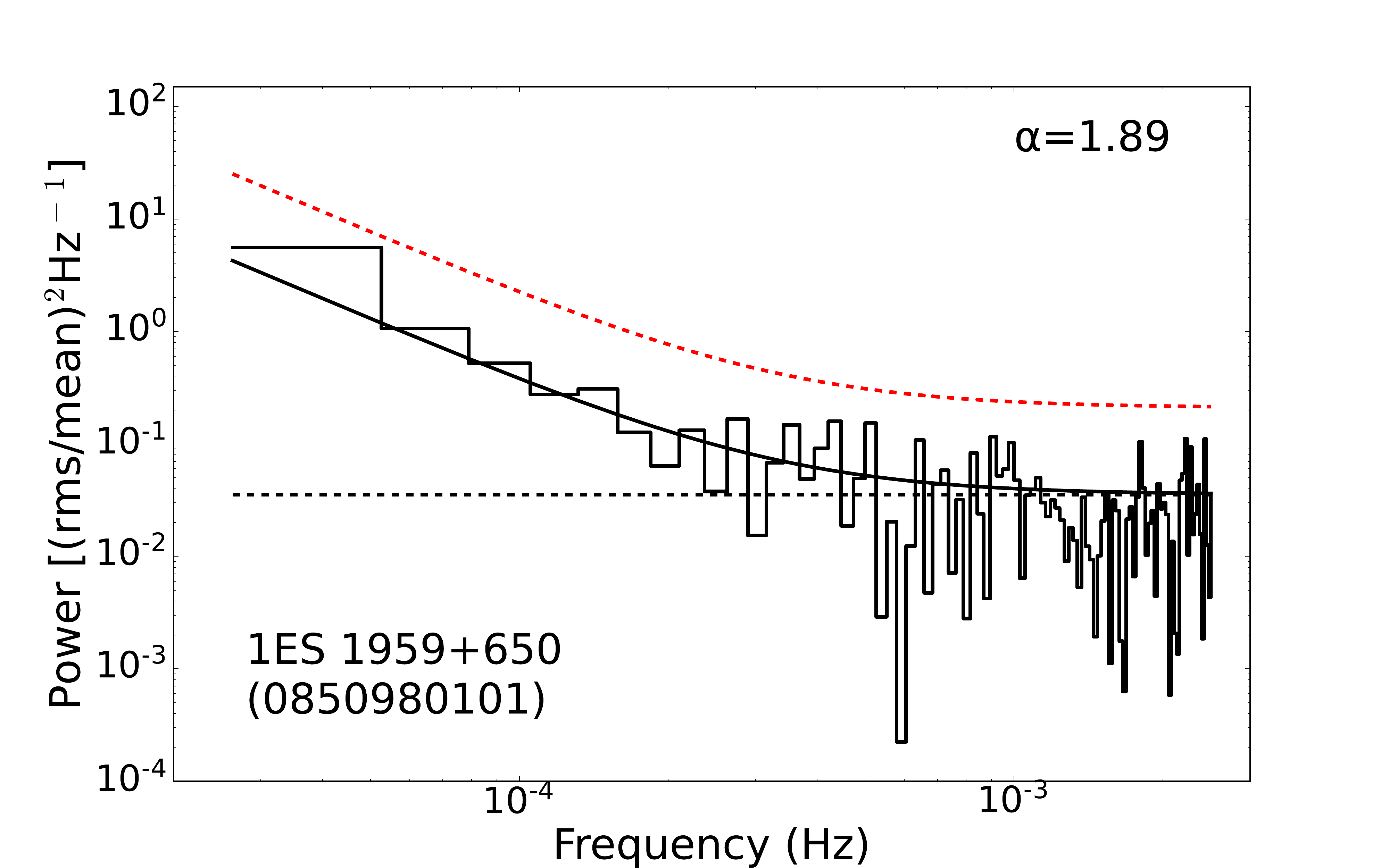}
\includegraphics[scale=0.2]{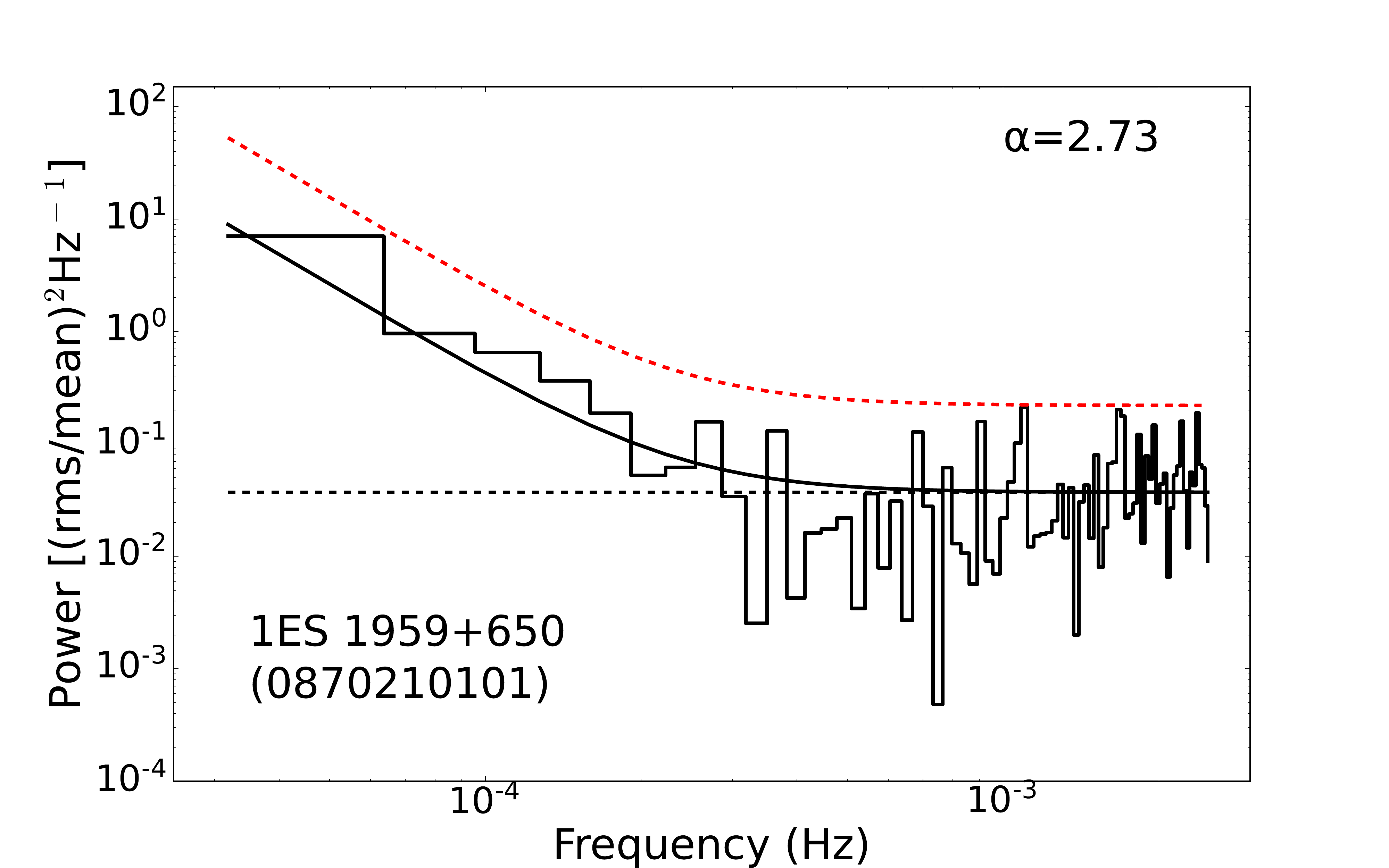}

\caption{Continued.}  
\end{figure*}

\end{document}